\documentclass[a4paper,epsf]{jpconf-app-fixed}
\usepackage[spanish,english]{babel}
\usepackage{graphicx}
\usepackage{epsfig,rotating,feynmp,amssymb,calc}

\input axodraw.sty
\makeatletter
\def\citer{\@ifnextchar [{\@tempswatrue\@citexr}{\@tempswafalse\@citexr[]}}

\def\@citexr[#1]#2{\if@filesw\immediate\write\@auxout{\string\citation{#2}}\fi
  \def\@citea{}\@cite{\@for\@citeb:=#2\do
    {\@citea\def\@citea{--\penalty\@m}\@ifundefined
       {b@\@citeb}{{\bf ?}\@warning
       {Citation `\@citeb' on page \thepage \space undefined}}%
\hbox{\csname b@\@citeb\endcsname}}}{#1}}
\makeatother

\newcommand{\beq}{\begin{eqnarray}}
\newcommand{\eeq}{\end{eqnarray}}
\def\lessim{\mathrel{\hbox{\rlap{\hbox{\lower4pt\hbox{$\sim$}}}\hbox{
$<$}}}}

\def\decimalpoint{.}
\newcommand{\gsim}{\raisebox{-0.13cm}{~\shortstack{$>$ \\[-0.07cm] $\sim$}}~}
\newcommand{\tgb}{\mbox{tg}\beta}
\newcommand{\ctgb}{\mbox{ctg}\beta}
\newcommand{\tg}{\mbox{tg}}
\newcommand{\non}{\nonumber}

\newcommand{\STS}{\vspace{1mm}}
\newcommand{\GS}{\vspace{5mm}}

\newcommand{\ee}{$e^+e^-$\ }

\newcommand{\demi}{1\! /\! 2}

\newcommand{\W}{{\it W} }

\newcommand{\SM}{Standard Model }

\newcommand{\ra}{\to }
\newcommand{\epem}{e^+e^- }
\newcommand{\cs}{cross section }

\newcommand{\Hs}{Higgs-strahlung }
\newcommand{\p}{particle }
\newcommand{\ps}{particles }
\newcommand{\cp}{coupling }
\newcommand{\cps}{couplings }

\newcommand{\sces}{scenarios }
\newcommand{\ssy}{supersymmetric }

\begin{document}

\selectlanguage{spanish}
\renewcommand{\tablename}{Tabla}
\decimalpoint

\thispagestyle{empty}
\begin{flushright}
  DESY 05-160\\
  LAPTH-1111/05\\
  PSI-PR-05-07
\end{flushright}
\vspace{1cm}
{\Large\raggedright\noindent \bf\par
Rompimiento de la Simetr\'{\i}a Electrod\'{e}bil y la
  F\'{\i}sica del Higgs: Conceptos  B\'{a}sicos}\\[2ex]
\begin{indented}
      \item[]\normalsize\raggedright
{ \bf M. Gomez-Bock$^1$, M. Mondrag\'on$^2$, M. M\"uhlleitner$^{3,4}$,\\ 
R. Noriega-Papaqui$^1$, I.~Pedraza$^1$, M. Spira$^3$, P.M. Zerwas$^5$}
\end{indented}
\begin{indented}
\item[]\rm
{ $^1$ Inst. de F\'{\i}sica ``LRT'', Benem\'erita Univ. Aut\'on. de Puebla, 
72570 Puebla, Pue, M\'exico \\
$^2$ Inst. de F\'{\i}sica, Univ. Nac. Auton. de M\'exico, 01000 M\'exico D.F., 
M\'exico \\
$^3$ Paul Scherrer Institut, CH-5232 Villigen PSI, Switzerland \\
$^4$ Laboratoire d'Annecy-Le-Vieux de Physique Th\'eorique, LAPTH,
Annecy-Le-Vieux, France \\ 
$^5$ Deutsches Elektronen-Synchrotron DESY, Hamburg, Germany}

\end{indented}


\begin{abstract}
Presentamos una introducci\'on a los conceptos
b\'asicos del rompimiento de la simetr\'{\i}a electrod\'ebil y la
f\'isica del Higgs dentro del Modelo Est\'andar y sus extensiones
supersim\'etricas. Se presenta tambi\'en una breve perspectiva general
de mecanismos alternativos del rompimiento de la simetr\'{\i}a.
Adem\'as de las bases te\'oricas, se discute el estado actual de
la f\'{\i}sica experimental del Higgs y sus implicaciones
para futuros experimentos en el LHC y en colisionadores lineales
$\epem$.
\end{abstract}


\section{Introducci\'on}

\phantom{h}
\paragraph{1.}

Revelar el mecanismo f\'{\i}sico responsable del rompimiento de las
simetr\'{\i}as electrod\'ebiles, es uno de los problemas principales en
la F\'{\i}sica de Part\'{\i}culas. Si las part\'{\i}culas
fundamentales - leptones, quarks y bosones de norma (gauge)- siguen
interactuando d\'ebilmente a altas eneg\'{\i}as, potencialmente cercanas
a la escala de Planck, el sector en el cual la simetr\'{\i}a
electrod\'ebil es rota debe contener uno o m\'as bosones escalares
fundamentales de Higgs con masas ligeras del orden de la escala del
rompimiento de la simetr\'{\i}a $v\sim 246$ GeV. La masa de las
part\'{\i}culas fundamentales son generadas a trav\'es de la interacci\'on
con un campo de fondo escalar de Higgs, el cual es diferente de cero
en su estado base \cite{1}.  De manera alternativa, el rompimiento de la
simetr\'\i a podr\'\i a ser generado din\'amicamente por nuevas fuerzas
fuertes caracterizadas por una escala de interacci\'on $\Lambda \sim 1$ TeV o
m\'as alta \cite{2,2A}. Si las simetr\'{\i}as globales de las
interacciones fuertes son rotas espont\'aneamente, los bosones de
Goldstone asociados pueden ser absorbidos por los campos de norma
(gauge fields), generando las masas de las part\'\i culas de norma.
Las masas de los leptones y quarks pueden ser generadas a trav\'es de
interacciones con un condensado de fermiones. Otros mecanismos de
rompimiento de las simetr\'{\i}as electrod\'ebiles est\'an asociados con
la din\'amica en el espacio de dimensiones extra a bajas energ\'{\i}as
\cite{RI2}.

\paragraph{2.}

Un mecanismo simple para describir el rompimiento de la simetr\'{\i}a
electrod\'ebil est\'a incorporado en el Modelo Est\'andar (SM) \cite{3}.
Para acoplar a todos los fen\'omenos observados, se introduce un campo
escalar complejo isodoblete; \'este adquiere un valor esperado del vac\'\i o
no nulo a trav\'es de sus auto-interacciones, rompiendo
espont\'aneamente la simetr\'\i a electrod\'ebil SU(2)$_I \times$ U(1)$_Y$ hasta la
simetr\'{\i}a electomagn\'etica U(1)$_{EM}$.  Las interacciones de los
bosones de norma y los fermiones con el campo de fondo generan las masas
de estas part\'{\i}culas. Una componente del campo escalar no se
absorbe en este proceso, manifiest\'andose como la part\'\i cula f\'\i sica
Higgs $H$.

La masa del bos\'on de Higgs es el \'unico par\'ametro desconocido en el
sector de rompimiento de simetr\'{\i}a del Modelo Est\'andar, mientras
que todos los acoplamientos est\'an fijos por las masas de las
part\'\i culas, una consecuencia del mecanismo de Higgs {\it sui generis}.
Sin embargo, la masa del bos\'on de Higgs est\'a acotada de dos maneras.
Dado que el auto-acoplamiento cu\'artico del campo de Higgs crece
indefinidamente conforme la energ\'{\i}a aumenta, un l\'{\i}mite
superior a la masa del Higgs puede ser derivado requiriendo que las
part\'{\i}culas del SM permanezcan d\'ebilmente interactuantes a escalas
de $\Lambda$ \cite{4}. Por otro lado,  l\'{\i}mites inferiores estrictos 
a la masa del Higgs provienen de pedir que el vac\'{\i}o del campo
electrod\'ebil sea estable \cite{5}. Si el Modelo Est\'andar es v\'alido
hasta la escala de Planck, la masa del Higgs del SM est\'a restringido a
un intervalo estrecho entre los $130$ y los $190$ GeV. Para las masas
del Higgs que est\'en fuera de este intervalo, se esperar\'{\i}a que
ocurrieran nuevos fen\'omenos f\'\i sicos a escalas $\Lambda$ entre $\sim 1$ TeV y la
escala de Planck. Para masas del Higgs del orden de $1$ TeV, la escala de las
nuevas interacciones fuertes ser\'{\i}a tan baja como $\sim 1$ TeV
\cite{4,6}.

Las observables electrod\'ebiles est\'an afectadas por la masa del Higgs a
trav\'es de correcciones radiativas \cite{7}. A pesar de la dependencia
logar\'{\i}tmica d\'ebil, la alta precisi\'on de los datos electrod\'ebiles,
{\it c.f.}  Fig.~\ref{fg:SMHiggs}, indican una preferencia por masas
ligeras del Higgs cercanas a $\sim 100$ GeV \cite{8}. A un 95\% del
nivel de confianza (CL), estos datos requieren un valor para la masa
de Higgs menor que $\sim 186$ GeV. Mediante la b\'usqueda directa de la
particula Higgs del SM, los experimentos del LEP han fijado un
l\'{\i}mite inferior de $M_H\gsim 114$ GeV sobre la masa del Higgs
\cite{9}. Dado que el bos\'on de Higgs no ha sido encontrado en LEP2, la
b\'usqueda continuar\'a en el Tevatron, que puede alcanzar masas de hasta
$\sim 140$ GeV \cite{11}. El colisionador de protones LHC abarca el
rango can\'onico completo de la masa del Higgs del Modelo Est\'andar
\cite{12}. Las propiedades de la part\'{\i}cula del Higgs pueden ser
analizadas con gran precisi\'on en colisionadores lineales $e^+ e^-$
\cite{13}, y as\'\i{} establecer el mecanismo de Higgs experimentalmente.
\begin{figure}[hbt]
\begin{center}
\hspace*{-1.1cm}
\epsfig{figure=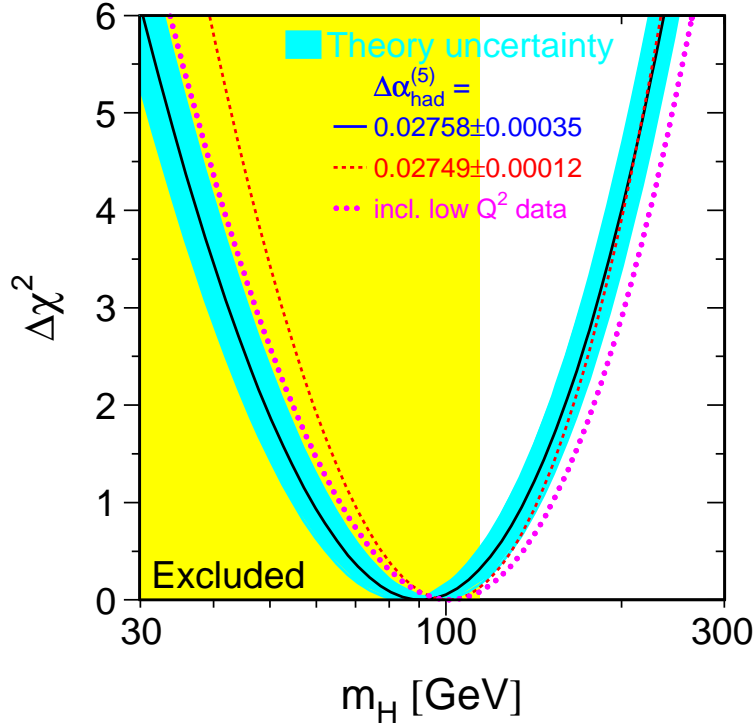,bbllx=17,bblly=36,bburx=564,bbury=564,width=10cm,clip=}
\end{center}
\vspace*{-0.4cm}

\caption[]{\label{esp-fg:SMHiggs}\it La curva $\Delta\chi^2$ 
derivada de mediciones electrod\'ebiles de precisi\'on $Q^2$-alta,
realizadas en el LEP y por SLD, CDF y D0, como funci\'on de la
masa del bos\'on de Higgs, suponiendo al Modelo Est\'andar como la
teor\'{\i}a de la naturaleza.}
\end{figure}

\paragraph{3.}
Si el Modelo Est\'andar puede encastrarse en una Teor\'{\i}a de 
Gran Unificaci\'on (GUT) a altas energ\'{\i}as, la escala natural
del rompimiento de la simetr\'{\i}a electrod\'ebil se
esperar\'{\i}a cercana a la escala de unificaci\'on $M_{GUT}$. La
Supersimetr\'{\i}a \cite{14} provee una soluci\'on a este
problema de jerarqu\'{\i}as. Las contribuciones cuadr\'aticamente
divergentes a las correcciones radiativas a la
masa del bos\'on escalar de Higgs son canceladas por la
interferencia destructiva entre los lazos (loops) fermi\'onicos y
bos\'onicos en las teor\'{\i}as supersim\'etricas \cite{15}. La
Extensi\'on M\'{\i}nima supersim\'etrica del Modelo Est\'andar
(MSSM) puede ser obtenida como una teor\'{\i}a efectiva a partir
de las teor\'{\i}as supersim\'etricas de gran unificaci\'on.
Una indicaci\'on fuerte para la realizaci\'on de este esquema f\'\i sico en la
naturaleza es el excelente acuerdo entre el valor del \'angulo de mezcla
electrod\'ebil $sin^2 \theta_W$ predicho por  la unificaci\'on de los acoplamientos de
norma y el valor medido experimentalmente.
 Si los acoplamientos de norma est\'an unificados en la
Teor\'{\i}a M\'{\i}nima Supersim\'etrica a una escala $M_{GUT} =
{\cal O}(10^{16}~\mbox{GeV})$, el valor predicho del \'angulo de mezcla
electrod\'ebil es $sin^2 \theta_W = 0.23120 \pm
0.0017$ \cite{16}  para el espectro de masas de las
part\'{\i}culas supersim\'etricas del orden de $M_Z$ a $1$ TeV.
Esta predicci\'on te\'orica concuerda muy bien con el resultado
experimental $sin^2 \theta_W^{exp} = 0.23120 \pm 0.0015$ \cite{8};
la diferencia entre los dos n\'umeros es menor que el 2 por mil.

En el MSSM, el sector de Higgs se establece mediante dos dobletes de
Higgs \cite{17}. Es necesario que sean dos para poder generar las
masas de los fermiones de tipo up y down en una teor\'{\i}a
supersim\'etrica y convertirla en una teor\'{\i}a sin anomal\'{\i}as. El
espectro de part\'{\i}culas de Higgs consiste en un quinteto de
estados: dos escalares neutrales CP-pares ($h,H$), un neutral
pseudoescalar CP-impar ($A$), y un par de bosones de Higgs cargados
($H^{\pm}$) \cite{19}.  Se espera que la masa de los bosones pesados de
Higgs $H,A,H^\pm$ sean del orden de $v$, pero se podr\'\i an extender hasta
el rango de TeV. Por contraste, dado que los acoplamientos propios
cu\'articos de Higgs est\'an determinados por los acoplamientos de norma,
la masa del bos\'on de Higgs m\'as ligero $h$ est\'a restringida muy
rigurosamente.  A nivel \'arbol, la masa ha sido predicha a ser 
menor a la masa del $Z$ \cite{19}. Las correcciones radiativas,
que aumentan como la cuarta potencia de la masa del top, recorren el
l\'\i mite superior a un valor entre $\sim 100$
GeV and $\sim 140$ GeV, dependiendo del par\'ametro $\tgb$, que es la
raz\'on de los valores esperados de los dos campos escalares
neutrales de Higgs.

Un l\'{\i}mite bajo general de 91 GeV ha sido establecido
experimentalemte para la part\'{\i}cula de Higgs $h$ por el LEP
\cite{9}.
La b\'usqueda de la masa de $h$ en exceso de $\sim 100$ GeV y la
b\'usqueda para el bos\'on pesado de Higgs continuan en el Tevatron,
el LHC y el colisionador lineal $e^+e^-$.

\paragraph{4.}
Un bos\'on ligero de Higgs tambi\'en puede ser generado como un
(pseudo-)bos\'on de Goldstone mediante el rompimiento de la simetr\'\i a global
de nuevas interacciones. Alternativamente a la supersimetr\'{\i}a, las
divergencias cuadr\'aticas podr\'\i an ser canceladas mediante los nuevos
compa\~neros de las part\'{\i}culas del Modelo Est\'andar que no difieran
en su caracter fermi\'onico/bos\'onico. Los esquemas de simetr\'{\i}a
restringen a los acoplamientos de tal forma que las cancelaciones se
logran de una manera natural. Tales escenarios se realizan en Modelos
de Higgs Peque\~nos (Little Higgs Models) \cite{2A} los cuales predicen
un gran conjunto de nuevas part\'{\i}culas SM dentro del rango de
masas de unos cuantos TeV's.

\paragraph{5.}
Las amplitudes de la dispersi\'on el\'astica de bosones vectoriales
masivos crecen indefinidamente con la energ\'{\i}a si son calculadas
en una expansi\'on perturbativa del acoplamiento d\'ebil de una
teor\'{\i}a de norma no-Abeliana. Como resultado, se viola la
unitaridad m\'as all\'a de una escala cr\'{\i}tica de energ\'{\i}a $\sim
1.2$ TeV.  Aparte de introducir un bos\'on ligero de Higgs, este
problema se puede tambi\'en resolver suponiendo que el bos\'on $W$ se
vuelve fuertemente interactuante a energ\'{\i}as de TeV, y por lo
tanto  amortiguando el aumento de las amplitudes de la dispersi\'on
el\'astica.  Naturalmente, las fuerzas fuertes entre los bosones $W$
pueden atribuirse a nuevas interacciones fundamentales caracterizadas
por un escala del orden de 1 TeV \cite{2}. Si la teor\'{\i}a
fundamental es invariante quiral globalmente, esta simetr\'{\i}a puede
romperse espont\'aneamente. Los bosones de Goldstone asociados con el
rompimiento espont\'aneo de la simetr\'{\i}a pueden ser absorbidos por 
bosones de norma para generar sus masas y para establecer los grados de
libertad longitudinales de sus funciones de onda.

Dado que los bosones $W$ longitudinalmente polarizados est\'an asociados
con modos de Goldstone del rompimiento de la simetr\'{\i}a quiral, las
amplitudes de dispersi\'on para el bos\'on $W_L$ pueden ser predichas para
altas energ\'{\i}as, mediante una expansi\'on sitem\'atica de la
energ\'{\i}a. El t\'ermino principal est\'a libre de par\'ametros, una
consecuencia del mecanismo de rompimiento de la simetr\'{\i}a quiral
{\it per se}, el cual es independiente de la estructura particular de
la teor\'{\i}a din\'amica. Los t\'erminos de orden superior en la
expansi\'on quiral sin embargo est\'an definidos por la estructura
detallada de la teor\'{\i}a fundamental.  Con el aumento de
energ\'{\i}a se espera que la expansi\'on quiral diverja y se podr\'\i an
generar nuevas resonancias en la dispersi\'on $WW$ a escalas de masa
entre 1 y 3 TeV's. Este esquema es an\'alogo a la din\'amica del pi\'on en
QCD, donde las amplitudes de umbral pueden ser predichas en una
expansi\'on quiral, mientras que a altas energ\'{\i}as las resonancias
vectoriales y escalares se forman en la dispersi\'on $\pi \pi$.  Este
escenario puede ser estudiado en los experimentos de dispersi\'on $WW$,
donde los bosones $W$ son radiados, como part\'{\i}culas cuasi-reales
\cite{22}, emitidas por quarks de alta energ\'\i a en el haz de protones
en el LHC \cite{12}, \citer{23,22A} o emitidas por electrones y
positrones en Colisionadores Lineales TeV \cite{13,24,24a}.

\paragraph{6.}
Tambi\'en en teor\'{\i}as con dimensiones espaciales extra, las
simetr\'{\i}as electrod\'ebiles pueden ser rotas sin introducir campos
escalares fundamentales adicionales, lo que lleva tambi\'en a
teor\'{\i}as sin Higgs. Dado que en las teor\'{\i}as de 5-dimensiones
las funciones de onda se expanden por una quinta componente, las
simetr\'{\i}as pueden ser rotas eligiendo apropiadamente las
condiciones a la frontera para esta componente del campo \cite{RI2}.
La componente escalar adicional del campo original de norma
penta-dimensional es absorbida para generar las torres masivas de
Kaluza-Klein de los campos de norma en cuatro dimensiones. El
intercambio adicional de estas torres en la dispersi\'on $WW$ disminuye
la amplitud de dispersi\'on del Modelo Est\'andar y permite en principio
extender la teor\'{\i}a a energ\'{\i}as mayores al l\'\i mite de
unitaridad de 1.2 TeV de los escenarios sin Higgs. Sin embargo, 
hasta el momento no es claro si modelos realistas de este tipo pueden ser
construidos  de tal forma que  den lugar a amplitudes de dispersi\'on
el\'astica $WW$ suficientemente peque\~nas para ser compatibles 
con la unitaridad perturbativa \cite{RI3}.

\paragraph{7.}
Este reporte est\'a dividido en tres partes. Una introducci\'on
b\'asica y un resumen de los principales resultados te\'oricos y
experimentales del sector de Higgs en el Modelo Est\'andar se
presentan en la siguiente secci\'on. Tambi\'en describiremos el
futuro de la b\'usqueda del Higgs en los colisionadores
hadr\'onicos y $e^+e^-$. De la misma forma en la secci\'on que le
sigue, se discutir\'a el espectro del Higgs de las teor\'{\i}as
supersim\'etricas. Finalmente, las principales
caracter\'{\i}sticas de las interacciones fuertes $WW$ y su
an\'alisis en los experimentos de dispersi\'on $WW$ se
presentar\'an en la secci\'on final.

S\'olo los elementos b\'asicos del rompimiento de la simetr\'ia
electrod\'ebil y el mecanismo de Higgs son examinados en este
reporte. Otros aspectos pueden ser encontrados en la referencia
\cite{24b} y los reportes reunidos en \cite{24A}, sobre los cuales
est\'a basado este reporte.

\newpage
\section{El Sector de Higgs del Modelo Est\'andar}

\phantom{h}
\subsection{El Mecanismo de Higgs}

\phantom{h} A energ\'\i as altas, la amplitud para la dispersi\'on el\'astica
de bosones $W$ masivos, $WW \to WW$, crece indefinidamente con la
energ\'{\i}a para part\'{\i}culas linealmente polarizadas
longitudinalmente, Fig.~\ref{fg:wwtoww}a.  Esta es una consecuencia
del crecimiento lineal de la funci\'on de onda longitudinal $W_L$, $\epsilon_L
= (p,0,0,E)/M_W$, con la energ\'{\i}a de la part\'{\i}cula. A pesar de
que el t\'ermino de la amplitud de dispersi\'on que aumenta con la cuarta
potencia de la energ\'{\i}a se cancela debido a la simetr\'{\i}a de
norma no-Abeliana, la amplitud permanece cuadr\'aticamente divergente en
la energ\'{\i}a.  Por otro lado, la unitaridad requiere que las
amplitudes de dispersi\'on el\'astica de las ondas parciales $J$  est\'en
acotadas por $\Re e A_J \leq 1/2$.  Aplicado a la amplitud asint\'otica 
de la onda $S$, $A_0 = G_F s/8\pi\sqrt{2}$, del canal de isospin-cero
$2W_L^+W_L^- + Z_L Z_L$, la cota \cite{25}
\begin{equation}
s \leq 4\pi\sqrt{2}/G_F \sim (1.2~\mbox{TeV})^2
\end{equation}
a la energ\'{\i}a del c.m. $\sqrt{s}$ puede ser derivada para la validez
de una  teor\'ia de bosones masivos de norma d\'ebilmente acoplados.
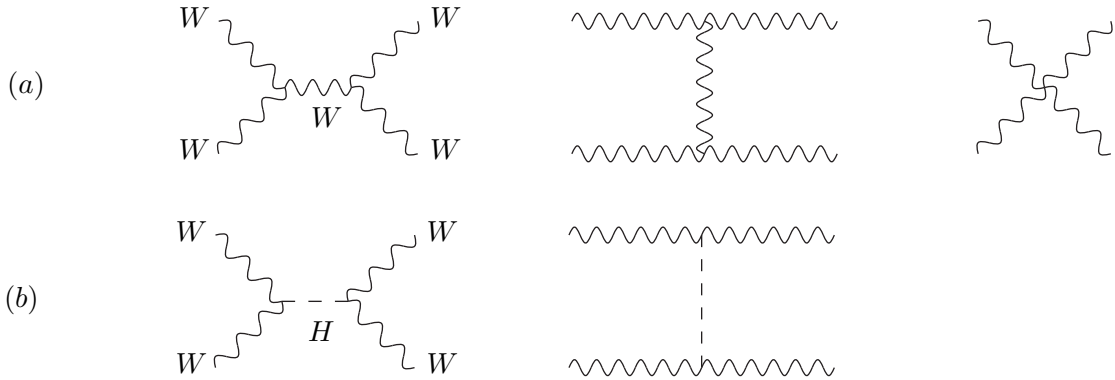
\begin{figure}[hbt]
\begin{center}
\begin{picture}(80,80)(50,-20)
\Photon(0,50)(25,25){3}{3} \Photon(0,0)(25,25){3}{3}
\Photon(25,25)(50,25){3}{3} \Photon(50,25)(75,50){3}{3}
\Photon(50,25)(75,0){3}{3} \put(-80,23){$(a)$} \put(-15,-2){$W$}
\put(-15,48){$W$} \put(80,-2){$W$} \put(80,48){$W$} \put(35,10){$W$}
\end{picture}
\begin{picture}(60,80)(0,-20)
\Photon(0,50)(100,50){3}{12} \Photon(0,0)(100,0){3}{12}
\Photon(50,50)(50,0){3}{6}
\end{picture}
\begin{picture}(60,80)(-90,-20)
\Photon(0,50)(25,25){3}{3} \Photon(0,0)(25,25){3}{3}
\Photon(25,25)(50,50){3}{3} \Photon(25,25)(50,0){3}{3}
\end{picture} \\
\begin{picture}(80,60)(83,0)
\Photon(0,50)(25,25){3}{3} \Photon(0,0)(25,25){3}{3}
\DashLine(25,25)(50,25){6} \Photon(50,25)(75,50){3}{3}
\Photon(50,25)(75,0){3}{3} \put(-80,23){$(b)$} \put(-15,-2){$W$}
\put(-15,48){$W$} \put(80,-2){$W$} \put(80,48){$W$} \put(35,10){$H$}
\end{picture}
\begin{picture}(60,60)(33,0)
\Photon(0,50)(100,50){3}{12} \Photon(0,0)(100,0){3}{12}
\DashLine(50,50)(50,0){5}
\end{picture}  \\
\end{center}
\caption[]{\it \label{esp-fg:wwtoww} Diagramas gen\'ericos para la
dispersi\'on el\'astica $WW$: (a) din\'amica norma-bos\'on pura, y (b)
el intercambio Higgs-bos\'on.}
\end{figure}
\noindent

Sin embargo, el aumento cuadr\'atico de la energ\'{\i}a puede ser
amortiguado si se intercambia una nueva part\'icula escalar,
Fig.~\ref{fg:wwtoww}b.  Para lograr la cancelaci\'on, el tama\~no
del acoplamiento debe estar dado por el producto del acoplamiento
de norma con la masa del bos\'on de norma.  Para altas energ\'{\i}as,
la amplitud $A'_0 = -G_F s/8\pi\sqrt{2}$ cancela exactamente la
divergencia cuadr\'atica de la amplitud  pura del bos\'on de norma 
$A_0$. Por lo tanto, la unitaridad puede restaurarse
introduciendo una  \underline{\it part\'{\i}cula de Higgs }
d\'ebilmente acoplada.  

De la misma forma, la divergencia lineal de
la amplitud  $A(f\bar f\to W_L W_L)\sim gm_f\sqrt{s}$ para la
aniquilaci\'on de un par fermi\'on--antifermi\'on a un par de
bos\'ones de norma longitudinalmente polarizados, puede ser aminorada
a\~nadiendo el intercambio del Higgs al intercambio de bos\'on de norma. 
En este caso la part\'{\i}cula de Higgs debe acoplarse
proporcionalmente a la masa $m_f$ del fermi\'on $f$.

Estas observaciones pueden ser resumidas en una regla\footnote{La
regla parece ser v\'alida a\'un si las teor\'ias en m\'as de cuatro
dimensiones son incluidas.}: {\it Una teor\'{\i}a de bosones de norma
 y  fermiones masivos que est\'an d\'ebilmente acoplados a muy altas
energ\'{\i}as, requiere, por unitaridad, la existencia de una
part\'{\i}cula de Higgs; la part\'{\i}cula de Higgs es una
part\'{\i}cula  escalar $0^+$ que se acopla a otras part\'{\i}culas
proporcionalmente a las masas de las part\'{\i}culas.}

La suposici\'on de que los acoplamientos de las part\'{\i}culas
fundamentales es d\'ebil hasta energ\'{\i}as altas se apoya
cualitativamente en la renormalizaci\'on perturbativa del \'angulo de
mezcla electrod\'ebil $\sin^2\theta_W$ a partir del valor de simetr\'\i a 3/8 a
la escala de GUT hasta $\sim 0.2$ a bajas energ\'\i as, el cual es
cercano al valor observado experimentalmente.\\

Estas ideas pueden expresarse en una forma matem\'atica elegante
interpretando a las interacciones electrod\'ebiles como una teor\'{\i}a
de norma con rompimiento espont\'aneo de la simetr\'{\i}a en el sector
escalar\footnote{El  mecanismo del rompimiento espont\'aneo de la
simetr\'{\i}a, incluyendo el teorema de Goldstone as\'{\i} como el
mecanismo de Higgs, son ejemplificados por el ilustrativo modelo
$O(3)$ $\sigma$ en el Ap\'endice A.}.  Tal teor\'{\i}a consiste en
campos fermi\'onicos, campos de norma y campos escalares acoplados por
las interacciones estandares de norma y las interacciones de Yukawa a
los otros campos. Adem\'as, una auto-interacci\'on
\begin{equation}
V = \frac{\lambda}{2} \left[ |\phi|^2 - \frac{v^2}{2} \right]^2
\label{esp-eq:potential}
\end{equation}
es introducida en el sector escalar, la cual conduce a un valor
diferente de cero para el estado base $v/\sqrt{2}$ del campo
escalar. Fijando la fase de la amplitud del vac\'{\i}o en cero, la
simetr\'{\i}a de norma se rompe espont\'aneamente  en el sector
escalar. Las interacciones del campo de norma con el campo escalar de
fondo, Fig.~\ref{fg:massgen}a, y las interacciones de  Yukawa
de los campo fermi\'onicos  con el campo de fondo,
Fig.~\ref{fg:massgen}b, recorren las masas de estos campos de un valor
cero a uno diferente de cero:
\begin{equation}
\begin{array}{lrclclcl}
\displaystyle (a) \hspace*{2.0cm} & \displaystyle \frac{1}{q^2} & \to
& \displaystyle \frac{1}{q^2} + \sum_j \frac{1}{q^2} \left[ \left(
\frac{gv}{\sqrt{2}} \right)^2 \frac{1}{q^2} \right]^j & = &
\displaystyle \frac{1}{q^2-M^2} & : & \displaystyle M^2 = g^2
\frac{v^2}{2} \\ \\ (b) & \displaystyle \frac{1}{\not \! q} & \to &
\displaystyle \frac{1}{\not \! q} + \sum_j \frac{1}{\not \! q} \left[
\frac{g_fv}{\sqrt{2}} \frac{1}{\not \! q} \right]^j & = &
\displaystyle \frac{1}{\not \! q-m_f} & : & \displaystyle m_f = g_f
\frac{v}{\sqrt{2}}
\end{array}
\end{equation}
Por lo tanto, en teor\'{\i}as con interacciones de norma y Yukawa, en
las cuales el campo escalar adquiere un valor diferente de cero para
el estado base, los acoplamientos son naturalmente proporcionales a
las masas. Esto garantiza la unitaridad de la teor\'\i a como se
discuti\'o anteriormente.  Estas teor\'{\i}as son renormalizables (como
resultado de la invariancia de norma, la cual est\'a solamente
disfrazada en la formulaci\'on unitaria que se ha adoptado hasta ahora),
y por tanto est\'an bien definidas y son matem\'aticamente consistentes.
\begin{figure}[hbt]
\begin{center}
\begin{picture}(60,10)(80,40)
\Photon(0,25)(50,25){3}{6} \LongArrow(65,25)(90,25) \put(-15,21){$V$}
\put(-15,50){$(a)$}
\end{picture}
\begin{picture}(60,10)(40,40)
\Photon(0,25)(50,25){3}{6} \put(60,23){$+$}
\end{picture}
\begin{picture}(60,10)(15,40)
\Photon(0,25)(50,25){3}{6} \DashLine(25,25)(12,50){3}
\DashLine(25,25)(38,50){3} \Line(9,53)(15,47) \Line(9,47)(15,53)
\Line(35,53)(41,47) \Line(35,47)(41,53) \put(45,45){$H$}
\put(70,23){$+$}
\end{picture}
\begin{picture}(60,10)(-10,40)
\Photon(0,25)(75,25){3}{9} \DashLine(20,25)(8,50){3}
\DashLine(20,25)(32,50){3} \DashLine(55,25)(43,50){3}
\DashLine(55,25)(67,50){3} \Line(5,53)(11,47) \Line(5,47)(11,53)
\Line(29,53)(35,47) \Line(29,47)(35,53) \Line(40,53)(46,47)
\Line(40,47)(46,53) \Line(64,53)(70,47) \Line(64,47)(70,53)
\put(90,23){$+ \cdots$}
\end{picture} \\
\begin{picture}(60,80)(80,20)
\ArrowLine(0,25)(50,25)
\LongArrow(65,25)(90,25)
\put(-15,23){$f$}
\put(-15,50){$(b)$}
\end{picture}
\begin{picture}(60,80)(40,20)
\ArrowLine(0,25)(50,25)
\put(65,23){$+$}
\end{picture}
\begin{picture}(60,80)(15,20)
\ArrowLine(0,25)(25,25)
\ArrowLine(25,25)(50,25)
\DashLine(25,25)(25,50){3}
\Line(22,53)(28,47)
\Line(22,47)(28,53)
\put(35,45){$H$}
\put(65,23){$+$}
\end{picture}
\begin{picture}(60,80)(-10,20)
\ArrowLine(0,25)(25,25)
\ArrowLine(25,25)(50,25)
\ArrowLine(50,25)(75,25)
\DashLine(25,25)(25,50){3}
\DashLine(50,25)(50,50){3}
\Line(22,53)(28,47)
\Line(22,47)(28,53)
\Line(47,53)(53,47)
\Line(47,47)(53,53)
\put(90,23){$+ \cdots$}
\end{picture}  \\
\end{center}
\caption[]{\it \label{esp-fg:massgen} Generador (a) de bosones de norma y
(b) masas de fermiones a trav\'es de interacciones con el campo
escalar de fondo.}
\end{figure}
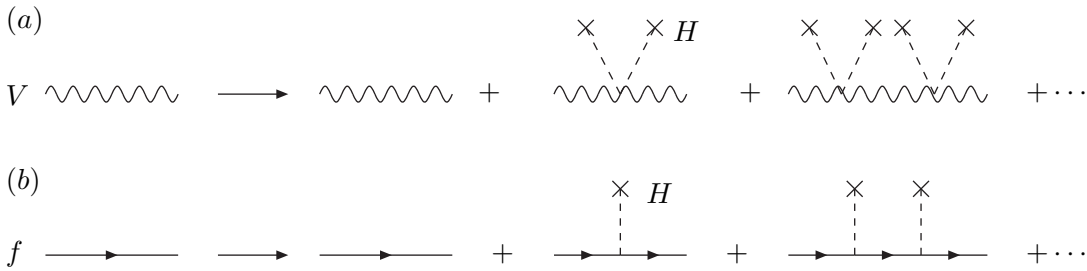

\subsection{El mecanismo de Higgs en el Modelo Est\'andar}

\phantom{h} Adem\'as de las partes de Yang--Mills y fermi\'onicas, el
Lagrangiano electrod\'ebil $SU_2 \times U_1$ incluye un campo escalar
isodoblete, $\phi$, acoplado a s\'\i{} mismo en el potencial $V$, cf. eq.
(\ref{eq:potential}), a los campos de norma a trav\'es de la derivada
covariante $iD = i\partial - g \vec{I} \vec{W} - g'YB$, y a los campos
fermi\'onicos up y down $u,d$ mediante las interacciones de Yukawa:
\begin{equation}
{\cal L}_0 = |D\phi|^2 - \frac{\lambda}{2} \left[ |\phi|^2 -
  \frac{v^2}{2} \right]^2 - g_d \bar d_L \phi d_R - g_u \bar u_L
\phi_c u_R + {\rm hc} ~.
\end{equation}
En la norma unitaria, el isodoblete $\phi$ es reemplazado por el campo
f\'{\i}sico de Higgs $H$, $\phi\to [0,(v+H)/\sqrt{2}]$, el cual describe
las fluctuaciones de la componente $I_3=-1/2$ alrededor del valor del
estado base $v/\sqrt{2}$. La escala $v$ del rompimiento de la
simetr\'{\i}a electrd\'ebil es fijada por la masa del $W$, la cual a su
vez puede re-expresarse por medio del acoplamiento fermi\'onico, $v =
1/\sqrt{\sqrt{2}G_F} \approx 246$ GeV.  El acoplamiento cu\'artico $\lambda$ y el
acoplamiento de Yukawa $g_f$ pueden ser re-expresados en t\'erminos de la
masa f\'isica del Higgs $M_H$ y las masas de los fermi\'on $m_f$:
\begin{eqnarray}
M_H^2 & = & \lambda v^2 \nonumber \\ m_f & = & g_f v / \sqrt{2}
\end{eqnarray}
respectivamente.

Dado que los acoplamientos de la part\'{\i}cula de Higgs con las
part\'iculas de norma, con los fermiones y con ellas mismas est\'an
dados por los acoplamientos de norma y las masas de las part\'{\i}culas,
el \'unico par\'ametro desconocido en el sector de Higgs (aparte de la
matr\'\i z de mezcla CKM) es la masa del Higgs. Cuando esta masa se
fija, todas las propiedades de la part\'{\i}cula del Higgs pueden
ser predichas, i.e. el tiempo de vida y las razones de desintegraci\'on
(branching ratios),
as\'{\i} como los mecanismos de producci\'on y las respectivas secciones
eficaces.

\subsubsection{La Masa del Higgs del Modelo Est\'andar \\ \\} 

Aun cuando la masa del bos\'on de Higgs no puede ser predicha en el
Modelo Est\'andar, l\'{\i}mites superior e inferior estrictos pueden ser
derivados de las condiciones de consistencia interna y extrapolaciones
del modelo a altas energ\'{\i}as.

El bos\'on de Higgs ha sido introducido como una part\'{\i}cula
fundamental para hacer a las amplitudes de dispersi\'on 2--2 que
involucran bosones $W$ polarizados longitudinalmente compatibles con
la unitaridad.  Basados en el principio general de incertidumbre de
tiempo-energ\'{\i}a, las part\'{\i}culas deben desacoplarse de un
sistema f\'{\i}sico si su energ\'{\i}a crece indefinidamente. La masa
de la part\'{\i}cula de Higgs debe por tanto estar obligada a
restablecer la unitaridad en el r\'egimen perturbativo. De la expansi\'on
asint\'otica de la amplitud de dispersi\'on el\'astica $W_L W_L$ onda $S$,
incluyendo $W$ y los intercambios de Higgs, $A(W_L W_L \to W_L W_L) \to
-G_F M_H^2/4\sqrt{2}\pi$, se sigue que \cite{25}
\begin{equation}
M_H^2 \leq 2\sqrt{2}\pi/G_F \sim (850~\mbox{GeV})^2 ~.
\end{equation}
Dentro de la formulaci\'on can\'onica del Modelo Est\'andar, las condiciones
de consistencia requieren por tanto una masa del Higgs por abajo de 1
TeV.\\

\begin{figure}[hbt]
\vspace*{-0.5cm}

\begin{center}
\begin{picture}(90,80)(60,-10)
\DashLine(0,50)(25,25){3}
\DashLine(0,0)(25,25){3}
\DashLine(50,50)(25,25){3}
\DashLine(50,0)(25,25){3}
\put(-15,45){$H$}
\put(-15,-5){$H$}
\put(55,-5){$H$}
\put(55,45){$H$}
\end{picture}
\begin{picture}(90,80)(10,-10)
\DashLine(0,50)(25,25){3}
\DashLine(0,0)(25,25){3}
\DashLine(75,50)(50,25){3}
\DashLine(75,0)(50,25){3}
\DashCArc(37.5,25)(12.5,0,360){3}
\put(-15,45){$H$}
\put(-15,-5){$H$}
\put(35,40){$H$}
\put(80,-5){$H$}
\put(80,45){$H$}
\end{picture}
\begin{picture}(50,80)(-40,2.5)
\DashLine(0,0)(25,25){3}
\DashLine(0,75)(25,50){3}
\DashLine(50,50)(75,75){3}
\DashLine(50,25)(75,0){3}
\ArrowLine(25,25)(50,25)
\ArrowLine(50,25)(50,50)
\ArrowLine(50,50)(25,50)
\ArrowLine(25,50)(25,25)
\put(-15,70){$H$}
\put(-15,-5){$H$}
\put(35,55){$t$}
\put(80,-5){$H$}
\put(80,70){$H$}
\end{picture}  \\
\setlength{\unitlength}{1pt}
\caption[]{\label{esp-fg:lambda} \it Diagramas generando la evoluci\'on de 
la auto-interacci\'on del Higgs $\lambda$.}
\end{center}
\end{figure}
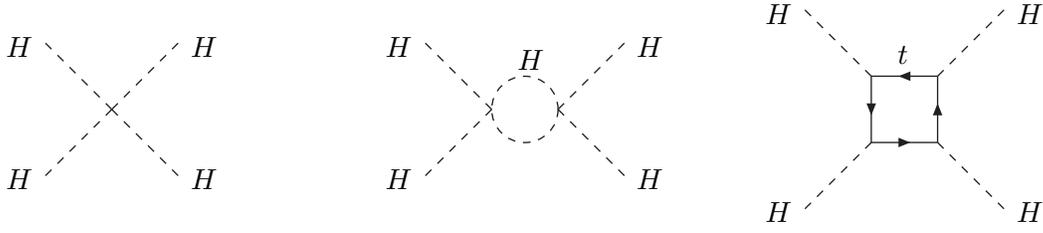

\begin{figure}[hbtp]

\vspace*{0.8cm}

\hspace*{2.0cm}
\begin{turn}{90}%
\epsfxsize=8.5cm \epsfbox{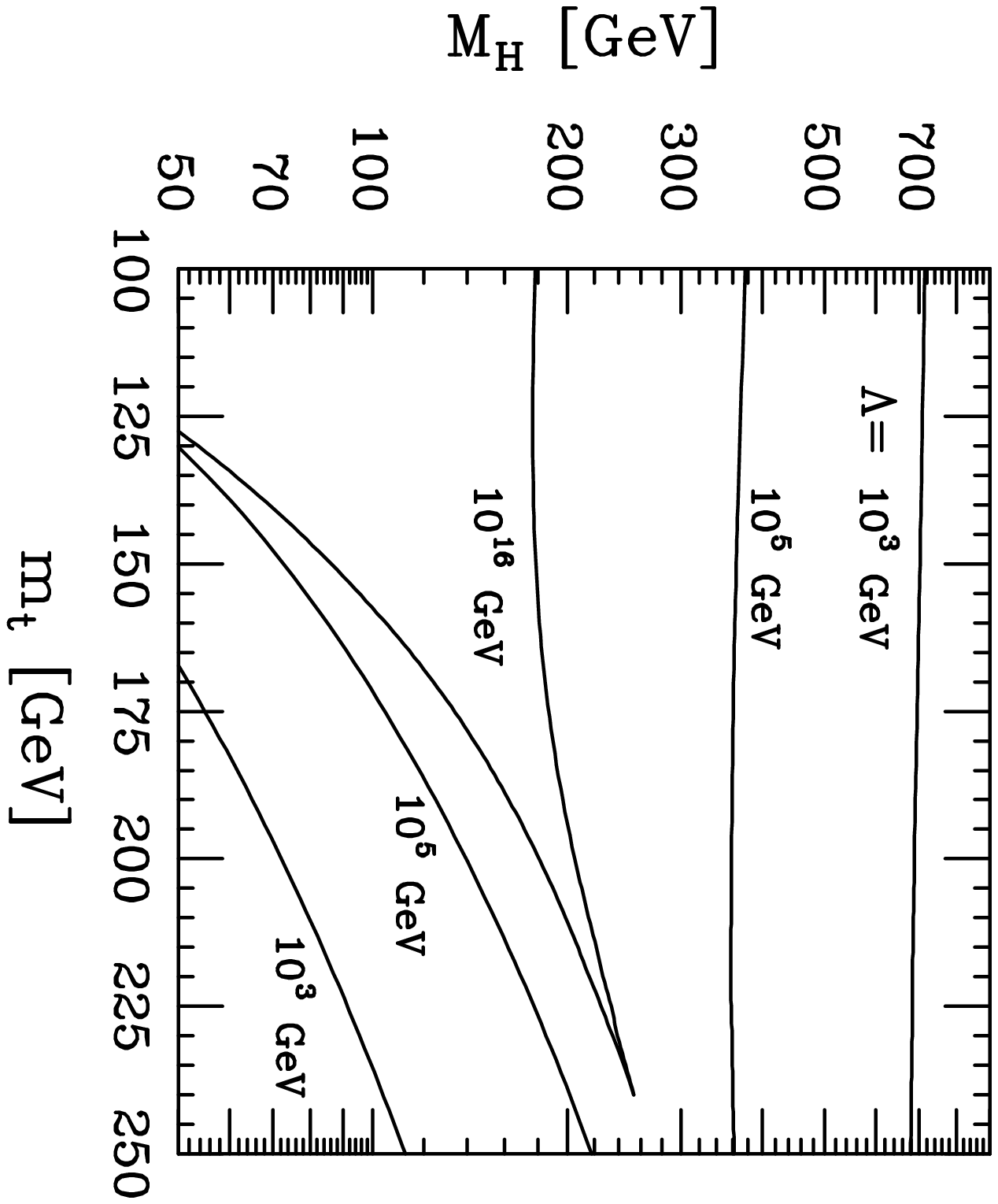}
\end{turn}
\vspace*{-0.2cm}

\caption[]{\label{esp-fg:triviality} \it L\'{\i}mites a la masa del bos\'on
Higgs en el SM. Aqu\'{\i} $\Lambda$ denota la escala de energ\'{\i}a a
la cual el sistema del bos\'on de Higgs del SM se volver\'\i a
fuertemente interactuante (l\'{\i}mite superior); el l\'{\i}mite
inferior proviene de pedir la estabilidad del
vac\'{\i}o. (Refs. \cite{4,5}.)}
\vspace*{-0.3cm}
\end{figure}

L\'{\i}mites muy restrictivos para el valor de la masa del Higgs del
Modelo Est\'andar provienen de la hip\'otesis sobre la escala de
energ\'{\i}a $\Lambda$ hasta la cual el Modelo Est\'andar puede extenderse
antes de que nuevos fen\'omenos de interacci\'on aparezcan. La clave para
estos l\'{\i}mites es la evoluci\'on del acoplamiento cu\'artico $\lambda$
con la energ\'{\i}a debida a fluctuaciones cu\'anticas \cite{4}. Las
contribuciones b\'asicas se presentan en la Fig.~\ref{fg:lambda}.  El lazo
del Higgs por s\'\i{} mismo da pie a un aumento indefinido del acoplamiento,
mientras que el lazo fermi\'onico-top quark, conforme la masa del top
aumenta, lleva al acoplamiento a valores peque\~nos, finalmente incluso
a valores abajo de cero. La variaci\'on del acoplamiento cu\'artico del
Higgs $\lambda$ y el acoplamiento de Yukawa top-Higgs $g_t$ con la
energ\'{\i}a, parametrizada por $t=\log \mu^2/v^2$, pueden ser escritos
como sigue \cite{4}
\begin{equation}
\begin{array}{rclcl} 
\displaystyle \frac{d\lambda}{dt} & = & \displaystyle \frac{3}{8\pi^2} 
\left[ \lambda^2 + \lambda g_t^2 - g_t^4 \right] & : & \displaystyle 
\lambda(v^2) = M_H^2/v^2 \\ \\ \displaystyle \frac{d g_t}{dt} & = & 
\displaystyle \frac{1}{32\pi^2} \left[ \frac{9}{2} g_t^3 - 8 g_t g_s^2 
\right] & : & \displaystyle g_t(v^2) = \sqrt{2}~m_f/v ~. 
\end{array}
\end{equation}

S\'olo las contribuciones principales de lazos de Higgs, top y
QCD son tomadas en cuenta. \\

Para masas del top moderadas, el acoplamiento cu\'artico $\lambda$ aumenta
indefinidamente, $\partial \lambda / \partial t \sim + \lambda^2$, y el acoplamiento se vuelve
fuerte poco antes de alcanzar el polo de Landau:
\begin{equation}
\lambda (\mu^2) = \frac{\lambda(v^2)}{1- \frac{3\lambda(v^2)}{8\pi^2} 
\log \frac{\mu^2}{v^2}} ~. 
\end{equation}
Reexpresando el valor inicial de $\lambda$ por la masa del Higgs, la
condici\'on $\lambda (\Lambda) < \infty$, puede ser traducida a un
\underline{l\'{\i}mite superior} para la masa del Higgs:
\begin{equation}
M_H^2 \leq \frac{8\pi^2 v^2}{3\log \Lambda^2/v^2} ~. 
\end{equation}

Este l\'{\i}mite para la masa est\'a relacionado logar\'{\i}tmicamente
con la energ\'{\i}a $\Lambda$ hasta la cual el Modelo Est\'andar se supone 
v\'alido.  El valor m\'aximo de $M_H$ para el corte m\'\i nimo $\Lambda \sim $
1~TeV est\'a dado por $\sim 750$ GeV. Esta cota es cercana a la
estimada $\sim 700$ GeV en los c\'alculos hechos con lattices para $\Lambda \sim
1$ TeV, los que permiten un control apropiado de
efectos no preturbativos cerca del l\'\i mite \cite{6}.\\

\begin{table}[hbt]
\renewcommand{\arraystretch}{1.5}
\begin{center}
\begin{tabular}{|l||l|} \hline
$\Lambda$ & $M_H$ \\ \hline \hline 1 TeV & 60 GeV $\lessim M_H \lessim
700$ GeV \\ $10^{19}$ GeV & 130 GeV $\lessim M_H \lessim 190$ GeV \\
\hline
\end{tabular}
\renewcommand{\arraystretch}{1.2}
\caption[]{\label{esp-tb:triviality} \it L\'{\i}mites de la masa del Higgs
para dos valores del corte $\Lambda$.}
\end{center}
\end{table}
Un \underline{l\'{\i}mite inferior} en la masa del Higgs puede ser
obtenido a partir de requerir la estabilidad del vac\'{\i}o
\cite{4,5}. Dado que las correcciones top-lazo reducen $\lambda$ para un
acoplamiento del Yukawa del top creciente, $\lambda$ se vuelve negativa si la
masa del top se se hace muy grande.  En ese caso, el potencial
de la energ\'{\i}a propia se volver\'\i a muy negativo y el 
estado base ya no ser\'\i a estable. Para evitar la
inestabilidad, la masa del Higgs debe de exceder un valor m\'{\i}nimo
para un valor dado de la masa del top. Este l\'{\i}mite inferior
depende del valor de corte $\Lambda$.\\

Para cualquier $\Lambda$ dada los valores permitidos de los pares
$(M_t,M_H)$ se muestran en Fig.~\ref{fg:triviality}. Los valores permitidos de
la masa del Higgs  est\'an reunidos en la Tabla \ref{tb:triviality}, 
para dos valores espec\'\i ficos del corte $\Lambda$.  Si el Modelo
Est\'andar se supone v\'alido hasta escalas de gran unificaci\'on, la masa del
Higgs est\'a restringida en el intervalo estrecho entre 130 y 
190~GeV. La observaci\'on de una masa del Higgs por arriba o por abajo de estos
valores requerir\'\i a una escala de nueva f\'{\i}sica por debajo de la
escala del GUT.

\subsubsection{Decaimientos de la part\'{\i}cula de Higgs \\ \\}
El perfil de la part\'{\i}cula de Higgs est\'a determinado
un\'{\i}vocamente si se fija la masa del Higgs. La intensidad de los
acoplamientos de Yukawa del bos\'on de Higgs a fermiones est\'a
dado por las masas de los fermiones $m_f$, y el acoplamiento
a los bosones de norma electrod\'ebiles $V=W,Z$, por sus masas $M_V$:

\begin{eqnarray}
g_{ffH} & = & \left[ \sqrt{2} G_F \right]^{1/2} m_f \\
g_{VVH} & = & 2 \left[ \sqrt{2} G_F \right]^{1/2} M_V^2 ~.
\nonumber
\end{eqnarray}

El ancho total de decaimiento y tiempo de vida, as\'{\i} como 
las razones de desintegraci\'on para canales de decaimiento
espec\'{\i}ficos, est\'an determinados por estos par\'ametros. La medici\'on
de las caracter\'{\i}sticas de decaimiento puede, por lo tanto, ser
explotada para establecer, experimentalmente, que los acoplamientos
del Higgs crecen con las masas de las part\'\i culas, una consecuencia
directa del mecanismo de Higgs {\it sui generis}.

Para las part\'{\i}culas de Higgs en el rango intermedio de masas
${\cal O}(M_Z) \leq M_H \leq 2M_Z$, los modos principales de decaimiento
son en pares $b\bar b$ y pares $WW,ZZ$, siendo virtual uno de los
bosones de norma por abajo del umbral respectivo. Arriba de los
umbrales de los pares $WW,ZZ$, las part\'{\i}culas de Higgs decaen
casi exclusivamente en estos dos canales, con una peque\~na contribuci\'on
de los decaimientos del top cerca del umbral del $t\bar t$.  Abajo de
los 140 GeV, los decaimientos $H\to \tau^+\tau^-, c\bar c$ y $gg$ son
tambi\'en importantes adem\'as del canal dominante $b\bar b$; los
decaimientos a $\gamma\gamma$, aunque de tasa suprimida, aportan no obstante
una se\~nal clara de 2-cuerpos para la formaci\'on de la part\'{\i}cula de
Higgs en este rango de masa.

\paragraph{(a) Decaimientos de Higgs a fermiones} ~\\[0.5cm]
El ancho parcial de los decaimientos de Higgs a pares de leptones y
quarks est\'a dado por \cite{26}
\begin{equation}
\Gamma (H\to f\bar f) = {\cal N}_c \frac{G_F}{4\sqrt{2}\pi}
m_f^2(M_H^2) M_H ~,
\end{equation}
siendo ${\cal N}_c = 1$ o 3 el factor de color. Cerca del umbral el
ancho parcial est\'a suprimido por el factor de onda-P, $\beta_f^3$
adicional, donde $\beta_f$ es la velocidad del fermi\'on. Asint\'oticamente,
el ancho fermi\'onico crece s\'olo linealmente con la masa del Higgs. La
mayor parte de las correcciones radiativas de QCD se pueden mapear
a la dependencia de escala de la masa del quark, evaluada en la masa
del Higgs.  Para $M_H\sim 100$ GeV los par\'ametros relevantes son $m_b
(M_H^2) \simeq 3$ GeV y $m_c (M_H^2) \simeq$ 0.6~GeV. La reducci\'on de la masa
efectiva del quark-$c$ sobrecompensa el factor de color en la raz\'on
entre los decaimientos a charm y a $\tau$ del bos\'on de Higgs. Las
correcciones residuales de QCD, $\sim 5.7 \times (\alpha_s/\pi)$, modifican los
anchos s\'olo ligeramente.

\paragraph{(b) Decaimientos del Higgs a pares debosones $WW$ y $ZZ$} ~\\[0.5cm]
Por arriba de los umbrales de los decaimientos $WW$ y $ZZ$, los anchos
parciales para estos canales pueden ser escritos como\cite{27}
\begin{equation}
\Gamma (H\to VV) = \delta_V \frac{G_F}{16\sqrt{2}\pi} M_H^3
(1-4x+12x^2) \beta_V ~,
\end{equation}
donde $x=M_V^2/M_H^2$ y $\delta_V = 2$ y 1 para $V=W$ y $Z$,
respectivamente. Para masas grandes del Higgs, los bosones vectoriales
est\'an polarizados longitudinalmente. Dado que las funciones de onda de
estos estados son lineales en la energ\'{\i}a, los anchos crecen como
la tercera potencia de la masa de Higgs. Por debajo del umbral de
dos bosones reales, la part\'{\i}cula de Higgs puede decaer en pares
$VV^*$, con uno de los bosones vectoriales siendo virtual. El ancho de
decaimiento est\'a dado en este caso \cite{28} por

\begin{equation}
\Gamma(H\to VV^*) = \frac{3G^2_F M_V^4}{16\pi^3}~M_H
R(x)~\delta'_V ~,
\end{equation}
donde $\delta'_W = 1$, $\delta'_Z = 7/12 - 10\sin^2\theta_W/9 + 40
\sin^4\theta_W/27$ y
\begin{displaymath}
R(x) =
\frac{3(1-8x+20x^2)}{(4x-1)^{1/2}}\arccos\left(\frac{3x-1}{2x^{3/2}}
\right) - \frac{1-x}{2x} (2-13x+47x^2) - \frac{3}{2} (1-6x+4x^2)
\log x ~.
\end{displaymath}

El canal $ZZ^*$ se vuelve relevante para masas del Higgs m\'as all\'a de
$\sim 140$ GeV. Por arriba del umbral, el canal de 4-leptones $H\to ZZ \to
4 \ell^\pm$ proporciona una se\~nal muy clara para los bosones de Higgs.
Tambi\'en el canal de decaimiento $WW$ demuestra su utilidad, a pesar
del escape de los neutrinos en decaimientos lept\'onicos del $W$, si el
canal $ZZ$ dentro de la capa de masa est\'a cerrado 
cinem\'aticamente.

\paragraph{(c) Decaimiento del Higgs a pares de $gg$ y $\gamma\gamma$} ~\\[0.5cm]

En el Modelo Est\'andar, los decaimientos glu\'onicos del Higgs son
mediados por lazos (loops) de quarks top y bottom, para los
decaimientos fot\'onicos, adem\'as son mediados por lazos del $W$.  Como
estos decaimientos son significantivos s\'olo muy por debajo de los
umbrales del top y del $W$,  son descritos por las expresiones
aproximadas \cite{29,30}
\begin{eqnarray}
\Gamma (H\to gg) & = & \frac{G_F
\alpha_s^2(M_H^2)}{36\sqrt{2}\pi^3}M_H^3 \left[ 1+
\left(\frac{95}{4} - \frac{7N_F}{6} \right) \frac{\alpha_s}{\pi}
\right] \label{esp-eq:htogg} \\ \nonumber \\
\Gamma (H\to \gamma\gamma) & = & \frac{G_F
\alpha^2}{128\sqrt{2}\pi^3}M_H^3 \left[  \frac{4}{3} {\cal N}_C
e_t^2 - 7 \right]^2 ~,
\end{eqnarray}
las cuales son v\'alidas en el l\'{\i}mite $M_H^2 \ll 4M_W^2, 4M_t^2$.
Las correcciones radiativas de QCD, las cuales incluyen los estados
finales $ggg$ y $gq\bar q$ en (\ref{eq:htogg}), son muy importantes;
estas correcciones incrementan el ancho parcial en alrededor de un
65\%. Aunque los decaimientos fot\'onicos del Higgs son muy raros,
ofrecen, no obstante, una se\~nal simple y atractiva para las
part\'{\i}culas de Higgs ya que llevan a s\'olo dos part\'{\i}culas
estables al estado final.

\paragraph{\underline{Digresi\'on:}}Los acoplamientos del Higgs mediados
por lazos pueden ser f\'acilmente 
calculados en el l\'{\i}mite en el cual la masa del Higgs es peque\~na
comparada con la masa del lazo, explotando un teorema a bajas
energ\'{\i}as para la amplitud externa del Higgs ${\cal A} (XH)$:
\begin{equation}
\lim_{p_H\to 0} {\cal A}(XH) = \frac{1}{v} \frac{\partial {\cal
A}(X)}{\partial
 \log m} ~.
\end{equation}
El teorema puede derivarse al observar que la inserci\'on de una
l\'{\i}nea externa de energ\'{\i}a cero del Higgs en un propagador
fermi\'onico, por ejemplo, es equivalente a la sustituci\'on
\begin{displaymath}
\frac{1}{\not\! p-m} \to \frac{1}{\not\! p-m} \frac{m}{v}
\frac{1}{\not\! p-m} = \frac{1}{v} \frac{\partial}{\partial \log
m} \frac{1}{\not\! p-m} ~.
\end{displaymath}
Las amplitudes para procesos incluyen una l\'{\i}nea externa del Higgs
pueden obtenerse as\'{\i} de la amplitud sin la l\'{\i}nea
externa del Higgs al tomar la derivada logar\'{\i}tmica. Si se aplica
al propagador del glu\'on en $Q^2=0$, $\Pi_{gg} \sim (\alpha_s/12\pi) GG \log
m$, la amplitud $Hgg$ puede derivarse f\'acilmente como ${\cal A}(Hgg) =
GG \alpha_s/(12\pi v)$~. Si se incluyen ordenes m\'as altos, el par\'ametro $m$
debe interpretarse como la masa desnuda.

\paragraph{(d) Resumen} ~\\[0.5cm]
Al sumar todos los posibles canales de decaimiento, obtenemos el ancho
total mostrado en la Fig.~\ref{fg:wtotbr}a. Hasta masas de 140 GeV,
la part\'{\i}cula del Higgs es muy estrecha, $\Gamma(H) \leq 10$ MeV. Despu\'es
de que los canales real y virtual de los bosones de norma se abren, el
estado r\'apidamente se ensancha, alcanzando un ancho de $\sim 1$ GeV en
el umbral de $ZZ$. El ancho no se puede medir directamente en la
regi\'on intermedia de masas en los colisionadores LHC o $e^+ e^-$. Sin
embargo, puede determinarse indirectamente; midiendo, por ejemplo, el
ancho parcial $\Gamma (H\to WW)$ a trav\'es proceso de fusi\'on $WW\to H$, y la
fracci\'on de desintegraci\'on $BR(H\to WW)$ en el proceso de decaimiento $H\to
WW$, el ancho total se sigue de la raz\'on de las dos observables.
Por arriba de una masa de $\sim 250$ GeV, el estado se ensancha lo
suficiente para ser resuelto experimentalmente.
\begin{figure}[hbtp]

\vspace*{0.5cm}
\hspace*{1.0cm}
\begin{turn}{-90}%
\epsfxsize=8.5cm \epsfbox{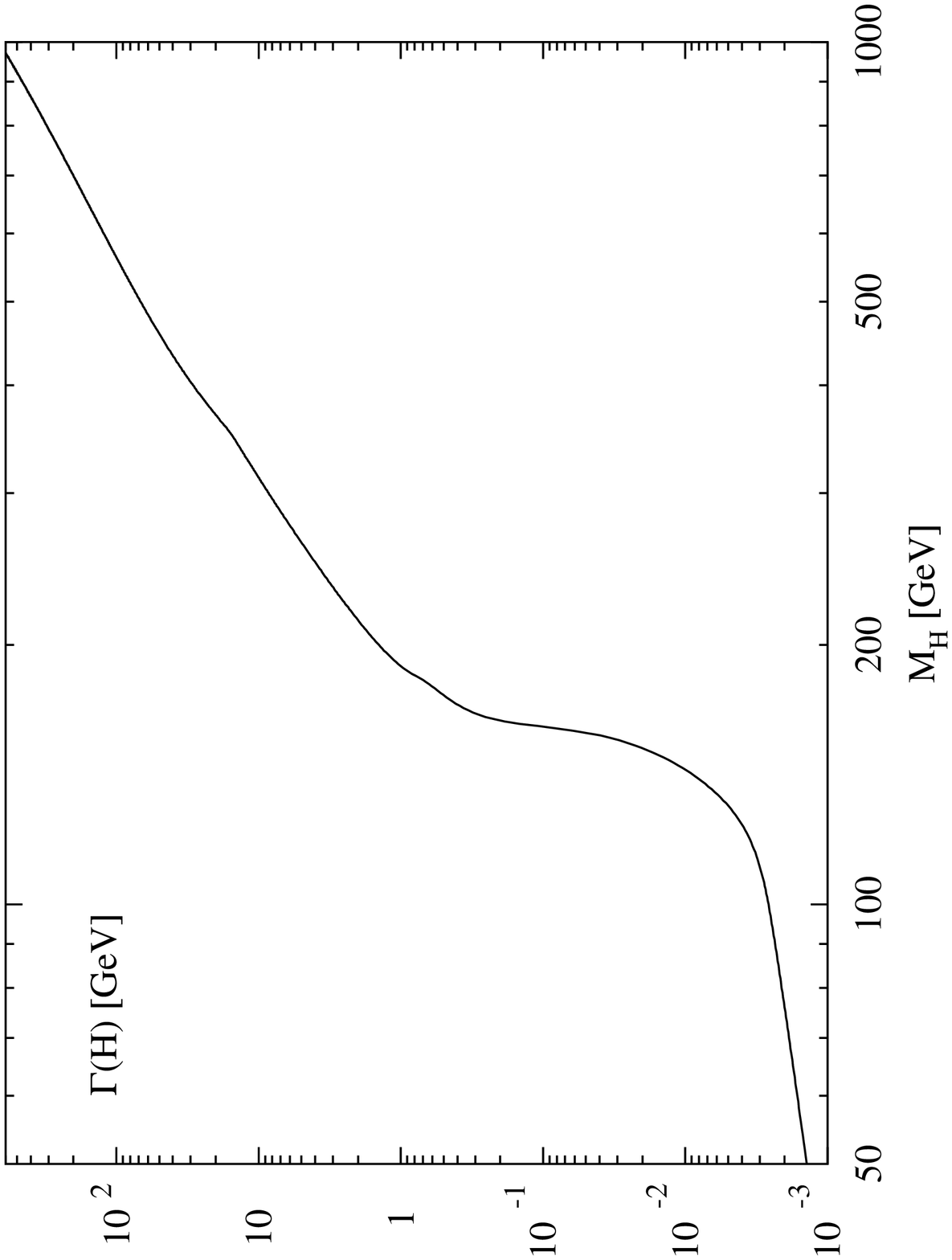}
\end{turn}

\vspace*{0.5cm}
\hspace*{1.0cm}
\begin{turn}{-90}%
\epsfxsize=8.5cm \epsfbox{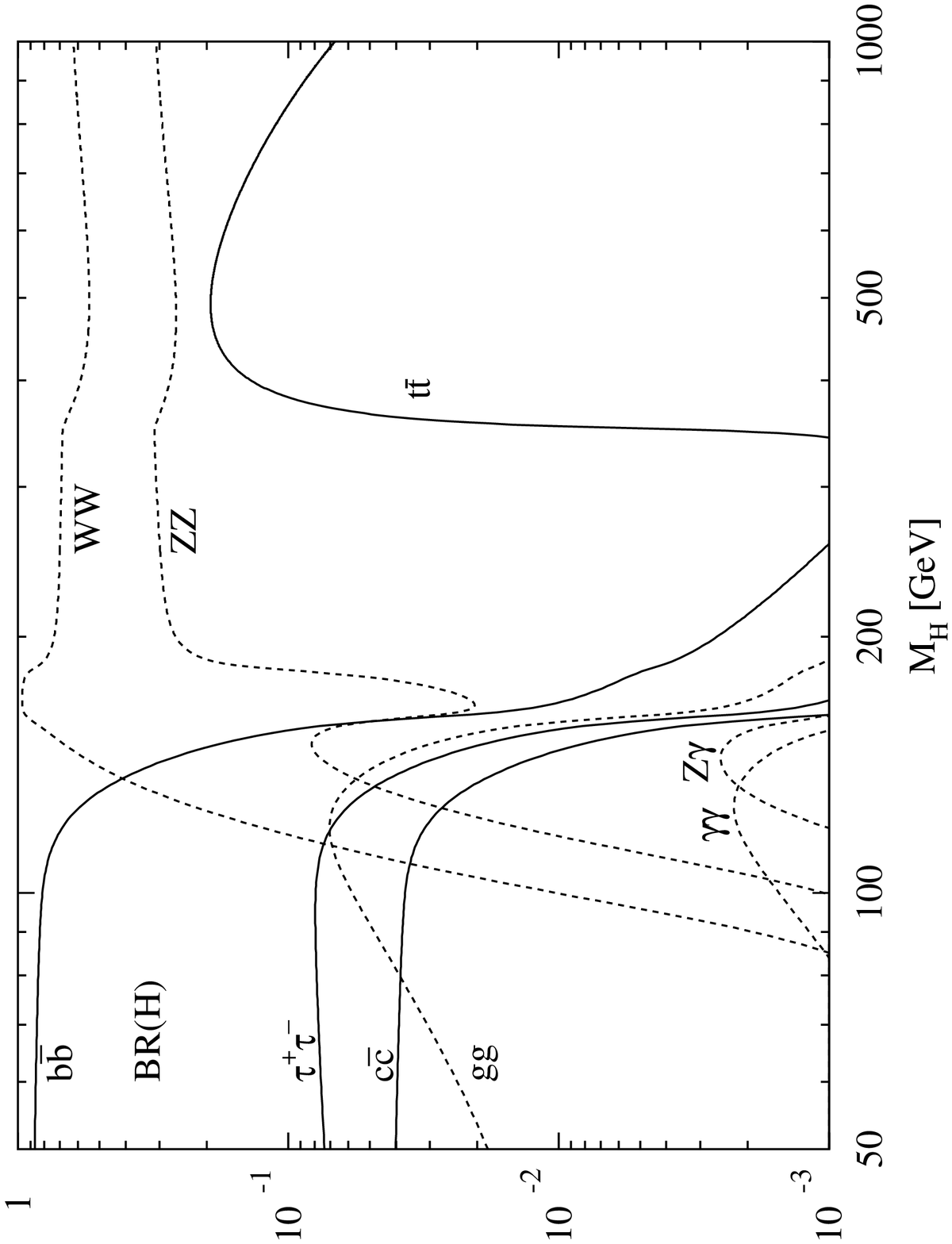}
\end{turn}
\vspace*{-0.0cm} \caption[]{\label{esp-fg:wtotbr} \it (a) Ancho total de
  decaimiento (en GeV) bos\'on de Higgs del SM como funci\'on de su masa.
  (b) Razones de desintegraci\'on (branching ratios) de los modos
  dominantes de decaimiento de 
  la part\'{\i}cula de Higgs del SM. Se toman en consideraci\'on todas
  las correcciones relevantes de orden m\'as alto.}

\end{figure}

Las razones de desintegraci\'on de los modos de decaimiento principales se
muestran en la Fig.~\ref{fg:wtotbr}b. Una gran variedad de canales
ser\'a accesible para masas del Higgs por debajo de 140 GeV. El modo
dominante es el decaimiento a $b\bar b$, aunque los decaimientos a
$c\bar c, \tau^+\tau^-$ y $gg$ ocurren tambi\'en a un nivel de porcentaje
considerable. A $M_H=$ 120~GeV, por ejemplo, las razones de
desintegraci\'on son de 68\% para $b\bar b$, 3.1\% para $c\bar c$, 6.9\%
para $\tau^+\tau^-$ y 7\% para $gg$.  Decaimientos a $\gamma\gamma$ se dan en un
nivel de 1 por mil.  Arriba de este valor de masa, el decaimiento del
bos\'on de Higgs a $W$'s se vuelve dominante, sobrepasando todos los
dem\'as canales si el modo de decaimiento a dos $W$'s reales es
cinem\'aticamente posible. Para masas del Higgs muy por arriba de los
umbrales, los decaimientos $ZZ$ y $WW$ se dan a un raz\'on de 1:2,
ligeramente modificados s\'olo justo por arriba del umbral de $t\bar t$.
Como el ancho crece como la tercera potencia de la masa, la
part\'{\i}cula de Higgs se vuelve muy ancha, $\Gamma(H) \sim \frac{1}{2}
M_H^3$ [TeV]. De hecho, para $M_H\sim 1$ TeV, el ancho alcanza $\sim
\frac{1}{2}$ TeV.

\subsection{Datos de Precisi\'on Electrod\'ebil: Una Estimaci\'on de la Masa
  del Higgs} 

\phantom{h} Evidencia indirecta del bos\'on ligero de Higgs se puede
derivar de las mediciones de alta precisi\'on de las observables
electrod\'ebiles en LEP y otros lugares. En realidad, el hecho de que el
Modelo Est\'andar sea renormalizable s\'olo despu\'es de incluir las
part\'\i culas top y Higgs en las correcciones de lazo, indica que
las observables electrod\'ebiles son sensibles a las masas de estas
part\'{\i}culas.

El acoplamiento de Fermi puede ser reescrito en t\'erminos del
acoplamiento d\'ebil y la masa del $W$; al orden m\'as bajo,
$G_F/\sqrt{2} = g^2/8M_W^2$. Despu\'es de sustituir el acoplamiento
electromagn\'etico $\alpha$, el \'angulo de mezcla electrod\'ebil y la masa del
$Z$ para el acoplamiento d\'ebil, y la masa del $W$, esta relaci\'on puede
reescribirse como
\begin{equation}
\frac{G_F}{\sqrt{2}} = \frac{2\pi\alpha}{\sin^2 2\theta_W M_Z^2}
[1+\Delta r_\alpha + \Delta r_t + \Delta r_H ] ~.
\end{equation}
Los t\'erminos $\Delta$ toman en cuenta las correcciones radiativas: $\Delta
r_\alpha$ describe el cambio en el acoplamiento electromagn\'etico $\alpha$ si
se eval\'ua en la escala $M_Z^2$ en lugar de en el momento cero; $\Delta
r_t$ denota las contribuciones del quark top (y bottom) a las masas de
los bosones $W$ y $Z$, las cuales son cuadr\'aticas en la masa del top.
Finalmente, $\Delta r_H$ toma en cuenta las contribuciones del Higgs virtual a
las masas; este t\'ermino depende s\'olo logar\'{\i}tmicamente \cite{7} en
la masa del Higgs al orden dominante:
\begin{equation}
\Delta r_H = \frac{G_F M_W^2}{8\sqrt{2}\pi^2} \frac{11}{3} \left[
\log \frac{M_H^2}{M_W^2} - \frac{5}{6} \right] \hspace{1cm} (M_H^2 \gg M_W^2)
~.
\end{equation}
El efecto de apantallamiento refleja el papel del campo de Higgs como
un regulador que hace a la teor\'{\i}a electrod\'ebil 
renormalizable.

Aunque la sensibilidad a la masa del Higgs es \'unicamente
logar\'{\i}tmica, el incremento en la precisi\'on de las medidas
de las observables electrod\'ebiles nos permite derivar
estimaciones interesantes y restricciones a la masa del Higgs
\cite{8}:
\begin{eqnarray}
M_H & = & 91^{+45}_{-32}~\mbox{GeV} \\
    & \lessim & 186 ~\mbox{GeV~~~(95\% CL)}  ~. \nonumber
\end{eqnarray}
Se puede concluir de estos n\'umeros que la formulaci\'on
can\'onica del Modelo Est\'andar incluyendo la existencia de un
bos\'on de Higgs ligero, es compatible con los datos
electrod\'ebiles. Sin embrago, mecanismos alternativos no se
pueden descartar si el sistema est\'a abierto a contribuciones de
\'areas de f\'{\i}sica m\'as all\'a del Modelo Est\'andar.

\subsection{Canales de Producci\'on de Higgs en colisionadores $e^+e^-$}

\phantom{h} El primer proceso que fue usado para buscar
directamente los bosones de Higgs sobre un rango de masas grande
fue el proceso Bjorken, $Z\to Z^* H, Z^* \to f\bar f$ \cite{34}.
Al explorar este canal de producci\'on se excluyeron valores para
la masa del bos\'on de Higgs menores a 65.4 GeV por los
experimentos en el LEP1. La b\'usqueda continu\'o al revertir los
papeles de los bosones $Z$ real y virtual en el continuo $e^+e^-$
en el LEP2.

EL principal mecanismo de producci\'on de los bosones de Higgs en
colisiones $e^+e^-$ son
\begin{eqnarray}
\mbox{Higgs-strahlung} & : & e^+e^- \to Z^* \to ZH \\ \nonumber \\
\mbox{$WW$ fusion}     & : & e^+e^- \to \bar \nu_e \nu_e (WW) \to \bar \nu_e
\nu_e H
\label{esp-eq:wwfusion}
\end{eqnarray}
En Higgs-strahlung \cite{30,34,35} el bos\'on de Higgs es emitido
de la l\'{\i}nea de bos\'on $Z$, mientras que la fusi\'on $WW$ es un
proceso de formaci\'on de los bosones de Higgs en la colisi\'on
de dos bosones $W$ cuasi-reales radiados de los rayos del
electr\'on y del positr\'on \cite{36}.

Como es evidente del an\'alisis subsecuente, el LEP2 pudo cubrir
el rango de masa del Higgs del SM hata alrededor de 114 GeV
\cite{9}. Los colisionadores lineales $e^+e^-$ de alta
energ\'{\i}a pueden cubrir el rango completo de la masa del Higgs,
el rango de masa intermedio lo cubre ya un colisionador de 500
GeV \cite{13}, el rango superior de masa se cubrir\'a en la
segunda fase de las m\'aquinas en las cuales se alcanzar\'a  una
energ\'{\i}a total de al menos 3~TeV \cite{38A}.

\paragraph{(a) Higgs-strahlung} ~\\[0.5cm]
La secci\'on eficaz para el Higgs-strahlung puede ser escrita en
una forma compacta como
\begin{equation}
\sigma (e^+e^- \to ZH) = \frac{G_F^2 M_Z^4}{96\pi s} \left[ v_e^2 + a_e^2
\right] \lambda^{1/2} \frac{\lambda + 12 M_Z^2/s}{\left[ 1- M_Z^2/s \right]^2}
~,
\end{equation}
donde $v_e = -1 + 4 \sin^2 \theta_W$ y $a_e=-1$ son las cargas $Z$ vector
y vector-axial del electr\'on, y $\lambda =
[1-(M_H+M_Z)^2/s] [1-(M_H-M_Z)^2/s]$ es la funci\'on usual de
espacio fase de dos part\'{\i}culas. La secci\'on eficaz es del
tama\~no de $\sigma \sim \alpha_W^2/s$, i.e. de segundo orden en
el acoplamiento d\'ebil, y se escala en el cuadrado de la
energ\'{\i}a. Contribuciones a m\'as alto orden a las secciones
eficaces est\'an bajo control te\'orico \cite{38B,38C}.

\begin{figure}[hbt]

\vspace*{-5.0cm}
\hspace*{-2.0cm}
\epsfxsize=20cm \epsfbox{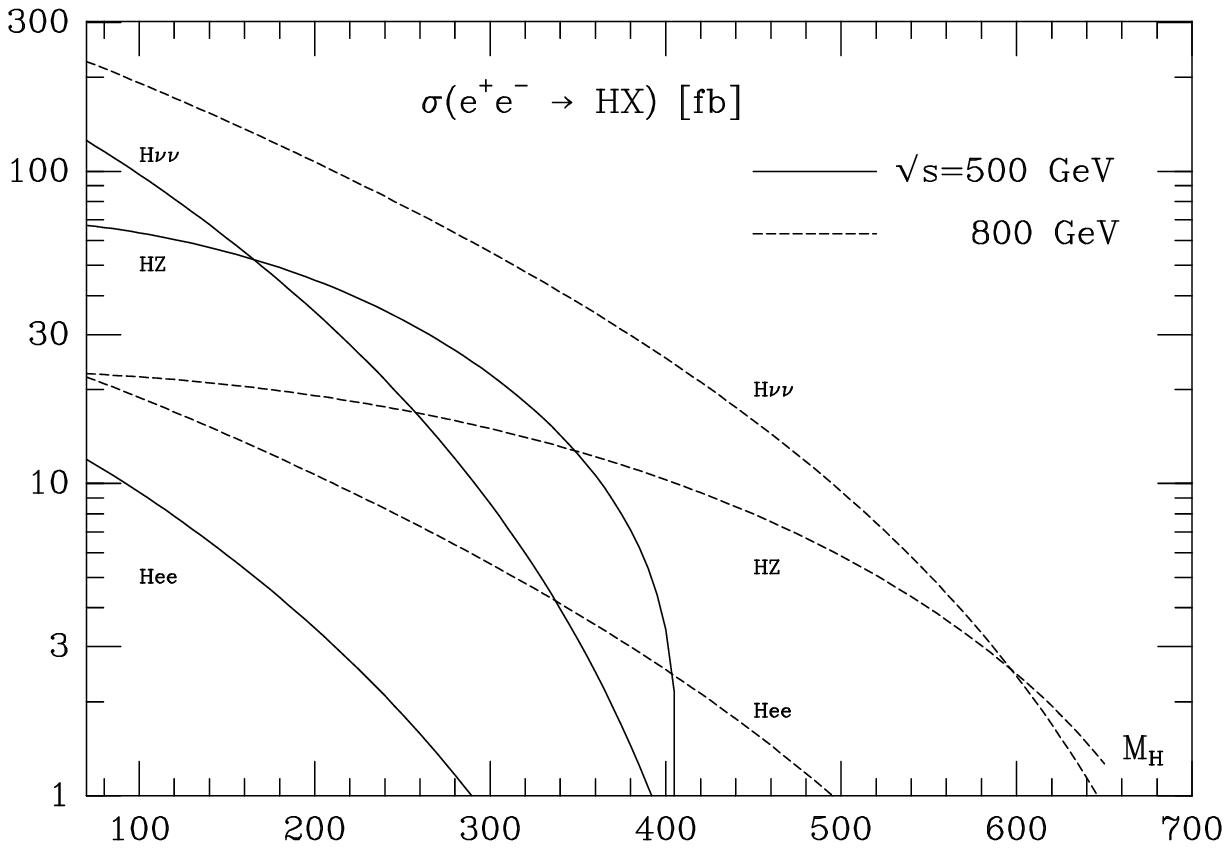}

\caption[]{\label{esp-fg:eehx} \it La secci\'on eficaz para la
producci\'on de bosones de Higgs del SM en el Higgs-strahlung
$e^+e^-\to ZH$, $WW/ZZ$ y fusi\'on $e^+e^- \to \bar \nu_e
\nu_e/e^+e^- H$; las curvas s\'olidas: $\sqrt{s}=500$ GeV, curvas
punteadas: $\sqrt{s}=800$ GeV.}
\end{figure}

Como la secci\'on eficaz se hace cero para energ\'{\i}as
asint\'oticas, el proceso de Higgs-strahlung es m\'as \'util para
la b\'usqueda de bosones de Higgs en el rango donde la
energ\'{\i}a del colisionador es del mismo orden que la masa del
Higgs, $\sqrt{s} \gsim {\cal O} (M_H)$. El tama\~no de la
secci\'on eficaz se ilustra en la Fig.~\ref{fg:eehx} para
energ\'{\i}as $\sqrt{s}=500$ GeV del colisionador lineal $e^+e^-$
como funci\'on de la masa del Higgs. Como la masa $Z$ de retroceso
en la reacci\'on de dos cuerpos $e^+e^- \to ZH$ es
mono-energ\'etica, la masa del bos\'on de Higgs puede ser
reconstruida a partir de la energ\'{\i}a del bos\'on $Z$, $M_H^2 =
s -2\sqrt{s}E_Z + M_Z^2$, sin la necesidad de analizar el producto
de los decaimientos del bos\'on del Higgs. Para los decaimientos
lept\'onicos del $Z$, t\'ecnicas de p\'erdida de masa proporcionan
una se\~nal muy clara, como se demuestra en la
Fig.~\ref{fg:zrecoil}.
\begin{figure}[hbt]
\begin{center}
\hspace*{-0.3cm}
\epsfig{figure=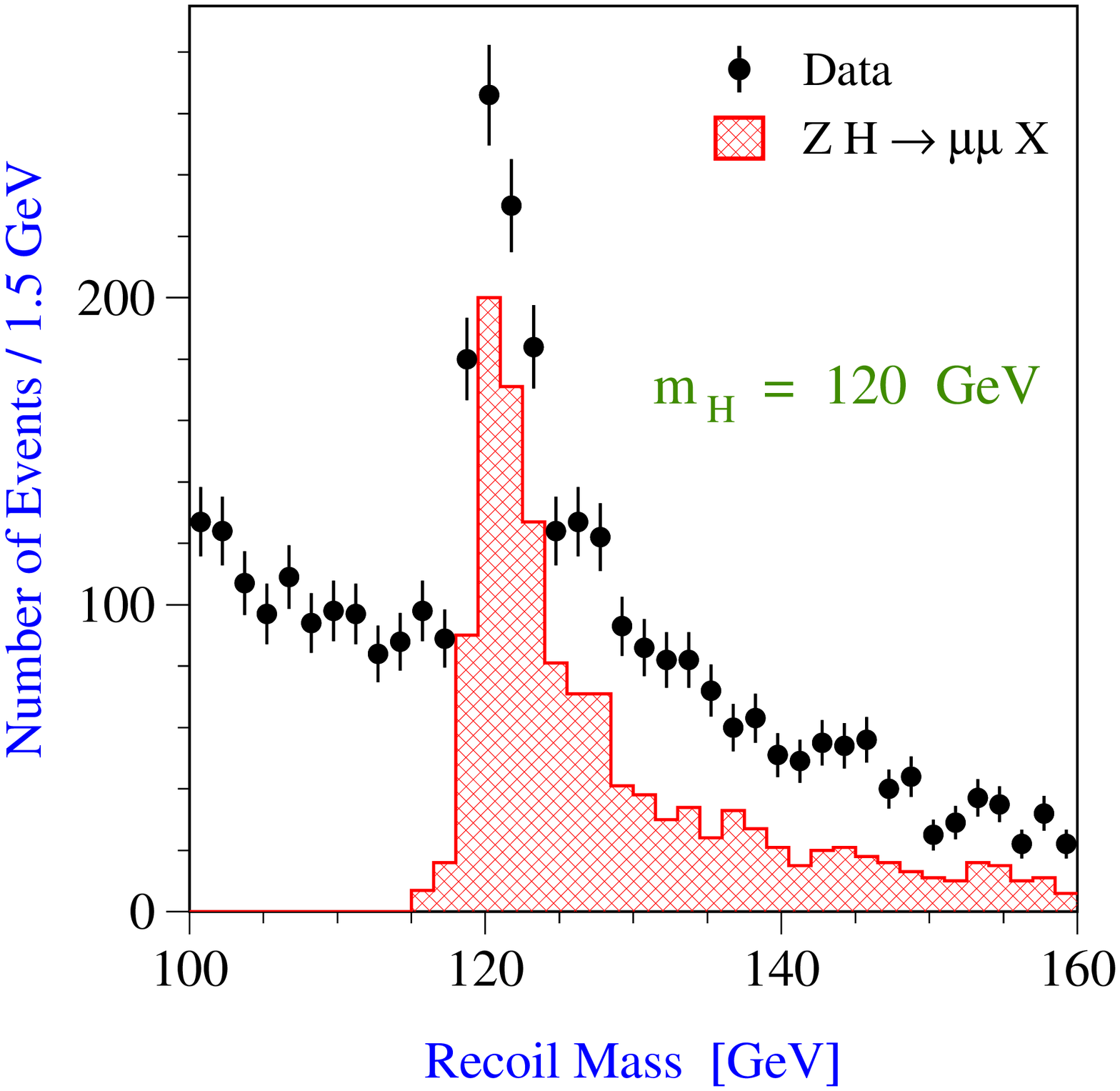,bbllx=0,bblly=0,bburx=567,bbury=561,width=10cm,clip=}
\end{center}
\vspace*{-0.4cm}

\caption[]{\label{esp-fg:zrecoil} \it La distribuci\'on de masa de
retroceso $\mu^+\mu^-$ en el proceso $e^+e^- \to H^0 Z\to X
\mu^+\mu^-$ para $M_H=120$~GeV y $\int {\cal L} = 500 fb^{-1}$ a
$\sqrt{s}=$ 350~GeV. Los puntos con barra de error son
simulaciones de Monte Carlo de la se\~nal de Higgs y se\~nales de
fondo. EL histograma sombreado representa s\'olo la se\~nal. Ref.
\cite{13}.}
\end{figure}

\paragraph{(b) Fusi\'on $WW$} ~\\[0.5cm]
Tambi\'en la secci\'on eficaz del proceso de fusi\'on
(\ref{eq:wwfusion}) puede ser puesta impl\'{\i}citamente en una 
forma compacta:
\begin{eqnarray}
\sigma (e^+e^-\to\bar \nu_e \nu_e H) & = & \frac{G_F^3 M_W^4}{4\sqrt{2}\pi^3}
\int_{\kappa_H}^1\int_x^1\frac{dx~dy}{[1+(y-x)/\kappa_W ]^2}f(x,y)
\\ \nonumber \\
f(x,y) & = & \left( \frac{2x}{y^3} - \frac{1+3x}{y^2} + \frac{2+x}{y} -1 \right)
\left[ \frac{z}{1+z} - \log (1+z) \right] + \frac{x}{y^3} \frac{z^2(1-y)}{1+z}
~,
\nonumber
\end{eqnarray}
con $\kappa_H=M_H^2/s$, $\kappa_W=M_W^2/s$ and
$z=y(x-\kappa_H)/(\kappa_Wx)$.

Dado que el proceso de fusi\'on es un proceso de intercambio del
canal-$t$, el tama\~no est\'a establecido por la longitud de onda
de Compton de $W$, suprimida, sin embargo, con respecto al
Higgs-strahlung por la tercera potencia del acoplamiento
electrod\'ebil, $\sigma \sim \alpha_W^3/M_W^2$. Como resultado, la
fusi\'on $W$ se convierte en el proceso de producci\'on principal
para las part\'{\i}culas de Higgs a energ\'{\i}as altas. A
energ\'{\i}as asint\'oticas la secci\'on eficaz se simplifica a
\begin{equation}
\sigma (e^+e^- \to \bar \nu_e \nu_e H) \to \frac{G_F^3 M_W^4}{4\sqrt{2}\pi^3}
\left[ \log\frac{s}{M_H^2} - 2 \right] ~.
\end{equation}
En este l\'\i mite, la fusi\'on de $W$ a bosones de Higgs
puede ser interpretado como un proceso de dos pasos: los bosones
$W$ son radiados como part\'{\i}culas cuasi-reales  a partir de
los electrones y positrones, $e \to \nu W$,
con la formaci\'on subsecuente de los bosones de Higgs en los
rayos $W$ que colisionan. Las correcciones electrod\'ebiles de
orden m\'as alto est\'an bajo control \cite{38C}.

En la Fig.~\ref{fg:eehx} se compara el tama\~no de la secci\'on
transversal de fusi\'on con el proceso de Higgs-strahlung. A
$\sqrt{s}=500$ GeV  las dos secciones eficaces son del mismo
orden, no obstante, el proceso de fusi\'on se vuelve cada vez
m\'as impotante con el incremento en la energ\'{\i}a.

\subsection{La Producci\'on de Higgs en Colisionadores Hadr\'onicos}

\phantom{h} Varios procesos pueden ser explotados para producir
part\'\i culas de Higgs en colisionadores hadr\'onicos \cite{24A,32}: \\[0.5cm]
\begin{tabular}{llll}
\hspace*{21mm} & fusi\'on de glu\'on & :              & $gg\to H$ \\ \\
& $WW,ZZ$ fusi\'on           & :    & $W^+ W^-, ZZ \to H$ \\ \\
& Higgs-strahlung emitido por $W,Z$ & :   & $q\bar q \to W,Z \to W,Z + H$ \\ \\
& Higgs bremsstrahlung emitido por top & : & $q\bar q, gg \to t\bar t + H$
\end{tabular} \\[0.5cm]

Mientras que la fusi\'on del glu\'on juega un papel dominante en
todo el rango de masa del Higgs  del Modelo Est\'andar, el
proceso de fusi\'on de $WW/ZZ$ se vuelve cada vez m\'as importante
al aumentar la masa del Higgs; sin embargo, tambi\'en juega un
papel importante en la b\'usqueda del bos\'on de Higgs el estudio
de sus propiedades en el rango intermedio de masas. Los \'ultimos
dos procesos de radiaci\'on son relevantes s\'olo para masas
ligeras del Higgs.

Las secciones eficaces de producci\'on en colisionadores
hadr\'onicos, en particular en el LHC, son bastante considerables  de
modo que en esta m\'aquina se puede producir una muestra grande de
part\'{\i}culas de Higgs del SM. Las dificultades experimentales
surgen del enorme n\'umero de eventos de fondo (background) que
aparecen junto con los eventos de se\~nales de Higgs. Este
problema ser\'a abordado ya sea al desencadenarse los decaimientos
lept\'onicos de $W,Z$ y $t$ en los procesos de radiaci\'on o
explotando el car\'acter de resonancia de los decaimientos del
Higgs $H\to \gamma\gamma$ y $H\to ZZ \to 4\ell^\pm$. De esta
manera, se espera que el Tevatron realice la b\'usqueda  de las
part\'{\i}culas de Higgs en el rango de masa que est\'a por arriba
del alcanzado en el LEP2 y hasta alrededor de 110 a 130 GeV
\cite{11}. Se espera que el LHC cubra el rango can\'onico completo
de la masa del Higgs $M_H \lessim 700$ GeV del Modelo Est\'andar
\cite{12}.

\paragraph{(a) Fusi\'on del glu\'on} ~\\[0.5cm]
El mecanismo de fusi\'on-glu\'on \cite{29,32,39A,39B}
\begin{displaymath}
pp \to gg \to H
\end{displaymath}
proporciona el mecanismo de producci\'on dominante de los bosones
de Higgs en el LHC en el rango completo de masas relevantes del
Higgs de hasta alredeor de 1 TeV. El acoplamiento del glu\'on al
bos\'on de Higgs en el SM est\'a mediado por lazos tiangulares de
los quarks top y bottom, cf. Fig.~\ref{fg:gghlodia}. Como el
acoplamiento de Yukawa de las part\'{\i}culas de Higgs a quarks
pesados crece con la masa del quark, equilibrando as\'{\i} la
disminuci\'on de la amplitud triangular, el factor de forma se
aproxima a un valor distinto de cero para masas grandes del lazo
del quark. [Si las masas de los quarks pesados m\'as all\'a de la
tercera generaci\'on fueran generadas s\'olamente por el mecanismo
de Higgs, estas part\'{\i}culas a\~nadir\'{\i}an al factor de
forma el mismo monto que el quark top en el l\'{\i}mite
asint\'otico de quark pesado.]
\begin{figure}[hbt]
\begin{center}
\setlength{\unitlength}{1pt}
\begin{picture}(180,90)(0,0)

\Gluon(0,20)(50,20){-3}{5}
\Gluon(0,80)(50,80){3}{5}
\ArrowLine(50,20)(50,80)
\ArrowLine(50,80)(100,50)
\ArrowLine(100,50)(50,20)
\DashLine(100,50)(150,50){5}
\put(155,46){$H$}
\put(25,46){$t,b$}
\put(-15,18){$g$}
\put(-15,78){$g$}

\end{picture}  \\
\setlength{\unitlength}{1pt} \caption[ ]{\label{esp-fg:gghlodia} \it
Diagrama que contribuye a la formaci\'on de los bosones de Higgs
en las colisiones glu\'on-glu\'on al m\'as bajo orden.}
\end{center}
\end{figure}
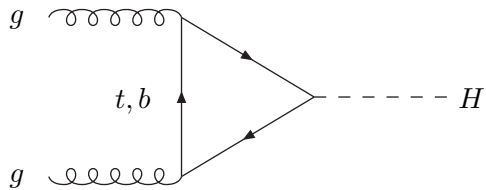\\

La secci\'on eficaz part\'onica, Fig.~\ref{fg:gghlodia}, puede ser
expresada por el ancho glu\'onico del bos\'on 
de Higgs a m\'as bajo orden \cite{32}:
\begin{eqnarray}
\hat \sigma_{LO} (gg\to H) & = & \sigma_0 M_H^2 \times BW(\hat{s}) \\
\sigma_0 & = & \frac{\pi^2}{8M_H^2} \Gamma_{LO} (H\to gg) =
\frac{G_F\alpha_s^2}{288\sqrt{2}\pi} \left| \sum_Q A_Q^H (\tau_Q) \right|^2 ~,
\nonumber
\end{eqnarray}
donde la variable de escalamiento est\'a definida como $\tau_Q =
4M_Q^2/M_H^2$ y $\hat s$ denota la energ\'{\i}a del c.m.
part\'onica al cuadrado. El factor de forma puede ser evaluado
f\'acilmente:
\begin{eqnarray}
A_Q^H (\tau_Q) & = & \frac{3}{2} \tau_Q \left[ 1+(1-\tau_Q) f(\tau_Q)
\right] \label{esp-eq:ftau} \\
f(\tau_Q) & = & \left\{ \begin{array}{ll}
\displaystyle \arcsin^2 \frac{1}{\sqrt{\tau_Q}} & \tau_Q \geq 1 \\
\displaystyle - \frac{1}{4} \left[ \log \frac{1+\sqrt{1-\tau_Q}}
{1-\sqrt{1-\tau_Q}} - i\pi \right]^2 & \tau_Q < 1
\end{array} \right. \nonumber
\end{eqnarray}

Para masas peque\~nas de los lazos el factor de forma se hace
cero, $A_Q^H(\tau_Q) \sim -3/8 \tau_Q [\log (\tau_Q/4)+i\pi]^2$,
mientras que para masas grandes de los lazos se aproxima a un
valor distinto de cero, $A_Q^H (\tau_Q) \to 1$. El t\'ermino
final $BW$ es la funci\'on de Breit-Wigner normalizada 
\beq
BW(\hat{s}) = \frac{M_H \Gamma_H/\pi}{[\hat{s}-M_H^2]^2 + M_H^2
\Gamma_H^2} 
\eeq 
acerc\'andose, en la aproximaci\'on de
anchura-estrecha (narrow-width) a una funci\'on $\delta$ en $\hat{s}=M_H^2$.

En la aproximaci\'on de anchura-estrecha, la secci\'on eficaz
hadr\'onica puede ponerse en la forma
\begin{equation}
\sigma_{LO} (pp\to H) = \sigma_0 \tau_H \frac{d{\cal L}^{gg}}{d\tau_H} ~,
\end{equation}
donde $d{\cal L}^{gg}/d\tau_H$ denota la luminosidad $gg$ del
colisionador $pp$, evaluado para la variable de Drell--Yan
$\tau_H = M_H^2/s$, donde $s$ es la energ\'{\i}a hadr\'onica total al cuadrado. \\

Las correcciones de QCD al proceso de fusi\'on del glu\'on
\cite{29,32,39B} son muy importantes. Estas estabilizan las
predicciones te\'oricas para la secci\'on eficaz cuando se var\'{\i}an las
escalas de renormalizaci\'on y de factorizaci\'on. M\'as a\'un, estas son
grandes y positivas, incrementando as\'{\i} la secci\'on eficaz de
producci\'on para los bosones de Higgs. Las correcciones de QCD consisten
en correcciones virtuales al proceso b\'asicos $gg\to H$, y a
correcciones reales debidas a la producci\'on asociada del bos\'on de Higgs
con partones sin masa, $gg\to Hg$ y $gq\to Hq,\, q\bar q\to Hg$. Estos
subprocesos contribuyen a la producci\'on del Higgs a orden de
${\cal O}(\alpha_s^3)$. Las correcciones virtuales reescalan la secci\'on
eficaz de fusi\'on al orden m\'as bajo con un coeficiente que depende s\'olo
en las razones de las masas del Higgs y del quark.  La radiaci\'on del
glu\'on conduce a estados finales de dos partones con energ\'{\i}a
invariante$\hat s \geq M_H^2$ en los canales
$gg, gq$ y $q\bar q$.\\

\begin{figure}[hbt]

\vspace*{0.4cm}
\hspace*{2.0cm}
\begin{turn}{-90}%
\epsfxsize=7cm \epsfbox{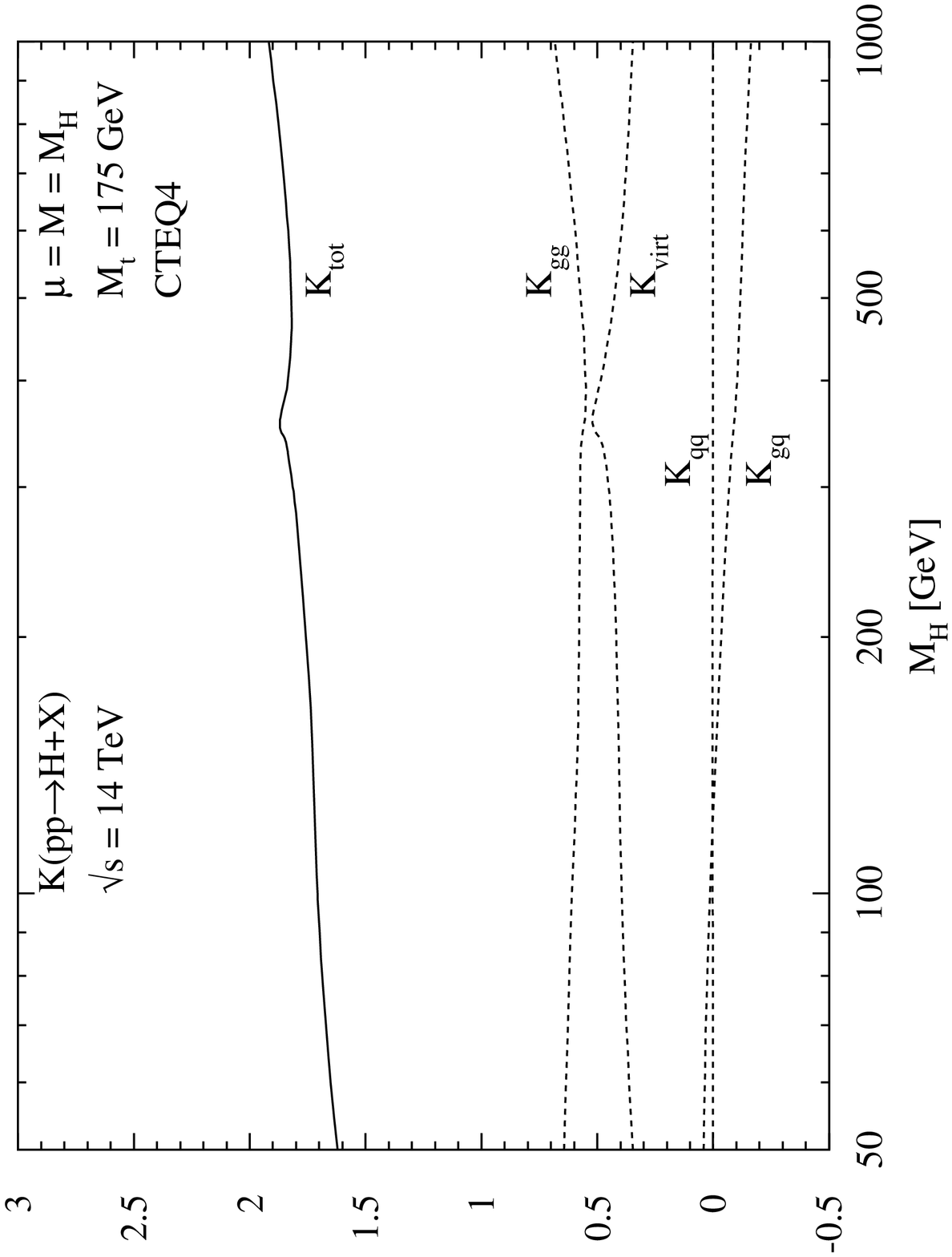}
\end{turn}
\vspace*{-0.2cm}

\caption[]{\label{esp-fg:gghk} \it Los factores K de la secci\'on
eficaz de la fusi\'on del glu\'on corregida por QCD $\sigma(pp \to
H+X)$ en el LHC con una energ\'{\i}a del c.m de $\sqrt{s}=14$ TeV.
Ls l\'{\i}neas punteadas muestran las contribuciones individuales
de las correcciones de QCD. Las escalas de renormalizaci\'on y
factorizaci\'on han sido identificadas con la masa del Higgs, y se
han adoptado densidades CTEQ4 del part\'on .}
\end{figure}

El tama\~no de las correcciones radiativas puede ser parametrizado
definiendo el factor $K$ como $K=\sigma_{NLO}/\sigma_{LO}$, en el cual todas
las cantidades son evaluadas en el numerador y denominador, en el
siguiente del primero (next-to-leading, NLO) y en el primer orden
respectivamente.  Los resultados de estos c\'alculos se muestran en la
Fig.~\ref{fg:gghk}. Las correciones virtuales $K_{virt}$ y las
correcciones reales $K_{gg}$ para las colisiones $gg$ son,
aparentemente del mismo tama\~no, y ambas son grandes y positivas; las
correcciones para las colisiones $q\bar q$ y las contribuciones
inel\'asticas de Compton $gq$ son menos importantes. Despu\'es de incluir
estas correcciones de QCD de orden m\'as alto, la dependencia de la
secci\'on eficaz en las escalas de renormalizaci\'on y factorizaci\'on son
sigificantemente reducidas de un nivel de ${\cal O}(100\%)$ hasta 
un nivel de alrededor de 20\%. Dependiendo s\'olo suavemente de la masa
del bos\'on de Higgs, el factor total $K$, $K_{tot}$, resulta estar
cerca de 2 \cite{29,32,39B,R}. Las contribuciones principales son
generadas por las correcciones virtuales y estados finales de
3-partones iniciados por estados iniciales $gg$. Se esperan grandes
correcciones NLO  para los procesos iniciados por gluones como
resultado de la carga de color grande.  Sin embargo, al estudiar los
ordenes siguientes en la correcci\'on en el l\'{\i}mite de masa grande
del top, las correcciones N$^{2}$LO generan \'unicamente un modesto
incremento adicinal del factor $K$, $\delta_2 K_{tot} \lessim 0.2$
\cite{x}.  Esto prueba que la expansi\'on es convergente, con las
correcciones m\'as importantes atribuidas a la contribuci\'on del orden
siguiente al primero NLO \cite{R}.

\begin{figure}[hbt]

\vspace*{0.5cm}
\hspace*{2.0cm}
\begin{turn}{-90}%
\epsfxsize=7cm \epsfbox{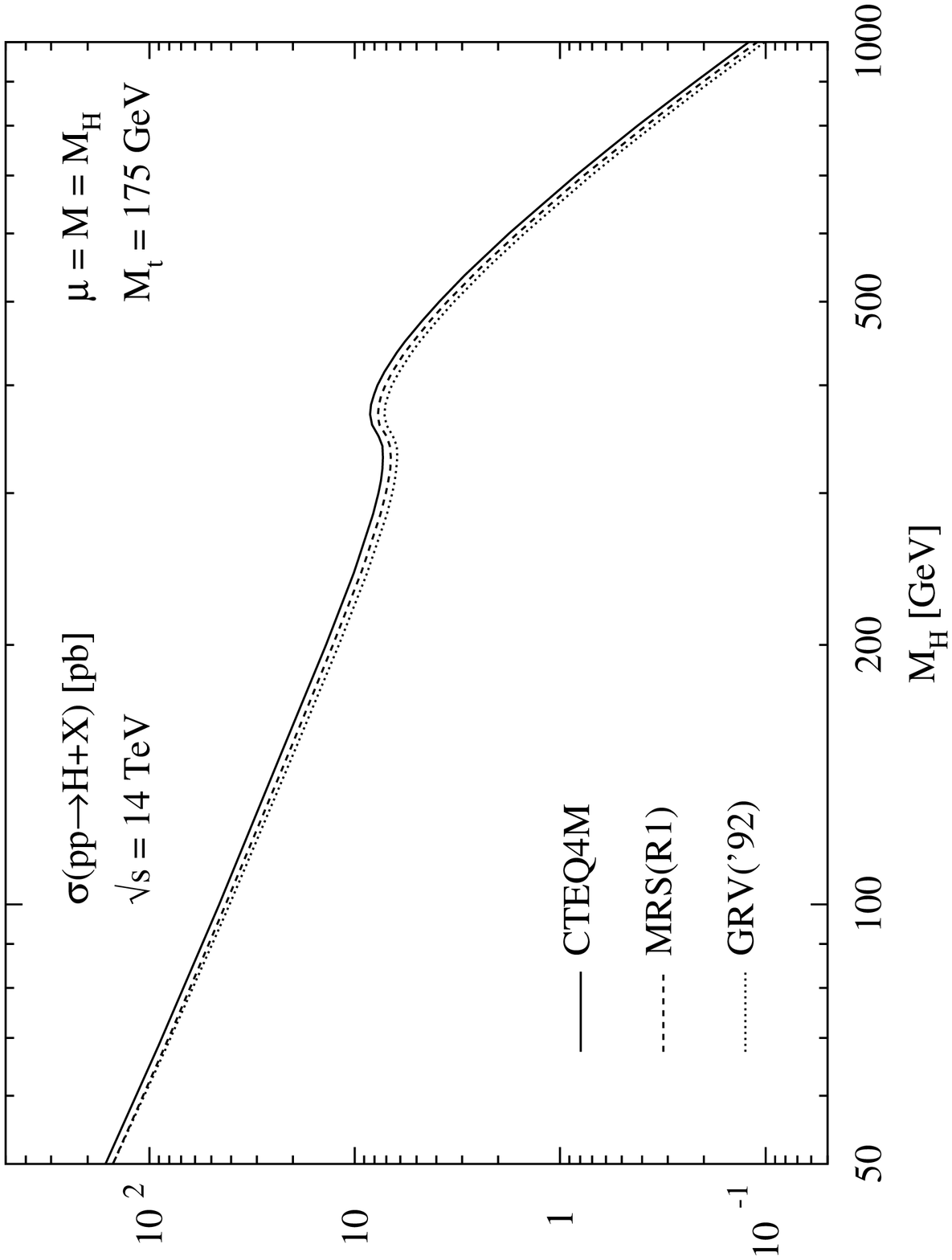}
\end{turn}
\vspace*{0.0cm}

\caption[]{\label{esp-fg:gghparton} \it La secci\'on eficaz para la
producci\'on de bosones de Higgs; se muestran tres diferentes
conjuntos de densidades del part\'on [CTEQ4M, MRS(R1) y
GRV('92)].}
\end{figure}

La predicci\'on te\'orica para la secci\'on eficaz de producci\'on de las
part\'{\i}culas de Higgs es presentada en la Fig.~\ref{fg:gghparton}
para el LHC como funci\'on de la masa del Higgs. La secci\'on eficaz
disminuye conforme la masa del Higgs aumenta. Esto es, en gran medida,
una consecuancia de la ca\'{\i}da abrubpa de la luminosidad $gg$ para
masas invariantes grandes.  La protuberancia en la secci\'on eficaz es
inducida por el umbral de $t\bar t$
en el tri\'angulo del top. La precisi\'on te\'orica total de este c\'alculo se
espera al nivel de 20 \'o  30\%.

\paragraph{(b) Fusi\'on del bos\'on vectorial} ~\\[0.5cm]
El segundo canal importante para la producci\'on de Higgs en el LHC es
la fusi\'on del bos\'on vectorial, $W^+W^- \to H$ \cite{23,22A}. Para masas
grandes del Higgs, este mecanismo se vuelve competitivo
para la fusi\'on del glu\'on; para masas intermedias la secci\'on eficaz es
menor por alrededor de un orden de magnitud.\\ 

Para masas grandes, los dos bosones electrod\'ebiles $W,Z$ que
forman el bos\'on de Higgs son predominantemente longitudinalmente
polarizados. A energ\'{\i}as altas, el espectro de part\'{\i}culas
equivalente de los bosones longitudinales $W,Z$ en haces de quark
est\'an dados por
\begin{eqnarray}
f^W_L (x) & = & \frac{G_F M_W^2}{2\sqrt{2}\pi^2} \frac{1-x}{x}
 \label{esp-eq:xyz} \\ \non \\
f^Z_L (x) & = & \frac{G_F M_Z^2}{2\sqrt{2}\pi^2}
\left[(I_3^q - 2e_q \sin^2\theta_W)^2 + (I_3^q)^2\right] \frac{1-x}{x} ~, \non
\end{eqnarray}
donde $x$ es la fracci\'on de la energ\'{\i}a transferida del
quark al bos\'on $W,Z$ en el proceso de escisi\'on (splitting) $q\to q +W/Z$.
De este espectro de part\'{\i}culas, las luminosidades de $WW$ y
$ZZ$ pueden f\'acilmente ser derivadas:
\begin{eqnarray}
\frac{d{\cal L}^{WW}}{d\tau_W} & = & \frac{G_F^2 M_W^4}{8\pi^4}
\left[ 2 - \frac{2}{\tau_W} -\frac{1+\tau_W}{\tau_W} \log \tau_W \right] \\
\non \\
\frac{d{\cal L}^{ZZ}}{d\tau_Z} & = & \frac{G_F^2 M_Z^4}{8\pi^4}
\left[(I_3^q - 2e_q \sin^2\theta_W)^2 + (I_3^q)^2\right]
\left[(I_3^{q'} - 2e_{q'} \sin^2\theta_W)^2 + (I_3^{q'})^2\right] \non \\
& & \hspace{1.5cm} \times \left[ 2 - \frac{2}{\tau_Z} -\frac{1+\tau_Z}{\tau_Z}
\log \tau_Z \right] \non
\end{eqnarray}
con la variable de Drell--Yan definida como $\tau_V = M_{VV}^2/s$.
La secci\'on eficaz para la producci\'on de Higgs en colisiones de
quark--quark est\'a dada por la convoluci\'on de las secciones
eficaces de partones $WW,ZZ \to H$ con luminosidades:
\begin{equation}
\hat \sigma(qq\to qqH) = \frac{d{\cal L}^{VV}}{d\tau_V} \sqrt{2} \pi G_F ~.
\label{esp-eq:vvhpart}
\end{equation}
La secci\'on eficaz hadr\'onica se obtiene finalmente al sumar la
seccion eficaz del part\'on (\ref{eq:vvhpart}) sobre el flujo de
todos las posibles pares de combinaciones de quark--quark y antiquark. \\

Ya que al orden m\'as bajo los remanentes del prot\'on son singletes de
color en los procesos de fusi\'on de $WW,ZZ$, no habr\'a intercambio de
color entre las dos l\'{\i}neas de quarks de las cuales los dos
bosones vectoriales son radiados. Como resultado, las correcciones
dominantes de QCD a estos procesos ya est\'an tomadas en cuenta por las
correcciones a las densidades part\'onicas del quark.\\

La secci\'on eficaz de la fusi\'on de $WW/ZZ$ para el bos\'on de Higgs en el
LHC se muestra en la Fig.~\ref{fg:lhcpro}. El proceso es aparentemente
muy importante para la b\'usqueda del bos\'on de Higgs en el rango
superior de la masa, donde la secci\'on eficaz se aproxima a valores
cercanos a la fusi\'on del glu\'on. Para masas intermedias, se acerca m\'as,
dentro de un orden de magnitud, a la secci\'on eficaz dominante de
fusi\'on del glu\'on.

\paragraph{(c) Emisi\'on radiativa de Higgs  por bosones vectoriales
  (Higgs-strahlung off top quarks)}  ~\\[0.5cm]
El proceso de radiaci\'on de Higgs (Higgs-strahlung) $q\bar q \to V^* \to
VH~(V=W,Z)$ es un mecanismo muy importante (Fig.~\ref{fg:lhcpro}) para
la b\'usqueda de bosones de Higgs ligeros en los colisionadores
hadr\'onicos Tevatron y LHC. Aunque la secci\'on eficaz es menor que la de
fusi\'on del glu\'on, los decaimientos lept\'onicos de los bosones
vectoriales electrod\'ebiles son extremadamente \'utiles para filtrar
eventos de se\~nales de Higgs de el enorme fondo. Como el
mecanismo din\'amico es el mismo que para los colisionadores $e^+e^-$,
excepto para el doblamiento con las densidades quark--antiquark, los
pasos intermedios del c\'alculo no necesitan ser mencionados m\'as aqu\'\i, y
simplemente se registran en la Fig.~\ref{fg:lhcpro} los valores
finales de las secciones eficaces para el Tevatron y el LHC.

\paragraph{(d) Higgs bremsstrahlung emitido por quarks top
  (Higgs bremsstrahlung off top quarks) } ~\\[0.5cm] 
Tambi\'en el proceso $gg,q\bar q \to t\bar t H$ es relevante
s\'olo para masas peque\~nas del Higgs, Fig.~\ref{fg:lhcpro}. La
expresi\'on anal\'{\i}tica para la secci\'on eficaz del part\'on,
incluso al orden m\'as bajo, es bastante complicada, de manera que
s\'olo se muestran los resultados finales para las secciones
eficaces del LHC en la Fig.~\ref{fg:lhcpro}. Correcciones a m\'as
alto orden han sido presentadas en la Ref.~\cite{z}.

Higgs-bremsstrahlung emitido por quarks top es tambi\'en un proceso
interesante para las mediciones del acoplamiento fundamental de Yukawa
$Htt$. La secci\'on eficaz $\sigma (pp\to t\bar t H)$ es directamente
proporcional al cuadrado de este acoplamiento fundamental.

\paragraph{\underline{Resumen.}} Una visi\'on general de las secciones
eficaces de producci\'on para las part\'{\i}culas de Higgs en el LHC
es presentada en la Fig.~\ref{fg:lhcpro}. Tres tipos de canales pueden
distinguirse. La fusi\'on de glu\'on de las part\'{\i}culas de Higgs es un
proceso universal, dominante sobre el rango entero de masas de Higgs
del SM. La emisi\'on radiativa de Higgs por bosones electrod\'ebiles $W,Z$
o quarks top es prominente para bosones de Higgs ligeros. El canal de
fusi\'on de $WW/ZZ$, en contraste, se vuelve cada vez m\'as importante en
la parte superior del rango de masas del Higgs del SM, aunque tambi\'en
prueba ser \'util en el rango intermedio de masa.

Las se\~nales para la b\'usqueda de part\'{\i}culas de Higgs son dictadas
por las razones de desintegraci\'on de decaimiento (decay branching
ratios). El la parte inferior del rango de masas intermedio, la
reconstrucci\'on de resonancias en estados finales $\gamma\gamma$ y $b\bar b$ de
jets puede ser explotada. En la parte superior del rango intermedio de
masas, los decaimientos a $ZZ^*$ y $WW^*$ son importantes, con los dos
bosones electrod\'ebiles decayendo lept\'onicamente. En el rango de masas
por arriba del umbral del decaimiento dentro de la capa de masa del
$ZZ$, los decaimientos a leptones cargados $H\to ZZ \to 4\ell^\pm$
proporcionan se\~nales muy valiosas.  S\'olo en el l\'\i mite superior del
rango cl\'asico de masas de Higgs del SM, decaimientos a neutrinos y
jets, generados en los decaimientos de $W$ y $Z$, completan las
t\'ecnicas de b\'usqueda.
\begin{figure}[hbt]

\vspace*{0.5cm}
\hspace*{0.0cm}
\begin{turn}{-90}%
\epsfxsize=10cm \epsfbox{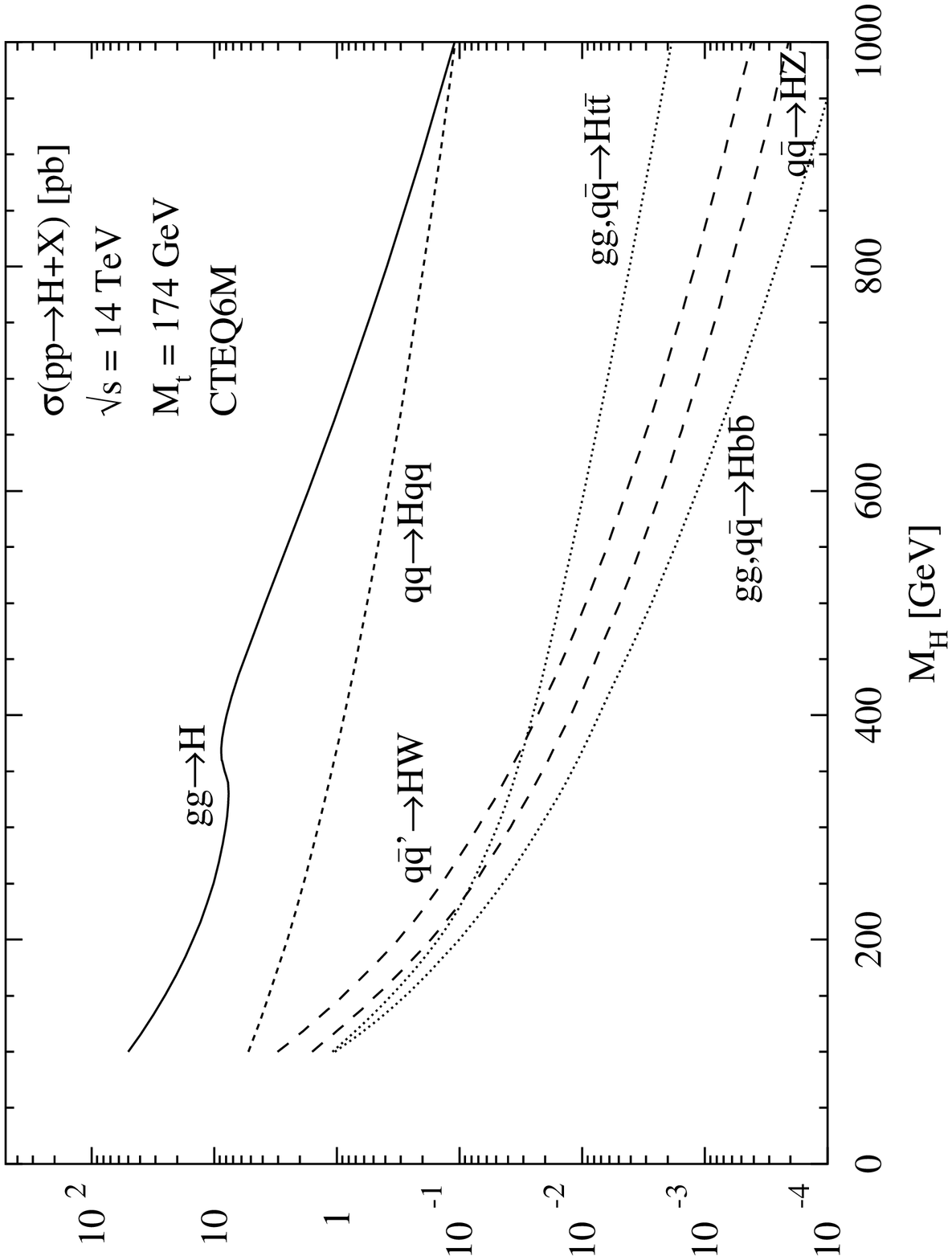}
\end{turn}
\vspace*{0.0cm}

\caption[]{\label{esp-fg:lhcpro} \it Las secciones eficaces de la
producci\'on de Higgs en el LHC para los varios mecanismos de
producci\'on como fusnciones de la masa del Higgs. Se muestran los
resultados con correcciones completas de QCD para la fusi\'on del
glu\'on $gg \to H$, para la fusi\'on del bos\'on vectorial $qq\to
VVqq \to Hqq$, el bremsstrahlung del bos\'on vectorial $q\bar q
\to V^* \to HV$ y la producci\'on asociada $gg,q\bar q \to Ht\bar
t, Hb\bar b$.}
\end{figure}


\subsection{ Perfil del Higgs en el ME}

\phantom{h} Para establecer experimentalmente el mecanismo de Higgs,
debe explorarse la naturaleza de esta part\'{\i}cula midiendo todas
sus caracter\'{\i}sticas, su masa y tiempo de vida, los n\'umeros
cu\'anticos externos esp\'{\i}n-paridad, los acoplamientos a los bosones
de norma y fermiones, y por \'ultimo pero no menos importante, los
auto-acoplamientos del Higgs.  Mientras que parte de este programa puede
ser realizado en LHC \cite{12,xx}, el perfil completo de la
part\'{\i}cula puede ser reconstruido a trav\'es de todo el rango de
masa en colisionadores $e^+ e^-$ \cite{13}.

\paragraph{(a) Masa} ~\\[0.5cm]
La masa de la part\'{\i}cula de Higgs puede ser medida juntando los
productos de decaimiento de la part\'{\i} en colisionadores de
hadrones y de $e^+e^-$. M\'as a\'un, en colisiones $e^+e^-$ el
Higgs-strahlung puede ser explotado para reconstruir muy precisamente
la masa a partir de la energ\'{\i}a de retroceso del $Z$ en los
procesos de dos cuerpos $e^+e^-\to ZH$. Se puede esperar una precisi\'on
total de $\delta M_H \sim 100$ MeV.

\paragraph{(b) Anchos/Tiempos de vida} ~\\[0.5cm]
El ancho del estado, i.e. el tiempo de vida de la part\'{\i}cula,
puede directamente ser medido por arriba del umbral de decaimiento de
$ZZ$ donde el ancho crece r\'apidamente. En la parte inferior del rango
intermedio de masa el ancho puede ser medido indirectamente \cite{13}
combinando la raz\'on de desintegraci\'on $H\to WW$ con la medida del ancho
parcial del $WW$, accesible a trav\'es de la secci\'on eficaz para la
fusi\'on del bos\'on $W$: $\Gamma_{tot} = \Gamma_{WW} / BR_{WW}$. Por lo tanto, el
ancho total para la part\'{\i}cula de Higgs puede ser determinado a
trav\'es de todo el rango de masa cuando los resultados del LHC y los
colisionadores $e^+e^-$ puedan ser combinados.

\paragraph{(c) Esp\'{\i}n-paridad} ~\\[0.5cm]
La distribici\'on angular de los bosones  $Z/H$ en el proceso de 
Higgs-strahlung es sensible al esp\'{\i}n y paridad de la
part\'{\i}cula de Higgs \cite{41}. Como la amplitud de
producci\'on est\'a dada por ${\cal A}(0^+) \sim
\vec{\epsilon}_{Z^*} \cdot \vec{\epsilon}_Z$, el bos\'on $Z$ es
producido en un estado de polarizaci\'on longitudinal a
energ\'{\i}as altas
 -- en concordancia con el teorema de equivalencia.
 Como resultado, la distribuci\'on angular
\begin{equation}
\frac{d\sigma}{d\cos\theta} \sim \sin^2 \theta + \frac{8M_Z^2}{\lambda s}
\end{equation}
se aproxima a la ley para esp\'{\i}n cero $\sin^2\theta$ asint\'oticamente.
Esto puede ser contrastado con la distriduci\'on $\sim 1 + \cos^2\theta$ para
estados de paridad negativa, la cual viene de amplitudes de
polarizaci\'on transversal ${\cal A}(0^-) \sim \vec{\epsilon}_{Z^*} \times \vec{\epsilon}_Z
\cdot \vec{k}_Z$.  Es tambi\'en caracter\'\i sticamente diferente de la
distribuci\'on del proceso de fondo $e^+e^- \to ZZ$, el cual, como
resultado del intercambio de $e$ en los canales $t/u$, tine un pico
estrecho en la direcci\'on adelante/atr\'as, Fig.~\ref{fg:spinpar}a.\\

De forma similar, el esp\'\i n cero de la part\'\i cula de Higgs puede ser
determinado de la distrubuci\'on isotr\'opica de los productos de
decaimiento. M\'as a\'un, la paridad puede ser medida observando las
correlaciones del esp\'{\i}n de los productos de decaimiento. De
acuerdo con el teorema de equivalencia, los \'angulos azimutales de los
planos de decaimiento en $H\to ZZ\to (\mu^+\mu^-) (\mu^+\mu^-)$ est\'an
asint\'oticamente descorrelacionados, $d\Gamma^+/d\phi_* \to 0$, para una
part\'{\i}cula $0^+$; en contraste con $d\Gamma^-/d\phi_* \to 1-\frac{1}{4}
\cos 2\phi_*$ para la distribuci\'on del \'angulo azimutal entre los planos
para el decaimiento de una part\'{\i}cula $0^-$~. La diferencia entre
las distribuciones angulares es una consecuencia de los diferentes
estados de polarizaci\'on de los bosones vectoriales en los dos casos.
Mientras estos se aproximan a estados con polarizaci\'on longitudinal 
para decaimientos escalares del Higgs, est\'an transversalmente
polarizados para  decaimientos de part\'{\i}culas
pseudo-escalares. \\

Un m\'etodo diferente para determinar el esp\'{i}n del bos\'on de Higgs es
porporcionado por la exploraci\'on del inicio de la curva de excitaci\'on
en el Higgs-strahlung \cite{MMM} $e^+e^- \to ZH$. Para el esp\'{\i}n
del Higgs $S_H = 0$ la curva de excitaci\'on se levanta abruptamente en
el umbral $\sim \sqrt{s - (M_H + M_Z)^2}$. Esta conducta es claramente
diferente de las excitaciones de esp\'{\i}n m\'as altas, las cuales se
elevan con una potencia $> 1$ en el factor de umbra.  Una
ambig\"{u}edad para estados con esp\'{\i}n/paridad $1^+$ y $2^+$ puede
ser resuelta evaluando tambi\'en la distrubuci\'on angular del Higgs y el
bos\'on $Z$ en el proceso Higgs-strahlung. La precisi\'on experimental
ser\'a suficiente para discriminar la asignaci\'on del esp\'{\i}n=0 al
bos\'on de Higgs de otras asignaciones, como se muestra en la
Fig.~\ref{fg:spinpar}b.

\begin{figure}[hbt]

\vspace*{0.5cm} \hspace*{0.0cm} \epsfxsize=8cm \epsfbox{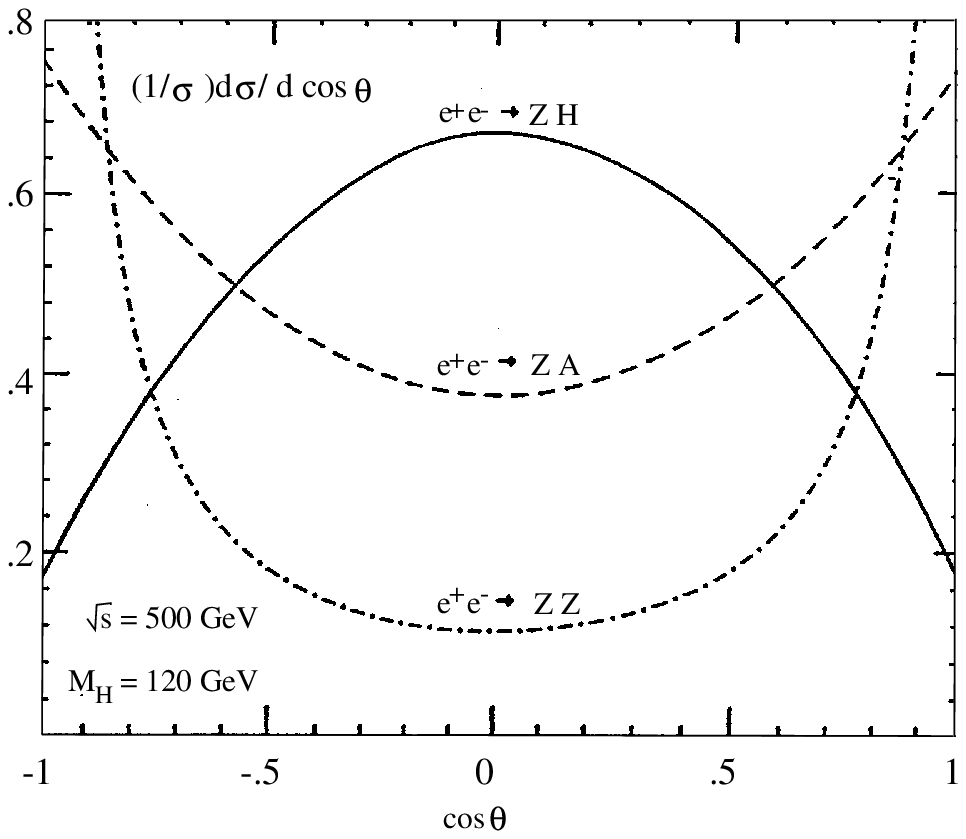}
\vspace*{0.0cm}

\vspace*{-6.3cm} \hspace*{9.0cm}
\epsfig{figure=spinexp.eps,bbllx=0,bblly=0,bburx=560,bbury=539,width=6.5cm,clip=}
\vspace*{0.0cm}

\caption[]{\it \label{esp-fg:spinpar} Izquierda:  Distribuci\'on
angular de los bosones $Z/H$ en el Higgs-strahlung, comparada con
la producci\'on de part\'{\i}culas pseudo-escalares y los estados
finales de fondo $ZZ$;
 Ref.~\cite{41}.
Derecha: Umbral de excitaci\'on del  Higgs-strahlung que discrimina 
el  spin=0 de otras asignaciones, Ref. \cite{MMM,exp}.}
\end{figure}

\paragraph{(d) Acoplamientos del  Higgs} ~\\[0.5cm]
Dado que las part\'{\i}culas fundamentales adquieren masa a trav\'es de
la interacci\'on con el campo de Higgs, la intensidad de los
acoplamientos del Higgs a fermiones y bosones de norma est\'a
establecida por las masas de las part\'{\i}culas. Ser\'a una tarea
experimental crucial medir estos acoplamientos, los cuales son
predichos un\'\i vocamente por la naturaleza misma del mecanismo de Higgs.\\

Los acoplamientos del Higgs a bosones de norma masivos pueden ser
determinados de la secciones eficaces de producci\'on en el Higgs-strahlung
y la fusi\'on $WW,ZZ$, con una precisi\'on esperada al nivel porcentual.
Para bosones de Higgs suficientemente pesados los anchos de decaimiento
pueden ser explotados para determinar los acoplamientos a bosones de
norma electrod\'ebiles. Para acoplamientos del Higgs a fermiones, las
razones de desintegraci\'on $H\to b\bar b, c\bar c, \tau^+\tau^-$ pueden ser
usadas en la parte inferior del rango intermedio de masa,
cf.~Fig.~\ref{fg:brmeas}; estas observables permiten la medici\'on
directa de los acoplamientos de Yukawa del Higgs.
\begin{figure}[hbt]
\begin{center}
\epsfig{figure=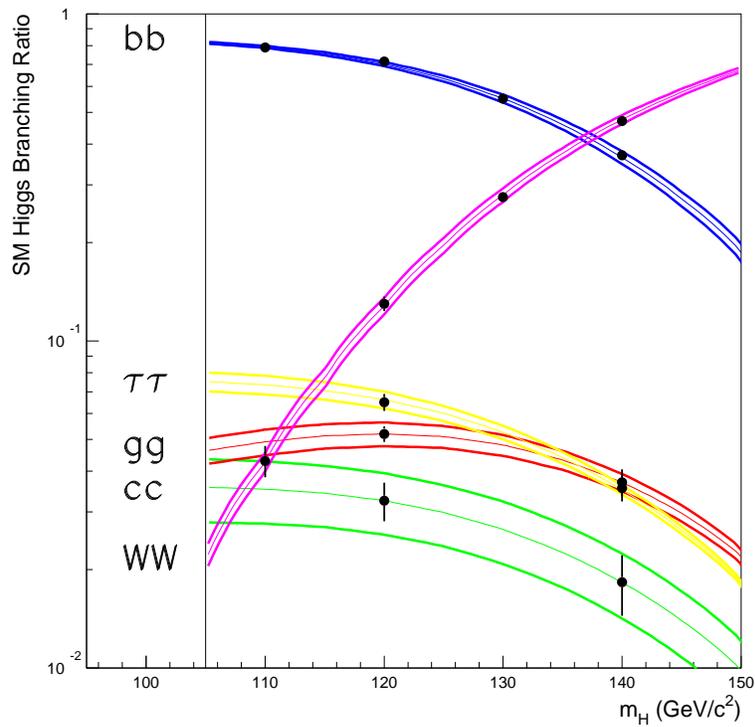,bbllx=0,bblly=18,bburx=521,bbury=517,width=10cm,clip=}
\end{center}
\vspace*{-0.2cm}

\caption[]{\it \label{esp-fg:brmeas} Razones de desintegraci\'on predichas para
  el bos\'on de Higgs del SM. Los puntos con barras de error muestran la
  precisi\'on esperada experimentalmente, mientras las l\'\i neas muestran
  las incertidumbres estimadas predichas en el SM. Ref.~\cite{13}.}
\end{figure}

Un acoplamiento particularmente interesante es el del Higgs a quarks
top. Dado que el quark top es por mucho el fermi\'on m\'as pesado del
Modelo Est\'andar, las irregularidades en la representac\'on est\'andar del
rompimiento de la simetr\'{\i} electrod\'ebil a trav\'es de un campo
fundamental de Higgs, pueden ser aparentes primero en este
acoplamiento. Por lo tanto el acoplamiento de Yukawa $Ht t$ podr\'\i a
eventualmente proveer claves esenciales de la naturaleza del mecanismo
de rompimiento de las simetr\'{\i}as electrod\'ebiles.

Los lazos del top mediando los procesos de producci\'on
$gg\to H$ y $\gamma\gamma\to H$ (y los correspondientes canales de
decaimiento) dan lugar a secciones eficaces y anchos parciales,
los cuales son proporcionales a la ra\'{\i}z cuadrada de los
acoplamientos de Yukawa del Higgs--top. Este acoplamiento de
Yukawa puede ser medido directamente, para la parte inferior del
rango intermedio de masa, en los procesos de bremsstrahlung $pp\to
t\bar t H$ y $e^+e^- \to t\bar t H$ \cite{44}. El bos\'on de Higgs
es radiado, exclusivamente en el primer proceso, en el segundo
proceso predominantemente, de los quarks top pesados. A\'un cuando
estos experimentos son  dif\'{\i}ciles debido a que las secciones
eficaces son peque\~nas [cf. Fig.~\ref{fg:eetth} para
colisionadores $e^+e^-$] y a la compleja topolog\'{\i}a del estado
final $b\bar bb\bar bW^+W^-$, este proceso es una herramienta
importante para explorar el mecanismo de rompimiento de la
simetr\'{\i}a electrod\'ebil. Para masas grandes del Higgs por arriba
del umbral $t\bar t$, el canal de decaimiento $H\to t\bar t$ puede
ser estudiado; en colisiones $e^+e^-$ la secci\'on eficaz de
$e^+e^- \to t\bar t Z$ aumenta a trav\'es de la reacci\'on $e^+e^-
\to ZH (\to t\bar t)$ \cite{45}. El intercambio de Higgs entre
quarks $t\bar t$ afecta tambi\'en la curva de excitaci\'on cerca del
umbral a un nivel de unos cuantos por ciento.

\begin{figure}[hbt]
  \vspace*{0.0cm} \hspace*{2.0cm} \epsfxsize=12cm
  \epsfbox{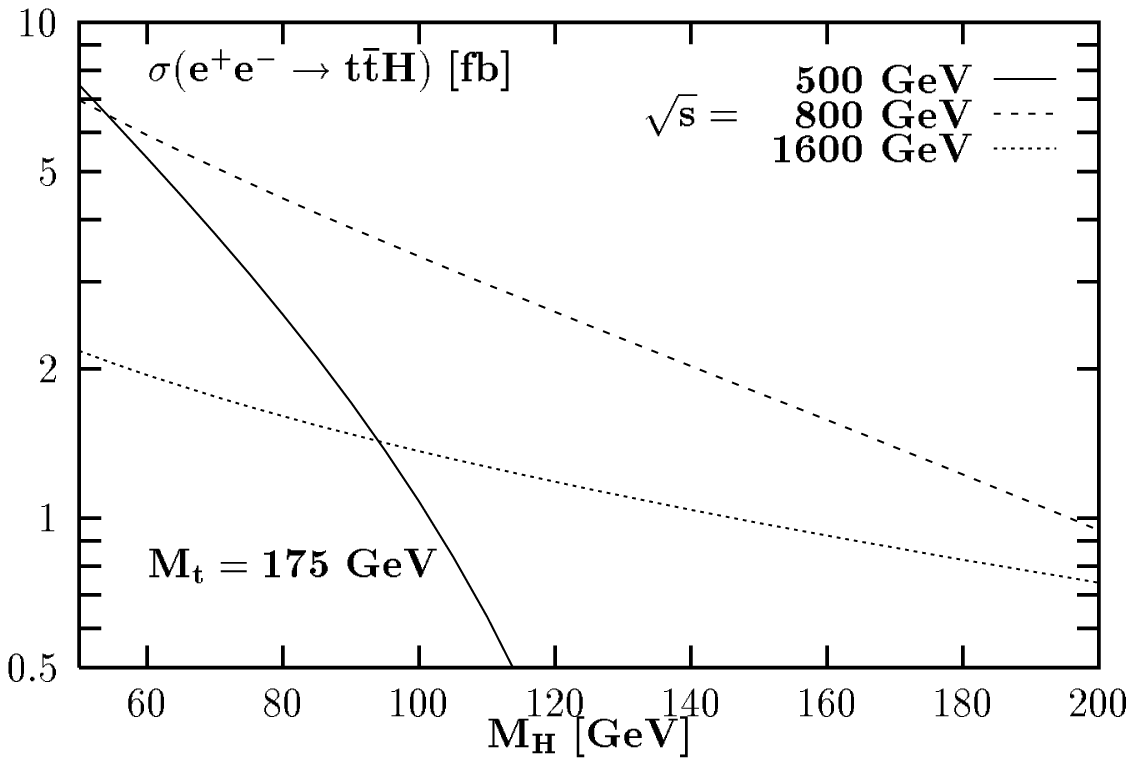} \vspace*{-0.5cm} \caption[]{\it
    \label{esp-fg:eetth} La secci\'on eficaz para bremsstrahlung de los
    bosones de Higgs del SM fuera por quarks top en los procesos de
    Yukawa $e^+e^-\to t\bar t H$. [La amplitud para la radiaci\'on por la
    l\'\i nea del bos\'on intermediario $Z$ es peque\~na.] Ref.
    \cite{44}.}
\end{figure}

S\'olo las razones de los acoplamientos del Higgs a bosones de norma
electrob\'ebil y fermiones pueden ser determinadas en el LHC sin
suposiciones auxiliares ya que solamente pueden ser medidos los
productos de las secciones eficaces de producci\'on y las razones de
desintegraci\'on de decaimiento. Sin embargo, en los colisionadores
lineales, las secciones eficaces de producci\'on pueden medirse
independientemente de los modos espec\'{\i}ficos de decaimiento del
Higgs, inclusive en Higgs-strahlung, por ejemplo. Esto puede ser
explotado para medir los acoplamientos del Higgs a los bosones $Z$ o
$W$ relativos a los cuales todos los dem\'as acoplamientos est\'an
escalados; y son determinados subsecuentemente por las razones de
desintegraci\'on. Las precisiones esperadas para algunos de los
acoplamientos estan reunidas en la Tabla~\ref{tab:muehll}.

\begin{table}[h]
\begin{center}
{\small
\begin{tabular}{|lll|}
\hline
Coupling & $M_H=120$~GeV & 140~GeV \\
\hline
$g_{HWW}$ & $\pm 0.012$ & $\pm 0.020$ \\
$g_{HZZ}$ & $\pm 0.012$ & $\pm 0.013$ \\
\hline
$g_{Htt}$ & $\pm 0.030$ & $\pm 0.061$ \\
$g_{Hbb}$ & $\pm 0.022$ & $\pm 0.022$ \\
$g_{Hcc}$ & $\pm 0.037$ & $\pm 0.102$ \\
\hline
$g_{H\tau\tau}$ & $\pm 0.033$ & $\pm 0.048$ \\
\hline
\end{tabular}
}
\end{center}
\caption{Las precisiones relativas en los acoplamientos de Higgs suponiendo
$\int\!{\cal L}=500$~fb$^{-1}$, $\sqrt{s}=500$~GeV ($\int\!{\cal L}=1$~ab$^{-1}$,
$\sqrt{s}=800$~GeV para $g_{Htt}$).} \label{esp-tab:muehll} \vspace*{-0.4cm}
\end{table}

\paragraph{(e) Auto-acoplamientos del Higgs} ~\\[0.5cm]
El mecanismo de Higgs, basado en un valor diferente de cero del
campo de Higgs en el vac\'{\i}o, debe finalmente ponerse de
manifiesto experimentalmente por la reconstrucci\'on del potencial
de interacci\'on que genera el campo diferente de cero en el
vac\'{\i}o. Este programa puede llevarse a cabo midiendo la
intensidad de los auto-acoplamientos trilineales y cu\'articos de
las part\'{\i}culas de Higgs:
\begin{eqnarray}
g_{H^3} & = & 3 \sqrt{\sqrt{2} G_F} M_H^2 \\ \non \\
g_{H^4} & = & 3 \sqrt{2} G_F M_H^2 ~.
\end{eqnarray}
Esta es una tarea dif\'{\i}cil ya que los procesos a ser
explotados est\'an suprimidos por acoplamientos  y
espacio fase peque\~nos. No obstante, el primer paso en este
problema puede ser resuelto en el LHC y en la fase de alta
energ\'{\i}a en colisionadores lineales $e^+e^-$ para
luminosidades suficientemente altas \cite{selfMMM}. La reacci\'on
m\'as conveniente en los colisionadores  $e^+e^-$ para la medida
de los acoplamientos trilineales de las masas del Higgs en el
rango de masas preferido te\'oricamente ${\cal
O}(100~\mbox{GeV})$, es el proceso doble de Higgs-strahlung
\begin{equation}
e^+e^- \to ZH \to ZHH
\end{equation}
en el cual, entre otros mecanismos, el estado final de dos
Higgs es generado por el intercambio de una part\'{\i}cula de
Higgs virtual de modo que este proceso es sensible al acoplamiento
trilineal $HHH$ en el potencial de Higgs, Fig.~\ref{fg:wwtohh}.
Como la secci\'on eficaz es s\'olo una fracci\'on de 1 fb, es
necesaria una luminosidad integrada de $\sim 1 ab^{-1}$ para
aislar los eventos en colisionadores lineales. 
Incertidumbres experimentales cercanas al 20\% se pueden esperar en
estas mediciones 
\cite{Rexp}. El acoplamiento cu\'artico $H^4$ parece ser 
accesible s\'olo a trav\'es de efectos de lazos en un futuro
pr\'oximo.\\

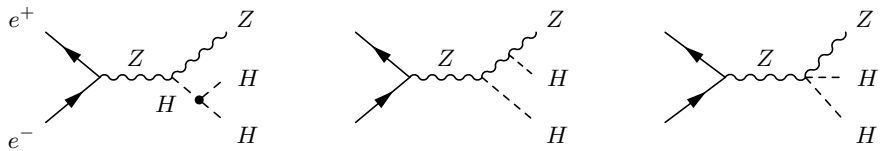
\begin{figure}[hbtp]
\begin{center}
\begin{fmffile}{fd}
{\footnotesize
\unitlength1mm
\begin{fmfshrink}{0.7}
\begin{fmfgraph*}(24,12)
  \fmfstraight
  \fmfleftn{i}{3} \fmfrightn{o}{3}
  \fmf{fermion}{i1,v1,i3}
  \fmflabel{$e^-$}{i1} \fmflabel{$e^+$}{i3}
  \fmf{bos\'on,lab=$Z$,lab.s=left,tens=3/2}{v1,v2}
  \fmf{bos\'on}{v2,o3} \fmflabel{$Z$}{o3}
  \fmf{phantom}{v2,o1}
  \fmffreeze
  \fmf{dashes,lab=$H$,lab.s=right}{v2,v3} \fmf{dashes}{v3,o1}
  \fmffreeze
  \fmf{dashes}{v3,o2}
  \fmflabel{$H$}{o2} \fmflabel{$H$}{o1}
  \fmfdot{v3}
\end{fmfgraph*}
\hspace{15mm}
\begin{fmfgraph*}(24,12)
  \fmfstraight
  \fmfleftn{i}{3} \fmfrightn{o}{3}
  \fmf{fermion}{i1,v1,i3}
  \fmf{bos\'on,lab=$Z$,lab.s=left,tens=3/2}{v1,v2}
  \fmf{dashes}{v2,o1} \fmflabel{$H$}{o1}
  \fmf{phantom}{v2,o3}
  \fmffreeze
  \fmf{bos\'on}{v2,v3,o3} \fmflabel{$Z$}{o3}
  \fmffreeze
  \fmf{dashes}{v3,o2}
  \fmflabel{$H$}{o2} \fmflabel{$H$}{o1}
\end{fmfgraph*}
\hspace{15mm}
\begin{fmfgraph*}(24,12)
  \fmfstraight
  \fmfleftn{i}{3} \fmfrightn{o}{3}
  \fmf{fermion}{i1,v1,i3}
  \fmf{bos\'on,lab=$Z$,lab.s=left,tens=3/2}{v1,v2}
  \fmf{dashes}{v2,o1} \fmflabel{$H$}{o1}
  \fmf{dashes}{v2,o2} \fmflabel{$H$}{o2}
  \fmf{bos\'on}{v2,o3} \fmflabel{$Z$}{o3}
\end{fmfgraph*}
\end{fmfshrink}
\\[0.5cm]
}
\end{fmffile}
\end{center}
\caption[]{\label{esp-fg:wwtohh} \it Diagramas gen\'ericos que contribuyen
al proceso doble de Higgs-strahlung $e^+e^- \to ZHH$.}
\end{figure}

{\it Para resumir}: los elementos esenciales del mecanismo de Higgs
pueden establecerse experimentalmente en el LHC y  colisionadores
TeV lineales $e^+e^-$.

\section{El Bos\'on de Higgs en Teor\'{\i}as Supersim\'etricas}

\phantom{h} Argumentos profundamente arraigados en el sector de
Higgs juegan un papel importante en la
introducci\'on de la supersimetr\'{\i}a como una simetr\'{\i}a
fundamental de la naturaleza \cite{14}. Esta es la \'unica
simetr\'{\i}a que correlaciona los grados de libertad bos\'onicos
y fermi\'onicos.
\paragraph{(a)}
La cancelaci\'on entre las contribuciones bos\'onicas y
fermi\'onicas a las correcciones radiativas de las masas de los
Higgs ligeros en teor\'{\i}as supersim\'etricas prove\'e una
soluci\'on al problema de la jerarqu\'{\i}a en el Modelo Est\'andar.
Si el Modelo Est\'andar se encastra en una teor\'{\i}a de gran
unificaci\'on, la gran brecha entre la escala alta de gran
unificaci\'on y la escala baja del rompimiento de la simetr\'{\i}a
electrod\'ebil se puede estabilizar de modo natural en 
teor\'{\i}as con  simetr\'{\i}as bos\'on--fermi\'on \cite{15,601}.
Denotando la masa desnuda del Higgs por $M_{H,0}^2$, las
correcciones radiativas debidas a los lazos de bos\'on vectorial en
el Modelo Est\'andar por $\delta M_{H,V}^2$, y a las contribuciones
de los gauginos, los compa\~neros supersim\'etricos fermi\'onicos,
por $\delta M_{\tilde H,\tilde V}^2$, la masa f\'{\i}sica del
Higgs est\'a dada por la suma $M_H^2 = M_{H,0}^2 + \delta
M_{H,V}^2 + \delta M_{\tilde H,\tilde V}^2$. La correcci\'on del 
bos\'on vectorial es cuadr\'aticamente divergente, $\delta M_{H,V}^2
\sim \alpha [\Lambda^2 - M^2]$, de tal manera que para una escala
de corte $\Lambda \sim \Lambda_{GUT}$ ser\'a necesario un
ajuste fino extremo entre la masa desnuda intr\'{\i}nseca  y las
fluctuaciones cu\'anticas radiativas para generar la masa del
Higgs del orden de $M_W$. Sin embargo, debido al principio de
Pauli, las contribuciones del gaugino fermi\'onico adicional en
teor\'{\i}as supersim\'etricas son justo del signo opuesto,
$\delta M_{\tilde H,\tilde V}^2\sim -\alpha [\Lambda^2-\tilde
M^2]$, de manera tal que los t\'erminos divergentes se cancelan
entre s\'{\i}\footnote{Las diferentes estad\'{\i}sticas para
bosones y fermiones son suficientes para obtener la cancelaci\'on
de las divergencias; sin embargo, ellas no son necesarias. Las
relaciones de simetr\'{\i}a entre los acoplamientos, como se dan
en Modelos de Higgs Peque\~no (Little Higgs Models), tambi\'en
pueden llevar a cancelaciones individualmente entre amplitudes
bos\'on-bos\'on o fermi\'on-fermi\'on.}. Como $\delta
M_H^2\sim\alpha [\tilde M^2-M^2]$, se evita cualquier ajuste fino
para masas de las part\'{\i}culas supersim\'etricas $\tilde M
\lessim {\cal O}(1$ TeV). As\'{\i}, dentro de este esquema de
simetr\'{\i}a el sector de Higgs es estable en el rango de bajas
energ\'{\i}as $M_H\sim M_W$ incluso en el contexto de escalas de
alta energ\'{\i}a de GUT. Este mecanismo conduce de manera natural
a la supersimetr\'{\i}a de bajas energ\'{\i}as.

\paragraph{(b)}
El concepto de supersimetr\'{\i}a est\'a apoyado fuertemente por
la exitosa predicci\'on del \'angulo de mezcla electrod\'ebil
obtenido en la versi\'on m\'\i nima de esta teor\'{\i}a \cite{16}. En \'esta, 
el espectro extendido de part\'{\i}culas lleva a la
evoluci\'on del \'angulo de mezcla electrod\'ebil del valor de
GUT de 3/8  hasta $\sin^2\theta_W = 0.2336 \pm 0.0017$ a bajas energ\'\i as,
donde el  error  incluye contribuciones desconocidas de umbral en
las escalas de baja y alta   masa supersim\'etrica. La
predicci\'on coincide con la medida experimental
$\sin^2\theta_W^{exp} = 0.23120 \pm 0.00015$ dentro de una
incertibumbre te\'orica menor al 2 por mil.

\paragraph{(c)}

Es conceptualmente muy interesante la interpretaci\'on del
mecanismo de Higgs en las teor\'{\i}as supersim\'etricas como un
efecto cu\'antico \cite{50A}. El rompimiento de la simetr\'{\i}a
electrod\'ebil $SU(2)_L \times U(1)_Y$ puede inducirse
radiativamente mientras se conservan la simetr\'\i a de norma
electromagn\'etica $U(1)_{EM}$  y la simetr\'\i a de norma de color
$SU(3)_C$ para masas del quark--top entre 150 y 200 GeV. Iniciando
con un conjunto de masas escalares univesales a la escala de GUT,
el par\'ametro de la masa al cuadrado del sector de Higgs toma
valores negativos a la escala electrod\'ebil, mientras que las
masas al cuadrado de los
sleptones y squarks se mantienen positivas.\\

El sector de Higgs de las teor\'{\i}as supersim\'etricas difiere
del Modelo Est\'andard en muchos aspectos \cite{17}. Para
preservar la supersimetr\'{\i}a y la invariancia de norma,  se deben
introducir al menos dos campos iso-dobletes, dej\'andonos con
un espectro de cinco o m\'as part\'{\i}culas de Higgs
f\'{\i}sicas. En la extensi\'on supersim\'etrica m\'{\i}nima del
Modelo Est\'andar (MSSM)las auto-interacciones del Higgs son
generadas por la acci\'on norma--escalar, de manera tal que los
acoplamientos cu\'articos est\'an relacionados a los
acoplamientos de norma en este escenario.  Esto lleva a l\'\i mites
fuertes \cite{19} para la masa del bos\'on de Higgs m\'as ligero, que debe
ser menor que alrededor de $140$ GeV [incluyendo correcciones radiativas].
Si se supone  que el sistema se mantiene d\'ebilmente
interactuante hasta escalas del orden de GUT o escala de Planck,
la masa se mantiene peque\~na, por razones muy parecidas a las
encontradas en el Modelo Est\'andar, incluso en teor\'{\i}as
supersim\'etricas m\'as complejas que involucran campos de Higgs
e interacciones de Yukawa adicionales. Las masas de los bosones de
Higgs pesados se espera que est\'en dentro del rango de la escala de
rompimiento de la simetr\'\i a electrod\'ebil hasta un orden de 1 TeV.

\subsection{El Sector de Higgs del MSSM}

\phantom{h} El espectro de part\'{\i}culas del MSSM \cite{14}
consiste de leptones, quarks y sus compa\~neros supersim\'etricos
escalares, y part\'{\i}culas de norma, part\'{\i}culas de Higgs y sus
compa\~neros de esp\'{\i}n-1/2. Los campos de 
mater\'{\i}a y de fuerzas est\'an acoplados en acciones 
supersim\'etricas e invariantes de norma:
\begin{equation}
\begin{array}{lrcll}
S = S_V + S_\phi + S_W: \hspace*{1cm}
& S_V    & = & \frac{1}{4} \int d^6 z \hat W_\alpha \hat W_\alpha
\hspace*{1cm} & \mbox{acci\'on de norma, gauge} ~, \\ \\
& S_\phi & = & \int d^8 z \hat \phi^* e^{gV} \hat \phi
& \mbox{acci\'on de materia} ~, \\ \\
& S_W    & = & \int d^6 z W[\hat \phi] & \mbox{superpotencial} ~.
\end{array}
\end{equation}
Descomponiendo los supercampos en sus componentes fermi\'onicas y
bos\'onicas, e integrando sobre las variables de Grassmann en
$z\to x$, se pueden derivar los siguientes Lagrangianos, los
cuales describen las interacciones de norma, materia y
campos de Higgs:
\begin{eqnarray*}
{\cal L}_V & = & -\frac{1}{4}F_{\mu\nu}F_{\mu\nu}+\ldots+\frac{1}{2}D^2 ~, \\ \\
{\cal L}_\phi & = & D_\mu \phi^* D_\mu \phi +\ldots+\frac{g}{2} D|\phi|^2  ~, \\ \\
{\cal L}_W & = & - \left| \frac{\partial W}{\partial \phi} \right|^2 ~.
\end{eqnarray*}
El campo $D$ es un campo auxiliar el cual no se propaga en el
espacio-tiempo y puede eliminarse aplicando las ecuaciones de
movimiento: $D=-\frac{g}{2} |\phi|^2$. Reinsertado en el
Lagrangiano, el auto-acoplamiento cu\'artico del campo escalar de
Higgs es generado:
\begin{equation}
{\cal L} [\phi^4] = -\frac{g^2}{8} |\phi^2|^2 ~.
\end{equation}
As\'\i, el acoplamiento cu\'artico de los campos de Higgs
est\'a dado, en la teor\'{\i}a supersim\'etrica m\'\i nima, por el
cuadrado del acoplamiento de norma.  A diferencia del caso del
Modelo Est\'andar, el acoplamiento cu\'artico no es un par\'ametro
libre.
Adem\'as, este acoplamiento es d\'ebil.\\

Dos campos dobletes de Higgs independientes $H_1$ y $H_2$ deben
introducirse en el superpotencial:
\begin{equation}
W = -\mu \epsilon_{ij} \hat H_1^i \hat H_2^j + \epsilon_{ij} [f_1 \hat H_1^i
\hat L^j \hat R + f_2 \hat H_1^i \hat Q^j \hat D +
f_2' \hat H_2^j \hat Q^i \hat U]
\end{equation}
para proporcionar masa a las part\'{\i}culas de tipo-down ($H_1$)
y a las par\'{\i}culas de tipo-up ($H_2$). A diferencia del Modelo
Est\'andar, el segundo campo de Higgs no puede ser identificado con
el conjugado de carga del primero, ya que $W$ debe ser
anal\'{\i}tico para preservar la supersimetr\'{\i}a. Adem\'as, los
campos del Higgsino asociados con un s\'olo campo de Higgs
generar\'{\i}an anomal\'{\i}as triangulares; \'estas se cancelan
si los dos dobletes conjugados se suman, y la invariancia de
norma cl\'asica de las interacciones no se destruye a nivel
cu\'antico. Integrando el superpotencial sobre las coordenadas de
Grassmann se genera la auto-energ\'{\i}a de Higgs supersim\'etrica
$V_0 = |\mu|^2 (|H_1|^2 + |H_2|^2)$. El rompimiento de la
supersimetr\'{\i}a puede ser incorporado en el sector de Higgs al
introducir t\'erminos de masa bilineales $\mu_{ij} H_i H_j$.
Sumados a la parte de la auto-energ\'{\i}a supersim\'etrica $H^2$ y la
parte cu\'artica $H^4$ generada por la acci\'on de norma, llevan
al siguiente potencial de Higgs
\begin{eqnarray}
V & = & m_1^2 H_1^{*i} H_1^i + m_2^2 H_2^{*i} H_2^i - m_{12}^2 (\epsilon_{ij}
H_1^i H_2^j + hc) \non \\ \non \\
& & + \frac{1}{8} (g^2 + g'^2) [H_1^{*i} H_1^i -
H_2^{*i} H_2^i]^2 + \frac{1}{2} |H_1^{*i} H_2^{*i}|^2 ~.
\end{eqnarray}
El potencial de Higgs incluye tres t\'erminos  bilineales de masa,
mientras que la intensidad del acoplamiento cu\'artico se
determina por los acoplamientos de norma $SU(2)_L$ y $U(1)_Y$ al
cuadrado. Los tres t\'erminos de masa son par\'ametros libres.

El potencial desarrolla un m\'{\i}nimo estable para $H_1 \to
[0,v_1]$ y $H_2\to [v_2,0]$, si se reunen las siguientes
condiciones:
\begin{equation}
m_1^2 +  m_2^2 >  2 | m^2_{12} |  \hspace*{0.5cm} \mbox{and}  \hspace*{0.5cm}
m_1^2    m_2^2  <  | m^2_{12} |^2 ~.
\end{equation}
Al expander los campos alrededor de los valores del estado base
$v_1$ y $v_2$,
\begin{equation}
\begin{array}{rclcl}
H_1^1 & = & & & H^+ \cos \beta + G^+ \sin \beta \\ \\
H_1^2 & = & v_1 & + & [H^0 \cos \alpha - h^0 \sin \alpha + i A^0 \sin \beta - i G^0
\cos \beta ]/\sqrt{2}
\end{array}
\end{equation}
y
\begin{equation}
\begin{array}{rclcl}
H_2^1 & = & v_2 & + & [H^0 \sin \alpha + h^0 \cos \alpha + i A^0 \cos \beta + i G^0
\sin \beta ]/\sqrt{2} \\ \\
H_2^2 & = & & & H^- \sin \beta - G^- \cos \beta ~,
\end{array}
\end{equation}
los eigenestados de masa est\'an dados por los estados neutros 
$h^0,H^0$ y $A^0$, lo cuales son pares e impares bajo
transformaciones de ${\cal CP}$; y por los estados cargados
$H^\pm$; los estados $G$ corresponden a los modos de Goldstone, los
cuales son absorbidos por los campos de norma para construir las
componenetes longitudinales. Despu\'es de introducir los tres
par\'ametros
\begin{eqnarray}
M_Z^2 & = & \frac{1}{2} (g^2 + g'^2) (v_1^2 + v_2^2) \non \\ \non \\
M_A^2 & = & m_{12}^2 \frac{v_1^2 + v_2^2}{v_1v_2} \non \\ \non \\
\tgb  & = & \frac{v_2}{v_1} ~,
\end{eqnarray}
la matriz de masa puede descomponerse en tres bloques de $2\times
2$, los cuales son f\'aciles de diagonalizar:
\begin{displaymath}
\begin{array}{ll}
\mbox{\bf masa pseudoescalar:} & M_A^2 \\ \\
\mbox{\bf masa cargada:} & M_\pm^2 = M_A^2+M_W^2 \\ \\
\mbox{\bf masa escalar:} & M_{h,H}^2 = \frac{1}{2} \left[ M_A^2 +
M_Z^2 \mp \sqrt{(M_A^2+M_Z^2)^2
- 4M_A^2M_Z^2 \cos^2 2\beta} \right] \\ \\
& \displaystyle \tg 2\alpha = \tg 2\beta \frac{M_A^2 + M_Z^2}{M_A^2 - M_Z^2}
\hspace*{0.5cm} \mbox{with} \hspace*{0.5cm} -\frac{\pi}{2} < \alpha < 0
\nonumber
\end{array}
\end{displaymath}

De las f\'ormulas de la masa, pueden derivarse dos desigualdades
importantes,
\begin{eqnarray}
M_h & \leq & M_Z, M_A \leq M_H \\ \nonumber \\
M_W & \leq & M_{H^\pm} ~,
\end{eqnarray}
las que, por construcci\'on, son v\'alidas en una aproximaci\'on de \'arbol.
Como resultado, se predice que la masa del escalar de Higgs m\'as ligero
est\'a acotada por la masa del $Z$, modulo correcciones
radiativas. Estos l\'{\i}mites provienen del hecho de que el
acoplamiento cu\'artico de los campos de Higgs est\'an determinados en el
MSSM por el tama\~no del acoplamiento de norma
al cuadrado. \\


\noindent
\underline{\bf Correcciones Radiativas en SUSY} \\[0.5cm]
Las relaciones a nivel \'arbol entre las masas de los Higgs son
fuertemente modificadas por correcciones radiativas que involucran
el espectro de part\'{\i}culas supersim\'etricas  del sector del
top \cite{50B}; cf.~Ref.~\cite{mhplot,66a} para recopilaciones
recientes. Estos efectos son proporcionales a la cuarta potencia
de la masa del top y al logaritmo de la masa del stop. El origen
de estas correcciones son las cancelaciones incompletas entre los
lazos del top virtual y  del stop, reflej\'ando el rompimiento
de la supersimetr\'{\i}a. M\'as a\'un, las relaciones de masa son
afectadas por la mezcla potencialmente grande entre $\tilde t_L$ y
$\tilde t_R$
debido al acoplamiento de Yukawa del top.\\

A primer orden en $M_t^4$ las correcciones radiativas se pueden
resumir en el par\'ametro
\begin{equation}
\epsilon = \frac{3G_F}{\sqrt{2}\pi^2}\frac{M_t^4}{\sin^2\beta}\log
\frac{M_{\tilde t_1}M_{\tilde t_2}}{M_t^2} ~.
\end{equation}
En esta aproximaci\'on, la masa del Higgs ligero $M_h$ puede
expresarse por $M_A$ y $\tgb$ en la siguiente forma compacta:
\begin{eqnarray*}
M^2_h & = & \frac{1}{2} \left[ M_A^2 + M_Z^2 + \epsilon \right.
\non \\
& & \left. - \sqrt{(M_A^2+M_Z^2+\epsilon)^2
-4 M_A^2M_Z^2 \cos^2 2\beta
-4\epsilon (M_A^2 \sin^2\beta + M_Z^2 \cos^2\beta)} \right]
\end{eqnarray*}
Las masas de los Higgs pesados  $M_H$ and $M_{H^\pm}$ se obtienen
de las reglas de suma
\begin{eqnarray*}
M_H^2 & = & M_A^2 + M_Z^2 - M_h^2 + \epsilon \non \\
M_{H^\pm}^2 & = & M_A^2 + M_W^2 ~.
\end{eqnarray*}
Finalmente, el par\'ametro de mezcla $\alpha$, el cual diagonaliza
la matriz de masa ${\cal CP}$-par, est\'a dada por la relaci\'on
mejorada radiativamente:
\begin{equation}
\tg 2 \alpha = \tg 2\beta \frac{M_A^2 + M_Z^2}{M_A^2 - M_Z^2 +
\epsilon/\cos 2\beta} ~.
\label{esp-eq:mssmalpha}
\end{equation}

Para masas grandes de $A$, las masas de las part\'{\i}culas Higgs
pesadas coinciden aproximadamente, $M_A\simeq M_H \simeq
M_{H^\pm}$, mientras que la masa del Higgs ligero  se acerca a un
valor asint\'otico peque\~no. El espectro para valores grandes de
$\tgb$ es bastante regular: para  valores peque\~nos de $M_A$ uno
encuentra que $\{ M_h\simeq M_A; M_H \simeq \mbox{const} \}$
\cite{intense}; para valores grandes de $M_A$ se encuentra la
relaci\'on opuesta $\{ M_h\simeq \mbox{const}, M_H \simeq
M_{H^\pm}\simeq M_A \}$,
cf.~Fig.~\ref{kdfig} la cual incluye correcciones radiativas.\\
\begin{figure}[hbt]
\begin{center}
\hspace*{-0.3cm}
\epsfig{figure=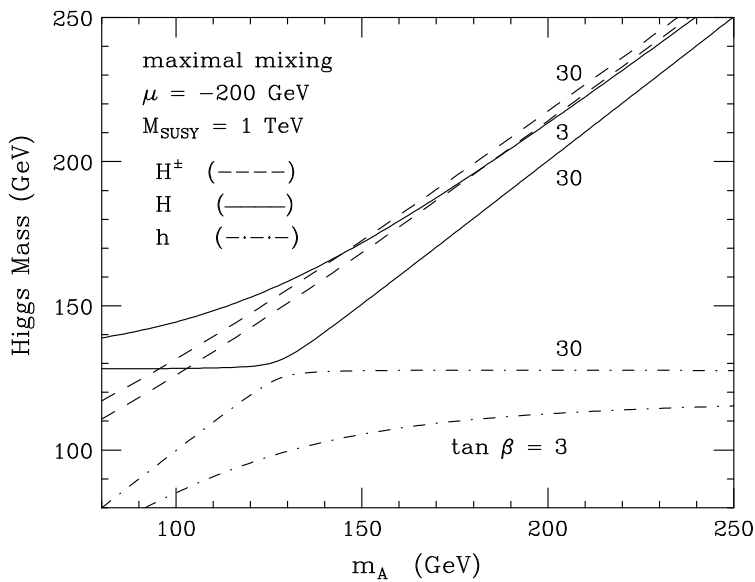,bbllx=80,bblly=205,bburx=556,bbury=570,width=10cm,clip=}
\end{center}
\vspace*{-0.4cm} \caption[]{\label{esp-kdfig} \it Las masas de los
bosones de Higgs del MSSM CP-par y cargados como una funci\'on de
$m_A$ para $\tan\beta=3$ y 30, incluyendo correcciones radiativas.
Ref.~\cite{66a}.}
\end{figure}

Mientras los efectos no dominantes de mezcla en las
relaciones de masa del Higgs son bastante complicados, el
impacto en la cota superior de la masa del Higgs ligero $M_h$ se
puede resumir de una forma sencilla:
\begin{equation}
M_h^2 \leq M_Z^2 \cos^2 2\beta + \delta M_t^2 + \delta M_X^2 ~.
\end{equation}
La principal contribuci\'on dada por el top est\'a relacionada con el
par\'ametro $\epsilon$,
\begin{equation}
\delta M_t^2 = \epsilon \sin^2\beta ~.
\end{equation}
La segunda contribuci\'on
\begin{equation}
\delta M_X^2 = \frac{3G_F M_t^4}{2\sqrt{2}\pi^2} X_t^2 \left[ 2
h(M_{\tilde t_1}^2, M_{\tilde t_2}^2) + X_t^2~g(M_{\tilde t_1}^2,
M_{\tilde t_2}^2) \right]
\end{equation}
depende del par\'ametro de mezcla
\begin{equation}
M_tX_t = M_t \left[A_t - \mu~\ctgb \right] ~,
\end{equation}
el cual acopla estados de quiralidad izquierda y derecha en la
matriz de masa del stop; $h,g$ son funciones  de las masas del
stop:
\begin{equation}
h = \frac{1}{a-b} \log \frac{a}{b} \hspace*{0.5cm} \mbox{and} \hspace*{0.5cm}
g = \frac{1}{(a-b)^2} \left[ 2 - \frac{a+b}{a-b} \log \frac{a}{b} \right] ~.
\end{equation}
Contribuciones subdominantes pueden ser reducidas esencialmente a
efectos de QCD de orden m\'as alto. Estas contribuciones pueden ser 
incorporadas efectivamente al interpretar el par\'ametro de masa del top
$M_t \to M_t(\mu_t)$ como la masa $\overline{\rm MS}$ del top evaluada
en la media geom\'etrica 
entre las masas del top y el estop, $\mu_t^2 = M_t M_{\tilde t}$.\\

Las cotas superiores en la masa del Higgs ligero se muestran en
Fig.~\ref{fg:mssmhiggs} como funci\'on de $\tg \beta$. Las curvas son el
resultado de los c\'alculos con y sin efectos de mezcla. Esto lleva a
que la cota superior general para una mezcla m\'axima est\'a dada por
$M_h\lessim 140$ GeV, incluyendo valores grandes de $\tgb$. El sector
del Higgs ligero no puede ser completamente cubierto por experimentos
del LEP2 debido al incremento que sufre el l\'{\i}mite de la masa con
la masa del top.
\begin{figure}[hbt]
\begin{center}
\hspace*{-0.3cm}
\epsfig{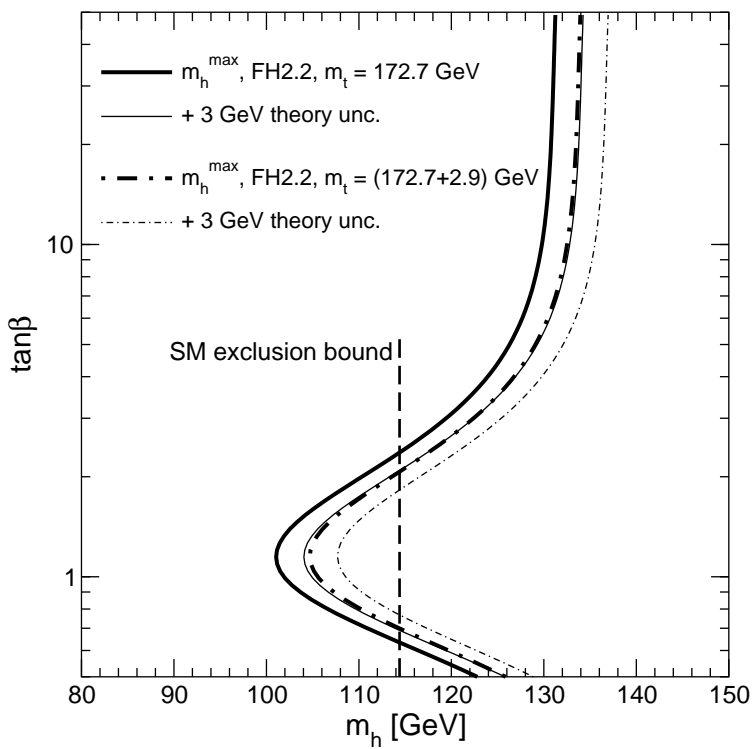}
\end{center}
\vspace*{-0.4cm} \caption[]{\label{esp-fg:mssmhiggs} \it Cotas
superiores a la masa del bos\'on de Higgs ligero como funci\'on
de $\tg\beta$ para varios escenarios de SUSY. Ref.
\cite{mhplot}.}
\end{figure}

\subsection{Acoplamientos SUSY del Higgs a Part\'{\i}culas del SM}

\phantom{h} El tama\~no de los acoplamientos de los Higgs del MSSM
con quarks, leptones y bosones de norma es similar al del Modelo
Est\'andar (SM), pero modificado por los \'angulos de mezcla
$\alpha$ and $\beta$. Normalizados a los valores del SM, estos
acoplamientos est\'an listados en la Tabla \ref{tb:hcoup}. El bos\'on
de Higgs pseudoescalar $A$ no se acopla a bosones de norma a nivel
\'arbol, pero el acoplamiento, compatible con la simetr\'{\i}a de 
${\cal CP}$, puede ser generado a nivel de lazos de m\'as alto
orden. Los bosones cargados de Higgs  se acoplan a los fermiones
up y down a trav\'es de las amplitudes quirales izquierda y
derecha $g_\pm = - \left[ g_t (1 \mp \gamma_5) + g_b (1 \pm
\gamma_5) \right]/\sqrt{2}$, donde $g_{t,b} = (\sqrt{2} G_F)^{1/2}
m_{t,b}$.
\begin{table}[hbt]
\renewcommand{\arraystretch}{1.5}
\begin{center}
\begin{tabular}{|lc||ccc|} \hline
\multicolumn{2}{|c||}{$\Phi$} & $g^\Phi_u$ & $g^\Phi_d$ &  $g^\Phi_V$ \\
\hline \hline
SM~ & $H$ & 1 & 1 & 1 \\ \hline
MSSM~ & $h$ & $\cos\alpha/\sin\beta$ & $-\sin\alpha/\cos\beta$ &
$\sin(\beta-\alpha)$ \\
& $H$ & $\sin\alpha/\sin\beta$ & $\cos\alpha/\cos\beta$ &
$\cos(\beta-\alpha)$ \\
& $A$ & $ 1/\tg\beta$ & $\tg\beta$ & 0 \\ \hline
\end{tabular}
\renewcommand{\arraystretch}{1.2}
\caption[]{\label{esp-tb:hcoup} \it Acoplamientos del Higgs a
fermiones y bos\'ons de norma [$V=W,Z$] en el MSSM relativos a los
acoplamientos del SM.}
\end{center}
\end{table}

Los acoplamientos modificados incorporan la renormalizaci\'on
debido a correcciones radiativas de SUSY, a primer orden en $M_t$,
si el \'angulo de mezcla $\alpha$ est\'a relacionado con $\beta$ y
con $M_A$ a trav\'es de la f\'ormula corregida
Eq.~(\ref{eq:mssmalpha}). Para valores grandes de $M_A$, en la
pr\'actica $M_A\gsim 200$ GeV, los acoplamientos del bos\'on de
Higgs ligero $h$ a los fermiones y a los bosones de norma  se
aproximan asint\'oticamente a los valores del SM. Esta es la
escencia del \underline{teorema de desacoplamiento} en el sector
de Higgs \cite{66AA}: Part\'{\i}culas con masas grandes deben
desacoplarse del sistema de part\'{\i}culas ligeras como
consecuencia del principio de incertidumbre de la mec\'anica cu\'antica.

\subsection{Decaimientos de las Part\'{\i}culas de Higgs}

\phantom{h} El \underline{\it bos\'on de Higgs neutro} m\'as
ligero $h$ decaer\'a principalmente en pares de fermiones ya que
su masa es menor a $\sim 140$ GeV, Fig.~\ref{fg:mssmbr}a (cf.
\cite{613A} para un resumen detallado). Este es en general,
tambi\'en el modo dominante de decaimiento para el bos\'on
pseudoescalar $A$. Para valores de $\tgb$ mayores a la unidad y
para masas menores que $\sim 140$ GeV, los principales modos de
decaimiento de los bosones neutros de Higgs son decaimientos a
pares de $b\bar b$ y $\tau^+\tau^-$; las razones de desintegraci\'on 
son del orden de $\sim 90\%$ y $8\%$, respectivamente. Los
decaimientos a pares de $c\bar c$ y gluones est\'an suprimidos,
especialmente para valores grandes de $\tgb$. Para masas grandes
de los bosones neutros de Higgs, se abre el canal de decaimiento
del top $H,A \to t\bar t$; aunque para valores grandes de $\tgb$
este modo permanece suprimido y los bosones neutros de Higgs
decaen casi exclusivamente en pares de $b\bar b$ y
$\tau^+\tau^-$. Si la masa es suficientemente grande, el bos\'on
de Higgs pesado ${\cal CP}$-par $H$ puede en principio decaer en
los bosones de norma d\'ebiles, $H\to WW,ZZ$.
Debido a que los anchos parciales de decaimiento son
proporcionales a $\cos^2(\beta - \alpha)$, en general est\'an
fuertemente suprimidos, y la se\~nal dorada $ZZ$ del
bos\'on pesado de Higgs en el Modelo Est\'andar se pierde en la
extensi\'on supersim\'etrica. Como resultado, los anchos totales
de los bosones de Higgs son mucho m\'as peque\~nos en teor\'{\i}as
supersim\'etricas que en el Modelo Est\'andar.
\begin{figure}[hbtp]

\vspace*{-2.5cm}
\hspace*{-4.5cm}
\begin{turn}{-90}%
\epsfxsize=16cm \epsfbox{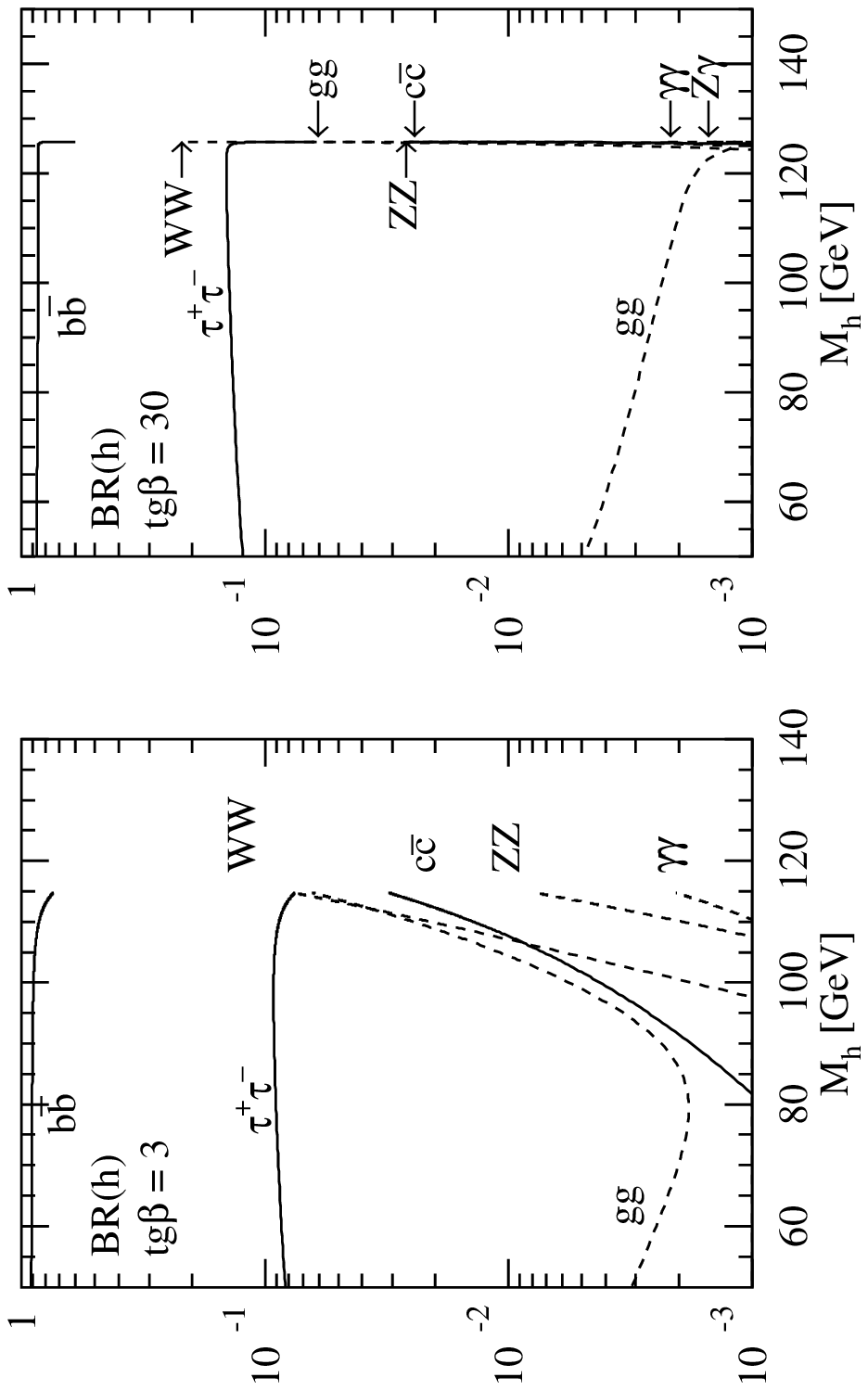}
\end{turn}
\vspace*{-4.2cm}

\centerline{\bf Fig.~\ref{fg:mssmbr}a}

\vspace*{-2.5cm}
\hspace*{-4.5cm}
\begin{turn}{-90}%
\epsfxsize=16cm \epsfbox{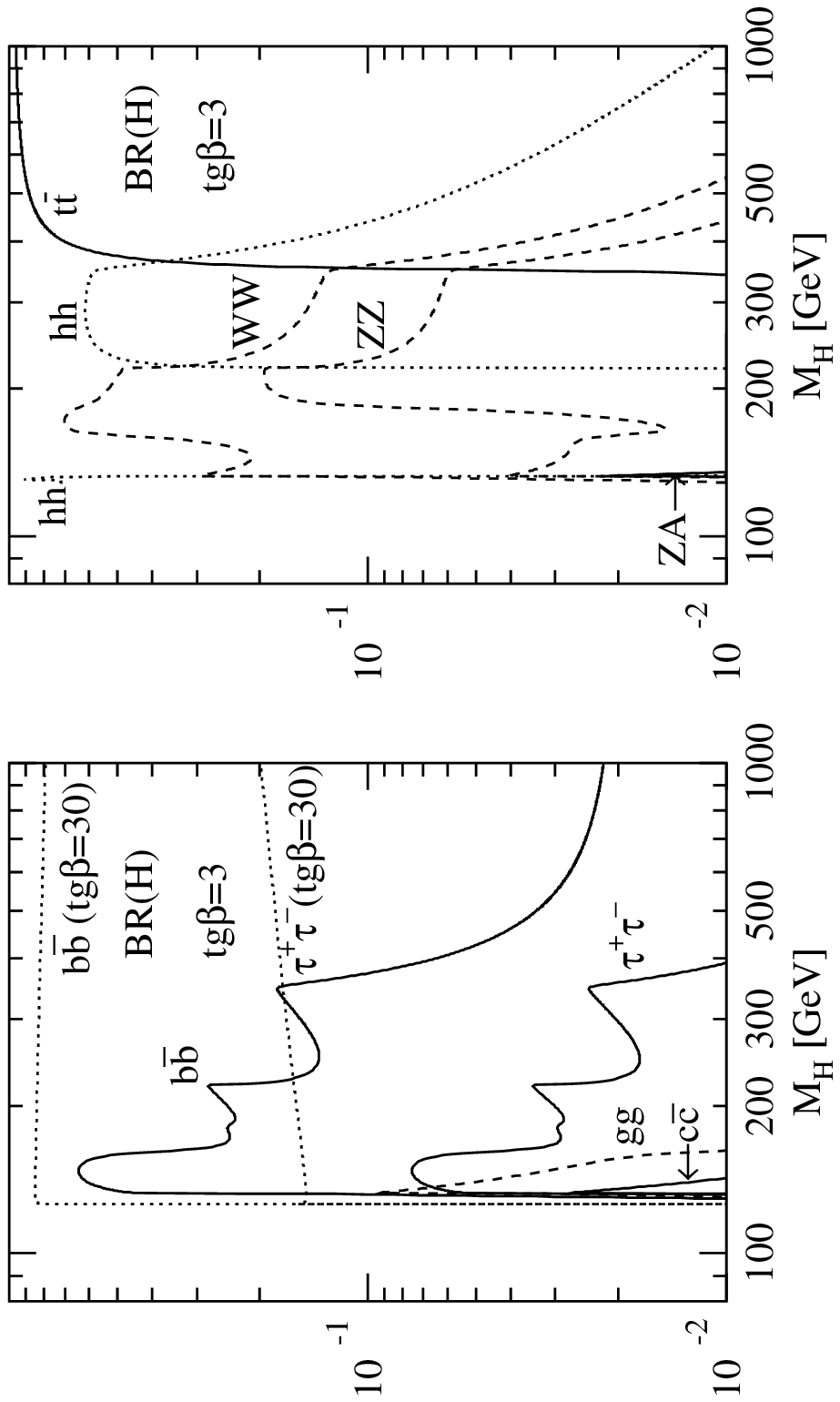}
\end{turn}
\vspace*{-4.2cm}

\centerline{\bf Fig.~\ref{fg:mssmbr}b}

\caption[]{\label{esp-fg:mssmbr} \it Razones de desintegraci\'on de los
bos\'ones de Higgs del MSSM $h, H, A, H^\pm$, para decaimientos no
supersim\'etricos como funci\'on de las
masas, para dos valores de $\tgb=3, 30$ y sin considerar mezcla. Se ha
escogido la masa com\'un de los squark como $M_S=1$ TeV.}
\end{figure}
\addtocounter{figure}{-1}
\begin{figure}[hbtp]

\vspace*{-2.5cm}
\hspace*{-4.5cm}
\begin{turn}{-90}%
\epsfxsize=16cm \epsfbox{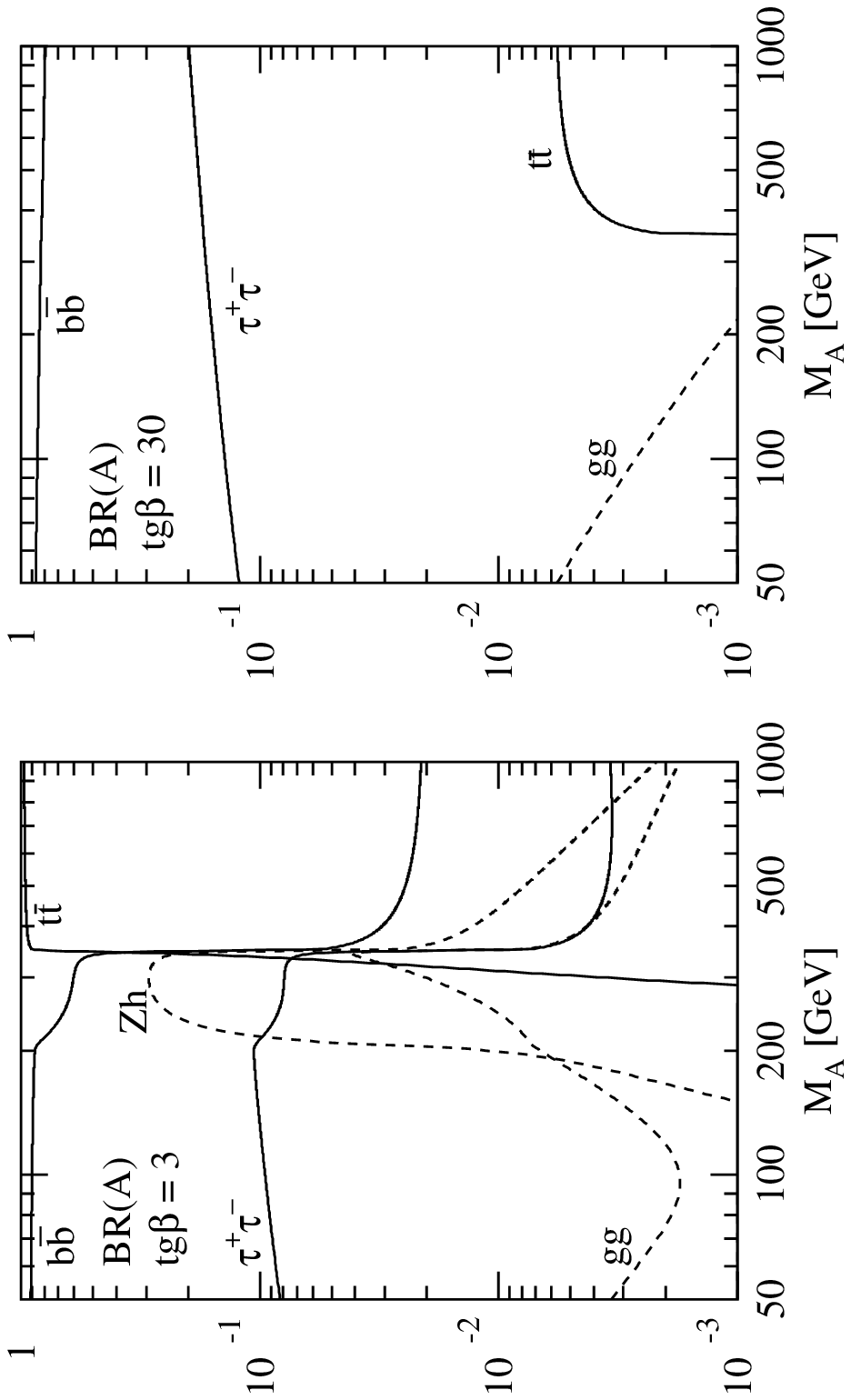}
\end{turn}
\vspace*{-4.2cm}

\centerline{\bf Fig.~\ref{fg:mssmbr}c}

\vspace*{-2.5cm}
\hspace*{-4.5cm}
\begin{turn}{-90}%
\epsfxsize=16cm \epsfbox{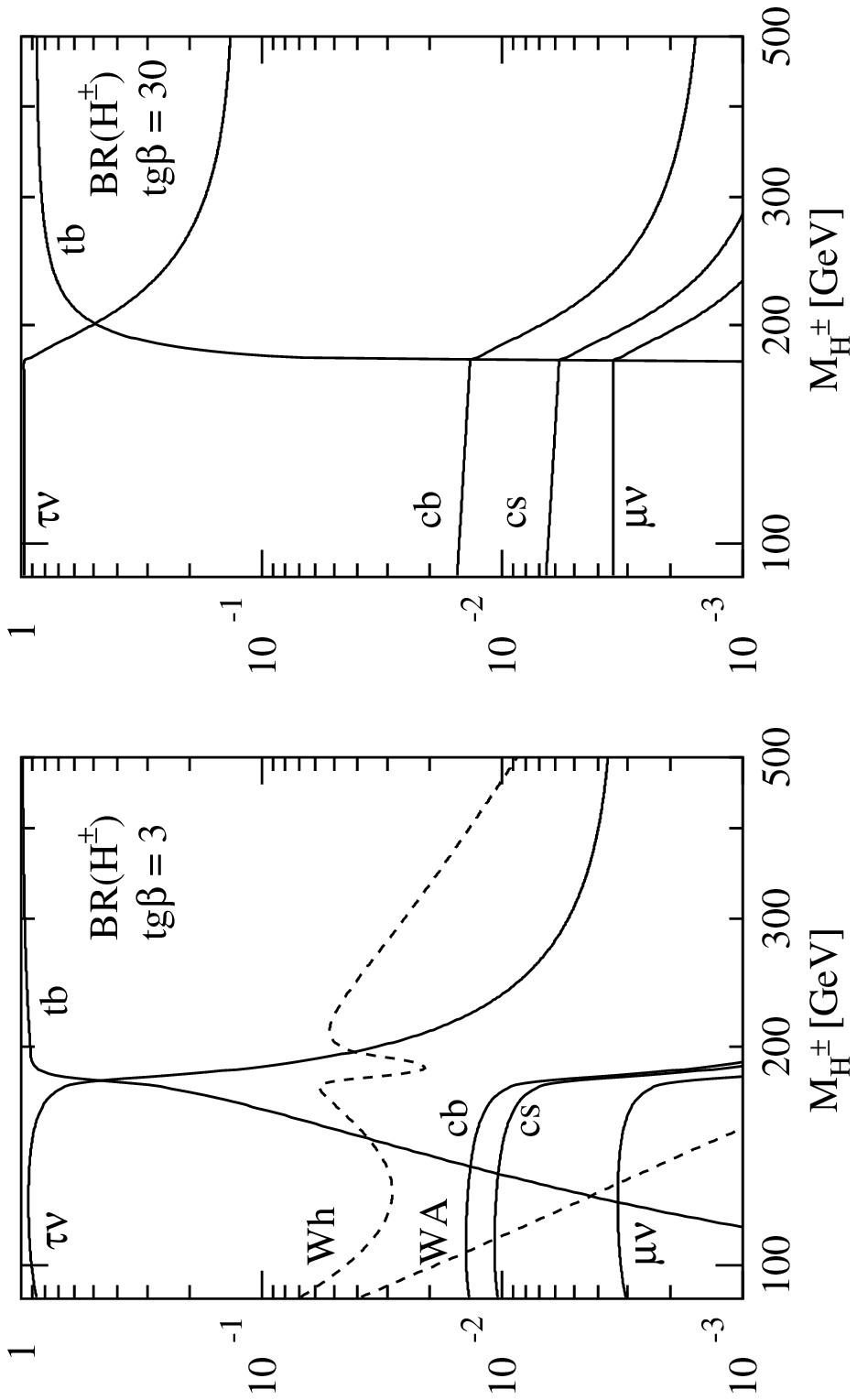}
\end{turn}
\vspace*{-4.2cm}

\centerline{\bf Fig.~\ref{fg:mssmbr}d}

\caption[]{\it Continaci\'on.}
\end{figure}

El bos\'on de Higgs  pesado  neutro $H$ puede tambi\'en decaer en
dos bosones de Higgs ligeros. Otros posibles canales son:
decaimientos en cascada del Higgs y decaimientos a part\'{\i}culas
supersim\'etricas \citer{614,616}, Fig.~\ref{fg:hcharneutsq}.
Adem\'as de los sfermiones ligeros, decaimientos del bos\'on de
Higgs a charginos y neutralinos eventuralmente podr\'\i an ser 
importantes. Estos nuevos canales son cinem\'aticamente
accesibles, al menos para los bosones de Higgs pesados $H,A$ and
$H^\pm$; de hecho, las razones de decaimiento pueden ser muy
grandes e incluso llegar a ser dominantes en algunas regiones del
espacio de par\'ametros del MSSM. Los decaimientos del $h$ a la
part\'{\i}cula supersim\'etrica m\'as ligera (LSP), neutralinos,
son tambi\'en importantes, excediendo el 50\% en algunas partes
del espacio  de par\'ametros. Estos decaimientos afectan
fuertemente las t\'ecnicas de investigaci\'on
experimentales.\\
\begin{figure}[hbt]

\vspace*{-2.5cm}
\hspace*{-4.5cm}
\begin{turn}{-90}%
\epsfxsize=16cm \epsfbox{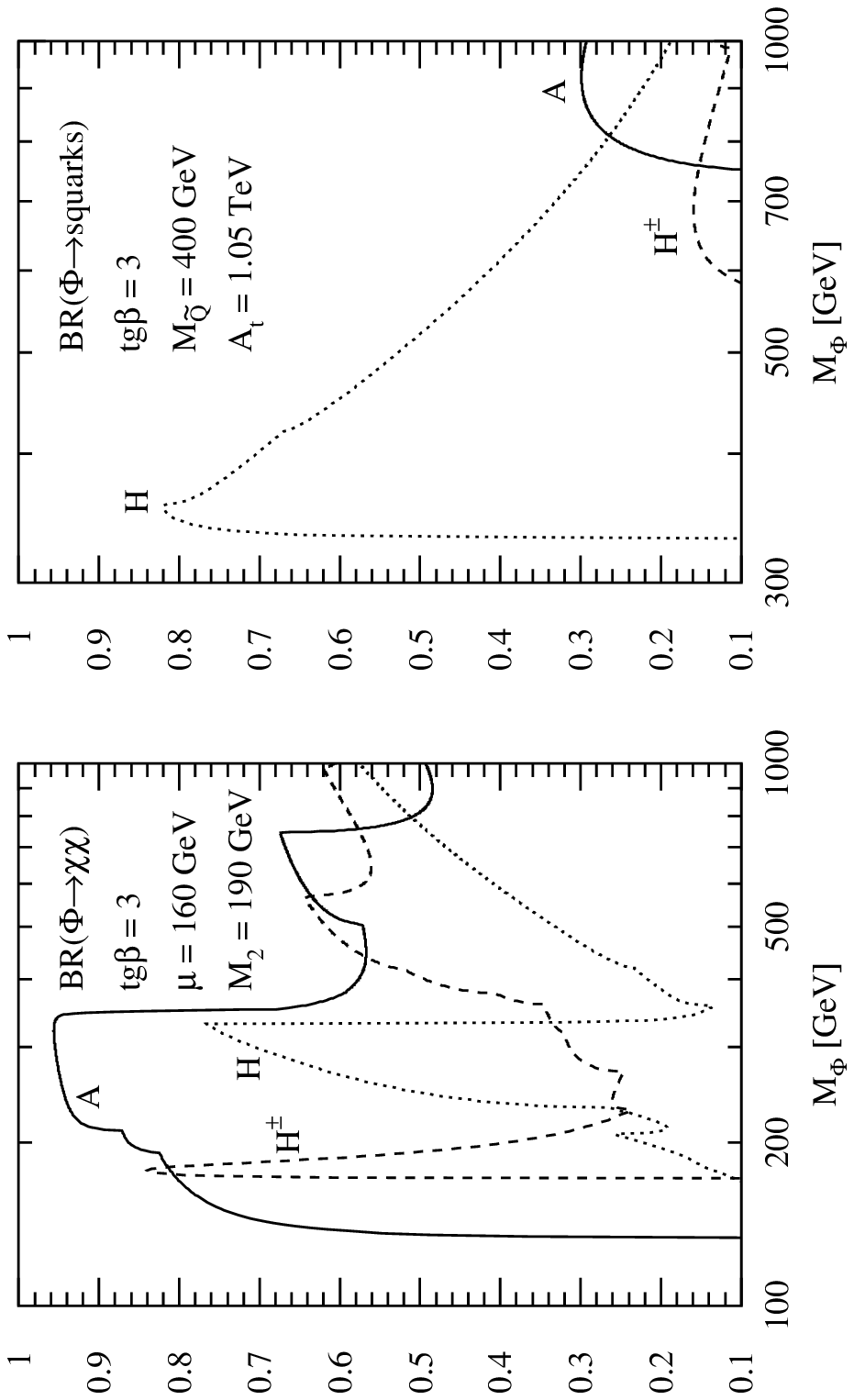}
\end{turn}
\vspace*{-4.2cm}

\caption[]{\label{esp-fg:hcharneutsq} \it Razones de desintegraci\'on de
los bosones de Higgs del MSSM $H,A,H^\pm$ en decaimientos a
charginos/neutralinos y squarks, como funci\'on de las masas del
Higgs para $\tgb=3$. Se han elegido los par\'ametros de mezcla
como $\mu=160$ GeV, $A_t=1.05$ TeV, $A_b=0$ y las masas de los
squarks de las primeras dos generaciones como
$M_{\widetilde{Q}}=400$ GeV. El par\'ametro de masa del gaugino se
toma como $M_2=190$ GeV.}
\end{figure}

Las \underline{\it part\'{\i}culas de Higgs cargadas} decaen a
fermiones, pero tambi\'en, si es permitido cinem\'aticamente, al
Higgs neutro m\'as ligero y al bos\'on $W$. Abajo de los umbrales
de $tb$ y $Wh$, las part\'{\i}culas de Higgs cargadas decaer\'an
primordialmente a pares $\tau \nu_\tau$ y $cs$, siendo dominante
el primero para $\tgb>1$. Para valores grandes de $M_{H^\pm}$, el
modo de decaimiento  top--bottom $H^+\to t\bar b$ se vuelve
dominante. En algunas partes del espacio de par\'ametros de SUSY,
los decaimientos a part\'{\i}culas supersim\'etricas podr\'{\i}an
exceder el 50\%.\\

Sumando los diversos modos de decaimiento, el ancho de todos los
cinco bosones de Higgs se mantiene muy estrecho, siendo del orden
de 10 GeV incluso para masas grandes.

\subsection{Producci\'on de Part\'{\i}culas Higgs de Supersim\'etricas en
  Colisiones $e^+e^-$} 

\phantom{h} La b\'usqueda de los bosones neutros de Higgs
supersim\'etricos en colisionadores lineales \ee  ser\'a una
extensi\'on directa de la b\'usqueda realizada en el LEP2, el cual
cubri\'o el rango de masas hasta los $\sim 100$~GeV para los
bos\'ones neutros de Higgs. Energ\'{\i}as mayores, un exceso en 
$\sqrt{s}$ de $250$~GeV, son requeridas para barrer todo el
espacio de par\'ametros del MSSM con valores desde moderados hasta
grandes de $\tgb$.

\GS Los principales mecanismos de producci\'on de \underline{\it
los bosones neutros de Higgs} en colisionadores \ee \cite{19, 615,
617} son los procesos \Hs y producci\'on de pares asociados,
as\'{\i} como tambi\'en los procesos de fusi\'on:

\begin{eqnarray}
(a) \ \ \mbox{Higgs--strahlung:} \hspace{1.5cm} \epem &
\stackrel{Z}{\longrightarrow} & Z+h/H \hspace{5cm}
\nonumber  \\
(b) \ \ \mbox{{\rm Producci\'on de pares:}} \hspace{8.6mm} \epem
& \stackrel{Z}{\longrightarrow} & A+h/H
\nonumber \\
(c) \ \ \mbox{{\rm Procesos de fusi\'on:}} \hspace{10.7mm} \ \epem
& \stackrel{WW}{\longrightarrow} & \overline{\nu_e} \ \nu_e \ +
h/H
\hspace{3.3cm} \nonumber  \\
\epem &
\stackrel{ZZ}{\longrightarrow} &  \epem + h/H  \nonumber
\end{eqnarray}

El bos\'on de Higgs ${\cal CP}$-impar $A$ no puede producirse en
procesos de fusi\'on a primer orden. Las secciones eficaces para
los cuatro \Hs y procesos de producci\'on de pares pueden ser
expresados como

\begin{eqnarray}
\sigma(\epem \ra Z + h/H) & =& \sin^2/\cos^2(\beta-\alpha) \ \sigma_{SM}
\nonumber \\
\sigma(\epem \ra A + h/H) & =& \cos^2/\sin^2(\beta-\alpha) \
\bar{\lambda} \  \sigma_{SM} ~,
\end{eqnarray}
donde $\sigma_{SM}$ es la secci\'on eficaz en el SM para el \Hs y
el coeficiente $\bar{\lambda} \sim \lambda^{3/2}_{Aj} /
\lambda^{\demi}_{Zj}$ que considera la supresi\'on de onda-$P$
$Ah/H$ en las secciones eficaces cerca del umbral.

\STS Las secciones eficaces para el Higgs-strahlung y para la
producci\'on de pares, tanto como para la producci\'on de los
bosones de Higgs neutros, ligero y pesado, $h$ y $H$, son
complementarios, ya sea que tengan cualquiera de los coeficientes
$\sin^2(\beta-\alpha)$ \'o $\cos^2(\beta-\alpha)$. Como resultado,
ya que $\sigma_{SM}$ es grande, al menos el bos\'on de Higgs
${\cal CP}$-par m\'as ligero debe ser detectado en los experimentos de
$e^+e^-$.

\begin{figure}[hbtp]
\begin{center}
\vspace*{5mm}
\hspace*{5mm}
\epsfig{file=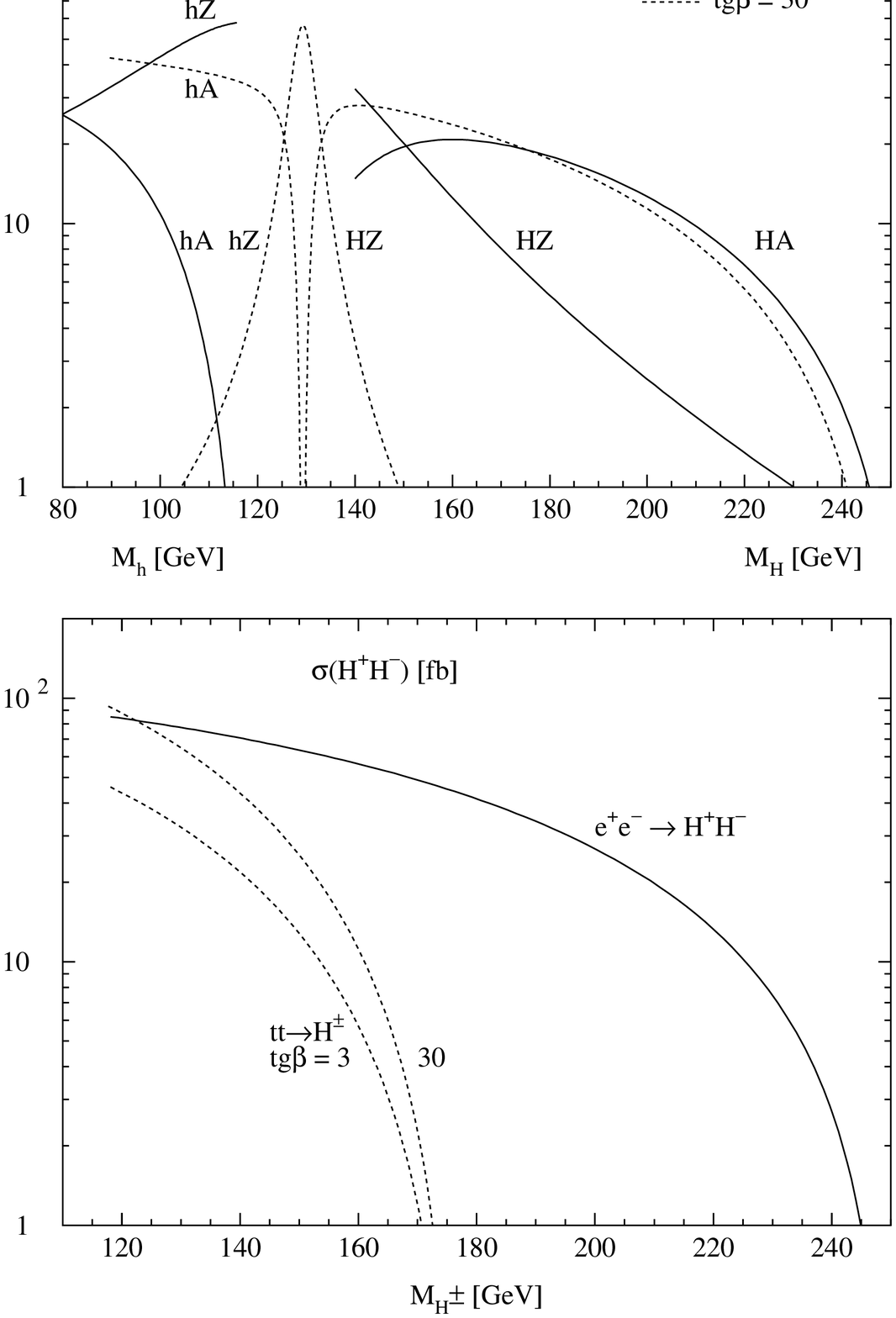,width=15cm}
\end{center}
\vspace{-2.5cm}
\caption[]{\it Secciones eficaces de la producci\'on de bosones de Higgs
  en el MSSM a $\sqrt{s} = 500$~GeV: Higgs-strahlung y producci\'on de
  pares; parte superior: bosones de Higgs neutros; inferior: bosones de Higgs
  cargados.  Ref. \protect\cite{613A}.
  \protect\label{esp-f604}\label{esp-prodcs}}
\end{figure}
\STS Ejemplos representativos de secciones eficaces para
mecanismos de producci\'on de los bosones de Higgs neutros est\'an
dados en la Fig.~\ref{f604}, como funci\'on de las masas de los
Higgs, para $\tgb= 3$ y 30.  La secci\'on eficaz para $hZ$ es
grande para $M_h$ cerca del valor m\'aximo permitido de $\tgb$; es
del orden de 50~fb, correspondiente a $\sim$ 2,500 eventos para
una luminosidad integrada de 50 fb$^{-1}$. En contraste, la
secci\'on eficaz para $HZ$ es grande si $M_h$ est\'a
suficientemente por abajo del valor m\'aximo [implicando que $M_H$ es
peque\~na]. Para $h$ y para una masa ligera de $H$, las se\~nales
consisten de un bos\'on $Z$ acompa\~nado por un par $b\bar{b}$ o
$\tau^+ \tau^-$. Estas se\~nales son f\'aciles de separar de las
se\~nales de fondo (background), el cual proviene principalmente
de la producci\'on de $ZZ$ si la masa del Higgs tiene un valor
cercano a $M_Z$. Para los canales asociados $\epem \to Ah$ y $AH$,
la situaci\'on es opuesta a la anterior: la secci\'on eficaz para
$Ah$ es grande para un $h$ ligero, mientras que la producci\'on
del par $AH$ es el mecanismo dominante en la regi\'on
complementaria para bosones $H$ y $A$ pesados. La suma de las dos
secciones eficaces decrece de $\sim 50$ a 10~fb si $M_A$ aumenta
de $\sim 50$ a 200~GeV en $\sqrt{s} = 500$~GeV. En la mayor parte
del espacio de par\'ametros, las se\~nales consisten de cuatro
quarks $b$ en el estado final, requiriendo que se prevea un
eficiente caracterizaci\'on o etiquetado de $b$-quark. Las
restricciones en la masa ayudar\'an a eliminar las se\~nales de
fondo de estados finales de jets de QCD y $ZZ$. Para el mecanismo
de fusi\'on del $WW$, las secciones eficaces son mayores que para
el mecanismo de Higgs-strahlung, si la masa del Higgs es
moderadamente peque\~na -- menor que 160~GeV en $\sqrt{s} = 500$
GeV. Sin embargo, como el estado final no puede ser completamente
reconstruido, es m\'as dif\'{\i}cil extraer la se\~nal. Como en el
caso de los procesos de \Hs, la producci\'on de bosones de Higgs
ligero $h$ y pesado $H$ tambi\'en se complementan entre ellos en
la fusi\'on de $WW$.

\GS Los \underline{\it bosones de Higgs cargados}, si son m\'as
ligeros que el quark top, pueden producirse en decaimientos del
top, $t \ra b + H^+$, con una raz\'on de desintegraci\'on que
var\'{\i}a entre $2\%$ y $20\%$ en la regi\'on cinem\'aticamente
permitida. Debido a que la secci\'on eficaz para la producci\'on
de top-pares es del orden de 0.5 pb en $\sqrt{s} = 500$~GeV, esto
corresponde a una producci\'on de 1,000 a 10,000 bosones de Higgs
cargados a una luminosidad de 50~fb$^{-1}$. Dado que para $\tgb$
mayores a uno, los bosones de Higgs cargados decaer\'an
principalmente a $\tau \nu_\tau$,  habr\'a un exceso de estados
finales $\tau$ sobre los estados finales $e, \mu$ en decaimientos
del $t$, un rompimiento aparente de la universalidad lept\'onica.
Para masas grandes del Higgs  el modo de decaimiento dominante es
el decaimiento al top  $H^+ \to t \overline{b}$. En este caso las
part\'{\i}culas de Higgs cargadas deben ser producidas a pares en
colisionadores \ee:
\[
              \epem \to H^+H^- ~.
\]
La secci\'on eficaz depende \'unicamente en la masa del Higgs
cargado. Es del orden de 100 fb para masas peque\~nas del Higgs en
$\sqrt{s} = 500$~GeV, pero desciende muy r\'apidamente debido a la
supresi\'on de onda-$P$ $\sim \beta^3$ cerca del umbral. Para
$M_{H^{\pm}} = 230$~GeV, la secci\'on eficaz cae a un nivel de
$\simeq 5\,$~fb. La secci\'on eficaz es considerablemente m\'as
grande para colisiones $\gamma \gamma$.

\GS \noindent
\underline{\bf Estrategias de B\'usqueda Experimental} \\[0.5cm]
Las estrategias de b\'usqueda de los bosones de Higgs cargados y
neutros han sido descritas en la Ref. \cite{13}. La situaci\'on
experimental en su conjunto puede sintetizarse en los siguientes
dos puntos:

\STS \noindent {\bf (i)} La part\'{\i}cula de Higgs ${\cal
CP}$-par m\'as ligera, $h$ puede ser detectada en el rango
completo del espacio de par\'ametros del MSSM, ya sea v\'{\i}a
Higgs-strahlung $\epem \to hZ$ o v\'{\i}a producci\'on $\epem
\to hA$. Esta conclusi\'on se mantiene cierta incluso a
una energ\'{\i}a del c.m. de 250 GeV, independientemente del valor
de la masa de los squarks; es tambi\'en v\'alido si los
decaimientos a neutralinos visibles y otras part\'{\i}culas SUSY
se dan en el sector de Higgs.

\STS \noindent {\bf (ii)} El \'area del espacio de par\'ametros
donde {\it  todos los bososnes de Higgs de SUSY} pueden ser
descubiertos en colisionadores \ee se caracteriza por $M_H, M_A
\lessim \frac{1}{2} \sqrt{s}$, independientemente de $\tgb$. Los
bosones de Higgs $h, H$ pueden producirse ya sea v\'{\i}a \Hs o en
producci\'on asociada de $Ah, AH$; los bososnes de Higgs cargados
se producir\'an en pares
$H^+H^-$. \\

La b\'usqueda del bos\'on de Higgs SUSY neutro m\'as ligero $h$ ha
sido uno de las tareas experimentales m\'as importantes del LEP2.
Valores de la masa del bos\'on pseudoescalar $A$ menor que
alrededor de 90 GeV ya han sido excluidos, independientemente de
$\tgb$, cf.~Fig.~\ref{fig:igo}. En escenarios del MSSM sin efectos
de mezcla, el rango completo de masa para la part\'{\i}cula de
Higgs ligera $h$ ya ha sido cubierta para $\tgb$ menor que
alrededor de 1.6; sin embargo, esta conclusi\'on no se cumple para
escenarios con fuertes efectos de mezcla \cite{mhplot}.
\begin{figure}[hbt]
\begin{center}
\hspace*{-0.3cm}
\epsfig{figure=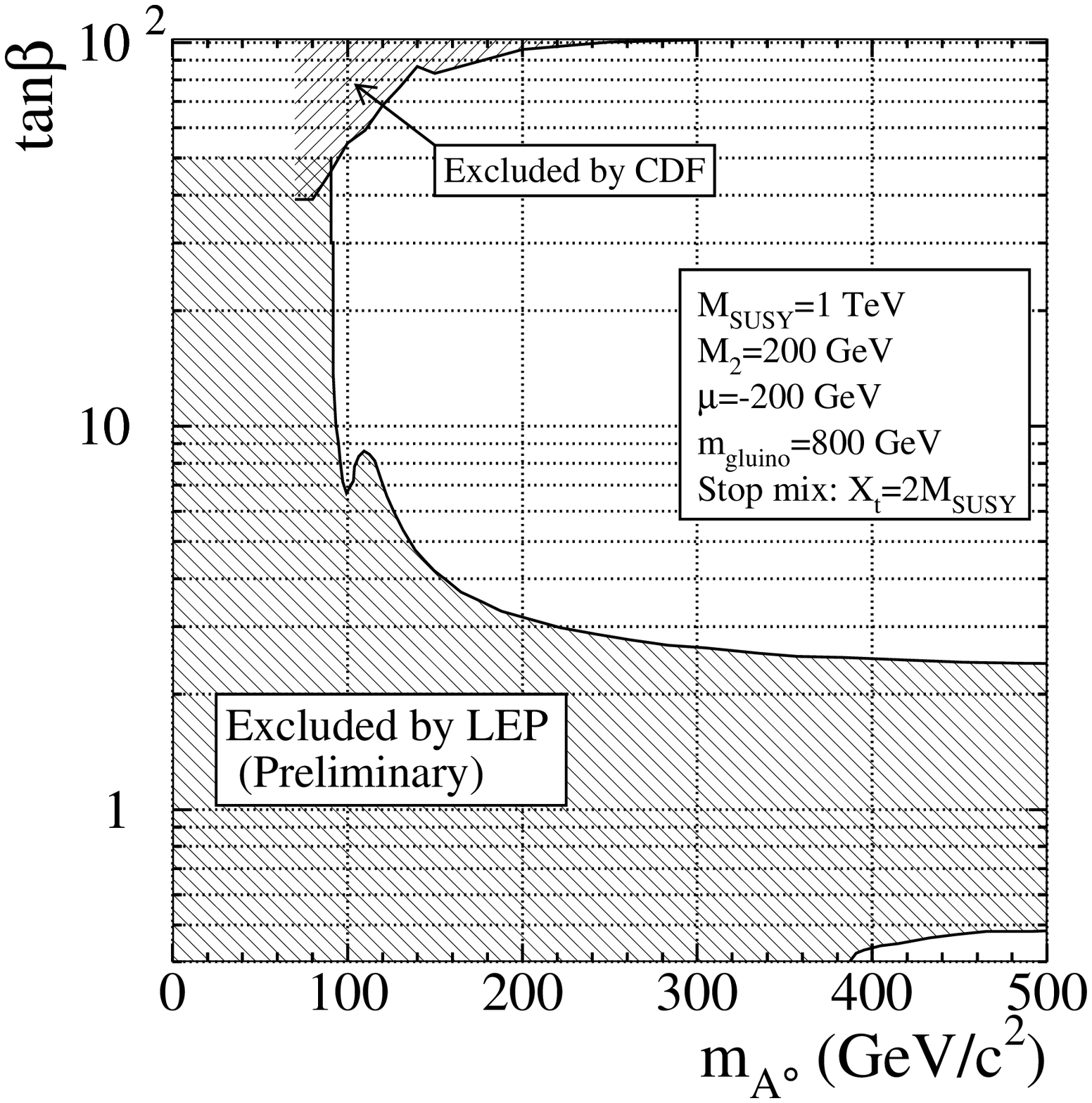,bbllx=16,bblly=5,bburx=531,bbury=524,width=10cm,clip=}
\end{center}
\vspace*{-0.4cm} \caption[]{\label{esp-fig:igo} \it El 95\% de las
cotas CL en $m_A$ y $\tan\beta$ para el escenario de referencia de
$m_h-max$ del LEP \cite{lep}. Se indica tambi\'en la zona excluida
de CDF en $\tan\beta$ grandes. Ref. \cite{cdf}.}
\end{figure}

\subsection{La Producci\'on de Part\'{\i}culas Higgs Supersim\'etricas en Colisiones
Hadr\'onicas}

\phantom{h} Los procesos b\'asicos de producci\'on de part\'{\i}culas de
Higgs en colisionadores hadr\'onicos \cite{24A,32,620B} son
esencialmente los mismos que en el Modelo Est\'andar.  Diferencias
importantes son, no obstante, generadas por los acoplamientos
modificados, el espectro de part\'{\i}culas extendido y la paridad
negativa del bos\'on $A$. Para $\tgb$ grandes el acoplamiento $hb\bar b$
es aumentado de tal manera que el lazo bottom-quark se vuelve
competitivo con respecto al lazo top-quark en el acoplamiento efectivo
$hgg$. M\'as a\'un, los lazos de squarks contribuir\'an a este
acoplamiento.\cite{sqloop}.\\ 

La secci\'on eficaz part\'onica $\sigma(gg\to \Phi)$ para la
fusi\'on de glu\'on de las part\'{\i}culas de Higgs puede ser expresada
por  acoplamientos $g$, en unidades de los correspondientes
acoplamientos del SM y de los factores de forma $A$; al orden m\'as
bajo \cite{32,sqloopqcd}:
\begin{eqnarray}
\hat\sigma^\Phi_{LO} (gg\to \Phi) & = & \sigma^\Phi_0 M_\Phi^2 \times
BW(\hat{s}) \\
\sigma^{h/H}_0 & = & \frac{G_{F}\alpha_{s}^{2}(\mu)}{128 \sqrt{2}\pi} \
\left| \sum_{Q} g_Q^{h/H} A_Q^{h/H} (\tau_{Q})
+ \sum_{\widetilde{Q}} g_{\widetilde{Q}}^{h/H} A_{\widetilde{Q}}^{h/H}
(\tau_{\widetilde{Q}}) \right|^{2} \nonumber \\
\sigma^A_0 & = & \frac{G_{F}\alpha_{s}^{2}(\mu)}{128 \sqrt{2}\pi} \
\left| \sum_{Q} g_Q^A A_Q^A (\tau_{Q}) \right|^{2} \nonumber
\end{eqnarray}
Mientras los acoplamientos de quarks han sido definidos en la
Tabla \ref{tb:hcoup}, los acoplamientos  de las part\'{\i}culas de
Higgs con squarks est\'an dados por
\begin{eqnarray}
g_{\tilde Q_{L,R}}^{h} & = & \frac{M_Q^2}{M_{\tilde Q}^2} g_Q^{h}
\mp \frac{M_Z^2}{M_{\tilde Q}^2} (I_3^Q - e_Q \sin^2 \theta_W)
\sin(\alpha + \beta) \nonumber \\ \nonumber \\
g_{\tilde Q_{L,R}}^{H} & = & \frac{M_Q^2}{M_{\tilde Q}^2} g_Q^{H}
\pm \frac{M_Z^2}{M_{\tilde Q}^2} (I_3^Q - e_Q \sin^2 \theta_W)
\cos(\alpha + \beta)
\end{eqnarray}
S\'olo la no-invariancia de ${\cal CP}$ permite contribuciones
diferentes de cero de los squarks a la producci\'on del bos\'on
pseudoescalar $A$. Los factores de forma pueden ser expresados en
t\'erminos de las funciones de escala $f(\tau_i=4M_i^2/M_\Phi^2)$,
cf. Eq. (\ref{eq:ftau}):
\begin{eqnarray}
A_Q^{h/H} (\tau) & = & \tau [1+(1-\tau) f(\tau)] \nonumber \\
A_Q^A (\tau) & = & \tau f(\tau) \nonumber \\
A_{\tilde Q}^{h/H} (\tau) & = & -\frac{1}{2}\tau [1-\tau f(\tau)] ~.
\end{eqnarray}
Para valores de $\tgb$ peque\~nos la contribuci\'on del lazo del top es
dominante, mientras que para $\tgb$ grandes el lazo del bottom aumenta
dr\'asticamente. Los lazos de squarks pueden ser muy
significativos para masas de squarks menores a $\sim 400$ GeV
\cite{sqloopqcd}.\\ 

Otros mecanismos de producci\'on para bosones de Higgs supersim\'etricos
como fusi\'on de bos\'on vectorial, Higgs-strahlung por los bosones $W,Z$
y Higgs-bremsstrahlung por quarks top y bottom, pueden tratarse en
analog\'{\i}a con los correspondientes procesos del SM.\\
\begin{figure}[hbtp]

\vspace*{0.3cm}
\hspace*{1.0cm}
\begin{turn}{-90}%
\epsfxsize=8.5cm \epsfbox{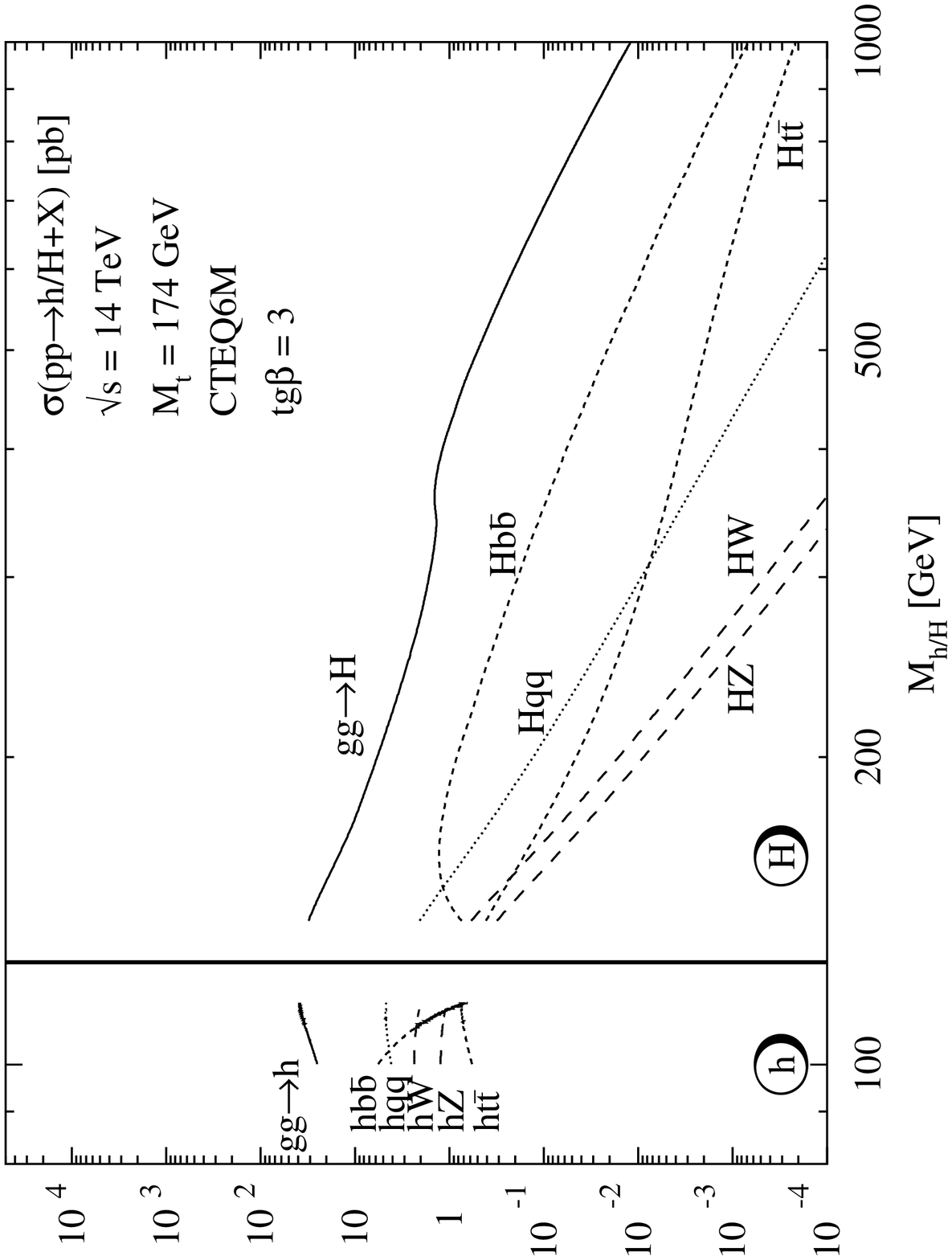}
\end{turn}
\vspace*{0.3cm}

\centerline{\bf Fig.~\ref{fg:mssmprohiggs}a}

\vspace*{0.2cm}
\hspace*{1.0cm}
\begin{turn}{-90}%
\epsfxsize=8.5cm \epsfbox{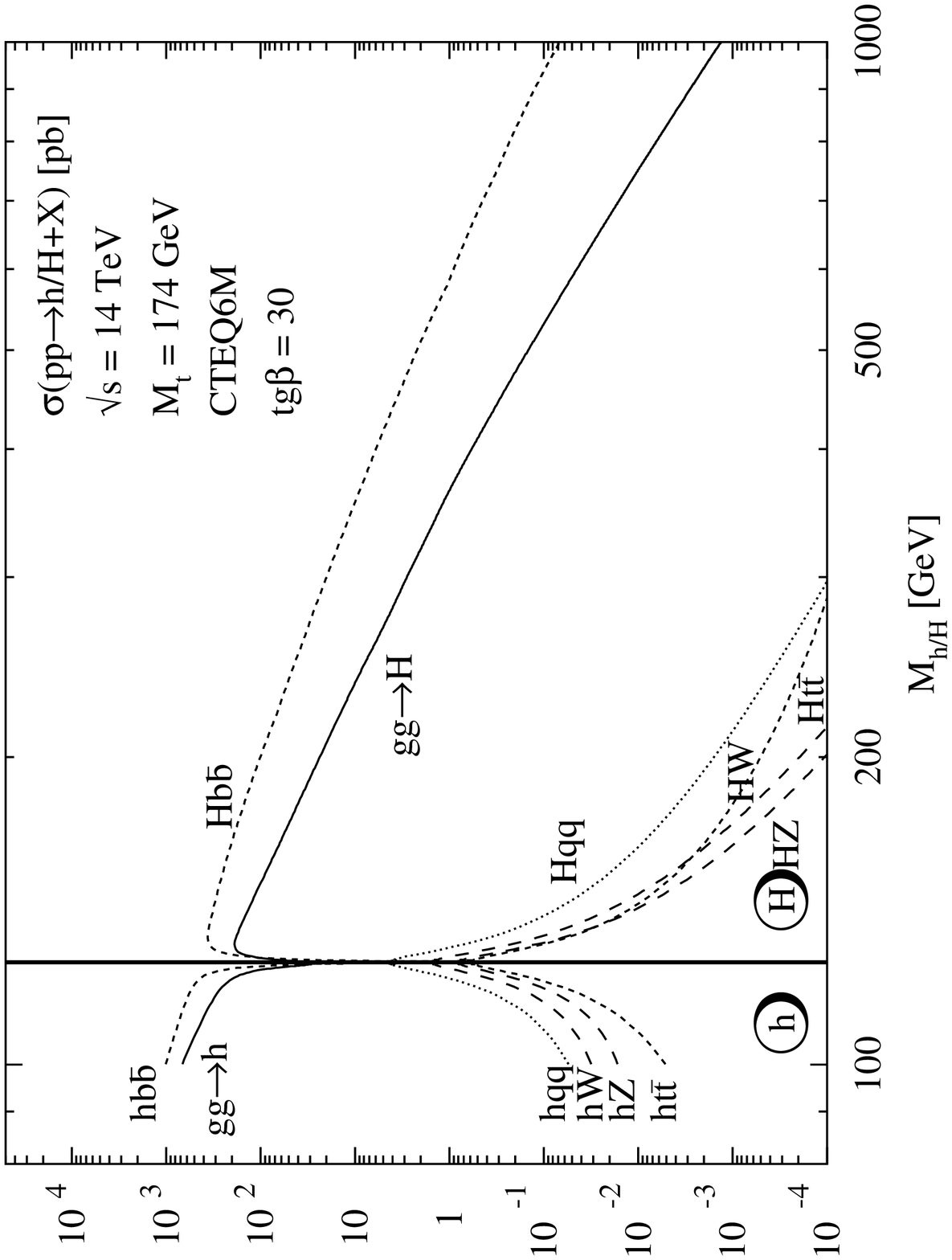}
\end{turn}
\vspace*{0.3cm}

\centerline{\bf Fig.~\ref{fg:mssmprohiggs}b}

\caption[]{\label{esp-fg:mssmprohiggs} \it Secciones eficaces de la
producci\'on del Higgs del MSSM neutro en el LHC para fusi\'on de
glu\'on $gg\to \Phi$, fusi\'on del bos\'on vectorial $qq\to qqVV
\to qqh/ qqH$, Higgs-strahlung $q\bar q\to V^*
\to hV/HV$ y la producci\'on asociada $gg,q\bar q \to b\bar b \Phi/
t\bar t \Phi$, incluyendo todas las correcciones conocidas de QCD.
(a) producci\'on de $h,H$ para $\tgb=3$, (b) producci\'on de $h,H$
para $\tgb=30$, (c) producci\'on de $A$ para $\tgb=3$, (d)
producci\'on de $A$ para $\tgb=30$.}
\end{figure}
\addtocounter{figure}{-1}
\begin{figure}[hbtp]

\vspace*{0.3cm}
\hspace*{1.0cm}
\begin{turn}{-90}%
\epsfxsize=8.5cm \epsfbox{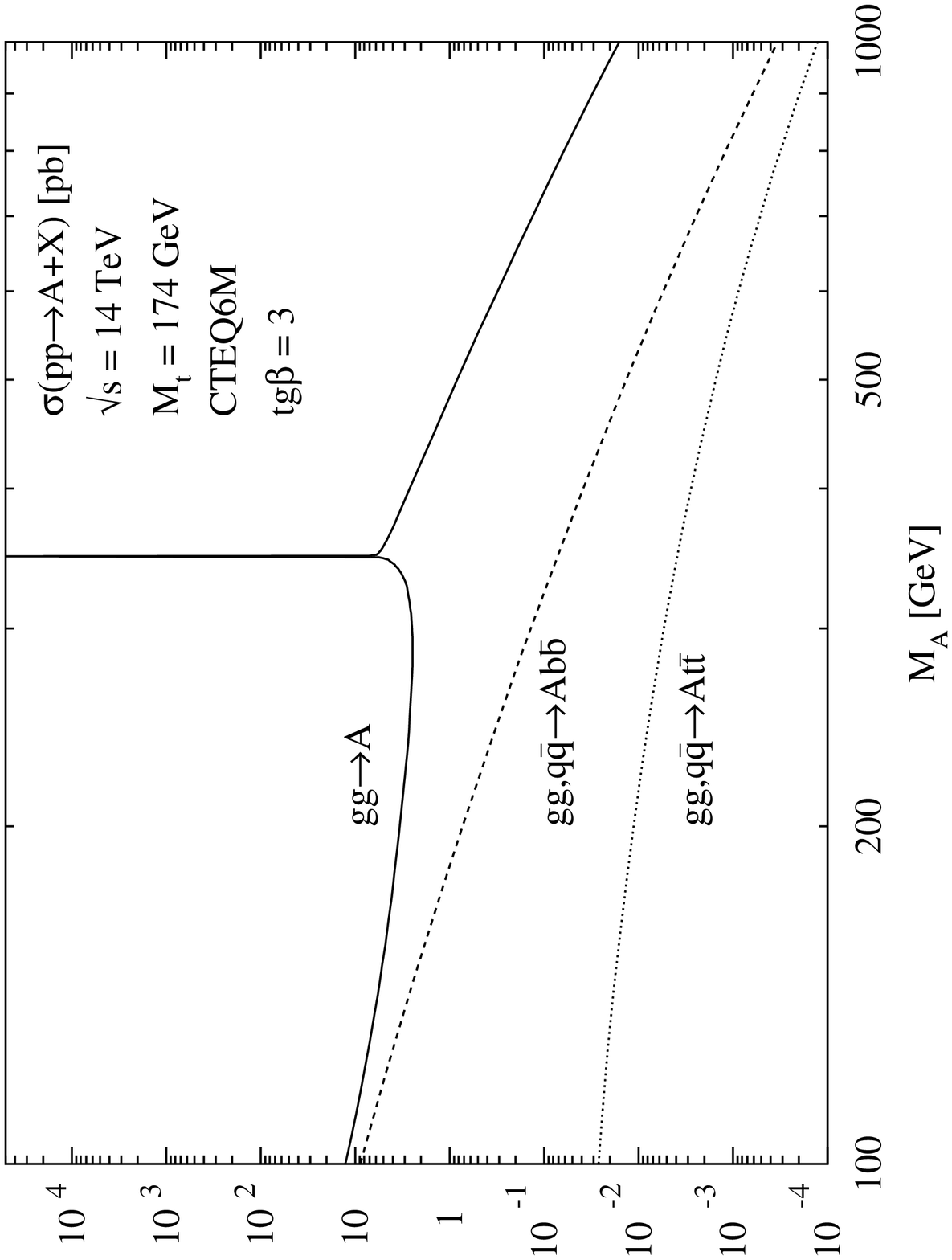}
\end{turn}
\vspace*{0.3cm}

\centerline{\bf Fig.~\ref{fg:mssmprohiggs}c}

\vspace*{0.2cm}
\hspace*{1.0cm}
\begin{turn}{-90}%
\epsfxsize=8.5cm \epsfbox{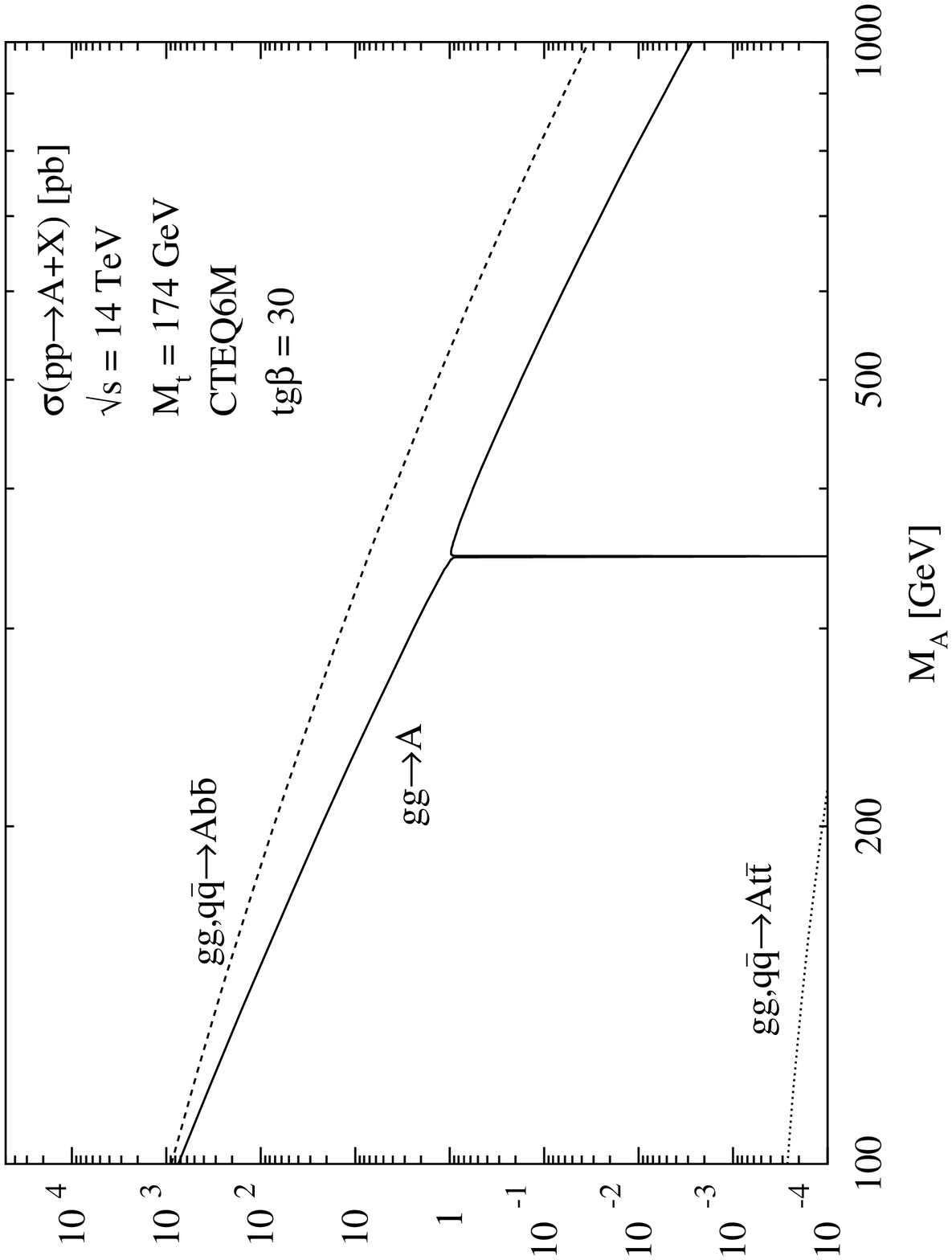}
\end{turn}
\vspace*{0.3cm}

\centerline{\bf Fig.~\ref{fg:mssmprohiggs}d}

\caption[]{\it Continuaci\'on.}
\end{figure}

Los datos extra\'{\i}dos del Tevatron en al canal $p \bar p \to b
\bar b \tau^+ \tau^-$ han sido explotados \cite{63A} para excluir
parte del espacio de par\'ametros supersim\'etrico del Higgs en el
plano $[ M_A, \tgb]$. En el rango interesante de $\tgb$ de entre
30 y 50, las masas para el pseudoescalar $M_A$ de hasta 150 a 190
GeV parecen estar excluidas.\\

Las secciones eficaces de los diversos mecanismos de producci\'on del
Higgs del MSSM en el LHC se muestran en las Figs.
\ref{fg:mssmprohiggs}a--d para dos valores representativos de
$\tgb = 3$ and 30, como funci\'on de la correspondiente masa del
Higgs. Las densidades del part\'on CTEQ6M han sido tomadas con
$\alpha_s(M_Z)=0.118$; las masas del top y del bottom se han
fijado en los valores $M_t=174$ GeV y $M_b=4.62$ GeV. Para el
Higgs bremsstrahlung por los quarks $t,b$, $pp \to Q\bar Q A
+X$, se han usado las densidades dominantes del part\'on 
CTEQ6L1. Para valores peque\~nos y moderados de $\tgb\lessim 10$ a
secci\'on eficaz de la fusi\'on-glu\'on prove\'e la secci\'on
eficaz del producci\'on dominante para la regi\'on completa de
masa del Higgs de hasta $M_\Phi\sim 1$ TeV. Sin embargo, para
$\tgb$ grandes, Higgs bremsstrahlung por quarks bottom, $pp\to
b\bar b \Phi+X$, domina sobre el mecanismo de fusi\'on-glu\'on ya
que los acoplamientos de Yukawa del bottom, en
este caso, aumentan dr\'asticamente.\\

La b\'usqueda del Higgs del MSSM en el LHC ser\'a m\'as complicada que la
b\'usqueda del Higgs del SM.
%
El resumen final se presenta en la Fig.~\ref{fg:atlascms}. Esta
gr\'afica exhibe una regi\'on dif\'{\i}cil para la b\'usqueda del
Higgs del MSSM en el LHC. Para $\tgb \sim 5$ y $M_A \sim 150$ GeV,
es necesaria la luminosidad total y la muestra total de datos de
los experimentos ATLAS y CMS en el LHC para cubrir la regi\'on
problem\'atica de los par\'ametros \cite{richter}. Por otro lado,
si no se encontrara exceso de eventos de Higgs por encima de los
procesos del fondo del SM m\'as all\'a de 2 desviaciones
est\'andar, los bosones de Higgs del MSSM se pueden excluir en un
95\% C.L. A pesar de que se espera que el espacio de par\'ametros
completo del Higgs supersim\'etrico sea finalmente cubierto por
experimentos del LHC, el conjunto total de los bosones de Higgs
individuales es accesible s\'olo en parte del espacio de
par\'ametros. M\'as a\'un, la b\'usqueda de las part\'{\i}culas de
Higgs pesadas $H,A$ es muy
dif\'{\i}cil por el fondo continuo de $t\bar t$ para masas $\gsim
500$ GeV.\\ 

\begin{figure}[hbtp]
\begin{center}
\epsfig{figure=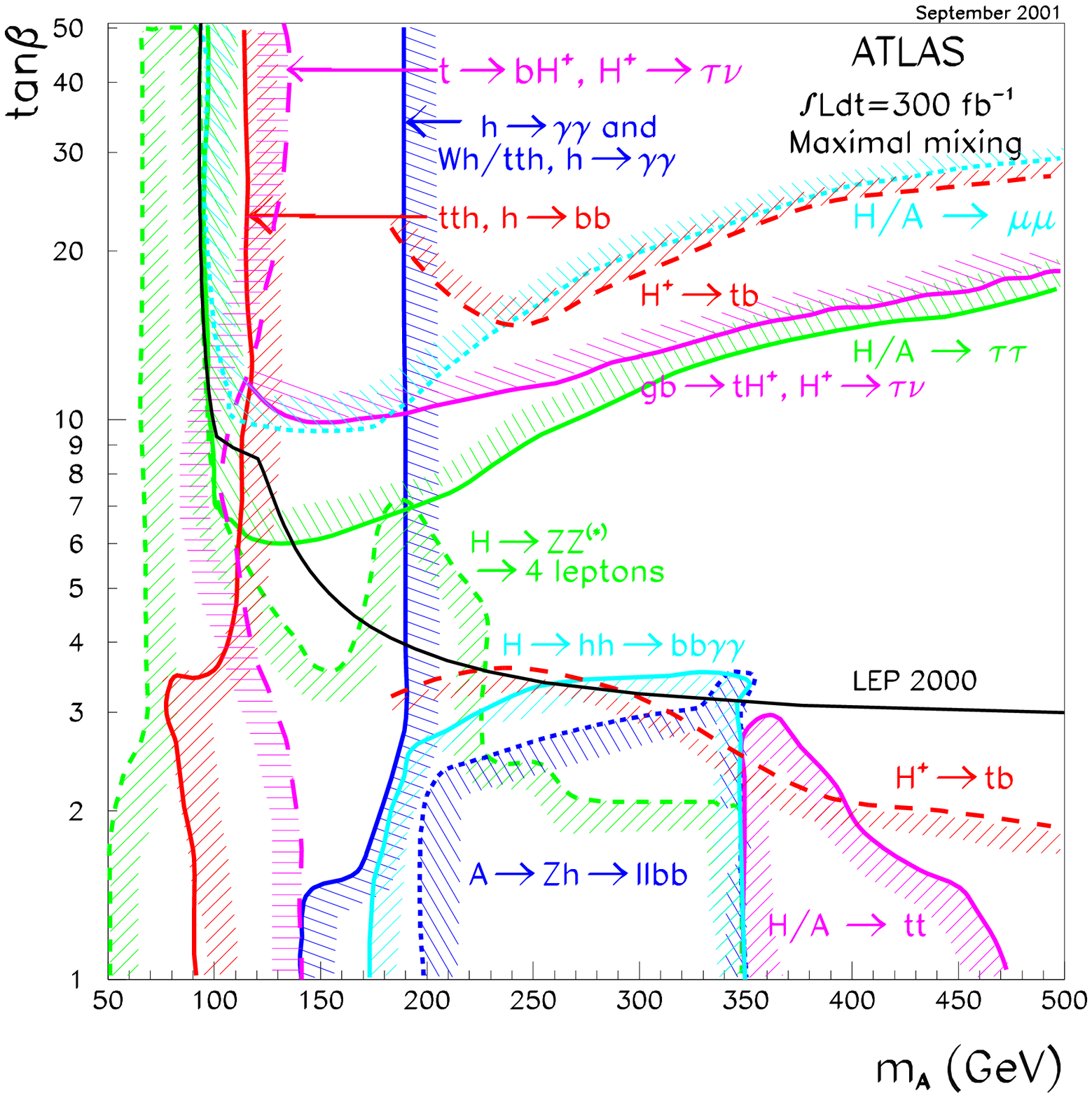,bbllx=0,bblly=0,bburx=523,bbury=504,width=10cm,clip=}
\end{center}
\caption[]{\label{esp-fg:atlascms} \it La sensibilidad de ATLAS 
para el descubrimiento del bos\'on de Higgs del MSSM en el caso de
 mezcla m\'axima. Las curvas de descubrimiento 5$\sigma$ se
muestran en el plano $(\tan\beta,m_A)$ para los canales
individuales y para una luminosidad integrada de 300 fb$^{-1}$.
Tambi\'en se muestra el correspondiente l\'{\i}mite del LEP.
[Tomado de la ref.~\cite{richter}.]}
\end{figure}

\subsection{Midiendo la Paridad de los bosones de Higgs}

\phantom{h} Una vez que los bosones de Higgs sean descubiertos, se
deben establecer las propiedades de las part\'{\i}culas. Aunado a
la reconstrucci\'on del potencial de Higgs supersim\'etrico
\cite{66A}, lo cual ser\'a un esfuerzo muy exigente, deben
establecerse los n\'umeros cu\'anticos externos, en particular la
paridad de las part\'{\i}culas
de Higgs escalar y pseudoescalar $H$ y $A$ \cite{618}.\\[-0.1cm]

Para masas grandes de $H,A$ los decaimientos $H,A\to t\bar t$ a
estados finales del top pueden ser usados para discriminar entre
las diferentes asignaciones de paridad \cite{618}. Por ejemplo,
los bosones $W^+$ y $W^-$ en los decaimientos $t$ y $\bar t$
tienden a emitirse antiparalelamente y paralelamente en el plano
perpendicular el eje  $t\bar t$:
\begin{equation}
\frac{d\Gamma^\pm}{d\phi_*} \propto 1 \mp \left( \frac{\pi}{4} \right)^2
\cos \phi_*
\end{equation}
para decaimientos $H$ y $A$, respectivamente. \\[-0.1cm]

Para masas ligeras de $H,A$, las colisiones $\gamma\gamma$ parecen
proporcionar una soluci\'on viable \cite{618}. La fusi\'on de las
part\'{\i}culas de Higgs en rayos de fotones linealmente
polarizados depende del \'angulo entre los vectores de
polarizaci\'on. Para part\'{\i}culas escalares $0^+$ la amplitud
de producci\'on es diferente de cero para vectores de
polarizaci\'on paralela, mientras que las part\'{\i}culas
pseudoescalares $0^-$ requieren vectores de polarizaci\'on
perpendiculares:
\begin{equation}
{\cal M}(H)^+  \sim  \vec{\epsilon}_1 \cdot \vec{\epsilon}_2
\hspace*{0.5cm} \mbox{y} \hspace*{0.5cm} {\cal M}(A)^-  \sim
\vec{\epsilon}_1 \times \vec{\epsilon}_2 ~.
\end{equation}
El montaje experimental para dispersi\'on hacia atr\'as de Compton
de luz laser se puede ajustar de tal manera que la polarizaci\'on
lineal de los rayos de fotones duros se aproxime a valores
cercanos al 100\%. Dependiendo de la paridad $\pm$ de la
resonancia producida, la asimetr\'{\i}a medida  para fotones de
polarizaci\'on paralela y perpendicular,
\begin{equation}
{\cal A} = \frac{\sigma_\parallel - \sigma_\perp}{\sigma_\parallel +
\sigma_\perp} ~,
\end{equation}
es o positiva o negativa.

\newpage
\subsection{Extensiones Supersim\'etricas No-M\'{\i}nimas}

\phantom{h} La extensi\'on supersim\'etrica m\'{\i}nima  del
Modelo Est\'andar puede parecer muy restrictiva para
teor\'{\i}as supersim\'etricas en general, en particular en el sector de
Higgs donde los acoplamientos cu\'articos se identifican con los
acoplamientos de norma. Sin embargo, resulta que el patr\'on de
masas del MSSM es bastante representativo si la teor\'{\i}a se
considera v\'alida hasta la escala de GUT -- la motivaci\'on para 
supersimetr\'{\i}a {\it sui generis}. Este patr\'on general ha
sido estudiado concienzudamente dentro de la extensi\'on que sigue
a la m\'{\i}nima (next-to-minimal extensi\'on): el MSSM, que
incorpora dos isodobletes de Higgs se extiende al introducir un
campo isosinglete adicional $N$. Esta extensi\'on lleva a un
modelo \citer{621,70A} que es generalmente referido como NMSSM.

\STS El singlete de Higgs adicional puede resolver el 
llamado problema-$\mu$ ($\mu$-problem) [i.e. $\mu \sim$
orden de $M_W$] al eliminar el par\'ametro del higgsino $\mu$ del 
potencial y reemplazarlo por el valor esperado (expectation value) del
vac\'{\i}o del campo $N$, el cual puede ser naturalmente
relacionado con los valores esperados del vac\'{\i}o
usuales de los campos isodobletes de Higgs. En este escenario el
superpotencial involucra a los dos acoplamientos
trilineales $H_1 H_2 N$ y $N^3$.  Las consecuencias de este sector
de Higgs extendido se destacar\'an en el contexto de (s)gran
unificaci\'on, incluyendo los t\'erminos de rompimiento suave
universales de la supersimetr\'{\i}a. \cite{622,70A}.\\

\GS El espectro de Higgs del NMSSM incluye, adem\'as del conjunto
m\'{\i}nimo de part\'{\i}culas de Higgs, una part\'{\i}cula
adicional de Higgs escalar y una pseudoescalar. Las
part\'{\i}culas de Higgs neutras son en general mezclas de
isodobletes que se acoplan a los bosones $W, Z$ y a los fermiones;
y del isosinglete, desacoplado del sector que no es de Higgs.
Las auto-interacciones trilineales contribuyen a las masas de las
part\'{\i}culas de Higgs; para el bos\'on de Higgs m\'as ligero de
cada especie:
\begin{eqnarray}
M^2 (h_1) & \leq & M^2_Z \cos^2 2\beta + \lambda^2 v^2 \sin^2 2 \beta \\
M^2 (A_1) & \leq & M^2 (A)   \nonumber \\
M^2 (H^{\pm}) & \leq & M^2 (W) + M^2 (A) - \lambda^2 v^2 \nonumber
\end{eqnarray}
En contraste con el modelo m\'{\i}nimo, la masa de la
part\'{\i}cula de Higgs cargada podr\'{\i}a ser m\'as peque\~na
que la masa del \W. Un ejemplo del espectro de masas se muestra en
la Fig.~\ref{fig:26}. Considerando que los acoplamientos
trilineales aumentan con la energ\'{\i}a, se pueden derivar cotas
superiores para la masa del bos\'on de Higgs m\'as ligero,
$h_1^0$, en analog\'{\i}a con el Modelo Est\'andar, y bajo la
suposici\'on de que la teor\'{\i}a es v\'alida hasta la escala de
GUT: $m(h_1^0) \lessim 140 $~GeV. As\'{\i}, a pesar de las
interacciones adicionales, el patr\'on distintivo de la extensi\'on
m\'{\i}nima permanece v\'alido incluso en escenarios
supersim\'etricos m\'as complicados. De hecho, la cota de la masa
de 140~GeV para la part\'{\i}cula de Higgs m\'as ligera es v\'alido en
casi todas las teor\'{\i}as supersim\'etricas \cite{623}. Si $h_1^0$ es
(casi)puramente isosinglete, se desacopla del sistema de bos\'on de norma y
fermiones y su papel lo toma la part\'{\i}cula de
Higgs siguiente con una componente de isodoblete grande,
implicando, nuevamente, la
validez de la cota de la masa.\\

\begin{figure}[hbt]
\begin{center}
\hspace*{-0.3cm}
\epsfig{figure=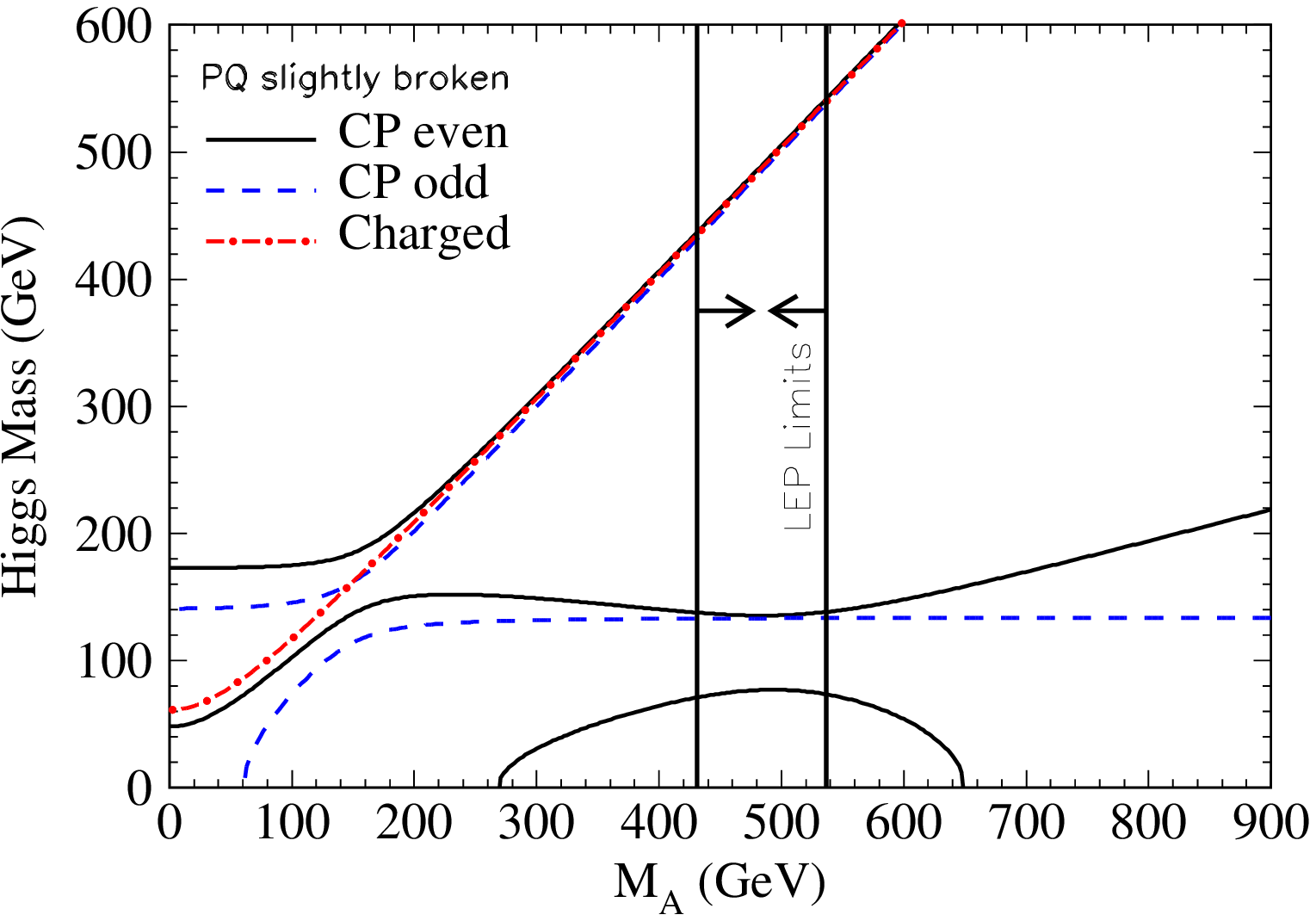,bbllx=0,bblly=14,bburx=439,bbury=320,width=10cm,clip=}
\end{center}
\vspace*{-0.4cm} \caption[]{\label{esp-fig:26} \it Las masas del
bos\'on de Higgs a un lazo como funci\'on de $M_A$ para
$\lambda=0.3$, $\kappa=0.1$, $v_s=3v$, $\tan\beta=3$ y
$A_\kappa=-100$~GeV. Las flechas denotan la region permitida por
las b\'usquedas del LEP con un 95\% de confianza. Ref.
\cite{70A}.}
\end{figure}

\STS Si la part\'{\i}cula de Higgs $h_1^0$ es primordialmente
isosinglete, el acoplamiento $ZZh_1^0$ es peque\~no y la part\'{\i}cula
no puede producirse por Higgs-strahlung. Sin embargo, en este caso,
$h_2^0$ es generalmente ligero y se
acopla con suficiente intensidad al bos\'on $Z$; de otra manera, $h_3^0$
juega este papel.\\ 

\STS {\it En suma}. Experimentos en colisionadores $e^+e^-$ no
est\'an en una situaci\'on de fracaso \cite{L141A} en la
detecci\'on de las part\'{\i}culas de Higgs en teor\'{\i}as
supersim\'etricas generales, incluso para energ\'{\i}as del c.m.
tan bajas como $\sqrt{s} \sim 300$ GeV.


\section{Rompimiento Din\'amico de la Simetr\'{\i}a}

\phantom{h}
\subsection{Modelos de Higgs Peque\~no}

\phantom{h} La interpretaci\'on del bos\'on de Higgs commo un
bos\'on (pseudo-)Goldstone ha sido una idea muy atractiva durante
mucho tiempo. El inter\'es en esta representaci\'on ha resurgido
dentro de escenarios de Higgs peque\~no \cite{2A} que
recientemente se han desarrollado para generar din\'amicamente el
rompimiento de la
simetr\'{\i}a electrod\'ebil por medio de nuevas interacciones fuertes.\\

Los modelos de Higgs peque\~no est\'an basados en un complejo
sistema de simetr\'{\i}as y mecanismos de rompimiento de
simetr\'{\i}a, para consultar una revisi\'on reciente ver
\cite{littlest}. Tres puntos son centrales para realizar
la idea:
\begin{itemize}
\item[(i)]  El campo de Higgs es un campo de Goldstone asociado con el
 rompimiento de una simetr\'{\i}a global $G$ a una escala de energ\'{\i}as
del orden de $\Lambda_s \sim 4 \pi f \sim$ 10 a 30 TeV, con $f$
caracterizando la escala del par\'ametro de rompimiento de
simetr\'{\i}a;
\item[(ii)] En el mismo paso, la simetr\'{\i}a de norma $G_0 \subset G$ se rompe
obteni\'endose el grupo de norma del Modelo Est\'andar $SU(2)
\times U(1)$, gener\'andose las masas para los bosones vectoriales
pesados y fermiones las cuales cancelan las divergencias
cuadr\'aticas est\'andar en las correcciones radiativas a la masa del
bos\'on de Higgs ligero. Como las masas de estas nuevas
part\'{\i}culas son generadas por el rompimiento de la
simetr\'{\i}a de norma $G_0$ son del tama\~no intermedio $M \sim g
f \sim 1$  a 3 TeV;
\item[(iii)] Los bosones Higgs adquieren finalmente una masa por 
  correcciones radiativas a la escala electrod\'ebil est\'andar del orden
  de $v \sim g^2 f / 4 \pi \sim$ 100 a 300 GeV.
\end{itemize}

Por lo tanto, en este modelo se encuentran tres escalas
caracter\'{\i}sticas: la escala de interacci\'on fuerte
$\Lambda_s$, la nueva escala de  masa $M$ y la escala de
rompimiento electrod\'ebil $v$, ordenadas en una cadena
jer\'arquica $\Lambda_s \gg M \gg v$. La masa del bos\'on de Higgs
ligero est\'a protegida a un valor peque\~no al requerir el
rompimiento colectivo de  dos simetr\'{\i}as. En contraste a la
simetr\'{\i}a bos\'on-fermi\'on que cancela las divergencias
cuadr\'aticas en supersimetr\'{\i}a, en los modelos de Higgs
peque\~no la cancelaci\'on opera individualmente en los sectores
bos\'onico y fermi\'onico, siendo asegurada por
las simetr\'{\i}as entre los acoplamientos de los campos
del SM y los nuevos campos con los campos de Higgs.\\

\noindent
\underline{Ejemplo: Modelo del Higgs M\'as Peque\~no}\\

\noindent Un ejemplo interesante  en donde estas ideas se llevan a
cabo es proporcionado por el ``Modelo del Higgs M\'as Peque\~no''
\cite{91A,91B}. El modelo est\'a formulado como un modelo de sigma
no lineal con un grupo de simetr\'{\i}a global $SU(5)$. Este grupo
se descompone a $SO(5)$ por el valor esperado del
vac\'{\i}o distinto de cero \beq \Sigma_0 =
crossdiag[{\scriptstyle '}\!\mathbb{I},1, {\scriptstyle
'}\!\mathbb{I}] \eeq del campo $\Sigma$. 
Suponiendo que el
subgrupo $[SU(2) \times U(1)]^2$ sea normado (gauged), el
rompimiento de la simetr\'{\i}a global conduce tambi\'en al
rompimiento de su grupo de norma descomponi\'endolo al grupo
$[SU(2) \times U(1)]$. El rompimiento de la simetr\'{\i}a global
genera $24 - 10 = 14$ bosones de Goldstone, cuatro de los cuales
son absorbidos por los bosones de norma asociados con el grupo de
norma roto. Los 10 bosones de Goldstone restantes, incorporados
en el campo $\Sigma$ \beq \Sigma = exp[2i\Pi/f]: \quad \Pi =
\left|\left|
\begin{array}{ccc} 0 & h^\dagger/\sqrt{2} &
\varphi^\dagger \\
h/\sqrt{2} & 0 & h^*/\sqrt{2} \\
\varphi & h^{\mathrm{T}}/\sqrt{2} & 0
\end{array} \right|\right|
\eeq son identificados como un iso-doblete $h$ que se convertir\'a
en el campo de Higgs ligero del Modelo Est\'andar y un triplete de
Higgs $\varphi$ que adquirir\'a una masa de orden $M$.\\

Los principios de construcci\'on fundamentales del modelo
deber\'an ser ilustrados al analizar cualitativamente los sectores
de norma y de Higgs. El sector del top, extendido por un nuevo
doblete pesado $[T_L, T_R]$, puede tratarse de un modo similar
despu\'es de introducir las interacciones apropiadas top-Higgs.\\

\noindent
\underline{Sector del Bos\'on Vectorial}\\

\noindent Al insertar los campos de norma $[SU(2) \times U(1)]^2$ en la
Lagrangiana de sigma, 
\beq {\cal L} = \frac{1}{2} \frac{f^2}{4}
\mathrm{Tr} | {\cal D}_\mu \Sigma |^2 \eeq 
con 
\beq {\cal D}_\mu \Sigma = \partial_\mu
\Sigma - i \sum_{j=1}^2 [ g_j (W_j \Sigma + \Sigma W_j^\mathrm{T}) + \{U(1)\} ] \eeq
los cuatro bosones vectoriales de la simetr\'{\i}a de norma rota
$[SU(2) \times U(1)]$ adquieren masa 
\beq M[W_H,Z_H,A_H] \sim g f \eeq
donde $W_H$ etc.,  denotan los campos de norma $W,Z$ y el fot\'on pesado.\\

Destaca el hecho de que el bos\'on de norma $W_H$ se acopla con el
signo opuesto al cuadrado del bos\'on de Higgs ligero comparado
con los bosones est\'andar W: \beq
{\cal L} &=& + \frac{g^2}{4} W^2 \,\mathrm{Tr} h^\dagger h \nonumber \\
&& - \frac{g^2}{4} W'^2 \,\mathrm{Tr} h^\dagger h + ... \eeq

Es por esto que las divergencias cuadr\'aticas de los dos
diagramas de lazo cerrados W y W' que acompa\~nan al campo de Higgs
ligero, se cancelan entre ellos  y, similarmente a los grados de
libertad supersim\'etricos, los bosones vectoriales nuevos
deber\'an tener masas que no excedan de 1 a 3 TeV para evitar
ajustes finos excesivos.\\

Hasta este punto, los bosones de norma del Modelo Est\'andar se matienen
sin masas; adquieren masa despu\'es de que el mecanismo de
rompimiento electrod\'ebil est\'andar se pone en operaci\'on.\\

\noindent
\underline{Sector de Higgs}\\

\noindent Hasta este nivel en la evoluci\'on de la teor\'{\i}a,
las simetr\'{\i}as globales impiden un potencial de Higgs
diferente de cero. S\'olo si las correcciones radiativas  se ponen
de manifiesto, entonces el mecanismo de Coleman-Weinberg genera el
potencial de Higgs que proporciona masas a los bosones de Higgs y
rompe la simetr\'{\i}a de norma del Modelo Est\'andar.\\

D\'andole al potencial de Higgs la forma 
\beq V = m_\varphi^2
\,\mathrm{Tr} \varphi^\dagger \varphi - \mu^2 h h^\dagger +
\lambda_4 (h h^\dagger)^2 \eeq 
el primer t\'ermino proporciona una
masa distinta de cero al bos\'on de Higgs $\varphi$, mientras que los
dos siguientes t\'erminos son responsables del rompimiento de la
simetr\'{\i}a en el sector de norma
del Modelo Est\'andar. \\

\noindent -- Truncando las contribuciones al potencial de
Coleman-Weinberg cuadr\'aticamente divergentes en $\Lambda_s$, las
masas al cuadardo de los [ahora] bosones de pseudo-Goldstone
$\varphi$ son del orden de \beq m_\varphi^2 \sim g^2 (\Lambda_s /
4 \pi )^2 \sim g^2 f^2~. \eeq 
As\'\i{} que los bosones de Higgs
pesados adquieren masas del orden
de los bosones vectoriales pesados.\\

\noindent -- El acoplamiento cu\'artico  del bos\'on de Higgs
ligero es del orden de $g^2$. Sin embargo, el coeficiente $\mu^2$
recibe \'unicamente contribuciones de las partes de un-lazo
logar\'{\i}tmicamente divergente y de dos-lazos cuadr\'aticamente
divergentes en el potencial de Coleman-Weinberg: \beq \mu^2 =
\mu_1^2 + \mu_2^2 : && \mu_1^2 \sim (\Lambda_s/4 \pi)^2 \log
\left(\Lambda_s^2/f^2\right)/16\pi^2
\sim f^2 \log \left(\Lambda_s^2/f^2\right)/16\pi^2 \nonumber\\
&& \mu_2^2 \sim \Lambda_s^2/(16\pi^2)^2 \sim f^2/16\pi^2 \eeq
Ambas contribuciones son naturalmente del orden de $f/4\pi$, i.e.
son  un orden de magnitud menor que la escala intermedia $M$ del
Higgs pesado y las masas vectoriales.\\

Resumiendo. Un bos\'on de Higgs ligero con masa del orden de 100
GeV puede ser generado en modelos de Higgs Peque\~no como un
bos\'on pseudo-Goldstone, y la masa peque\~na est\'a protegida
contra correcciones radiativas grandes individualmente en los
sectores bos\'onico y fermi\'onico.
\\

\noindent
\underline{Fenomenolog\'{\i}a}\\

\noindent De tales escenarios surgen muchas predicciones que
pueden verificarse experimentalmente.

Lo m\'as importante, el espectro de los nuevos bosones vectoriales
pesados y fermiones deber\'an observarse con masas en el rango
intermedio de 1 a algunos TeV en el LHC o en colisionadores
lineales $e^+e^-$ TeV/multi-TeV.

Sin embargo, el modelo puede ya ser verificado analizando los
datos de precisi\'on existentes extraidos del LEP y otros. El
impacto de los nuevos grados de libertad en los modelos de Higgs
Peque\~no debe mantenerse lo suficientemente reducido para no
arruinar el \'exito de las correcciones radiativas al incluir
s\'olo el bos\'on de Higgs ligero en la descripci\'on de los
datos. Esto lleva a restringir el par\'ametro $f$ a un orden de 3
a 5 TeV, Fig.~\ref{fig:kilian}. As\'{\i} la teor\'{\i}a es
compatible con los datos de precisi\'on actuales, pero solo
marginalmente y parece ya estar restringida.

\begin{figure}[hbt]
\begin{center}
\epsfig{figure=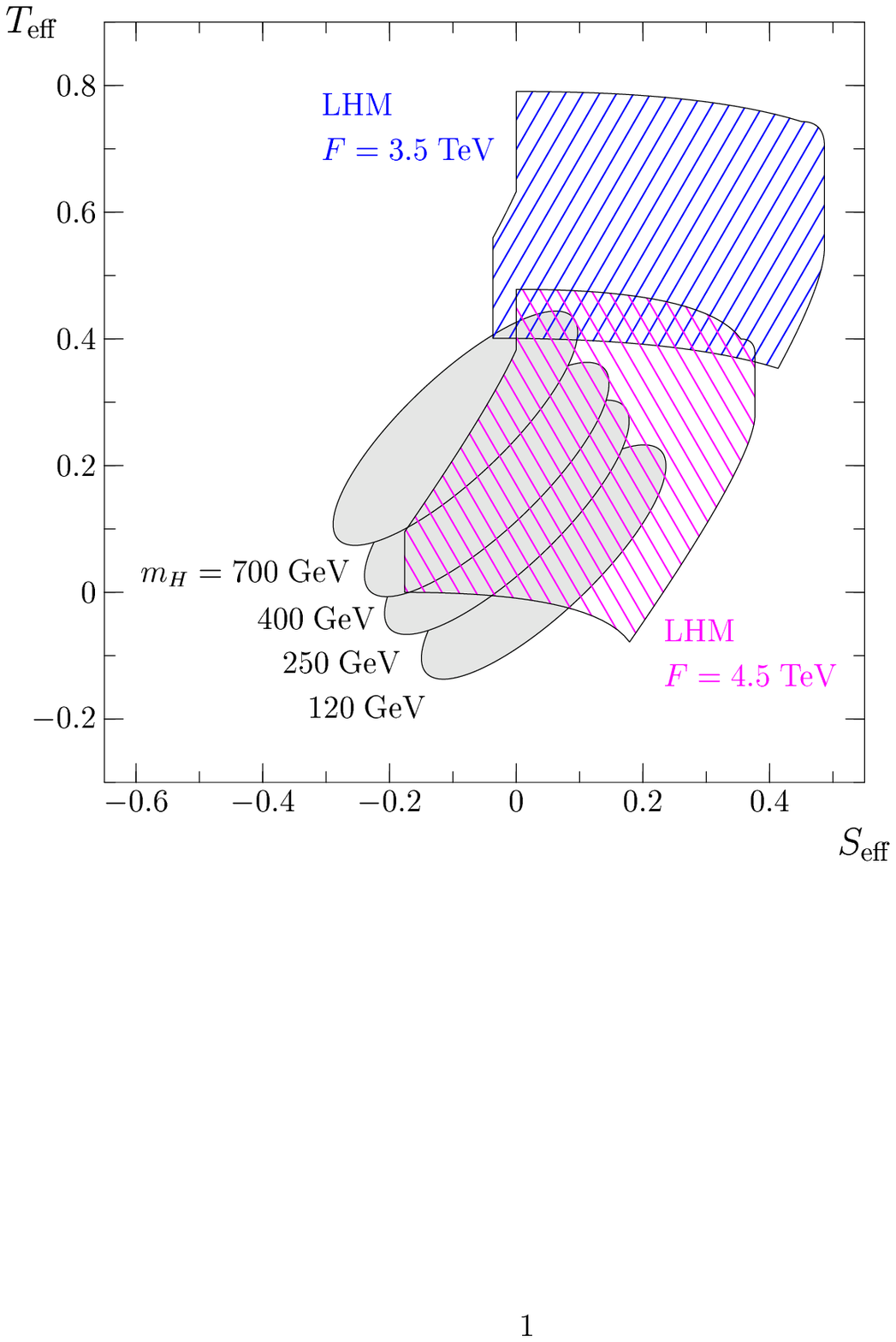,bbllx=114,bblly=296,bburx=451,bbury=620,width=10cm,clip=}
\end{center}
\vspace*{-0.4cm}

\caption[]{\label{esp-fig:kilian} \it Predicciones  de los
par\'ametros de precisi\'on $S,T$ para el modelo del Higgs M\'as
Peque\~no con asignaciones de carga est\'andares $U(1)$. Las
elipses sombreadas son el 68 \% de los contornos de exclusi\'on
los cuales provienen de los datos de precisi\'on electrod\'ebiles,
suponiendo cuatro masas diferentes de los Higgs. Las \'areas
rayadas son los rangos de los par\'ametros permitidos del modelo
del Higgs M\'as Peque\~no para dos valores diferentes de la escala
$F$. Los l\'{\i}mites de las interacciones de  contacto se han
tomado en cuenta. Ref. \cite{kilfigure}.}
\end{figure}

\subsection{Bosones $W$ Fuertemente Interactuantes}

\phantom{h} El mecanismo de Higgs est\'a basado en el concepto te\'orico
del rompimiento espont\'aneo de la simetr\'{\i}a \cite{1}. En la
formulaci\'on can\'onica, adoptada por el Modelo Est\'andar, un campo
escalar {\it fundamental} de cuatro componentes se introduce, el cual es
dotado con una auto-interacci\'on tal que el campo adquiere un
valor del estado base diferente de cero. La direcci\'on espec\'{\i}fica
en el isoespacio, el cual se distingue por la soluci\'on del estado
base, rompe la invariancia del isoesp\'\i n  de la interacci\'on
espont\'aneamente\footnote{ Mantenemos el lenguaje com\'un tambi\'en en
  el contexto de teor\'\i as de norma, a\'un cuando la simetr\'\i a de norma no
  est\'a rota en el sentido estricto}. La interacci\'on de
los campos de norma con el campo escalar en el estado base, genera las
masas de estos campos. Los grados de libertad longitudinales de los
campos de norma se construyen al absorber los modos Goldstone, los
cuales est\'an asociados con el rompimiento espont\'aneo de las
simetr\'{\i}as electrod\'ebiles en el sector del campo escalar.  Los
fermiones adquieren masa a trav\'es de las interacciones de Yukawa con
el estado base del campo. Mientras tres de las componentes escalares son
absorbidas por los campos de norma, un grado de libertad se manifiesta
como una part\'{\i}cula f\'\i sica, el bos\'on de Higgs. El
intercambio de esta part\'{\i}cula en las amplitudes de dispersi\'on,
incluyendo campos de norma longitudinales y campos de fermiones masivos,
garantiza la unitaridad de la teor\'\i a hasta energ\'{\i}as asint\'oticas.

En la alternativa a este escenario basado en el campo fundamental
de Higgs, el rompimiento espont\'aneo de la simetr\'{\i}a es
generado {\it din\'amicamente } \cite{2}. Un sistema de nuevos
fermiones es introducido el cual interact\'ua fuertemente a la
escala del orden de 1 TeV. En el estado base de dicho sistema un
condensado escalar de pares fermi\'on-antifermi\'on podr\'{\i}a
formarse. Generalmente se espera que tal proceso se lleve a cabo
en cualquier teor\'{\i}a de norma no Abeliana de las nuevas
interacciones fuertes [y que se d\'e en QCD, por ejemplo]. Como el
condensado escalar rompe  la simetr\'{\i}a quiral del sistema
fermi\'onico, se formar\'an campos de Goldstone, y estos
pueden ser absorbidos por los campos de norma electrod\'ebiles
para construir las componentes longitudinales y las masas de los
campos de norma. Nuevas interacciones de norma deben ser
introducidas, las cuales acoplan los leptones y los quarks del
Modelo Est\'andar a los nuevos fermiones de tal manera que se
generen las masas de los leptones y quarks a trav\'es de las
interacciones con el estado base del condesado
fermi\'on-antifermi\'on. En el sector de bajas energ\'{\i}as de la
teor\'{\i}a electrod\'ebil, la aproximaci\'on del campo de Higgs
fundamental y la alternativa din\'amica son equivalentes. Sin
embargo, las dos teor\'{\i}as son fundamentalmente diferentes a
energ\'{\i}as altas. Mientras que la unitaridad  de la
teor\'{\i}a de norma electrod\'ebil se garantiza por el
intercambio de un part\'{\i}cula de Higgs escalar en procesos de
dispersi\'on, la unitaridad se restablece a altas energ\'{\i}as en
la teor\'{\i}a din\'amica a trav\'es de interacciones fuertes no
perturbativas entre las part\'{\i}culas. Como las componentes de
los campos de norma longitudinales son equivalentes a los campos
de Goldstone asociados con la teor\'{\i}a microsc\'opica, sus
interacciones fuertes a altas energ\'{\i}as son transferidas a los
bos\'ones de norma electrod\'ebiles. Puesto que por unitaridad, la
amplitud de dispersi\'on de la onda $S$ de los bosones $W, Z$
polarizados longitudinalmente en el canal isoescalar $(2W^+W^- +
ZZ) / \sqrt{3}$, $a^0_0 = \sqrt{2} G_F s/ 16 \pi$, est\'a acotado
por 1/2, la escala caracter\'{\i}stica de las nuevas interacciones
fuertes debe estar cercana a 1.2 TeV. Entonces, cerca de la
energ\'{\i}a cr\'{\i}tica de 1 TeV, los bosones $W,Z$ interact\'uan
fuertemente entre ellos. Las teor\'{\i}as tecnicolor proporcionan
una forma elaborada de dichos escenarios.

\subsubsection{Bases Te\'oricas\\ \\}
Los escenarios f\'{\i}sicos del rompimiento din\'amico de la
simetr\'{\i}a pueden estar basados en teor\'{\i}as de nueva
interacci\'on fuerte, las cuales extienden el espectro de
part\'{\i}culas de materia y de las interacciones m\'as all\'a de
los grados de libertad tomados en cuenta en el Modelo Est\'andar.
Si las nuevas interacciones fuertes son invariantes bajo
transformaciones de un grupo de simetr\'{\i}a quiral $SU(2) \times
SU(2)$, la invariancia quiral generalmente es rota
espont\'aneamente al grupo de isoesp\'{\i}n custodial diagonal
$SU(2)$. Este proceso est\'a asociado con la formaci\'on de un
condensado quiral en el estado base y con la existencia  de tres
bosones Goldstone sin masa.

\begin{figure}[hbt]
\begin{center}
\begin{picture}(60,10)(90,30)
\Photon(0,25)(50,25){3}{6}
\LongArrow(60,25)(75,25)
\put(-10,21){$V$}
\end{picture}
\begin{picture}(60,10)(70,30)
\Photon(0,25)(50,25){3}{6}
\put(55,21){$+$}
\end{picture}
\begin{picture}(60,10)(55,30)
\Photon(0,25)(25,25){3}{3}
\Photon(50,25)(75,25){3}{3}
\Line(25,24)(50,24)
\Line(25,26)(50,26)
\GCirc(25,25){5}{0.5}
\GCirc(50,25){5}{0.5}
\put(80,21){$+$}
\put(35,30){$G$}
\end{picture}
\begin{picture}(60,10)(20,30)
\Photon(0,25)(25,25){3}{3}
\Photon(50,25)(75,25){3}{3}
\Photon(100,25)(125,25){3}{3}
\Line(25,24)(50,24)
\Line(25,26)(50,26)
\Line(75,24)(100,24)
\Line(75,26)(100,26)
\GCirc(25,25){5}{0.5}
\GCirc(50,25){5}{0.5}
\GCirc(75,25){5}{0.5}
\GCirc(100,25){5}{0.5}
\put(130,21){$\cdots$}
\put(35,30){$G$}
\put(85,30){$G$}
\end{picture}
\end{center}
\caption[]{\label{esp-fg:gaugemass} \it Generando masas de los bosones
de norma (V) a trav\'es de la interacci\'on con los bosones de
Goldstone (G).}
\end{figure}

Los bosones de Goldstone pueden ser absorbidos por los campos de
norma, generando estados longitudinales y masas distintas de cero
para los bosones de norma, como se muestra en la
Fig.~\ref{fg:gaugemass}. Sumando las series geom\'etricas de las
transiciones bos\'on de Goldstone-bos\'on vectorial en el
propagador, lleva a un cambio en el polo de la masa:
\begin{eqnarray}
\frac{1}{q^2} & \to & \frac{1}{q^2} + \frac{1}{q^2} q_\mu \frac{g^2 F^2/2}{q^2}
q_\mu \frac{1}{q^2} + \frac{1}{q^2} \left[ \frac{g^2 F^2}{2} \frac{1}{q^2}
\right]^2 + \cdots \nonumber \\
& \to & \frac{1}{q^2-M^2}
\end{eqnarray}
El acoplamiento entre bosones de norma y bosones de Goldstone ha sido
definido como $ig F/\sqrt{2} q_\mu$. La masa generada por los
campos de norma est\'a relacionada a este acoplamiento por
\begin{equation}
M^2 = \frac{1}{2} g^2 F^2 ~.
\end{equation}
El valor num\'erico del acoplamiento $F$ debe coincidir con $v=246$ GeV.\\

La simetr\'{\i}a custodial $SU(2)$ restante garantiza que el
par\'ametro $\rho$, que es la intensidad relativa entre los
acoplamientos $NC$ y $CC$, sea uno. Denotando los elementos de la
matriz de masa $W/B$ por
\begin{equation}
\begin{array}{rclcrcl}
\langle W^i | {\cal M}^2 | W^j \rangle & = & \displaystyle \frac{1}{2} g^2
F^2 \delta_{ij}
& \hspace*{1cm} & \langle W^3 | {\cal M}^2 | B \rangle & = & \langle B |
{\cal M}^2 | W^3 \rangle \\ \\
\langle B | {\cal M}^2 | B \rangle & = & \displaystyle \frac{1}{2} g'^2 F^2 &
& & = & \displaystyle \frac{1}{2} gg' F^2
\end{array}
\end{equation}
la universalidad del acoplamiento $F$ conduce a la raz\'on
$M_W^2/M_Z^2 = g^2/(g^2+g'^2) = \cos^2\theta_W$ de los
eigenvalores de masa, equivalente a $\rho=1$.\\

Puesto que las funciones de onda de los bosones vectoriales
polarizados longitudinalmente aumentan con la energ\'{\i}a, las
componentes del campo longitudinales son los grados de libertad
dominantes a altas energ\'{\i}as. Estos estados pueden, sin
embargo, para energ\'{\i}as asint\'oticas identificarse con los
bosones de Goldstone absorbidos. Esta equivalencia \cite{75} es
aparente en la norma de 't Hooft--Feynman donde, para
energ\'{\i}as asint\'oticas,
\begin{equation}
\epsilon_\mu^L W_\mu \to k_\mu W_\mu \sim M^2 \Phi ~.
\end{equation}
Por consiguiente, la din\'amica de los bosones de norma puede ser
identificada a altas enreg\'{\i}as con la din\'amica de los
campos de Goldstone escalares. Una representaci\'on elegante de
los campos de Goldstone $\vec{G}$ en este contexto est\'a dada por
la forma exponencial
\begin{equation}
U = \exp [-i \vec{G} \vec{\tau}/v ] ~,
\end{equation}
la cual corresponde a un campo matricial $SU(2)$.\\

El Lagrangiano del sistema de bosones fuertemente interactuantes
consiste en tal escenario para la parte de Yang-Mills  ${\cal
L}_{YM}$ y las interacciones de los campos de Goldstone ${\cal
L}_G$,
\begin{equation}
{\cal L}={\cal L}_{YM}+{\cal L}_G ~.
\end{equation}
La parte de Yang--Mills es escrita en la forma usual ${\cal
L}_{YM} = -\frac{1}{4} {\rm Tr} [W_{\mu\nu} W_{\mu\nu} +
B_{\mu\nu} B_{\mu\nu} ]$. La interaccion de los campos de
Goldstone pueden ser sistem\'aticamente expandida en teor\'{\i}as
quirales en las derivadas de los campos, correspondiendo a
expansiones en potencias de la energ\'{\i}a para amplitudes de
dispersi\'on \cite{76}:
\begin{equation}
{\cal L}_G = {\cal L}_0 + \sum_{dim=4} {\cal L}_i + \cdots
\end{equation}
Denotando la derivada covariante del SM para los campos de
Goldstone como
\begin{equation}
D_\mu U = \partial_\mu U - i g W_\mu U + i g' B_\mu U ~,
\end{equation}
el t\'ermino principal ${\cal L}_0$, el cual es de dimensi\'on = 2,
est\'a dado por
\begin{equation}
{\cal L}_0 = \frac{v^2}{4} {\rm Tr} [ D_\mu U^+ D_\mu U ] ~.
\end{equation}
Este t\'ermino genera las masas de los bosones de norma $W,Z$:
$M_W^2 = \frac{1}{4} g^2 v^2$ y $M_Z^2 = \frac{1}{4} (g^2+g'^2)
v^2$. El \'unico par\'ametro en esta parte de la interacci\'on es
$v$, el cual sin embargo, se fija de manera \'unica por el valor
experimental de la masa del $W$; de modo que las amplitudes
predichas por el t\'ermino principal en la expansi\'on quiral
puede efectivamente ser
considerado como par\'ametro libre.\\

 La componente al siguiente orden
en la expansi\'on con dimensi\'on = 4 consiste de diez t\'erminos
individuales. Si la simetr\'{\i}a custodial $SU(2)$ se impone,
s\'olo quedan dos t\'erminos, los cuales no afectan a los
propagadores ni a los v\'ertices de 3 bosones, pero s\'{\i} a los
de 4 bosones. Introduciendo el campo vectorial $V_\mu$ como
\begin{equation}
V_\mu = U^+ D_\mu U
\end{equation}
estos dos t\'erminos est\'an dados por las densidades de
interacci\'on
\begin{equation}
{\cal L}_4  =  \alpha_4 \left[Tr V_\mu V_\nu \right]^2
\hspace*{0.5cm} \mbox{y} \hspace*{0.5cm} {\cal L}_5  =  \alpha_5
\left[Tr V_\mu V_\mu \right]^2
\end{equation}

Los dos coeficientes $\alpha_4,\alpha_5$ son par\'ametros libres
que debe ser ajustados experimentalmente con datos de la
dispersi\'on $WW$.

Ordenes mayores en la expansi\'on quiral dan lugar a una
expansi\'on de la energ\'{\i}a de las amplitudes de dispersi\'on
de la forma ${\cal A} = \sum c_n (s/v^2)^n$. Esta serie
divergir\'a a energ\'{\i}as para las cuales las resonancias de las
teor\'{\i}as de nuevas interacciones fuertes pueden formarse en
colisiones de $WW$: $0^+$ `tipo-Higgs', resonancias $1^-$
`tipo-$\rho$', etc. Las masas de estos estados de resonancia se
esperan en el rango $M_R \sim 4\pi v$ donde la expansi\'on de lazo
quiral diverge, i.e. entre alrededor de 1 a 3 TeV.

\subsubsection{Un Ejemplo: Teor\'{\i}as Tecnicolor\\ \\}
Un ejemplo simple para tales escenarios es proporcionado por las
teor\'{\i}as de Tecnicolor, ver e.g. Ref.~\cite{94A}. \'Estas son
construidas con patrones similares a QCD pero caracterizadas por
una escala $\Lambda_{TC}$ en el rango de TeV de manera que la
interacci\'on se
vuelve fuerte ya a cortas distancias, del orden de $10^{-17}$~cm. \\

Los grados de libertad b\'asicos en la versi\'on m\'as simple son
un conjunto quiral de fermiones sin masa $[(U,D)_L;U_R,D_R]$ que
interact\'uan con campos de norma tecnicolor. La simetr\'{\i}a
quiral $SU(2)_L\times SU(2)_R$ de esta teor\'{\i}a se rompe a la
simetr\'{\i}a vectorial diagonal $SU(2)_{L+R}$ por la formacion de
condesados de vac\'{\i}o  $\langle\bar{U} U\rangle =\langle\bar{D}
D\rangle = {\cal O}(\Lambda^3_{TC})$. El rompimiento de la
simetr\'{\i}a quiral genera tres bosones de Goldstone sin masa GB
$\sim \bar{Q} i \gamma_5 \stackrel{\to}{\tau} Q$, que
pueden ser absorbidos por campos de norma del Modelo Est\'andar
para construir los estados masivos con $M_W \sim 100$~GeV. De la
cadena 
\beq 
M_W = \frac{1}{2} g F \quad \mathrm{y} \quad F \sim
\Lambda_{TC} / 4 \pi 
\eeq  
se estima que el par\'ametro $F$ sea
del orden de 1 TeV
en tanto que $\Lambda_{TC}$ debe estar en el rango de 10 TeV. \\

Mientras que el sector de norma electrod\'ebil puede ser formulado
consistentemente en esta marco, generar las masas de los fermiones
conduce a dificultades severas. Como las interacciones de norma
acoplan s\'olo las componentes del campo izquierdo-izquierdo y
derecho-derecho, un cambio en la helicidad del operador de masa
izquierdo-derecho $\bar{f}_L f_R$ no es generado para los fermiones
del Modelo Est\'andar. Para resolver este problema, se deben introducir
nuevas interacciones entre fermiones del SM y fermiones del TC
[Tecnicolor Extendido ETC] de manera que la helicidad pueda cambiar a
trav\'es del condesado de ETC en el vac\'{\i}o. Las masas del SM
predichas de esta manera son del orden $m_f \sim g^2_E \Lambda^3_{ETC}/M_E^2$~,
siendo $g_E$ el acoplamiento de la teor\'{\i}a de norma de Tecnicolor
extendida y $M_E$ es la masa de los campos de norma del ETC. Sin
embargo, estimaciones de $M_E$ llevan a conflictos si uno intenta
reconciliar el tama\~no de la escala requerida para generar la masa del
top, del orden de TeV, con la supresi\'on de procesos de cambio de sabor, como
las oscilaciones $K\bar{K}$,
las cuales requieren un tama\~no del orden de PeV. \\

De este modo, la realizaci\'on m\'as simple  de las teor\'{\i}as de
tecnicolor sufre de  conflictos internos en el sector
fermi\'onico. Modelos te\'oricos m\'as complicados son necesarios
para reconciliar estas estimaciones conflictivas \cite{94A}. No
obstante, la idea de generar el rompimiento de la simetr\'{\i}a
electrod\'ebil din\'amicamente es  te\'oricamente
atractiva y un escenario en principio interesante.

\subsection{$WW$ Dispersi\'on en colisionadores de Alta Energ\'{\i}a}

\phantom{h} Independientemente de la forma espec\'{\i}fica en que
se realiza el rompimiento din\'amico de la simetr\'{\i}a, se han
desarrollado herramientas te\'oricas que pueden ser
\'utiles para investigar  estos escenarios de una manera
bastante general. Las amplitudes de dispersi\'on (cuasi-)
el\'asticas de 2--2 $WW$ pueden expresarse a altas energ\'{\i}as
por una amplitud maestra $A(s,t,u)$, la cual depende de las tres
variables de Mandelstam de los procesos de dispersi\'on:
\begin{eqnarray}
A(W^+ W^- \to ZZ) & = & A(s,t,u) \\
A(W^+ W^- \to W^+ W^-) & = & A(s,t,u) + A(t,s,u) \nonumber \\
A(ZZ \to ZZ) & = & A(s,t,u) + A(t,s,u) + A(u,s,t) \nonumber \\
A(W^- W^- \to W^- W^-) & = & A(t,s,u) + A(u,s,t) ~. \nonumber
\end{eqnarray} ~\\

Al orden m\'as bajo en la expansi\'on quiral, ${\cal L} \to {\cal
L}_{YM} + {\cal L}_0$, la amplitud maestra est\'a dada, en la
forma libre de par\'ametros, por la energ\'{\i}a al cuadrado $s$:
\begin{equation}
A(s,t,u) \to \frac{s}{v^2} ~.
\end{equation}
Esta representaci\'on es v\'alida para energ\'{\i}as $s \gg M_W^2$
pero abajo de la nueva regi\'on de resonancia, i.e. en la
pr\'actica, a energ\'{\i}as de $\sqrt{s}={\cal O}(1~\mbox{TeV})$.
Denotando la longitud de dispersi\'on para el canal que lleva
isoesp\'{\i}n $I$ y momento angular $J$ por $a_{IJ}$, los \'unicos
canales de dispersi\'on distinta de cero que predice el t\'ermino
principal de la expansi\'on quiral corresponden a
\begin{eqnarray}
a_{00} & = & + \frac{s}{16\pi v^2} \\
a_{11}   & = & + \frac{s}{96\pi v^2} \nonumber \\
a_{20}   & = & - \frac{s}{32\pi v^2} ~.
\end{eqnarray}
Mientras que el canal ex\'otico $I=2$ es repulsivo, los canales
$I=J=0$ y $I=J=1$ son atractivos, indicando la formaci\'on de
resonancias no-fundamentales tipo Higgs y tipo $\rho$.\\

Tomando en consideraci\'on los t\'erminos  siguientes al dominante en
la expansi\'on quiral (next-to-leading terms), la amplitud maestra
resulta ser \cite{24}
\begin{equation}
A(s,t,u) = \frac{s}{v^2} + \alpha_4 \frac{4(t^2+u^2)}{v^4}
+ \alpha_5 \frac{8s^2}{v^4} + \cdots ~,
\end{equation}
incluyendo los dos par\'ametros $\alpha_4$ y $\alpha_5$.\\

Al incrementar la energ\'{\i}a, las amplitudes se aproximar\'an al
\'area de resonancia. En esa \'area, el car\'acter quiral de la
teor\'{\i}a no proporciona m\'as principios de gu\'\i a en
la construcci\'on de amplitudes de dispersi\'on. En su lugar, se
deben introducir hip\'otesis {\it ad-hoc} para definir la
naturaleza de las resonancias, ver e.g. Ref. \cite{24a}. Un
ejemplo t\'{\i}pico lo da la
\begin{eqnarray}
\hspace*{-0.5cm}\mbox{\bf{Resonancia escalar acoplada
quiralmente:}}\quad A & = & \frac{s}{v^2} - \frac{g_s^2 s^2}{v^2}
\frac{1}{s-M_S^2 - iM_S \Gamma_S}
\\
& & \mbox{con}~~~\Gamma_S = \frac{3g_s^2 M_S^3}{32 \pi v^2}
\nonumber
\end{eqnarray}

Para energ\'{\i}as peque\~nas, la amplitud de dispersi\'on es
reducida a la forma quiral principal $s/v^2$. En la regi\'on de
resonancia es descrita por dos par\'ametros, la masa y el ancho de
resonancia.
Las amplitudes se interpolan entre las dos regiones de una manera suave.\\

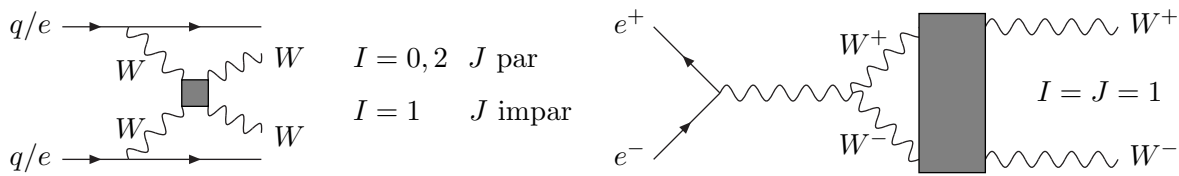
\begin{figure}[hbt]
\begin{center}
\begin{picture}(60,50)(140,0)
\ArrowLine(0,50)(25,50) \ArrowLine(25,50)(75,50)
\ArrowLine(0,0)(25,0) \ArrowLine(25,0)(75,0)
\Photon(25,50)(45,30){-3}{3} \Photon(25,0)(45,20){3}{3}
\Photon(55,20)(75,10){-3}{3} \Photon(55,30)(75,40){3}{3}
\GBox(45,20)(55,30){0.5} \put(-20,48){$q/e$} \put(-20,-2){$q/e$}
\put(20,8){$W$} \put(20,30){$W$} \put(80,35){$W$} \put(80,5){$W$}
\put(110,35){$I=0,2~~J~\mbox{par}$}
\put(110,15){$I=1~~~~~J~\mbox{impar}$}
\end{picture}
\begin{picture}(60,50)(-20,0)
\ArrowLine(25,25)(0,50)
\ArrowLine(0,0)(25,25)
\Photon(25,25)(75,25){3}{5}
\Photon(75,25)(100,50){3}{4}
\Photon(75,25)(100,0){3}{4}
\Photon(125,0)(175,0){3}{5}
\Photon(125,50)(175,50){3}{5}
\GBox(100,-5)(125,55){0.5}
\put(-15,48){$e^+$}
\put(-15,-2){$e^-$}
\put(70,2){$W^-$}
\put(70,40){$W^+$}
\put(180,48){$W^+$}
\put(180,-2){$W^-$}
\put(145,22){$I=J=1$}
\end{picture}
\end{center}
\caption[]{\label{esp-fg:qqtoqqww} \it Dispersi\'on y redispersi\'on
de $WW$ a altas energ\'{\i}as en el LHC y en colisionadores TeV
lineales $e^+e^-$.}
\end{figure}

\begin{figure}[hbtp]
\begin{center}
\vspace*{-5.5cm}

\hspace*{-3.5cm}
\epsfig{file=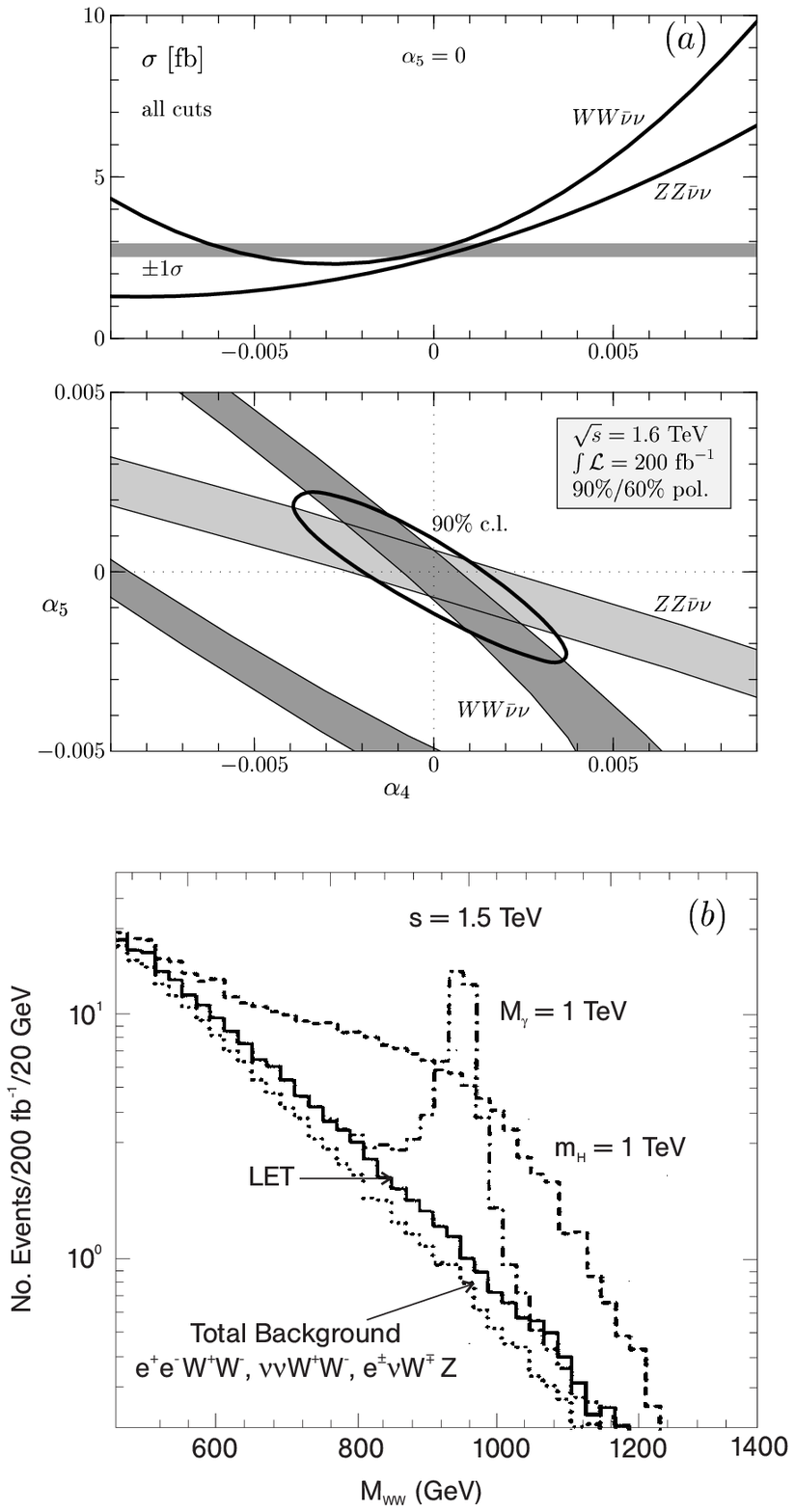,width=22.0cm,angle=0}
\vspace*{-8.5cm}

\end{center}
\caption[]{\it
  Parte superior: La sensibilidad a los par\'ametros de expansi\'on
  en modelos electrod\'ebiles quirales de la dispersi\'on $WW \to WW$ y $WW \to
  ZZ$ en el umbral de la interacci\'on fuerte;
  Ref. \protect\cite{24}.
Parte inferior: La distribuci\'on de la energ\'{\i}a invariante de
$WW$ en $e^+e^-
  \to \overline{\nu} \nu WW$ para modelos de resonancia escalares y
   vectoriales [$M_H, M_V$ = 1 TeV];
  Ref. \protect\cite{24a}.
\protect\label{esp-17tt}\label{esp-PKB}
}
\end{figure}
\begin{figure}[hbtp]

\vspace*{-2.0cm}
\hspace*{-2.0cm}
\epsfxsize=20cm \epsfbox{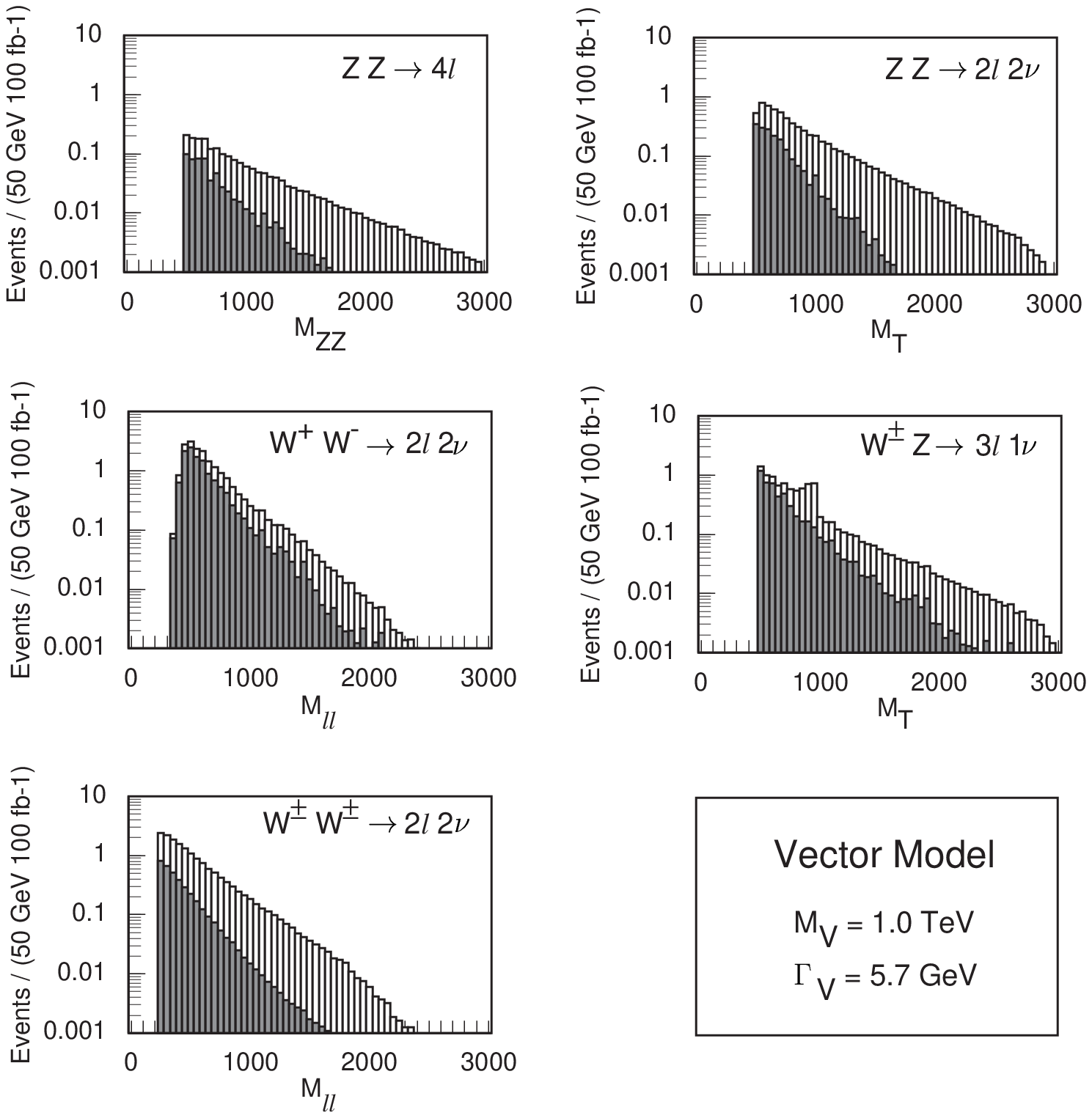}
\vspace*{-11cm}

\caption[]{\label{esp-fg:vvto4l} \it Distribuciones de masa invariante
para estados ``dorados'' finales puramente lept\'onicos  que
surgen de los procesos $pp\to ZZX \to 4\ell X, pp\to ZZX\to 2\ell
2\nu X, pp\to W^+W^-X, pp\to W^\pm ZX$ y $pp\to W^\pm W^\pm X$,
para el LHC (la masa en GeV). La se\~nal se grafica arriba del
fondo ya sumado. Se muestran las distribuciones para vectores
acoplados quiralmente con $M_V=1$ TeV, $\Gamma_V=5.7$ GeV; Ref.
\protect\cite{23B}.}
\end{figure}

La dispersi\'on de $WW$ puede estudiarse en el LHC y en el colisionador
TeV lineal $e^+e^-$. A altas energ\'{\i}as, rayos de $W$ equivalentes
acompa\~nan a los rayos de quark y electr\'on/positr\'on
(Fig.~\ref{fg:qqtoqqww}) en los procesos de fragmentaci\'on $pp\to qq \to
qqWW$ y $ee\to \nu\nu WW$; el espectro de los bosones $W$ polarizados
longitudinalmente est\'an dados en la Eq. (\ref{eq:xyz}). En el ambiente
hadr\'onico de LHC los estados finales del bos\'on $W$ s\'olo pueden ser
observados en decaimientos lept\'onicos. La reconstrucci\'on de la
resonancia no es entonces posible para estados finales $W$ cargados.
Sin ambargo, el ambiente limpio de los colisionadores de $e^+e^-$
permitir\'a la reconstrucci\'on de resonancias de decaimientos a pares de
jet. Los resultados de tres simulaciones experimentales se exhiben en
la Fig.~\ref{PKB}. En la Fig.~\ref{PKB} se muestra la sensibilidad a
los par\'ametros $\alpha_4,\alpha_5$ de la expansi\'on quiral para la dispersi\'on
$WW$ en colisionadores $e^+e^-$ \cite{24}. Los resultados de estos
an\'alisis pueden reinterpretarse como sensibilidad a la predicci\'on de
los par\'ametros libres de la expansi\'on quiral, correspondiendo a un
error de cerca del 10\% en el primer t\'ermino de la amplitud maestra
$s/v^2$. Estos experimentos prueban el concepto b\'asico de rompimiento
de simetr\'{\i}a din\'amico a trav\'es del rompimiento espont\'aneo de la
simetr\'{\i}a. La producci\'on de una resonancia bos\'on vectorial de masa
$M_V=1$ TeV se ejemplifica en la Fig.~\ref{PKB}b \cite{24a}. Las
energ\'{\i}as invariantes lept\'onicas esperadas de los estados finales
de la dispersi\'on $WW$ en el LHC son comparadas en el
modelo vectorial con las se\~nales de fondo en la Fig.~\ref{fg:vvto4l}
\cite{23B}.\\ 

Un segundo poderoso m\'etodo  mide la dispersi\'on el\'astica $W^+W^-
\to W^+W^-$ en el canal $I=1, J=1$. La redispersi\'on de los
bosones $W^+W^-$ producidos en la aniquilaci\'on $e^+e^-$, cf.
Fig.~\ref{fg:qqtoqqww}, depende, a energ\'{\i}as altas, de la fase
$\delta_{11}$ de dispersi\'on de $WW$ \cite{78}. La amplitud de
producci\'on $F = F_{LO} \times R$ es el producto del diagrama
perturbativo a m\'as bajo orden con la amplitud de redispersi\'on
de Mushkelishvili--Omn\`es ${\cal R}_{11}$,
\begin{equation}
{\cal R}_{11} = \exp \frac{s}{\pi} \int \frac{ds'}{s'}
\frac{\delta_{11}(s')}{s'-s-i\epsilon} ~,
\end{equation}
la cual est\'a determinada por el cambio en la fase $\delta_{11}$
de $WW$, obtenida de $I = J = 1$. La eficacia de este m\'etodo se
deriva del hecho de que toda la energ\'{\i}a del colisionador
$e^+e^-$ es transferida al sistema $WW$ [mientras que una
fracci\'on mayor de la energ\'{\i}a se pierde en la
fragmentaci\'on de $e \to \nu W$ si la dispersi\'on de $WW$ es
estudiada en el proceso $ee\to \nu\nu WW$]. Simulaciones
detalladas \cite{78} han demostrado que este proceso es sensible a
las masas del bos\'on vectorial de hasta
cerca de $M_V \lessim 6$ TeV en teor\'{\i}as tipo tecnicolor. \\

El an\'alisis experimental de los par\'ametros $\alpha$ en el
colisionador lineal $e^+e^-$ en la primera fase con energ\'{\i}as
de hasta $\sim 1$~TeV puede ser reinterpretado de la siguiente
forma. Asociando los par\'ametros $\alpha$ con nuevas escalas de
interacci\'on fuerte, $\Lambda_\star \sim M_W/\sqrt{\alpha}$, cotas superiores para
  $\Lambda_\star$ de $\sim 3$~TeV pueder ser exploradas en la dispersi\'on $WW$.
 As\'{\i} pues, este instrumento
permite cubrir toda la regi\'on umbral $\lessim 4\pi v \sim 3$~TeV
de las nuevas interacciones fuertes. El canal de producci\'on
de$W^+W^-$ en colisiones $e^+e^-$  permite probar indirectamente
un rango que alcanza incluso los 10~TeV. Si se descubriera una
nueva escala $\Lambda_\star$ de hasta $\sim 3$~TeV, se
podr\'{\i}an buscar nuevas resonancias $WW$ en el LHC mientras que
CLIC podr\'{\i}a investigar potencialmente nuevos estados de
resonancia de masas cercanas a 5 TeV.


\section{Sinopsis}

\phantom{h} El mecanismo de rompimiento de la simetr\'{\i}a
elctrod\'ebil puede quedar establecido en los colisionadores
$p\bar p/pp$ y $e^+e^-$ actuales o en la nueva generaci\'on de
estos:
\begin{itemize}
\item[$\star$] La existencia de un bos\'on de Higgs fundamental ligero;
\item[$\star$] El perfil de la part\'{\i}cula de Higgs puede ser reconstruido,
revelando as\'{\i} la naturaleza f\'{\i}sica 
del subyacente mecanismo del rompimiento de la simetr\'{\i}a electrod\'ebil;
\item[$\star$] Pueden realizarse an\'alisis sobre la dispersi\'on WW fuerte
si el rompimiento de la simetr\'{\i}a es de naturaleza din\'amica
y generada por nuevas interacciones fuertes.
\end{itemize}
M\'as a\'un, dependiendo de la respuesta experimental a estas
preguntas, el sector electrod\'ebil proporcionar\'a la plataforma
para extrapolaciones en \'areas f\'{\i}sicas m\'as all\'a del
Modelo Est\'andar: ya sea para el sector supersim\'etrico de bajas
energ\'{\i}as o, alternativamente, para una teor\'{\i}a de nuevas
interacciones fuertes a una escala caracter\'{\i}stica del orden
de 1 TeV y m\'as all\'a.

\section*{Agradecimientos}

P.M Zerwas est\'a muy agradecido con los organizadores, A.Bashir,
J.Erler y M.Mondrg\'on, por la invitaci\'on a la XI Escuela Mexicana de
Part\'\i culas y Campos, Xalapa (Veracruz) 2004.  Reconoce con
agradecimiento la cooperaci\'on de sus co-autores en la escritura de
este reporte, en particular de los estudiantes mexicanos y de Myriam
Mondrag\'on, los cuales prepararon la versi\'on en espa\~nol.
\\\\
Este trabajo fue parcialmente apoyado por los proyectos
PAPIIT-IN116202 y Conacyt 42026-F.  \newpage

\appendix
\def\thesection{Ap\'endice \Alph{section}}   

\section{El Modelo $\sigma$ O(3)}

\phantom{h} Un modelo transparente pero, a la vez, suficientemente
complejo para estudiar todos los aspectos del rompimiento de la
simetr\'{\i}a electrod\'ebil, es el modelo $\sigma$ O(3). Al
empezar por la versi\'on est\'andar, de un n\'umero de variantes
se puede desarrollar la idea del rompimiento espont\'aneo de la
simetr\'{\i}a y el teorema de Goldstone, mientras que normando
(gauging) la teor\'{\i}a conduce al fen\'omeno del Higgs. Esta
evoluci\'on ser\'a descrita
paso a paso en las siguientes tres subsecciones. \\

El modelo $\sigma$ O(3) incluye un triplete de componentes de
campo:
\beq 
\sigma = (\sigma_1,\sigma_2,\sigma_3) 
\eeq
Si el potencial de auto-interacci\'on del campo depende
\'unicamente en la intensidad global del campo, la teor\'{\i}a,
descrita por el Lagrangiano
 \beq 
{\cal L} = \frac{1}{2} (\partial \sigma)^2 -
V(\sigma^2) 
\eeq 
es invariante rotacional O(3). Estas
iso-rotaciones son generadas por la transformaci\'on 
\beq 
\sigma \to e^{i\alpha t}\sigma \quad \mathrm{con} \quad t_{ik}^j = -i
\epsilon_{ijk} 
\eeq
Esta transformaci\'on corresponde a una rotaci\'on alrededor del
eje $\alpha = (\alpha_1,\alpha_2,\alpha_3)$. Eligiendo una
interacci\'on cu\'artica para el potencial, la teor\'{\i}a es
renormalizable y por lo tanto bien definida.

\subsection{Teor\'{\i}a ``Normal'':}

\phantom{h}
 Si el potencial cu\'artico $V$ se escoge a que sea
cf.~Fig.~\ref{fig:pot}, 
\beq 
V(\sigma^2) = \lambda^2 (\sigma^2 + \mu^2)^2 
\eeq 
el espectro de part\'{\i}culas y las interacciones
pueden f\'acilmente derivarse a partir de la siguiente forma
 \beq
V(\sigma^2) = 2\lambda^2\mu^2\sigma^2 + \lambda^2 \sigma^4 +
\mathrm{const.} \eeq
\begin{figure}[hbt]
\begin{center}
\epsfig{figure=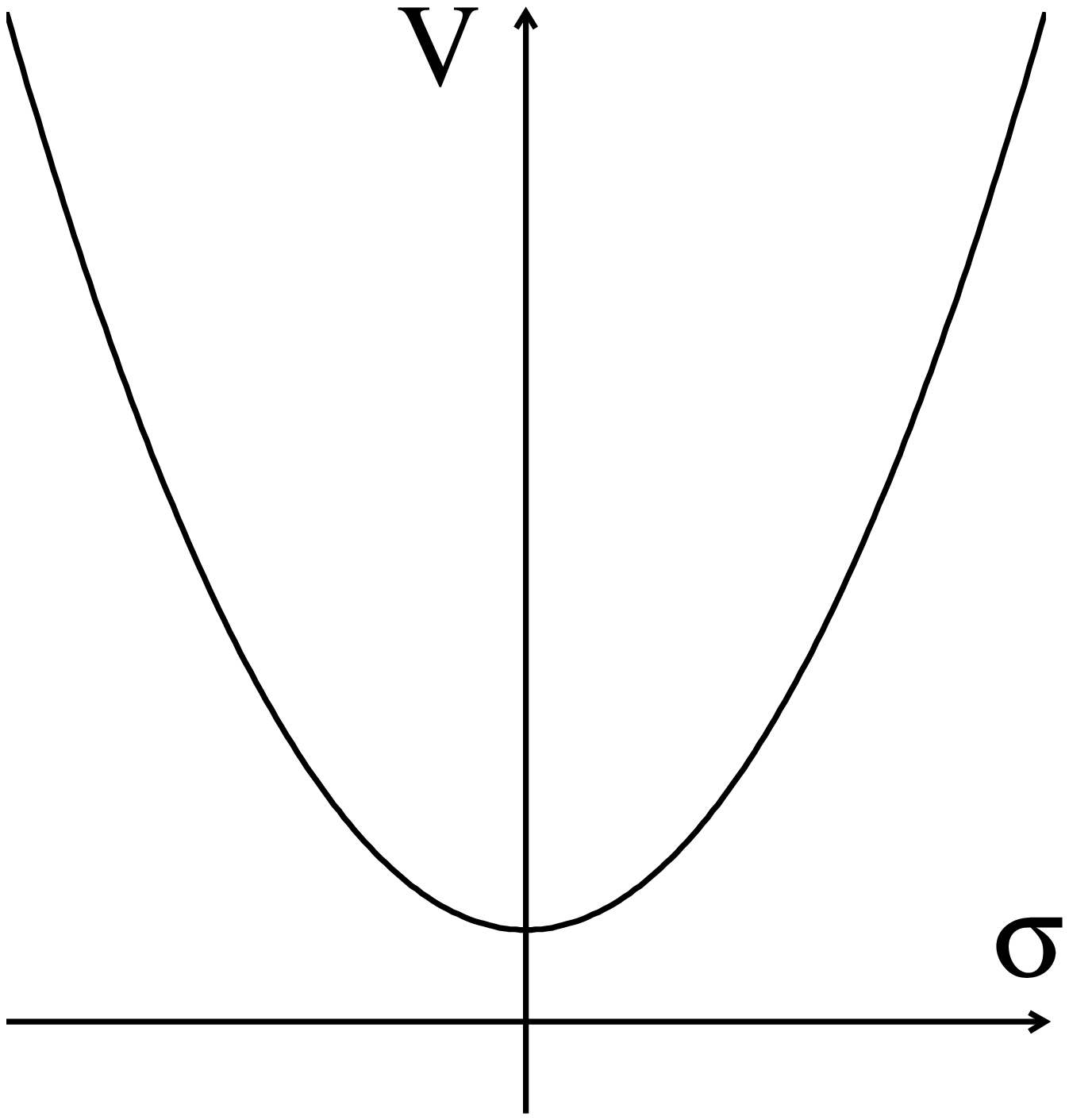,width=4cm}
\end{center}
\vspace*{-0.4cm}
\caption{\it \label{esp-fig:pot}}
\end{figure}

\noindent El t\'ermino de campo bilineal describe tres masas
degeneradas 
\beq 
m(\sigma_1) = m(\sigma_2) = m(\sigma_3) = 2\lambda\mu 
\eeq 
correspondiendo a tres grados de libertad de
part\'{\i}culas f\'{\i}sicas. Los campos interact\'uan a trav\'es
del segundo t\'ermino cu\'artico. El estado base del sistema se
obtiene para la intesidad cero del campo:
 \beq \sigma^0 = (0,0,0) \eeq
Esta teor\'{\i}a describe un sistema de part\'{\i}culas est\'andar
en el cual el estado base preserva la invariancia rotacional del
Lagrangiano. Por lo que el Lagrangiano y la soluci\'on a la
ecuaci\'on del campo odedecen el mismo grado de simetr\'{\i}a.

\subsection{Rompimiento Espont\'aneo de la Simetr\'{\i}a y el Teorema de
  Goldstone:} 

\phantom{h} Sin embargo, si el signo del par\'ametro de masas el
potencial cambia a valores negativos,
 \beq 
V(\sigma^2) = \lambda^2(\sigma^2-\mu^2)^2 
\eeq 
el estado base es un estado de
intensidad de campo diferente de cero,
cf.~Fig.~\ref{fig:pothiggs}. Fijando el eje del estado base de tal
manera que
 \beq \sigma^0 = (0,0,v) \quad \mathrm{con} \quad v=\mu
\eeq 
la invariancia rotacional original O(3) del Lagrangiano ya no
es respetada por la soluci\'on del estado base el cual distingue
una direcci\'on espec\'{\i}fica del iso-espacio. Sin embargo, no
hay un principio que determine la direcci\'on arbitraria del
estado base en el iso-espacio. Dicho fen\'omeno, en el cual las
soluciones de las ecuaciones de campo no obedecen la simetr\'{\i}a
del Lagrangiano, es generalmente llamado
``rompimiento espont\'aneo de la simetr\'{\i}a''.\\
\begin{figure}[hbt]
\begin{center}
\epsfig{figure=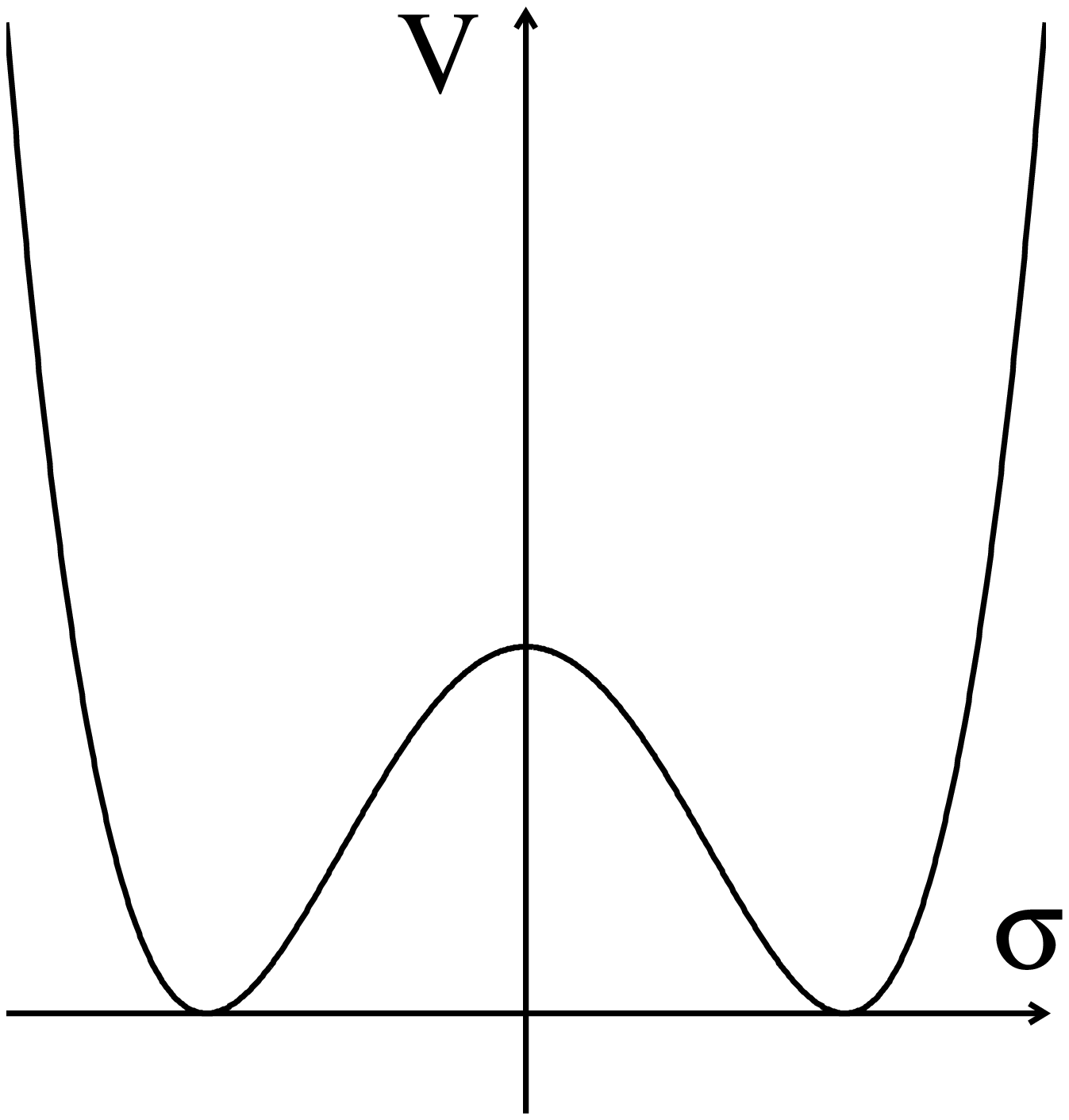,width=4cm}
\end{center}
\vspace*{-0.4cm}
\caption{\it \label{esp-fig:pothiggs}}
\end{figure}

Expandiendo el campo $\sigma$ alrededor del estado base, 
\beq
\sigma = (\sigma_1', \sigma_2', v+\sigma_3') 
\eeq 
emerge una teor\'{\i}a efectiva para los nuevos grados de libertad
din\'amicos $\sigma_1', \sigma_2'$ y $\sigma_3'$. Evaluando el
potencial para los nuevos campos, 
\beq 
V = 4v^2\lambda^2\sigma_3^{'2} + 4 v \lambda^2 \sigma_3' (\sigma_1^{'2}
+ \sigma_2^{'2} + \sigma_3^{'2}) + \lambda^2 (\sigma_1^{'2} +
\sigma_2^{'2} + \sigma_3^{'2})^2 
\eeq 
dos part\'{\i}culas sin masa m\'as una part\'{\i}cula masiva
corresponden a los t\'erminos de campo bilineales: 
\beq
m(\sigma_1')=m(\sigma_2')=0 \nonumber\\
m(\sigma_3')=2\sqrt{2}\lambda v \neq 0
\eeq
A las dos part\'{\i}culas sin masa se les llama bosones de Goldstone,
Ref.~\cite{appcite}.\\ 

Los bosones de Goldstone y las part\'{\i}culas masivas
interact\'uan entre ellas por medio de los t\'erminos trilineales
del potencial
efectivo, adem\'as de los t\'erminos cu\'articos est\'andar. \\

La simetr\'{\i}a de la teor\'{\i}a efectiva se reduce, de la
invariancia rotacional O(3) original a la invariancia O(2)
restringida a
rotaciones alrededor del eje del estado base. \\

Este modelo $\sigma$ es s\'olo un simple ejemplo del m\'as general \\

{\it \noindent\underline{Teorema de Goldstone:}\\
Si N es la dimensi\'on del grupo de simetr\'{\i}a del Lagrangiano
b\'asico, pero la simetr\'{\i}a de la soluci\'on del estado base
se reduce a una dimensi\'on M, entonces la teor\'{\i}a incluye
(N-M) bosones escalares de Goldstone sin masa.}\\

Por cada grado de libertad de la simetr\'{\i}a que es destruido,
aparece en el espectro una part\'{\i}cula sin masa. Un ejemplo muy
famoso de este teorema son los tres piones casi sin masa que
emergen del rompimiento espont\'aneo de la simetr\'{\i}a de
isoesp\'{\i}n quiral en QCD.

\subsection{El mecanismo de Higgs}

\phantom{h}

El mecanismo de Higgs proporciona masa a los bosones vectoriales
en teor\'{\i}as de norma sin destruir la renormalizabilidad de la
teor\'{\i}a. Si las masas fueran introducidas a mano, la 
invariancia de norma que garantiza la renormalizabilidad, se
destruir\'{\i}a por
los t\'erminos de masa en el Lagrangiano.\\

La simetr\'{\i}a global de isoesp\'{\i}n del modelo $\sigma$ O(3)
puede ser extendida a una simetr\'{\i}a local al introducir un
iso-triplete $W$ de campos de norma acoplados de forma m\'{\i}nima
al campo $\sigma$. Introduciendo la derivada covariante  
\beq
\partial_\mu \sigma \to \partial_\mu \sigma + igtW\sigma
\eeq 
en el Lagrangiano
\beq 
{\cal L} = \frac{1}{2} [ (\partial +
igtW)\sigma]^2 - V(\sigma^2) + {\cal L}_{kin}(W) 
\eeq
la teor\'{\i}a es invariante bajo transformaciones de norma local
\beq \sigma \to e^{i\alpha t} \sigma \quad \mathrm{con} \quad
\alpha = \alpha(x) 
\eeq
con la transformaci\'on de materia complementada por la
transformaci\'on usual del campo de norma no abeliano. El
Lagrangiano normado (gauged) incluye la parte cin\'etica de norma,
la parte cin\'etica de $\sigma$ y la interacci\'on $\sigma$-norma, as\'{\i} como
tambi\'en el potencial. 

\begin{itemize}
\item[--] Si el potencial $\sigma$ es justo el potencial
est\'andar, la teor\'{\i}a es una teor\'{\i}a de norma de
Yang-Mills no abeliana con un triplete de part\'{\i}culas $\sigma$
interactuando de la manera  est\'andar con los campos tripeltes de norma
$W$.\\ 

\item[--] Sin embargo, si el potencial se escoge del tipo
$V=\lambda^2 (\sigma^2-\mu^2)^2$, que en el modelo $\sigma$
conduce al rompimiento espont\'aneo de la simetr\'{\i}a, el
contenido f\'{\i}sico campo/part\'{\i}cula de la teor\'{\i}a
cambia dram\'aticamente [un fen\'omeno similar a la teor\'{\i}a no
normada (non-gauged theory)].
\end{itemize}

Parametrizando el campo-triplete $\sigma$ a trav\'es de una
rotaci\'on del campo alrededor del eje del estado base , \beq
\sigma = e^{i\Theta t/v} (\sigma^0 + \eta) 
\eeq 
con 
\beq 
\sigma^0= (0,0,v)\,;\quad \eta = (0,0,\eta)\,; \quad \Theta =
(\Theta_1,\Theta_2,0) 
\eeq
las componentes $\Theta$ de $\sigma$ perpendiculares al eje del
estado base pueden ser removidas por la transformaci\'on de norma
$\sigma \to exp[-i\Theta t/v] \sigma$ suplementada por la
transformaci\'on correspondiente del campo de norma. Manteniendo
la notaci\'on original para los campos transformados de norma, el
nuevo Lagrangiano para los grados de libertad f\'{\i}sicos est\'a
dado por
 \beq {\cal L} = \frac{1}{2} [(\partial + igWt)(\sigma^0
+\eta)]^2 - V([\sigma^0 + \eta]^2) + {\cal L}_{kin}(W) 
\eeq
Despu\'es de escribir el Lagrangian resultante de la teor\'{\i}a
efectiva como
 \beq {\cal L} = {\cal L}_{kin}(W) + \frac{1}{4} g^2
v^2 ( W_1^2 + W_2^2) + \frac{1}{2} (\partial\eta)^2 - V + {\cal
L}_{int} (\eta, W) \eeq 
el contenido f\'{\i}sico
part\'{\i}cula/campo se pone de manifiesto:

\begin{itemize}
\item[--] un campo vectorial sin masa $W_3$ correspondiente a la invariancia
rotacional residual alrededor del eje-3 del estado base;\\

\item[--] dos campos $W$ masivos $W_1$ y $W_2$ perpendiculares al eje
  del estado base con masas que est\'an determinadas por la intensidad
  del campo $v$ del estado base $\sigma$ y la constante de acoplamiento de norma
  $g$.  Estos dos campos masivos corresponden a los grados de libertad
  de la simetr\'{\i}a que fueron rotos espont\'aneamente en el modelo
  $\sigma$   no-normado;\\

\item[--] los bosones de Goldstone han desaparecido del espectro,
  absorbidos para formar los grados longitudinales de los bosones
  de norma masivos;\\ 

\item[--] un bos\'on de Higgs escalar real $\eta$.
\end{itemize}

Este ejemplo puede extenderse f\'acilmente, paralelamente al
teorema de Goldstone, para describir la forma general de\\

{\it \noindent \underline{El mecanismo de Higgs}

\noindent Si N es la dimensi\'on del grupo de simetr\'{\i}a del
Lagrangiano original, M la dimensi\'on del grupo de simetr\'{\i}a
que deja invariante el estado base de n campos escalares, entonces
la teor\'{\i}a f\'{\i}sica consiste en M campos vectoriales sin
masa, (N-M) campos vectoriales masivos, y n-(N-M) campos de Higgs
escalares.}

\newpage

\renewcommand\thefigure{\arabic{figure}}
\setcounter{figure}{0}
\setcounter{section}{0}
\setcounter{equation}{0}
\setcounter{footnote}{0}

\setcounter{table}{0}

\renewcommand{\thesection}{\arabic{section}}
\renewcommand{\thetable}{\arabic{table}}

\renewcommand{\theequation}{\arabic{equation}}
\selectlanguage{english}

\thispagestyle{empty}
{\Large\raggedright\noindent \bf\par
 Electroweak Symmetry Breaking and Higgs Physics: Basic Concepts}\\[2ex]
\begin{indented}
      \item[]\normalsize\raggedright
{ \bf M. Gomez-Bock$^1$, M. Mondrag\'on$^2$, M. M\"uhlleitner$^{3,4}$, 
R.~Noriega-Papaqui$^1$, I.~Pedraza$^1$, M. Spira$^3$, P.M. Zerwas$^5$}
\end{indented}
\begin{indented}
\item[]\rm
{ $^1$ Inst. de F\'{\i}sica ``LRT'', Benemerita Univ. Auton. de Puebla, 
72570 Puebla, Pue, Mexico \\
$^2$ Inst. de F\'{\i}sica, Univ. Nac. Auton. de Mexico, 01000 Mexico D.F., 
Mexico \\
$^3$ Paul Scherrer Institut, CH-5232 Villigen PSI, Switzerland \\
$^4$ Laboratoire d'Annecy-Le-Vieux de Physique Th\'eorique, LAPTH,
Annecy-Le-Vieux, France \\ 
$^5$ Deutsches Elektronen-Synchrotron DESY, Hamburg, Germany}

\end{indented}


\begin{abstract}
We present an introduction to the basic concepts of electroweak symmetry 
breaking and Higgs physics within the Standard Model and its sypersymmetric 
extensions. A brief overview will also be given on alternative mechanisms of 
symmetry breaking. In addition to the theoretical basis, the present 
experimental status of Higgs physics and implications for future experiments 
at the LHC and $\epem$ linear colliders are discussed.
\end{abstract}


\section{Introduction}

\phantom{h}
\paragraph{1.}
Revealing  the physical mechanism 
responsible for the breaking of 
electroweak symmetries, is one of the key
problems in particle physics. If the fundamental
particles -- leptons, quarks and gauge
bosons -- remain weakly interacting up to
very high energies, potentially close to the Planck scale, 
the sector in which
the electroweak symmetry is broken
must contain one or more fundamental
scalar Higgs bosons with light masses
of the order of the symmetry-breaking
scale $v\sim 
246$ GeV. The masses of the fundamental
particles are generated through the
interaction with the scalar background
Higgs field, which is  non-zero in the ground
state \cite{1}. Alternatively, the symmetry
breaking could be generated dynamically
by new strong forces characterized by
an interaction scale $\Lambda \sim
1$ TeV and beyond \cite{2,2A}. If global symmetries of
the strong interactions are broken
spontaneously, the associated Goldstone
bosons can be absorbed by the gauge
fields, generating the masses of the
gauge particles. The masses of leptons
and quarks can be generated through
interactions with the fermion condensate.
Other breaking mechanisms of the electroweak symmetries are associated
with the dynamics in extra space dimensions at low energies \cite{RI2}.

\paragraph{2.}
A simple mechanism for the breaking of
the electroweak symmetry is incorporated in 
the Standard Model (SM) \cite{3}. To
accommodate all observed phenomena, a
complex isodoublet scalar field is introduced; 
this acquires a non-vanishing vacuum expectation
value through self-interactions, breaking spontaneously the electroweak
symmetry SU(2)$_I\times$ U(1)$_Y$
down to the electromagnetic U(1)$_{EM}$
symmetry. The interactions of the gauge
bosons and fermions with the background
field generate the masses of these
particles. One scalar field component
is not absorbed in this process,
manifesting itself as the physical
Higgs particle $H$.

The mass of the Higgs boson is the
only unknown parameter in the symmetry-breaking 
sector of the Standard
Model, while all couplings are fixed
by the masses of the particles, a
consequence of the Higgs mechanism
{\it sui generis}. However, the mass of the Higgs
boson is constrained in two ways. Since
the quartic self-coupling of the Higgs
field grows indefinitely with rising
energy, an upper limit on the Higgs
mass can be derived from demanding
the SM particles to remain weakly
interacting up to a scale $\Lambda$
\cite{4}. On the other hand, stringent
lower bounds on the Higgs mass follow
from requiring the electroweak vacuum
to be stable \cite{5}. If the Standard
Model is valid up to scales near the 
Planck scale, the SM Higgs mass is
restricted to a narrow window between
130 and 190 GeV. For Higgs masses
either above or below this window, new
physical phenomena are expected to
occur at a scale $\Lambda$ between
$\sim 1$ TeV and the Planck scale. For Higgs
masses of order 1 TeV, the scale of
new strong interactions would be as
low as $\sim 1$ TeV \cite{4,6}. 

\begin{figure}[hbt]
\begin{center}
\hspace*{-1.1cm}
\epsfig{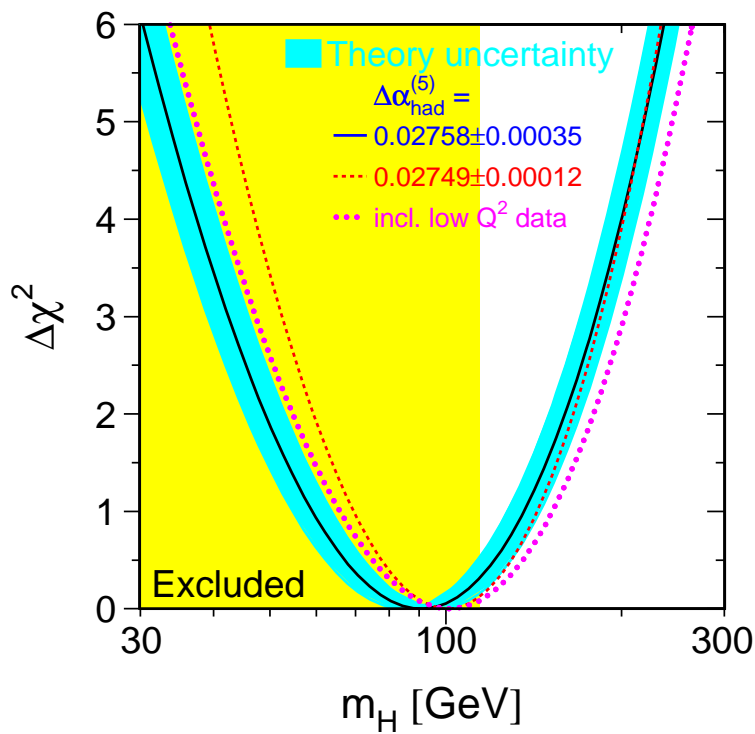}
\end{center}
\vspace*{-0.4cm}

\caption[]{\label{fg:SMHiggs}\it The $\Delta\chi^2$ curve 
derived from high-$Q^2$ precision 
electroweak measurements, performed at LEP and by SLD, CDF and D0, as a 
function of the Higgs boson mass, assuming the Standard Model to be the 
correct theory of nature.}
\end{figure}

The electroweak observables are affected
by the Higgs mass through radiative
corrections \cite{7}. Despite  the weak 
logarithmic dependence, the high-precision
electroweak data, {\it c.f.} Fig.~\ref{fg:SMHiggs}, indicate a preference
for light Higgs masses close to
$\sim 100$ GeV \cite{8}. At the 95\% CL, they
require a value of the Higgs mass less than $\sim 186$~GeV.
By searching directly for the SM Higgs particle,
the LEP experiments have set a lower
limit of $M_H\gsim 114$ GeV
 on the Higgs mass \cite{9}. Since the
Higgs boson has not been found at LEP2, 
the search will continue at the
Tevatron, which may reach masses up to $\sim 140$ GeV
\cite{11}. The proton collider LHC can sweep
the entire canonical Higgs mass range
of the Standard Model \cite{12}. The properties
of the Higgs particle can be analysed very
accurately at $e^+e^-$ linear colliders \cite{13}, thus establishing
the Higgs mechanism experimentally.

\paragraph{3.}
If the Standard Model is embedded in a 
Grand Unified Theory (GUT) at high energies,
the natural scale of electroweak symmetry
breaking would be expected close to the
unification scale $M_{GUT}$.
Supersymmetry \cite{14} provides a solution of
this hierarchy problem. The quadratically
divergent contributions to the radiative
corrections of the scalar Higgs boson
mass are cancelled by the destructive
interference between bosonic
and fermionic loops in sypersymmetric theories \cite{15}. The Minimal
Supersymmetric extension of the Standard
Model (MSSM) can be derived as an effective
theory from supersymmetric grand unified
theories. A strong indication for the realization
of this physical picture in nature is the
excellent agreement between the value of
the electroweak mixing angle $\sin^2 \theta_W$
predicted by the unification of the gauge
couplings, and the experimentally measured value. If the
gauge couplings are unified in the
minimal supersymmetric theory at a scale
$M_{GUT} = {\cal O}(10^{16}~\mbox{GeV})$, 
the electroweak mixing angle is predicted
to be $\sin^2\theta_W = 0.2336 \pm 0.0017$
\cite{16} for a mass spectrum of the supersymmetric
particles of order $M_Z$ to 1 TeV.
This theoretical prediction is matched very well by 
 the experimental result
$\sin^2\theta_W^{exp} = 0.23120 \pm 0.00015$
\cite{8}; the difference between the two numbers
is less than 2 per-mille.

In the MSSM, the Higgs sector is built up 
by two Higgs doublets \cite{17}. The doubling is
necessary to generate masses for up- and
down-type fermions in a supersymmetric
theory and to render the theory anomaly-free. 
The Higgs particle spectrum consists
of a quintet of states: two CP-even
scalar neutral ($h,H$), one CP-odd pseudoscalar
neutral ($A$), and a pair of charged ($H^\pm$)
Higgs bosons \cite{19}. The masses of the heavy
Higgs bosons, $H,A,H^\pm$, are expected to be of order $v$, 
but they may extend up to the TeV range. By contrast,
since the quartic Higgs self-couplings are 
determined by the gauge couplings, the mass
of the lightest Higgs boson $h$ is constrained
very stringently. At tree level, the mass
has been predicted to be smaller than the
$Z$ mass \cite{19}. Radiative corrections,
increasing as the fourth power of the
top mass, shift the upper limit to a value
between $\sim 100$ GeV and $\sim 140$
GeV, depending on the parameter $\tgb$,
the ratio of the vacuum expectation values
of the two neutral scalar Higgs fields.

A general lower bound of 91 GeV has been 
experimentally established
for the Higgs particle $h$ at
LEP \cite{9}. 
The search for $h$
masses in excess of $\sim 100$ GeV
and the search for the heavy Higgs bosons
continues at the Tevatron, the LHC and $e^+e^-$
linear colliders. 

\paragraph{4.}
A light Higgs boson may also be generated as a (pseudo-)Goldstone boson by 
breaking the global symmetry of new interactions.
Alternatively to supersymmetry, the quadratic divergencies could be
cancelled by new partners of the Standard Model particles that do not differ 
in the fermionic/bosonic character. Symmetry schemes constrain the couplings 
in such a way that the cancellations are achieved in a natural way. Such 
scenarios are realized in Little Higgs Models \cite{2A} which predict a large 
ensemble of new SM-type particles in the mass range of a few TeV.

\paragraph{5.}
Elastic-scattering amplitudes of massive
vector bosons grow indefinitely with
energy if they are calculated in a
perturbative expansion in the weak coupling
of a non-Abelian gauge theory. As a
result, they violate the unitarity beyond
a critical energy scale of 
$\sim 1.2$ TeV. Apart from
introducing a light Higgs boson, this problem can
also be solved by assuming the $W$ boson to become 
strongly interacting at TeV energies,
thus damping the rise of the elastic-scattering 
amplitudes. Naturally, the strong
forces between the $W$ bosons may be traced
back to new fundamental interactions
characterized by a scale of order 1 TeV \cite{2}.
If the underlying theory is globally
chiral-invariant, this  symmetry may be broken
spontaneously. The Goldstone bosons 
associated with the spontaneous 
breaking of the symmetry can be absorbed by  gauge
bosons to generate their masses and to build
up the longitudinal degrees of freedom of their wave functions.

Since the longitudinally polarized $W$ bosons
are associated with the Goldstone modes
of chiral symmetry breaking, the scattering
amplitudes of the $W_L$
bosons can be predicted for high energies
by a systematic expansion in the energy.
The leading term is parameter-free, a
consequence of the chiral symmetry-breaking
mechanism {\it per se}, which is independent of
the particular structure of the dynamical theory. The 
higher-order terms in the chiral expansion however 
are defined by the detailed structure 
of the underlying theory. With rising
energy the chiral expansion is expected to diverge
and new resonances may be generated in
$WW$ scattering at mass scales between 1
and 3 TeV. This picture is analogous to
pion dynamics in QCD, where the threshold
amplitudes can be predicted in a chiral
expansion, while at higher energies vector
and scalar resonances are formed in $\pi \pi$
scattering.

Such a scenario can be studied in $WW$
scattering experiments, where the $W$ bosons
are radiated, as quasi-real particles \cite{22},
off high-energy quarks in the proton
beams of the LHC \cite{12}, \citer{23,22A} or off electrons
and positrons in TeV linear colliders \cite{13,24,24a}.

\paragraph{6.}
Also in theories with extra space dimensions, the electroweak symmetries can 
be broken without introducing additional fundamental scalar fields, leading 
also to higgsless theories. Since in 5-dimensional theories the wave-functions 
are expanded by a fifth component, the symmetries can be broken by applying 
appropriately chosen boundary conditions to this field component \cite{RI2}. 
This additional scalar component of the original 5-dimensional gauge field is 
absorbed to generate the massive Kaluza-Klein towers of the gauge fields in 
four dimensions. The additional exchange of these towers in $WW$ scattering 
damps the scattering amplitude of the Standard Model and allows in principle 
to extend the theory to energies beyond the 1.2 TeV unitarity bound of 
higgsless scenarios. However, it is presently unclear whether realistic models 
of this type can be constructed that give rise to small enough elastic $WW$ 
scattering amplitudes compatible with perturbative unitarity \cite{RI3}.

\paragraph{7.}
This report is divided into three parts.
A basic introduction and a summary of the
main theoretical and experimental results
will be presented in the next section on
the Higgs sector of the Standard Model.
Also the search for the Higgs particle
at future $e^+e^-$
and hadron colliders will be described. In
the same way, the Higgs spectrum of 
supersymmetric theories will be discussed
in the following section. Finally, the main 
features of strong $W$ interactions and their
analysis in $WW$ scattering experiments will
be presented in the last section.

Only the basic elements of electroweak symmetry
breaking and Higgs mechanism can be reviewed  
in this report. Other aspects may be traced back
from Ref. \cite{24b} and the reports collected in Ref. \cite{24A} on which 
these lectures are based.

\section{The Higgs Sector of the Standard Model}

\phantom{h}
\subsection{The Higgs Mechanism}

\begin{figure}[hbt]
\begin{center}
\begin{picture}(80,80)(50,-20)
\Photon(0,50)(25,25){3}{3}
\Photon(0,0)(25,25){3}{3}
\Photon(25,25)(50,25){3}{3}
\Photon(50,25)(75,50){3}{3}
\Photon(50,25)(75,0){3}{3}
\put(-80,23){$(a)$}
\put(-15,-2){$W$}
\put(-15,48){$W$}
\put(80,-2){$W$}
\put(80,48){$W$}
\put(35,10){$W$}
\end{picture}
\begin{picture}(60,80)(0,-20)
\Photon(0,50)(100,50){3}{12}
\Photon(0,0)(100,0){3}{12}
\Photon(50,50)(50,0){3}{6}
\end{picture}
\begin{picture}(60,80)(-90,-20)
\Photon(0,50)(25,25){3}{3}
\Photon(0,0)(25,25){3}{3}
\Photon(25,25)(50,50){3}{3}
\Photon(25,25)(50,0){3}{3}
\end{picture} \\
\begin{picture}(80,60)(83,0)
\Photon(0,50)(25,25){3}{3}
\Photon(0,0)(25,25){3}{3}
\DashLine(25,25)(50,25){6}
\Photon(50,25)(75,50){3}{3}
\Photon(50,25)(75,0){3}{3}
\put(-80,23){$(b)$}
\put(-15,-2){$W$}
\put(-15,48){$W$}
\put(80,-2){$W$}
\put(80,48){$W$}
\put(35,10){$H$}
\end{picture}
\begin{picture}(60,60)(33,0)
\Photon(0,50)(100,50){3}{12}
\Photon(0,0)(100,0){3}{12}
\DashLine(50,50)(50,0){5}
\end{picture}  \\
\end{center}
\caption[]{\it \label{fg:wwtoww} Generic diagrams of elastic $WW$ scattering:
(a) pure gauge-boson dynamics, and (b) Higgs-boson exchange.}
\end{figure}
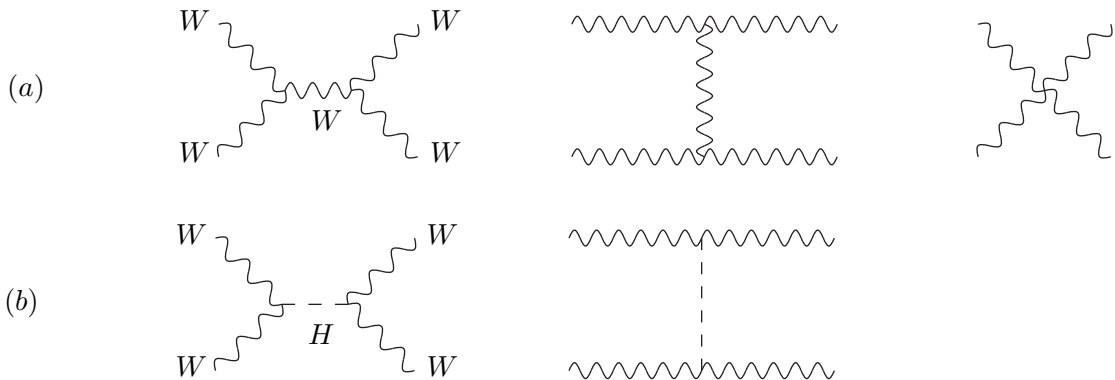

\phantom{h}
At high energies, the amplitude for the elastic 
scattering of massive $W$ bosons, $WW \to WW$, 
grows indefinitely with energy for longitudinally 
polarized particles, Fig.~\ref{fg:wwtoww}a. This is a consequence
of the linear rise of the longitudinal $W_L$ wave function, 
$\epsilon_L = (p,0,0,E)/M_W$,
with the energy of the particle. Even though the term of 
the scattering amplitude rising as the fourth power in the energy
is cancelled by virtue of the non-Abelian 
gauge symmetry, the amplitude remains quadratically
divergent in the energy. On the other hand, 
unitarity requires elastic-scattering 
amplitudes of partial waves $J$ to be bounded by
$\Re e A_J \leq 1/2$.
Applied to the asymptotic $S$-wave amplitude
$A_0 = G_F s/8\pi\sqrt{2}$ of the isospin-zero channel
$2W_L^+W_L^- + Z_L Z_L$,  the bound  \cite{25}
\begin{equation}
s \leq 4\pi\sqrt{2}/G_F \sim (1.2~\mbox{TeV})^2
\end{equation}
on the c.m. energy $\sqrt{s}$ can be derived for 
the validity of a theory of weakly 
coupled massive gauge bosons.
\noindent

However, the quadratic rise in the energy can be damped by 
exchanging a new scalar particle, Fig.~\ref{fg:wwtoww}b. 
To achieve the 
cancellation, the size of the coupling must be given 
by the product of the gauge coupling with the gauge
boson mass. For high energies, the
amplitude $A'_0 = -G_F s/8\pi\sqrt{2}$
cancels exactly the quadratic divergence of the pure 
gauge-boson amplitude $A_0$.
Thus, unitarity can be restored by introducing a weakly
coupled \underline{\it Higgs particle}.

In the same way, the linear divergence of the amplitude 
$A(f\bar f\to W_L W_L)\sim gm_f\sqrt{s}$
for the annihilation of a fermion--antifermion pair to 
a pair of longitudinally polarized gauge bosons,
can be damped by adding the Higgs exchange to 
the gauge-boson exchange. In this case
the Higgs particle must couple proportionally
to the mass $m_f$ of the fermion $f$.

These observations can be summarized in a rule\footnote{The rule 
appears to remain valid even if theories in more than four space-time 
dimensions are included.}:
{\it A theory of massive gauge bosons and fermions
that are weakly coupled up to very high 
energies, requires, by unitarity, the existence 
of a Higgs particle; the Higgs particle is a
scalar $0^+$ particle that couples to other particles 
proportionally to the masses of the particles.}

The assumption that the couplings of the 
fundamental particles are weak up to very
high energies is qualitatively supported
by the perturbative renormalization of the
electroweak mixing angle $\sin^2\theta_W$
from the symmetry value 3/8 at the GUT scale
down to $\sim 0.2$, 
which is close to the experimentally observed
value at low energies.\\

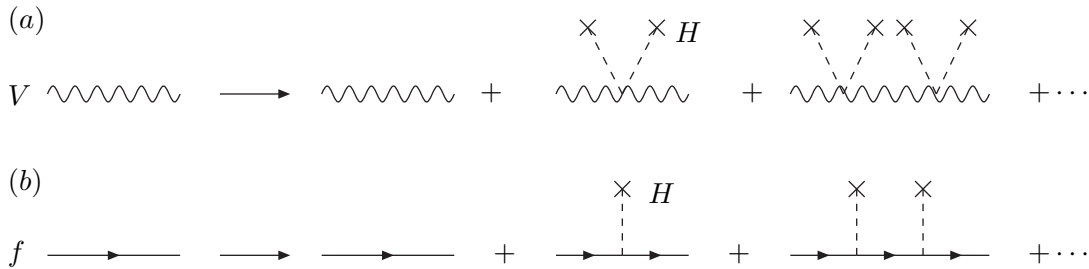
\begin{figure}[hbt]
\begin{center}
\begin{picture}(60,10)(80,40)
\Photon(0,25)(50,25){3}{6}
\LongArrow(65,25)(90,25)
\put(-15,21){$V$}
\put(-15,50){$(a)$}
\end{picture}
\begin{picture}(60,10)(40,40)
\Photon(0,25)(50,25){3}{6}
\put(60,23){$+$}
\end{picture}
\begin{picture}(60,10)(15,40)
\Photon(0,25)(50,25){3}{6}
\DashLine(25,25)(12,50){3}
\DashLine(25,25)(38,50){3}
\Line(9,53)(15,47)
\Line(9,47)(15,53)
\Line(35,53)(41,47)
\Line(35,47)(41,53)
\put(45,45){$H$}
\put(70,23){$+$}
\end{picture}
\begin{picture}(60,10)(-10,40)
\Photon(0,25)(75,25){3}{9}
\DashLine(20,25)(8,50){3}
\DashLine(20,25)(32,50){3}
\DashLine(55,25)(43,50){3}
\DashLine(55,25)(67,50){3}
\Line(5,53)(11,47)
\Line(5,47)(11,53)
\Line(29,53)(35,47)
\Line(29,47)(35,53)
\Line(40,53)(46,47)
\Line(40,47)(46,53)
\Line(64,53)(70,47)
\Line(64,47)(70,53)
\put(90,23){$+ \cdots$}
\end{picture} \\
\begin{picture}(60,80)(80,20)
\ArrowLine(0,25)(50,25)
\LongArrow(65,25)(90,25)
\put(-15,23){$f$}
\put(-15,50){$(b)$}
\end{picture}
\begin{picture}(60,80)(40,20)
\ArrowLine(0,25)(50,25)
\put(65,23){$+$}
\end{picture}
\begin{picture}(60,80)(15,20)
\ArrowLine(0,25)(25,25)
\ArrowLine(25,25)(50,25)
\DashLine(25,25)(25,50){3}
\Line(22,53)(28,47)
\Line(22,47)(28,53)
\put(35,45){$H$}
\put(65,23){$+$}
\end{picture}
\begin{picture}(60,80)(-10,20)
\ArrowLine(0,25)(25,25)
\ArrowLine(25,25)(50,25)
\ArrowLine(50,25)(75,25)
\DashLine(25,25)(25,50){3}
\DashLine(50,25)(50,50){3}
\Line(22,53)(28,47)
\Line(22,47)(28,53)
\Line(47,53)(53,47)
\Line(47,47)(53,53)
\put(90,23){$+ \cdots$}
\end{picture}  \\
\end{center}
\caption[]{\it \label{fg:massgen} Generating (a) gauge boson and (b)
fermion masses through interactions with the scalar background field.}
\end{figure}

These ideas can be cast into an elegant
mathematical form by interpreting the electroweak
interactions as a gauge theory with spontaneous
symmetry breaking in the scalar sector\footnote{The mechanism of spontaneous
symmetry breaking, including the Goldstone theorem as well as the Higgs 
mechanism, are exemplified for the illustrative $O(3)$ $\sigma$ model in 
Appendix A.}. 
Such a theory consists of fermion fields, gauge
fields and a scalar field coupled by the 
standard gauge interactions and Yukawa interactions
to the other fields. Moreover, a self-interaction
\begin{equation}
V = \frac{\lambda}{2} \left[ |\phi|^2 - \frac{v^2}{2} \right]^2
\label{eq:potential}
\end{equation}
is introduced in the scalar sector, which leads to 
a non-zero ground-state value $v/\sqrt{2}$
of the scalar field. By fixing the phase of the 
vacuum amplitude to  zero, the gauge symmetry
is spontaneously broken in the scalar sector. Interactions of the 
gauge fields with the scalar background field,
Fig.~\ref{fg:massgen}a, and Yukawa interactions of the 
fermion fields with the background field, Fig.~\ref{fg:massgen}b, 
shift the masses of these fields from 
zero to non-zero values:
\begin{equation}
\begin{array}{lrclclcl}
\displaystyle
(a) \hspace*{2.0cm} &
\displaystyle
\frac{1}{q^2} & \to & \displaystyle \frac{1}{q^2} + \sum_j \frac{1}{q^2}
\left[ \left( \frac{gv}{\sqrt{2}} \right)^2 \frac{1}{q^2} \right]^j & = &
\displaystyle \frac{1}{q^2-M^2} & : & \displaystyle M^2 = g^2 \frac{v^2}{2}
\\ \\
(b) &
\displaystyle
\frac{1}{\not \! q} & \to & \displaystyle \frac{1}{\not \! q} +
\sum_j \frac{1}{\not \! q} \left[ \frac{g_fv}{\sqrt{2}} \frac{1}{\not
\! q} \right]^j & = & \displaystyle \frac{1}{\not \! q-m_f} & : &
\displaystyle m_f = g_f \frac{v}{\sqrt{2}} 
\end{array}
\end{equation}
Thus, in theories with gauge and Yukawa interactions,
in which the scalar field acquires a non-zero
ground-state value, the couplings are naturally
proportional to the masses. This ensures the
unitarity of the theory as discussed before.
These theories are renormalizable (as a result
of the gauge invariance, which is only disguised 
in the unitary formulation adopted so far), and 
thus they are well-defined and mathematically 
consistent.

\subsection{The Higgs Mechanism in the Standard Model}

\phantom{h}
Besides the Yang--Mills and the fermion parts, the
electroweak $SU_2 \times U_1$
Lagrangian includes a scalar isodoublet field
$\phi$, coupled to itself in the potential $V$,
cf. eq. (\ref{eq:potential}),
to the gauge fields through the covariant derivative
$iD = i\partial - g \vec{I} \vec{W} - g'YB$,
and to the up and down fermion fields $u,d$
through Yukawa interactions:
\begin{equation}
{\cal L}_0 = |D\phi|^2 - \frac{\lambda}{2} \left[ |\phi|^2
- \frac{v^2}{2} \right]^2 - g_d \bar d_L \phi d_R - g_u \bar u_L
\phi_c u_R + {\rm hc} ~.
\end{equation}
In the unitary gauge, the isodoublet $\phi$
is replaced by the physical Higgs field
$H$, $\phi\to [0,(v+H)/\sqrt{2}]$,
which describes the fluctuation of the $I_3=-1/2$
component about the ground-state 
value $v/\sqrt{2}$. The scale $v$
of the electroweak symmetry breaking is fixed
by the $W$ mass, which in turn can be reexpressed by the
Fermi coupling, $v = 1/\sqrt{\sqrt{2}G_F} \approx 246$ GeV.
The quartic coupling $\lambda$
and the Yukawa couplings $g_f$
can be reexpressed in terms of the physical
Higgs mass $M_H$ and the fermion masses $m_f$: 
\begin{eqnarray}
M_H^2 & = & \lambda v^2 \nonumber \\
m_f & = & g_f v / \sqrt{2}
\end{eqnarray}
respectively.

Since the couplings of the Higgs particle to gauge particles, 
to fermions and to itself are given by the gauge couplings
and the masses of the particles, the only unknown
parameter in the Higgs sector (apart from the CKM 
mixing matrix) is the Higgs mass. When this mass is fixed,
all properties of the Higgs particle can be predicted,
i.e. the lifetime and decay branching ratios, as 
well as the production mechanisms and the corresponding
cross sections.

\subsubsection{The SM Higgs Mass \\ \\}

Even though the mass of the Higgs boson cannot be 
predicted in the Standard Model, stringent upper
and lower bounds can nevertheless be derived from
internal consistency conditions and extrapolations
of the model to high energies.

The Higgs boson has been introduced as a fundamental
particle to render 2--2 scattering amplitudes involving
longitudinally polarized $W$
bosons compatible with unitarity. Based on the general
principle of time-energy uncertainty, particles must
decouple from a physical system if their mass grows
indefinitely. The mass of the Higgs particle must 
therefore be bounded to restore unitarity in the 
perturbative regime. From the asymptotic expansion of 
the elastic $W_L W_L$ $S$-wave scattering amplitude including 
$W$ and Higgs exchanges, $A(W_L W_L \to W_L W_L) \to -G_F M_H^2/4\sqrt{2}\pi$,
it follows \cite{25} that
\begin{equation}
M_H^2 \leq 2\sqrt{2}\pi/G_F \sim (850~\mbox{GeV})^2 ~.
\end{equation}
Within the canonical formulation of the Standard Model,
consistency conditions  therefore require a Higgs mass below 1 TeV.\\

\begin{figure}[hbt]
\vspace*{-0.5cm}

\begin{center}
\begin{picture}(90,80)(60,-10)
\DashLine(0,50)(25,25){3}
\DashLine(0,0)(25,25){3}
\DashLine(50,50)(25,25){3}
\DashLine(50,0)(25,25){3}
\put(-15,45){$H$}
\put(-15,-5){$H$}
\put(55,-5){$H$}
\put(55,45){$H$}
\end{picture}
\begin{picture}(90,80)(10,-10)
\DashLine(0,50)(25,25){3}
\DashLine(0,0)(25,25){3}
\DashLine(75,50)(50,25){3}
\DashLine(75,0)(50,25){3}
\DashCArc(37.5,25)(12.5,0,360){3}
\put(-15,45){$H$}
\put(-15,-5){$H$}
\put(35,40){$H$}
\put(80,-5){$H$}
\put(80,45){$H$}
\end{picture}
\begin{picture}(50,80)(-40,2.5)
\DashLine(0,0)(25,25){3}
\DashLine(0,75)(25,50){3}
\DashLine(50,50)(75,75){3}
\DashLine(50,25)(75,0){3}
\ArrowLine(25,25)(50,25)
\ArrowLine(50,25)(50,50)
\ArrowLine(50,50)(25,50)
\ArrowLine(25,50)(25,25)
\put(-15,70){$H$}
\put(-15,-5){$H$}
\put(35,55){$t$}
\put(80,-5){$H$}
\put(80,70){$H$}
\end{picture}  \\
\setlength{\unitlength}{1pt}
\caption[]{\label{fg:lambda} \it Diagrams generating the evolution of
the Higgs self-interaction $\lambda$.}
\end{center}
\end{figure}
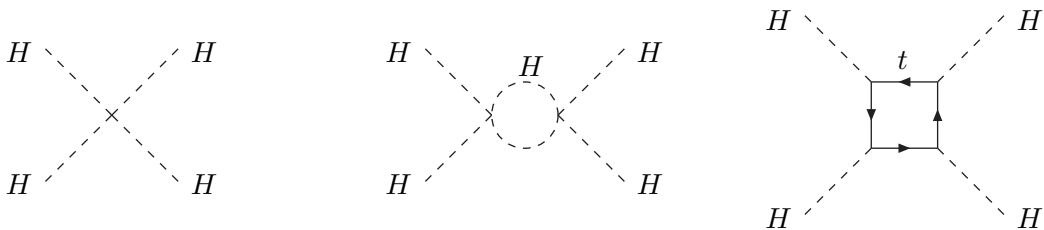

\begin{figure}[hbtp]

\vspace*{0.8cm}

\hspace*{2.0cm}
\begin{turn}{90}%
\epsfxsize=8.5cm \epsfbox{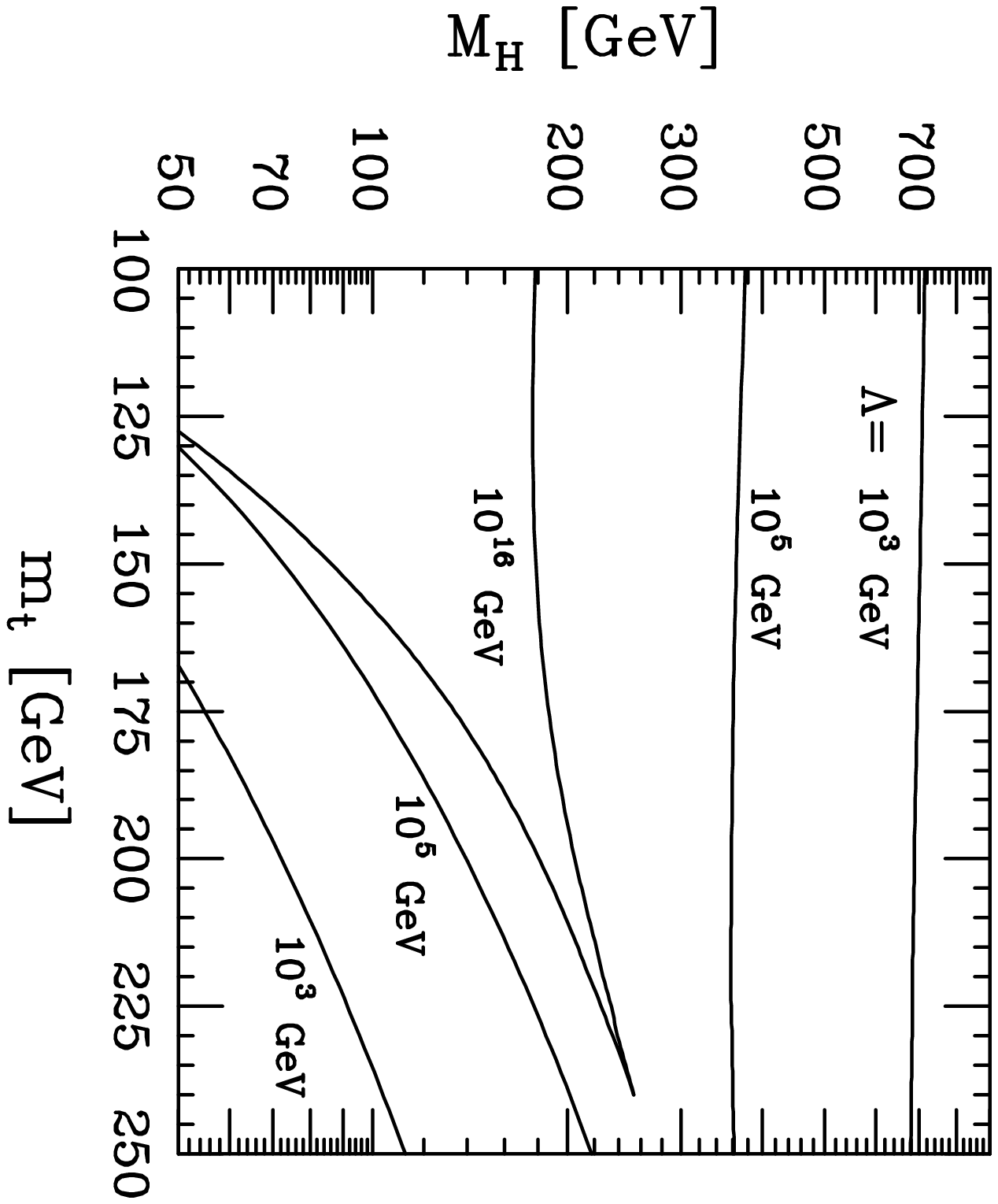}
\end{turn}
\vspace*{-0.2cm}

\caption[]{\label{fg:triviality} \it Bounds on the mass of the Higgs boson in
the SM. Here $\Lambda$ denotes the energy scale at which the Higgs-boson system of
the SM would become strongly interacting (upper bound); the lower bound follows
from the requirement of vacuum stability. (Refs. \cite{4,5}.)}
\vspace*{-0.3cm}
\end{figure}

Quite restrictive bounds on the value of the SM Higgs
mass follow from the hypothesis on the 
energy scale $\Lambda$
up to which the Standard Model can be extended before
new strong interaction phenomena emerge. The 
key to these bounds is the evolution of the quartic
coupling $\lambda$
with the energy 
due to quantum fluctuations \cite{4}. The basic
contributions are depicted in Fig.~\ref{fg:lambda}. The
Higgs loop itself gives rise to an indefinite increase
of the coupling while the fermionic top-quark
loop, with increasing top mass, drives the coupling
to smaller values, finally even to values below zero.
The variation of the quartic Higgs coupling
$\lambda$ and the top-Higgs Yukawa coupling $g_t$
with energy, parametrized by $t=\log \mu^2/v^2$,
may be written as \cite{4}
\begin{equation}
\begin{array}{rclcl}
\displaystyle \frac{d\lambda}{dt} & = & \displaystyle \frac{3}{8\pi^2}
\left[ \lambda^2 + \lambda g_t^2 - g_t^4 \right] & : & \displaystyle
\lambda(v^2) = M_H^2/v^2 \\ \\
\displaystyle \frac{d g_t}{dt} & = & \displaystyle \frac{1}{32\pi^2}
\left[ \frac{9}{2} g_t^3 - 8 g_t g_s^2 \right] & : & \displaystyle
g_t(v^2) = \sqrt{2}~m_f/v ~.
\end{array}
\end{equation}
Only the leading contributions from Higgs, top
and QCD loops are taken into account.\\

For moderate top masses, the quartic coupling $\lambda$
rises indefinitely, $\partial \lambda / \partial t \sim + \lambda^2$,
and the coupling becomes strong shortly before 
reaching the Landau pole:
\begin{equation}
\lambda (\mu^2) = \frac{\lambda(v^2)}{1- \frac{3\lambda(v^2)}{8\pi^2} \log
\frac{\mu^2}{v^2}} ~.
\end{equation}
Reexpressing the initial value of $\lambda$
by the Higgs mass, the condition $\lambda (\Lambda) < \infty$,
can be translated to an \underline{upper bound} on the Higgs
mass: 
\begin{equation}
M_H^2 \leq \frac{8\pi^2 v^2}{3\log \Lambda^2/v^2} ~.
\end{equation}
This mass bound is related logarithmically to the energy $\Lambda$
up to which the Standard Model is assumed to be valid.
The maximal value of $M_H$ for the minimal cut-off $\Lambda \sim $ 1~TeV
is given by $\sim 750$ GeV. This bound is close to the estimate
of $\sim 700$ GeV
in lattice calculations for $\Lambda \sim 1$ TeV,
which allow  proper control of non-perturbative 
effects near the boundary \cite{6}.\\

\begin{table}[hbt]
\renewcommand{\arraystretch}{1.5}
\begin{center}
\begin{tabular}{|l||l|} \hline
$\Lambda$ & $M_H$ \\ \hline \hline
1 TeV & 60 GeV $\lessim M_H \lessim 700$ GeV \\
$10^{19}$ GeV & 130 GeV $\lessim M_H \lessim 190$ GeV \\ \hline
\end{tabular}
\renewcommand{\arraystretch}{1.2}
\caption[]{\label{tb:triviality}
\it Higgs mass bounds for two values of the cut-off $\Lambda$.}
\end{center}
\end{table}
A \underline{lower bound} on the Higgs mass can be derived from 
the requirement of vacuum stability \cite{4,5}. Since
top-loop corrections reduce $\lambda$
for increasing top-Yukawa coupling, $\lambda$
becomes negative if the top mass becomes too large. 
In such a case, the self-energy potential would 
become deep negative and the ground state would no longer 
be stable. To avoid the instability, the Higgs
mass must exceed a minimal value for a given top 
mass. This lower bound depends on the cut-off value $\Lambda$.\\

For any given $\Lambda$ the allowed values of $(M_t,M_H)$ pairs are shown
in Fig.~\ref{fg:triviality}. The allowed Higgs mass values are
collected in Table \ref{tb:triviality} for two specific cut-off values
$\Lambda$.  If the Standard Model is assumed to be valid up to the scale of
grand unification, the Higgs mass is restricted to a narrow window
between 59 and 136~GeV. The observation of a Higgs mass above or below
this window would demand a new physics scale below the GUT scale.

\subsubsection{Decays of the Higgs Particle \\ \\}
The profile of the Higgs particle is uniquely
determined if the Higgs mass is fixed. The
strength of the Yukawa couplings of the Higgs
boson to fermions is set by the fermion masses $m_f$,
and the coupling to the electroweak gauge bosons
$V=W,Z$ by their masses $M_V$:
\begin{eqnarray}
g_{ffH} & = & \left[ \sqrt{2} G_F \right]^{1/2} m_f \\
g_{VVH} & = & 2 \left[ \sqrt{2} G_F \right]^{1/2} M_V^2 ~. \nonumber
\end{eqnarray}
The total decay width and lifetime, as well as the
 branching ratios for specific decay channels, are 
determined by these parameters. The measurement
of the decay characteristics can therefore
by exploited to establish, experimentally, 
that Higgs couplings grow with the masses
of the particles, a direct consequence of
the Higgs mechanism {\it sui generis}.

For Higgs particles in the intermediate mass range
${\cal O}(M_Z) \leq M_H \leq 2M_Z$, 
the main decay modes are decays into 
$b\bar b$ pairs and $WW,ZZ$
pairs,  one of the gauge bosons being virtual
below the respective threshold. Above the
$WW,ZZ$ pair thresholds, the Higgs particles decay almost
exclusively into these two channels, with a small
admixture of top decays near the $t\bar t$ threshold.
Below 140 GeV, the decays $H\to \tau^+\tau^-, c\bar c$
and $gg$ are also important besides the dominating
$b\bar b$ channel; $\gamma\gamma$
decays, though suppressed in rate, nevertheless provide 
 a clear 2-body signature for the
formation of Higgs particles in this mass range.

\paragraph{(a) Higgs decays to fermions} ~\\[0.5cm]
The partial width of Higgs decays to lepton
and quark pairs is given by \cite{26}
\begin{equation}
\Gamma (H\to f\bar f) = {\cal N}_c \frac{G_F}{4\sqrt{2}\pi} m_f^2(M_H^2) M_H
~,
\end{equation}
${\cal N}_c = 1$ or 3 being the color factor. Near the threshold the partial
width is suppressed by the additional P-wave factor $\beta_f^3$, where 
$\beta_f$
is the fermion velocity. Asymptotically, the
fermionic width grows only linearly with the
Higgs mass.
The bulk of QCD radiative corrections can be 
mapped into the scale dependence of the quark mass,
evaluated at the Higgs mass. For $M_H\sim 100$ GeV
the relevant parameters are $m_b (M_H^2) \simeq 3$ GeV and
$m_c (M_H^2) \simeq$ 0.6~GeV.
The reduction of the effective $c$-quark mass
overcompensates the color factor in the ratio 
between charm and $\tau$
decays of Higgs bosons. The residual QCD corrections,
$\sim 5.7 \times (\alpha_s/\pi)$, modify the widths only slightly. 

\paragraph{(b) Higgs decays to $WW$ and $ZZ$ boson pairs} ~\\[0.5cm]
Above the $WW$ and $ZZ$ decay thresholds, the partial widths for these 
channels may be written as \cite{27}
\begin{equation}
\Gamma (H\to VV) = \delta_V \frac{G_F}{16\sqrt{2}\pi} M_H^3 (1-4x+12x^2)
\beta_V ~,
\end{equation}
where $x=M_V^2/M_H^2$ and $\delta_V = 2$ and 1 for $V=W$ and $Z$, 
respectively. For large Higgs masses, the vector bosons
are longitudinally polarized. Since the wave functions
of these states are linear in the energy, the widths
grow as the third power of the Higgs mass. Below the
threshold for two real bosons, the Higgs particle can
decay into $VV^*$
pairs, one of the vector bosons being virtual. The partial 
width is given in this case \cite{28} by
\begin{equation}
\Gamma(H\to VV^*) = \frac{3G^2_F M_V^4}{16\pi^3}~M_H  R(x)~\delta'_V ~,
\end{equation}
where $\delta'_W = 1$, $\delta'_Z = 7/12 - 10\sin^2\theta_W/9 + 40
\sin^4\theta_W/27$ and
\begin{displaymath}
R(x) = \frac{3(1-8x+20x^2)}{(4x-1)^{1/2}}\arccos\left(\frac{3x-1}{2x^{3/2}}
\right) - \frac{1-x}{2x} (2-13x+47x^2)
- \frac{3}{2} (1-6x+4x^2) \log x ~.
\end{displaymath}
The $ZZ^*$ channel becomes relevant for Higgs masses beyond 
$\sim 140$ GeV.  Above the threshold, the 4-lepton channel 
$H\to ZZ \to 4 \ell^\pm$
provides a very clear signal for Higgs bosons. Despite of escaping 
neutrinos in leptonic $W$ decays, also the $WW$ decay channel proves useful 
if the on-shell $ZZ$ channel is still closed kinematically.

\paragraph{(c) Higgs decays to $gg$ and $\gamma\gamma$ pairs} ~\\[0.5cm]
In the Standard Model, gluonic Higgs decays are 
mediated by top- and bottom-quark loops, photonic decays in addition by 
$W$ loops. Since these decay modes are significant
only far below the top and $W$
thresholds, they are described by the approximate
expressions \cite{29,30}
\begin{eqnarray}
\Gamma (H\to gg) & = & \frac{G_F \alpha_s^2(M_H^2)}{36\sqrt{2}\pi^3}M_H^3
\left[ 1+ \left(\frac{95}{4} - \frac{7N_F}{6} \right) \frac{\alpha_s}{\pi}
\right] \label{eq:htogg} \\ \nonumber \\
\Gamma (H\to \gamma\gamma) & = & \frac{G_F \alpha^2}{128\sqrt{2}\pi^3}M_H^3
\left[  \frac{4}{3} {\cal N}_C e_t^2 - 7 \right]^2 ~,
\end{eqnarray}
which are valid in the limit $M_H^2 \ll 4M_W^2, 4M_t^2$.
The QCD radiative corrections, which include the $ggg$ and $gq\bar q$
final states in (\ref{eq:htogg}), are very important; they
increase the partial width by about 65\%. Even
though photonic Higgs decays are very rare, 
they nevertheless offer a simple and 
attractive signature for Higgs particles
by leading to just two stable particles 
in the final state.

\paragraph{\underline{Digression:}} Loop-mediated Higgs couplings can 
easily be calculated in the limit in which the  Higgs
mass is small compared to the loop mass, by exploiting a low-energy 
theorem \citer{29,32} for the external Higgs amplitude ${\cal A} (XH)$:
\begin{equation}
\lim_{p_H\to 0} {\cal A}(XH) = \frac{1}{v} \frac{\partial {\cal A}(X)}{\partial
 \log m} ~.
\end{equation}
The theorem can be derived by observing that the
insertion of an external zero-energy Higgs line into a 
fermionic propagator, for instance, is equivalent
to the substitution
\begin{displaymath}
\frac{1}{\not\! p-m} \to \frac{1}{\not\! p-m} \frac{m}{v} \frac{1}{\not\! p-m}
= \frac{1}{v} \frac{\partial}{\partial \log m} \frac{1}{\not\! p-m} ~.
\end{displaymath}
The amplitudes for processes including an
external Higgs line can therefore be obtained
from the amplitude without the external Higgs
line by taking the logarithmic derivative.
If applied to the gluon propagator at $Q^2=0$, $\Pi_{gg} \sim 
(\alpha_s/12\pi)
GG \log m$, the $Hgg$ amplitude can easily be derived as
${\cal A}(Hgg) = GG \alpha_s/(12\pi v)$. If higher 
orders are included, the parameter $m$ must be interpreted
as bare mass.

\begin{figure}[hbtp]

\vspace*{0.5cm}
\hspace*{1.0cm}
\begin{turn}{-90}%
\epsfxsize=8.5cm \epsfbox{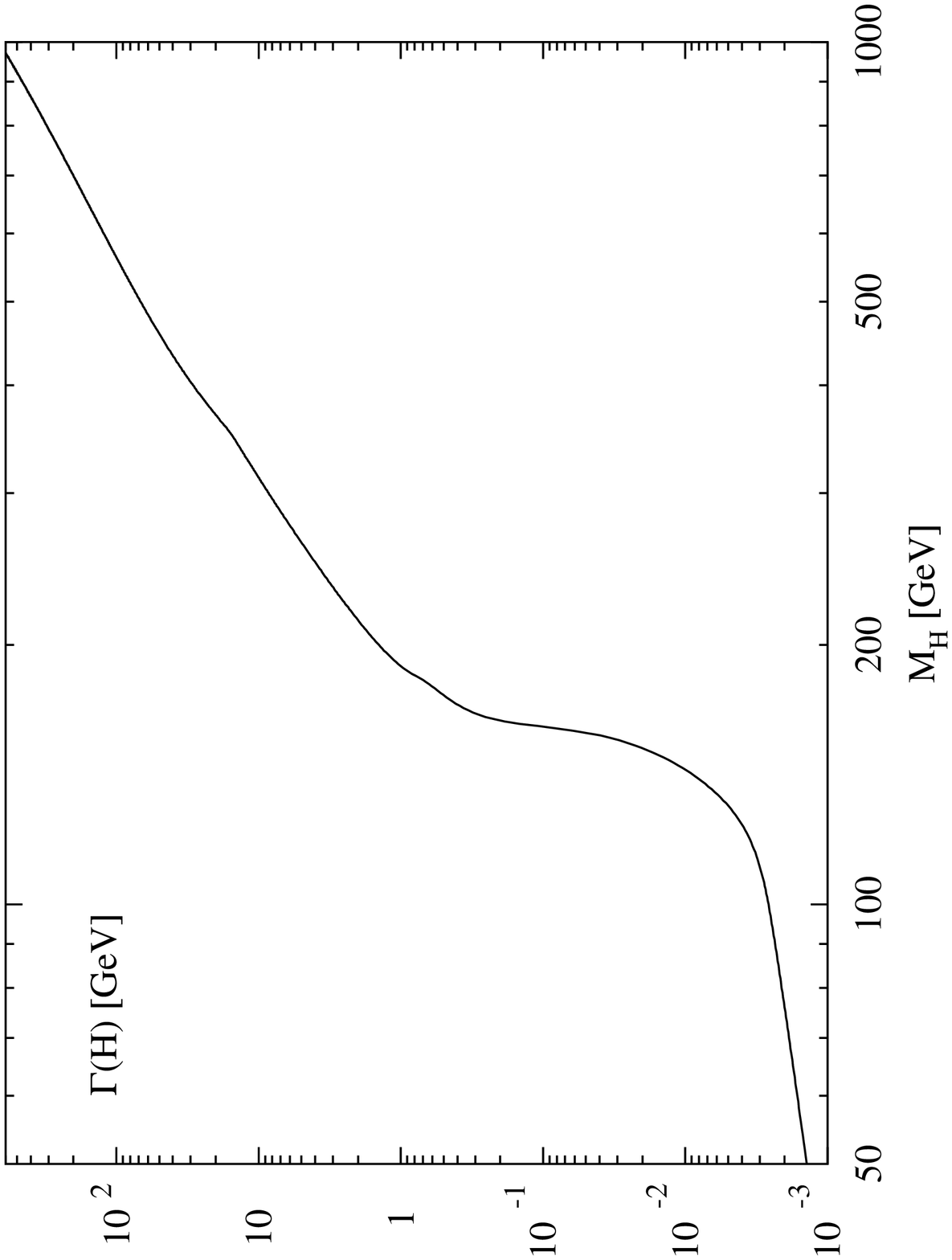}
\end{turn}

\vspace*{0.5cm}
\hspace*{1.0cm}
\begin{turn}{-90}%
\epsfxsize=8.5cm \epsfbox{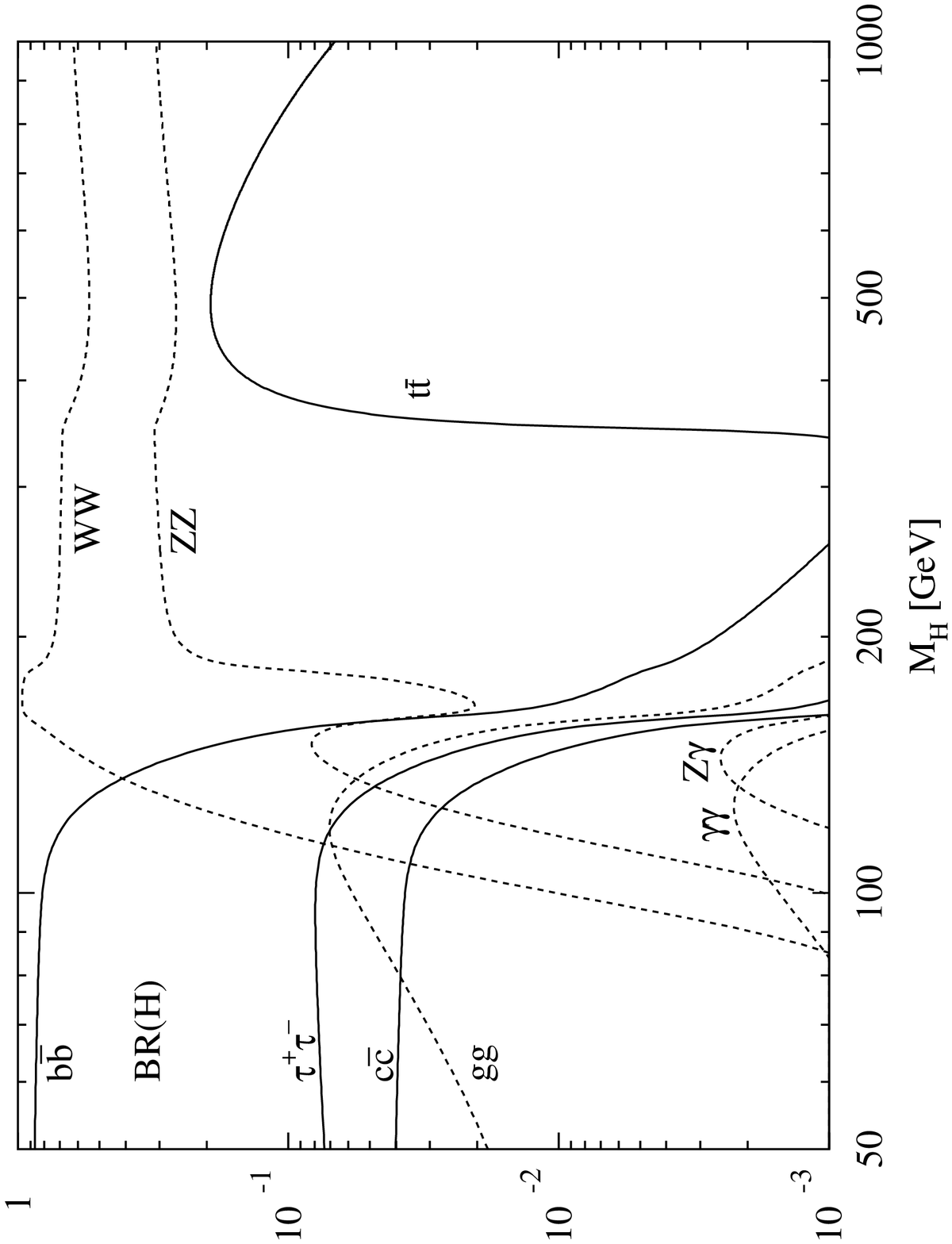}
\end{turn}
\vspace*{-0.0cm}

\caption[]{\label{fg:wtotbr} \it (a) Total decay width (in GeV) of the SM
Higgs boson as a function of its mass. (b) Branching ratios of the
dominant decay modes of the SM Higgs particle. All relevant higher-order
corrections are taken into account.}
\end{figure}

\paragraph{(d) Summary} ~\\[0.5cm]
By adding up all possible decay channels, we obtain 
the total width shown in Fig.~\ref{fg:wtotbr}a. Up to masses of 
140 GeV, the Higgs particle is very narrow, $\Gamma(H) \leq 10$ MeV.
After the real and virtual gauge-boson
channels open up, the state rapidly becomes  wider,
reaching a width of $\sim 1$ GeV at the $ZZ$
threshold. The width cannot be measured directly
in the intermediate mass region at the LHC or $e^+ e^-$
colliders. However it can be determined indirectly; measuring, for
example, the
partial width $\Gamma (H\to WW)$ through the fusion process $WW\to H$, and
the branching fraction $BR(H\to WW)$ in the decay process $H\to WW$, the total 
width follows from the ratio of the two observables. 
Above a mass of $\sim 250$ GeV,
the state becomes wide enough to be resolved experimentally.

The branching ratios of the main decay modes are
displayed in Fig.~\ref{fg:wtotbr}b. A large variety of channels
will be accessible for Higgs masses below 140 GeV. 
The dominant mode is $b\bar b$ decays, yet
$c\bar c, \tau^+\tau^-$ and $gg$ decays 
still occur at a level of several per cent.
At $M_H=$ 120~GeV for instance, the branching ratios are 68\% for
$b\bar b$, 3.1\% for $c\bar c$, 6.9\% for $\tau^+\tau^-$ and 7\% for $gg$.
$\gamma\gamma$
decays occur at a level of 1 per mille. Above this mass value, 
the Higgs boson decay into $W$'s
becomes dominant, overwhelming all other channels if 
the decay mode into two real $W$'s is kinematically possible.
For Higgs masses far above the thresholds, 
$ZZ$ and $WW$ decays occur at a ratio of 1:2, slightly modified only
just above the $t\bar t$ threshold. Since the width grows as the third
power of the mass, the Higgs particle becomes very wide, $\Gamma(H) \sim
\frac{1}{2} M_H^3$ [TeV]. In fact, for $M_H\sim 1$ TeV,
the width reaches $\sim \frac{1}{2}$ TeV.

\subsection{Electroweak Precision Data: Estimate of the Higgs Mass}  

\phantom{h}
Indirect evidence for a light Higgs 
boson can be derived from the
high-precision measurements of electroweak 
observables at LEP and elsewhere. Indeed, the
fact that the Standard Model is renormalizable only 
after including the top and Higgs particles in 
the loop corrections, indicates that the
electroweak observables are sensitive to the
masses of these particles.

The Fermi coupling can be rewritten in terms of 
the weak coupling and the $W$ mass; at lowest order, 
$G_F/\sqrt{2} = g^2/8M_W^2$.
After substituting the electromagnetic coupling $\alpha$, 
the electroweak mixing angle and the $Z$
mass for the weak coupling, and the $W$
mass, this relation can be rewritten as
\begin{equation}
\frac{G_F}{\sqrt{2}} = \frac{2\pi\alpha}{\sin^2 2\theta_W M_Z^2}
[1+\Delta r_\alpha + \Delta r_t + \Delta r_H ] ~.
\end{equation}
The $\Delta$ terms take account of the radiative corrections:
$\Delta r_\alpha$ describes the shift in the electromagnetic coupling 
$\alpha$ if evaluated at the scale $M_Z^2$ instead of zero-momentum;
$\Delta r_t$ denotes the top (and bottom) quark contributions to the
$W$ and $Z$ masses, which are quadratic in the top mass. Finally,
$\Delta r_H$ accounts for the virtual Higgs contributions to the masses;
this term depends only logarithmically \cite{7} on
the Higgs mass at leading order: 
\begin{equation}
\Delta r_H = \frac{G_F M_W^2}{8\sqrt{2}\pi^2} \frac{11}{3} \left[
\log \frac{M_H^2}{M_W^2} - \frac{5}{6} \right] \hspace{1cm} (M_H^2 \gg M_W^2)
~.
\end{equation}
The screening effect reflects the role of the Higgs
field as a regulator that renders the electroweak theory
renormalizable.

Although the sensitivity on the Higgs mass is 
only logarithmic, the increasing precision in the 
measurement of the electroweak observables allows 
us to derive interesting estimates and constraints
on the Higgs mass \cite{8}, {\it c.f.} Fig.~\ref{fg:SMHiggs}:
\begin{eqnarray}
M_H & = & 91^{+45}_{-32}~\mbox{GeV} \\
    & \lessim & 186 ~\mbox{GeV~~~(95\% CL)}  ~. \nonumber
\end{eqnarray}
It may be concluded from these numbers that the 
canonical formulation of the Standard Model
including the existence of a light Higgs boson,
is compatible with the electroweak data. However,
alternative mechanisms cannot be ruled out if the 
system is opened up to contributions from physics 
areas beyond the Standard Model.

\subsection{Higgs Production Channels at $e^+e^-$ Colliders}

\phantom{h}
The first process that was used to search
directly for Higgs bosons over a large mass range,
was the Bjorken process, $Z\to Z^* H, Z^* \to f\bar f$ \cite{34}.
By exploring this production channel,
Higgs bosons with masses less than 65.4 GeV 
were excluded by the LEP1 experiments.
The search continued by reversing the
role of the real and virtual $Z$ bosons in the 
$e^+e^-$ continuum at LEP2.

The  main production mechanisms for Higgs
bosons in $e^+e^-$ collisions are
\begin{eqnarray}
\mbox{Higgs-strahlung} & : & e^+e^- \to Z^* \to ZH \\
\mbox{$WW$ fusion}     & : & e^+e^- \to \bar \nu_e \nu_e (WW) \to \bar \nu_e
\nu_e H
\label{eq:wwfusion}
\end{eqnarray}
In Higgs-strahlung \cite{30,34,35} the Higgs boson is emitted
from the $Z$-boson line, while $WW$ fusion is a formation
process of Higgs bosons in the collision of two quasi-real 
$W$ bosons radiated off the electron and positron beams \cite{36}.

As evident from the subsequent analyses, LEP2 could cover
the SM Higgs mass range up to about 114 GeV
\cite{9}. The high-energy $e^+e^-$
linear colliders can cover the entire Higgs
mass range, the intermediate mass range already at a 500 GeV collider 
\cite{13}, the upper mass range in the second phase of the machines in which 
they will reach a total energy of at least 3~TeV \cite{38A}.

\paragraph{(a) Higgs-strahlung} ~\\[0.5cm]
The cross section for Higgs-strahlung can be
written in a compact form as 
\begin{equation}
\sigma (e^+e^- \to ZH) = \frac{G_F^2 M_Z^4}{96\pi s} \left[ v_e^2 + a_e^2
\right] \lambda^{1/2} \frac{\lambda + 12 M_Z^2/s}{\left[ 1- M_Z^2/s \right]^2}
~, 
\end{equation}
where $v_e = -1 + 4 \sin^2 \theta_W$ and $a_e=-1$
are the vector and axial-vector $Z$
charges of the electron and $\lambda = [1-(M_H+M_Z)^2/s] [1-(M_H-M_Z)^2/s]$
is the usual two-particle phase-space
function. The cross section is of the size $\sigma \sim \alpha_W^2/s$,
i.e. of second order in the weak coupling, and 
it scales in the squared energy. Higher order contributions to the cross 
sections are under theoretical control \cite{38B,38C}.

\begin{figure}[hbt]

\vspace*{-5.0cm}
\hspace*{-2.0cm}
\epsfxsize=20cm \epsfbox{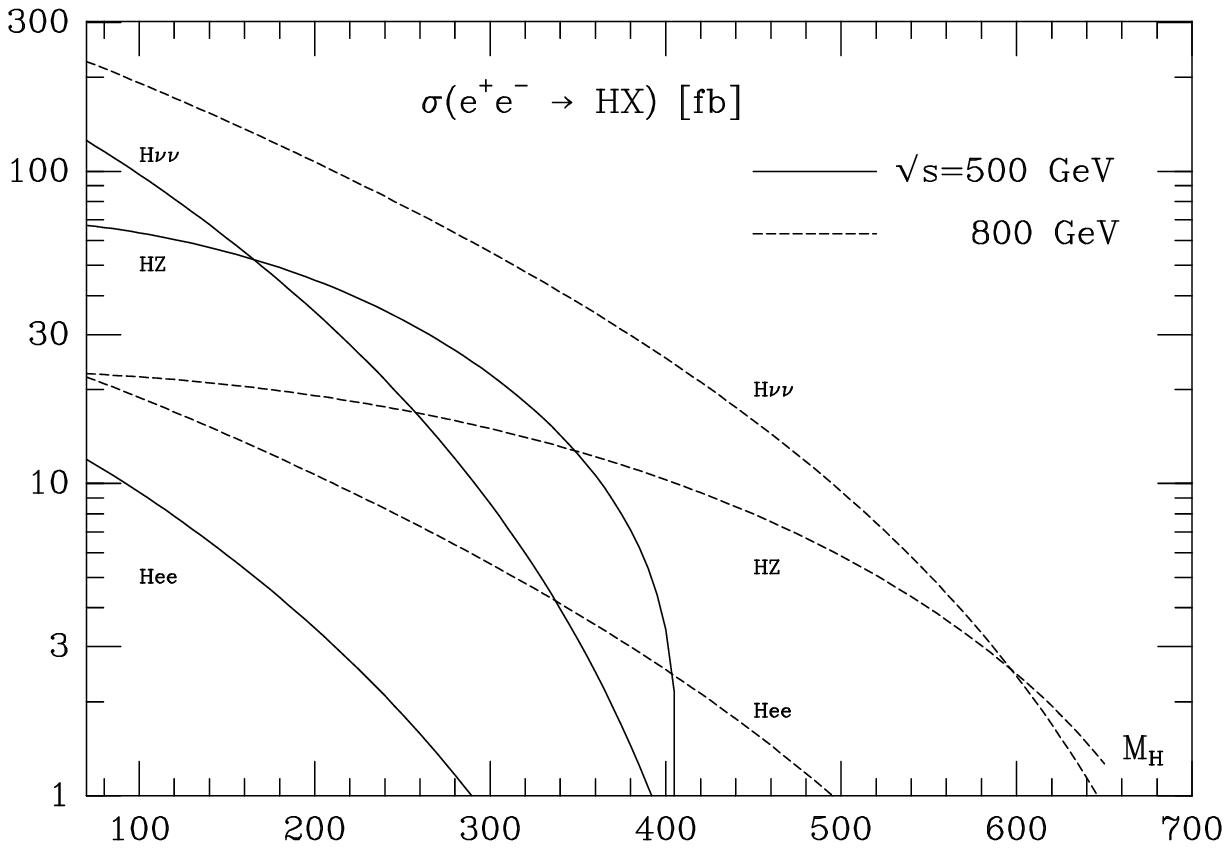}

\caption[]{\label{fg:eehx} \it The cross section for the production of SM
Higgs bosons in Higgs-strahlung $e^+e^-\to ZH$ and $WW/ZZ$ fusion $e^+e^- \to
\bar \nu_e \nu_e/e^+e^- H$; solid curves: $\sqrt{s}=500$ GeV, dashed curves:
$\sqrt{s}=800$ GeV.}
\end{figure}
\begin{figure}[hbt]
\begin{center}
\hspace*{-0.3cm}
\epsfig{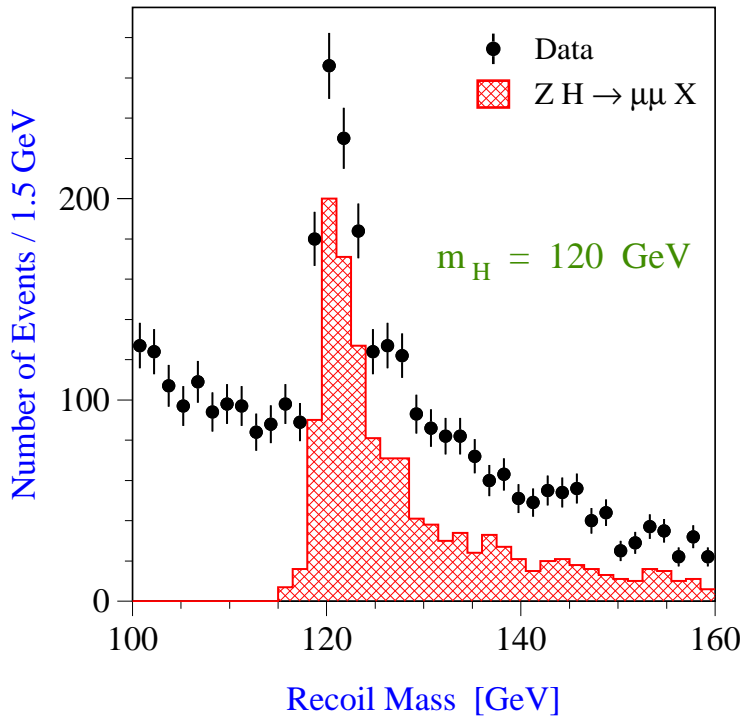}
\end{center}
\vspace*{-0.4cm}

\caption[]{\label{fg:zrecoil} \it The $\mu^+\mu^-$ recoil mass distribution in 
the process $e^+e^- \to H^0 Z\to X \mu^+\mu^-$ for $M_H=120$~GeV and
$\int {\cal L} = 500 fb^{-1}$ at $\sqrt{s}=$ 350~GeV. The dots with error
bars are Monte Carlo simulations of the Higgs signal and the background. The 
shaded histogram represents the signal only. Ref. \cite{13}.}
\end{figure}

Since the cross section vanishes for asymptotic 
energies, the Higgs-strahlung process is most
useful for searching Higgs bosons in the
range where the collider energy is of the 
same order as the Higgs mass, $\sqrt{s} \gsim {\cal O} (M_H)$.
The size of the cross section is illustrated 
in Fig.~\ref{fg:eehx} for the energy $\sqrt{s}=500$ GeV of 
$e^+e^-$ linear colliders as a function of the Higgs mass.
Since the recoiling $Z$ mass in the two-body reaction
$e^+e^- \to ZH$
is mono-energetic, the mass of the Higgs boson
can be reconstructed from the energy of the
$Z$ boson, $M_H^2 = s -2\sqrt{s}E_Z + M_Z^2$,
without any need of analyzing the decay products
of the Higgs boson. For leptonic $Z$
decays, missing-mass techniques provide a 
very clear signal, as demonstrated in Fig.~\ref{fg:zrecoil}.

\paragraph{(b) $WW$ fusion} ~\\[0.5cm]
Also the cross section for the fusion process (\ref{eq:wwfusion})
can be cast implicitly into a compact form:
\begin{eqnarray}
\sigma (e^+e^-\to\bar \nu_e \nu_e H) & = & \frac{G_F^3 M_W^4}{4\sqrt{2}\pi^3}
\int_{\kappa_H}^1\int_x^1\frac{dx~dy}{[1+(y-x)/\kappa_W ]^2}f(x,y)
\\ \nonumber \\
f(x,y) & = & \left( \frac{2x}{y^3} - \frac{1+3x}{y^2} + \frac{2+x}{y} -1 \right)
\left[ \frac{z}{1+z} - \log (1+z) \right] + \frac{x}{y^3} \frac{z^2(1-y)}{1+z}
~, 
\nonumber
\end{eqnarray}
with $\kappa_H=M_H^2/s$, $\kappa_W=M_W^2/s$ and $z=y(x-\kappa_H)/(\kappa_Wx)$.

Since the fusion process is a 
$t$-channel exchange process, the size is set by the 
$W$ Compton wavelength, suppressed however with 
respect to Higgs-strahlung by the third power
of the electroweak coupling, $\sigma \sim \alpha_W^3/M_W^2$.
As a result, $W$
fusion becomes the leading production process
for Higgs particles at high energies. At
asymptotic energies the cross section simplifies to
\begin{equation}
\sigma (e^+e^- \to \bar \nu_e \nu_e H) \to \frac{G_F^3 M_W^4}{4\sqrt{2}\pi^3}
\left[ \log\frac{s}{M_H^2} - 2 \right] ~.
\end{equation}
In this limit, $W$
fusion to Higgs bosons can be interpreted as a
two-step process: the $W$
bosons are radiated as quasi-real particles 
from electrons and positrons, $e \to \nu W$, 
with the Higgs bosons formed subsequently in the 
colliding $W$ beams. The electroweak higher order corrections are under 
control \cite{38C}.

The size of the 
fusion cross section is compared with Higgs-strahlung
in Fig.~\ref{fg:eehx}. At $\sqrt{s}=500$ GeV
the two cross sections are of the same order, yet the
fusion process becomes increasingly important with 
rising energy.

\subsection{Higgs Production at Hadron Colliders}

\phantom{h}
Several processes can be exploited to produce 
Higgs particles in hadron colliders \cite{24A,32}: \\[0.5cm]
\begin{tabular}{llll}
\hspace*{21mm} & gluon fusion & :              & $gg\to H$ \\ \\
& $WW,ZZ$ fusion           & :    & $W^+ W^-, ZZ \to H$ \\ \\
& Higgs-strahlung off $W,Z$ & :   & $q\bar q \to W,Z \to W,Z + H$ \\ \\
& Higgs bremsstrahlung off top & : & $q\bar q, gg \to t\bar t + H$
\end{tabular} \\[0.5cm]
While gluon fusion plays a dominant role 
throughout the entire Higgs mass range of the
Standard Model, the $WW/ZZ$
fusion process becomes increasingly important
with rising Higgs mass; however, it plays also an important role in the 
search for the Higgs boson and the study of its properties in the intermediate 
mass range. The last two radiation 
processes are relevant only for light Higgs masses.

The production cross sections at hadron colliders, at
the LHC in particular, are quite sizeable so that a 
large sample of SM Higgs particles can be produced
in this machine. Experimental difficulties 
arise from the huge number  of background events
that come along with the Higgs signal events.
This problem will be tackled by either
triggering on leptonic decays of $W,Z$ and $t$
in the radiation processes or by exploiting
the resonance character of the Higgs decays
$H\to \gamma\gamma$ and $H\to ZZ \to 4\ell^\pm$.
In this way, the Tevatron is expected to 
search for Higgs particles in the mass range
above that of LEP2 up to about 110 to 130 GeV \cite{11}. 
The LHC is expected to cover the entire canonical
Higgs mass range $M_H \lessim 700$ GeV
of the Standard Model \cite{12}.

\paragraph{(a) Gluon fusion} ~\\[0.5cm]
The gluon-fusion mechanism \cite{29,32,39A,39B}
\begin{displaymath}
pp \to gg \to H
\end{displaymath}
provides the dominant production mechanism of Higgs 
bosons at the LHC in the entire relevant Higgs
mass range up to about 1 TeV. The gluon coupling
to the Higgs boson in the SM is mediated by
triangular loops of top and bottom quarks, 
cf. Fig.~\ref{fg:gghlodia}.
Since the Yukawa coupling of the Higgs particle
to heavy quarks grows with the quark mass, thus
balancing the decrease of the triangle amplitude, the 
form factor approaches a non-zero value for large 
loop-quark masses. [If the masses of heavy quarks
beyond the third generation were generated solely
by the Higgs mechanism, these particles
would add the same amount to the form factor as
the top quark in the asymptotic heavy-quark limit.]
\begin{figure}[hbt]
\begin{center}
\setlength{\unitlength}{1pt}
\begin{picture}(180,90)(0,0)

\Gluon(0,20)(50,20){-3}{5}
\Gluon(0,80)(50,80){3}{5}
\ArrowLine(50,20)(50,80)
\ArrowLine(50,80)(100,50)
\ArrowLine(100,50)(50,20)
\DashLine(100,50)(150,50){5}
\put(155,46){$H$}
\put(25,46){$t,b$}
\put(-15,18){$g$}
\put(-15,78){$g$}

\end{picture}  \\
\setlength{\unitlength}{1pt}
\caption[ ]{\label{fg:gghlodia} \it Diagram contributing to the
formation of Higgs bosons in gluon-gluon collisions
at lowest order.}
\end{center}
\end{figure}
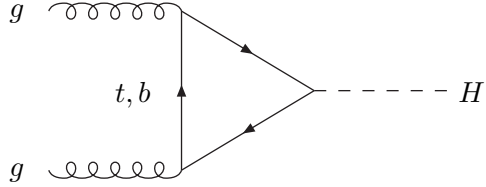\\

The partonic cross section, Fig.~\ref{fg:gghlodia}, can be expressed 
by the gluonic width of the Higgs boson at 
lowest order \cite{32}:
\begin{eqnarray}
\hat \sigma_{LO} (gg\to H) & = & \sigma_0 M_H^2 \times BW(\hat{s}) \\
\sigma_0 & = & \frac{\pi^2}{8M_H^2} \Gamma_{LO} (H\to gg) =
\frac{G_F\alpha_s^2}{288\sqrt{2}\pi} \left| \sum_Q A_Q^H (\tau_Q) \right|^2 ~,
\nonumber
\end{eqnarray}
where the scaling variable is defined as $\tau_Q = 4M_Q^2/M_H^2$ and $\hat s$
denotes the partonic c.m. energy squared. The
form factor can easily be evaluated:
\begin{eqnarray}
A_Q^H (\tau_Q) & = & \frac{3}{2} \tau_Q \left[ 1+(1-\tau_Q) f(\tau_Q)
\right] \label{eq:ftau} \\
f(\tau_Q) & = & \left\{ \begin{array}{ll}
\displaystyle \arcsin^2 \frac{1}{\sqrt{\tau_Q}} & \tau_Q \geq 1 \\
\displaystyle - \frac{1}{4} \left[ \log \frac{1+\sqrt{1-\tau_Q}}
{1-\sqrt{1-\tau_Q}} - i\pi \right]^2 & \tau_Q < 1
\end{array} \right. \nonumber
\end{eqnarray}
For small loop masses the form factor vanishes, $A_Q^H(\tau_Q) \sim -3/8 \tau_Q
[\log (\tau_Q/4)+i\pi]^2$,
while for large loop masses it approaches a non-zero value,
$A_Q^H (\tau_Q) \to 1$. The final term $BW$ is the normalized Breit-Wigner 
function
\beq
BW(\hat{s}) = \frac{M_H \Gamma_H/\pi}{[\hat{s}-M_H^2]^2 + M_H^2 \Gamma_H^2}
\eeq
approaching in the narrow-width approximation a $\delta$-function at 
$\hat{s}=M_H^2$.

In the narrow-width approximation, the hadronic
cross section can be cast into the form
\begin{equation}
\sigma_{LO} (pp\to H) = \sigma_0 \tau_H \frac{d{\cal L}^{gg}}{d\tau_H} ~,
\end{equation}
with $d{\cal L}^{gg}/d\tau_H$ denoting the $gg$ luminosity of the 
$pp$ collider, evaluated for the Drell--Yan variable
$\tau_H = M_H^2/s$, where $s$ is the total hadronic energy squared. \\

%
%
%
%

The QCD corrections to the gluon fusion 
process \cite{29,32,39B} are very important. They
stabilize the theoretical predictions for the 
cross section when the renormalization and
factorization scales are varied. Moreover,
they are large and positive, thus increasing the
production cross section for Higgs bosons. 
The QCD corrections consist of virtual 
corrections to the basic process $gg\to H$, 
and of real corrections due to the associated
production of the Higgs boson with massless
partons, $gg\to Hg$ and $gq\to Hq,\, q\bar q\to Hg$.
These subprocesses contribute to Higgs production
at ${\cal O}(\alpha_s^3)$.
The virtual corrections rescale the lowest-order
fusion cross section with a coefficient that depends
only on the ratios of the Higgs and quark masses. 
Gluon radiation leads to two-parton final states
with invariant energy $\hat s \geq M_H^2$ in the 
$gg, gq$ and $q\bar q$ channels.\\

\begin{figure}[hbt]

\vspace*{0.4cm}
\hspace*{2.0cm}
\begin{turn}{-90}%
\epsfxsize=7cm \epsfbox{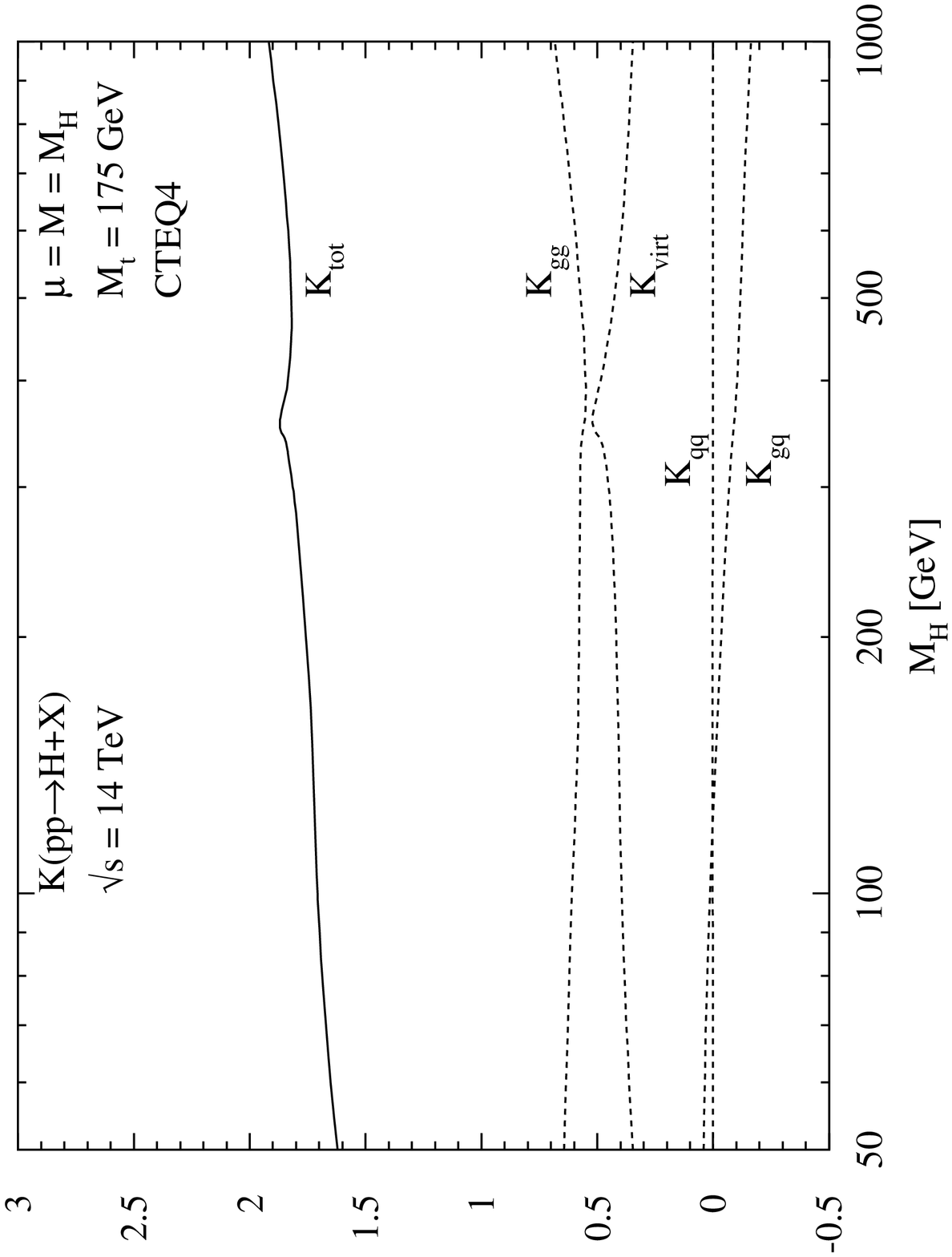}
\end{turn}
\vspace*{-0.2cm}

\caption[]{\label{fg:gghk} \it K factors of the QCD-corrected gluon-fusion
cross section $\sigma(pp \to H+X)$ at the LHC with c.m.~energy $\sqrt{s}=14$
TeV. The dashed lines show the individual contributions of 
the QCD corrections. The renormalization and
factorization scales have been identified with the Higgs mass,  
and  CTEQ4 parton densities have been adopted.}
\end{figure}
The size of the radiative corrections can be parametrized
by defining the $K$ factor as $K=\sigma_{NLO}/\sigma_{LO}$, in which
all quantities are evaluated in the  numerator and 
denominator in next-to-leading and leading order,
respectively. The results of this calculation are
shown in Fig.~\ref{fg:gghk}. The virtual corrections 
$K_{virt}$ and the real corrections $K_{gg}$ for the $gg$ collisions 
 are apparently of the same size, and both are large
and positive; the corrections for $q\bar q$
collisions and the $gq$
inelastic Compton contributions are less important.
After including these higher-order QCD corrections,
the dependence of the cross section on the renormalization
and factorization scales is significantly reduced
from a level of ${\cal O}(100\%)$ down to a level of about 20\%.
Depending only mildly on the Higgs bosons mass, the overall $K$ factor, 
$K_{tot}$, turns out to be close to 2 \cite{29,32,39B,R}.
The main contributions are 
generated by the virtual corrections and the 3-parton final states initiated by 
$gg$ initial states. Large NLO corrections are expected for 
gluon-initiated processes as a result of the large color charge. However, by 
studying the next order of corrections in the large top-mass limit, the 
N$^{2}$LO corrections generate only a modest additional increase of the $K$ 
factor, $\delta_2 K_{tot} \lessim 0.2$ \cite{x}. This proves the expansion to 
be convergent with the most important correction to be attributed to the 
next-to leading order contribution \cite{R}.

\begin{figure}[hbt]

\vspace*{0.5cm}
\hspace*{2.0cm}
\begin{turn}{-90}%
\epsfxsize=7cm \epsfbox{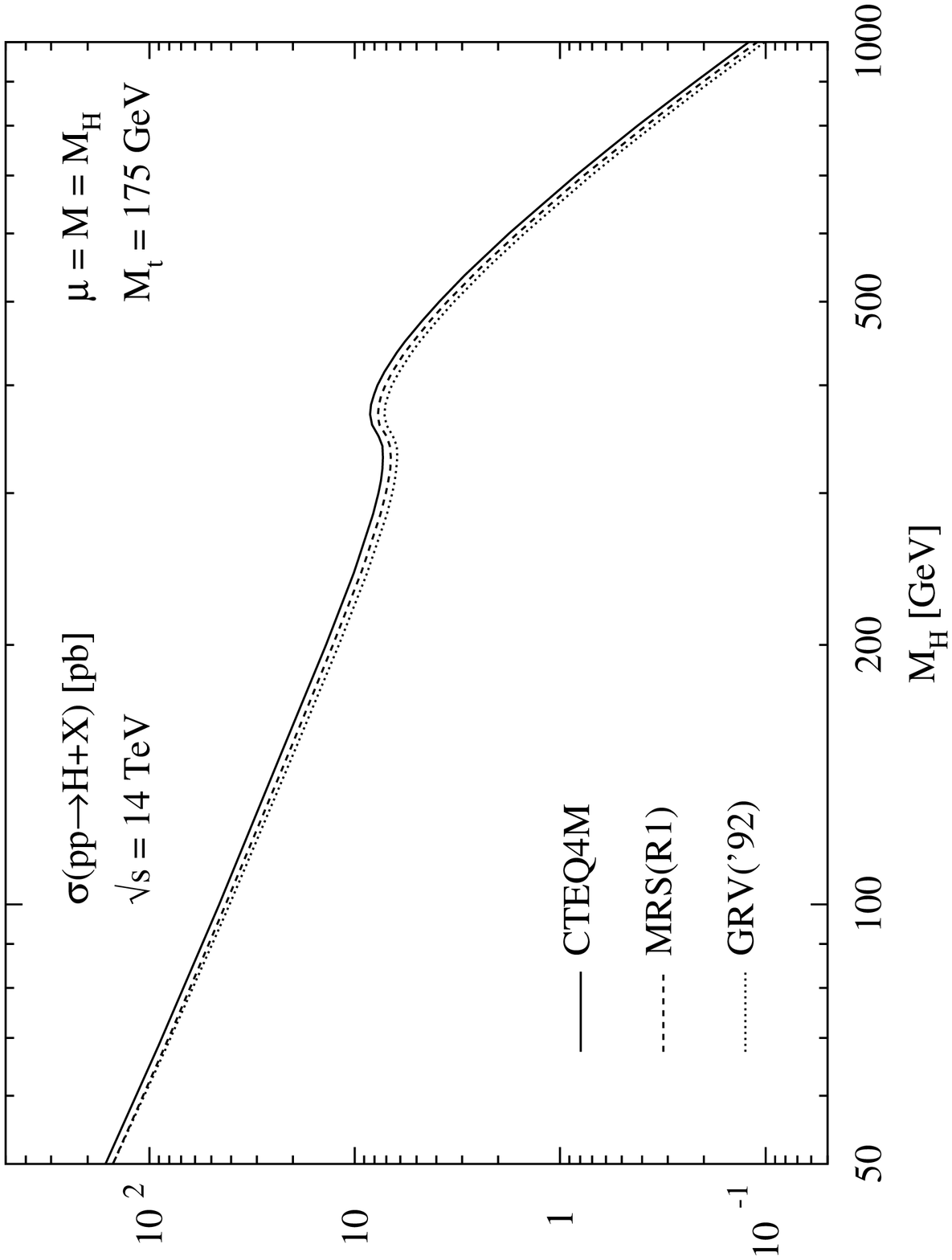}
\end{turn}
\vspace*{0.0cm}

\caption[]{\label{fg:gghparton} \it The cross 
section for the production of Higgs bosons;  three
different sets of parton densities are shown
[CTEQ4M, MRS(R1) and GRV('92)].}
\end{figure}
The theoretical prediction for the production cross 
section  of Higgs particles 
is presented in Fig.~\ref{fg:gghparton} for the LHC as a 
function of the Higgs mass.
The cross section decreases with increasing Higgs mass.
This is, to a large extent, a consequence of the
sharply falling $gg$
luminosity for large invariant masses. The bump in 
the cross section is induced by the $t\bar t$
threshold
in the top triangle. The overall theoretical 
accuracy of this calculation is expected to be at
a level of 20 to 30\%.  

\paragraph{(b) Vector-boson fusion} ~\\[0.5cm]
The second important channel for Higgs production at the
LHC is vector-boson fusion, $W^+W^- \to H$
\cite{23,22A}. For large Higgs masses this mechanism becomes
competitive to gluon fusion; for intermediate
masses the cross section is smaller by about an
order of magnitude.\\

For large Higgs masses, the two electroweak bosons $W,Z$
that form the Higgs boson are predominantly 
longitudinally polarized. At high energies, the 
equivalent particle spectra of the longitudinal $W,Z$
bosons in quark beams are given by
\begin{eqnarray}
f^W_L (x) & = & \frac{G_F M_W^2}{2\sqrt{2}\pi^2} \frac{1-x}{x} 
 \label{eq:xyz} \\ \non \\
f^Z_L (x) & = & \frac{G_F M_Z^2}{2\sqrt{2}\pi^2}
\left[(I_3^q - 2e_q \sin^2\theta_W)^2 + (I_3^q)^2\right] \frac{1-x}{x} ~, \non
\end{eqnarray}
where $x$ is the fraction of energy transferred from the quark
to the $W,Z$ boson in the splitting process
$q\to q +W/Z$. From these particle spectra, the $WW$ and $ZZ$
luminosities can easily be derived:
\begin{eqnarray}
\frac{d{\cal L}^{WW}}{d\tau_W} & = & \frac{G_F^2 M_W^4}{8\pi^4}
\left[ 2 - \frac{2}{\tau_W} -\frac{1+\tau_W}{\tau_W} \log \tau_W \right] \\
\non \\
\frac{d{\cal L}^{ZZ}}{d\tau_Z} & = & \frac{G_F^2 M_Z^4}{8\pi^4}
\left[(I_3^q - 2e_q \sin^2\theta_W)^2 + (I_3^q)^2\right]
\left[(I_3^{q'} - 2e_{q'} \sin^2\theta_W)^2 + (I_3^{q'})^2\right] \non \\
& & \hspace{1.5cm} \times \left[ 2 - \frac{2}{\tau_Z} -\frac{1+\tau_Z}{\tau_Z}
\log \tau_Z \right] \non
\end{eqnarray}
with the Drell--Yan variable defined as  $\tau_V = M_{VV}^2/s$.
The cross section for Higgs production in quark--quark
collisions is given by the convolution of the parton cross sections 
$WW,ZZ \to H$ with the luminosities:
\begin{equation}
\hat \sigma(qq\to qqH) = \frac{d{\cal L}^{VV}}{d\tau_V} \sqrt{2} \pi G_F ~.
\label{eq:vvhpart}
\end{equation}
The hadronic cross section is finally obtained by
summing the parton cross section (\ref{eq:vvhpart}) 
over the flux of all possible pairs
of quark--quark and antiquark combinations. \\

Since to lowest order the proton remnants are
color singlets in the $WW,ZZ$
fusion processes, no color will be exchanged between the
two quark lines from which the two vector bosons are
radiated. As a result, the leading QCD corrections to
these processes are already accounted for
by the corrections to the quark parton densities.\\

\begin{figure}[hbt]

\vspace*{0.5cm}
\hspace*{0.0cm}
\begin{turn}{-90}%
\epsfxsize=10cm \epsfbox{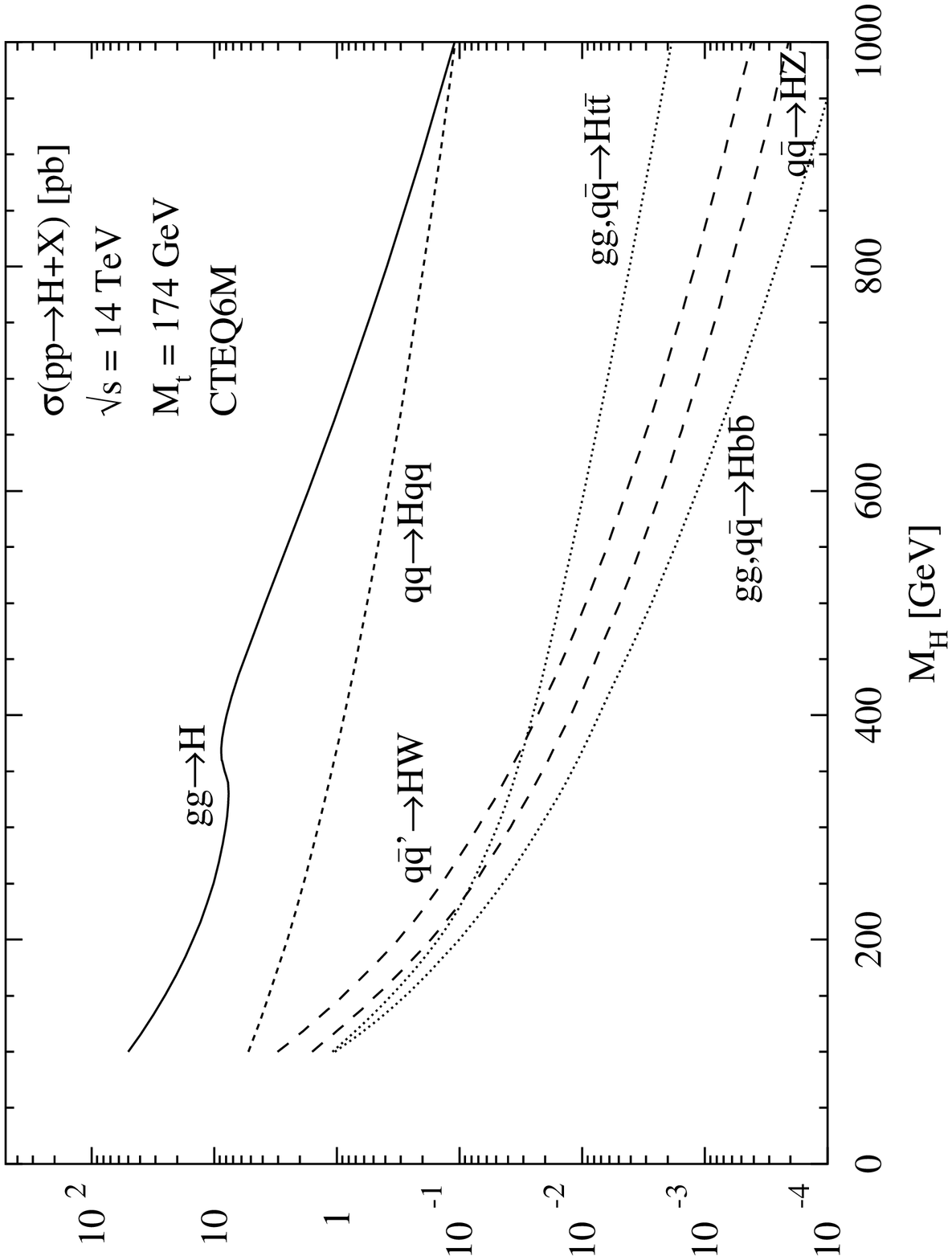}
\end{turn}
\vspace*{0.0cm}

\caption[]{\label{fg:lhcpro} \it Higgs production cross sections at the LHC
 for the various production mechanisms as a function of the
Higgs mass. The full QCD-corrected results for the gluon fusion $gg
\to H$, vector-boson fusion $qq\to VVqq \to Hqq$, vector-boson bremsstrahlung
$q\bar q \to V^* \to HV$ and associated production $gg,q\bar q \to Ht\bar t,
Hb\bar b$ are shown.}
\end{figure}

The $WW/ZZ$ fusion cross section for Higgs bosons at the LHC
is shown in Fig.~\ref{fg:lhcpro}. The process is apparently
very important for the search of the Higgs boson in the upper mass range, 
where the cross section approaches values close to gluon fusion. For 
intermediate masses, it comes close within an order of magnitude to the 
leading gluon fusion cross section.

\paragraph{(c) Higgs-strahlung off vector bosons} ~\\[0.5cm]
Higgs-strahlung $q\bar q \to V^* \to VH~(V=W,Z)$
is a very important mechanism (Fig.~\ref{fg:lhcpro}) for the 
search of light Higgs bosons at the hadron colliders
Tevatron and LHC. Though the cross section is 
smaller than for gluon fusion, leptonic decays
of the electroweak vector bosons are
extremely useful to filter Higgs signal events
out of the huge background. Since the dynamical
mechanism is the same as for $e^+e^-$
colliders, except for the  folding with
the quark--antiquark densities, intermediate steps of the
 calculation need not be noted here anymore, and merely  
the final values 
of the cross sections for the Tevatron and the 
LHC are recorded in Fig.~\ref{fg:lhcpro}.

\paragraph{(d) Higgs bremsstrahlung off top quarks} ~\\[0.5cm]
Also the process $gg,q\bar q \to t\bar t H$
is relevant only for small Higgs masses, Fig.~\ref{fg:lhcpro}.
The analytical expression for the parton cross
section, even at lowest order, is quite involved, 
so that just the final results for the LHC
cross section are shown in Fig.~\ref{fg:lhcpro}.
Higher order corrections have been presented in Ref.~\cite{z}.

Higgs bremsstrahlung off top quarks is also an interesting
process for measurements of the fundamental $Htt$
Yukawa coupling. The cross section $\sigma (pp\to t\bar t H)$
is directly proportional to the square of
this fundamental coupling.

\paragraph{\underline{Summary.}} An overview of the production cross
sections for Higgs particles at the LHC
is presented in Fig.~\ref{fg:lhcpro}. Three classes
of channels can be distinguished. The gluon fusion of Higgs particles
is a universal process, dominant over the 
entire SM Higgs mass range. Higgs-strahlung
off electroweak $W,Z$
bosons or top quarks is prominent for light
Higgs bosons. The $WW/ZZ$
fusion channel, by contrast, becomes increasingly
important in the upper part of the SM Higgs
mass range, though it proves also useful in the intermediate 
mass range.

The signatures for the search for Higgs particles are
dictated by the decay branching ratios. In the
lower part of the intermediate mass range, resonance
reconstruction in $\gamma\gamma$ final states and $b\bar b$
jets can be exploited. In the upper part of the
intermediate mass range, decays to $ZZ^*$ and $WW^*$
are important, with the two electroweak bosons 
decaying leptonically. In the mass range above
the on-shell $ZZ$ decay threshold, the charged-lepton decays
$H\to ZZ \to 4\ell^\pm$ provide  gold-plated signatures. Only at the
upper end of the classical SM Higgs mass range,
 decays to neutrinos and jets,  
generated in $W$ and $Z$ decays, complete the search techniques.

\subsection{The Profile of the SM Higgs Particle}

\phantom{h}
To establish the Higgs mechanism
experimentally, the nature of this particle
must be explored by measuring all its
characteristics, the mass and lifetime,
the external quantum numbers spin-parity,
the couplings to gauge bosons and fermions,
and last but not least, the Higgs self-couplings.
While part of this program
can be realized at the LHC \cite{12,xx}, the complete
profile of the particle can be reconstructed
across the entire mass range in $e^+ e^-$ colliders \cite{13}.

\paragraph{(a) Mass} ~\\[0.5cm]
The mass of the Higgs particle can be
measured by collecting the decay products
of the particle at hadron and $e^+e^-$ colliders. Moreover, in
$e^+e^-$ collisions Higgs-strahlung can be exploited
to reconstruct the mass very precisely from
the $Z$ recoil energy in the two-body
process $e^+e^-\to ZH$.
 An overall
accuracy of about $\delta M_H \sim 100$ MeV can be expected.

\paragraph{(b) Width/lifetime} ~\\[0.5cm]
The width of the state, i.e. the lifetime of
the particle, can be measured directly
above the $ZZ$ decay threshold where the
width grows rapidly. In the lower part of
the intermediate mass range the width can be
measured indirectly \cite{13}, by combining the branching
ratio for $H\to WW$ with the measurement
of the partial $WW$ width, accessible through the cross 
section for $W$ boson fusion:
$\Gamma_{tot} = \Gamma_{WW} / BR_{WW}$.
Thus, the total width of the Higgs particle can be
determined throughout the entire  mass
range when the experimental results from the LHC and
$e^+e^-$ colliders can be combined. 

\begin{figure}[hbt]
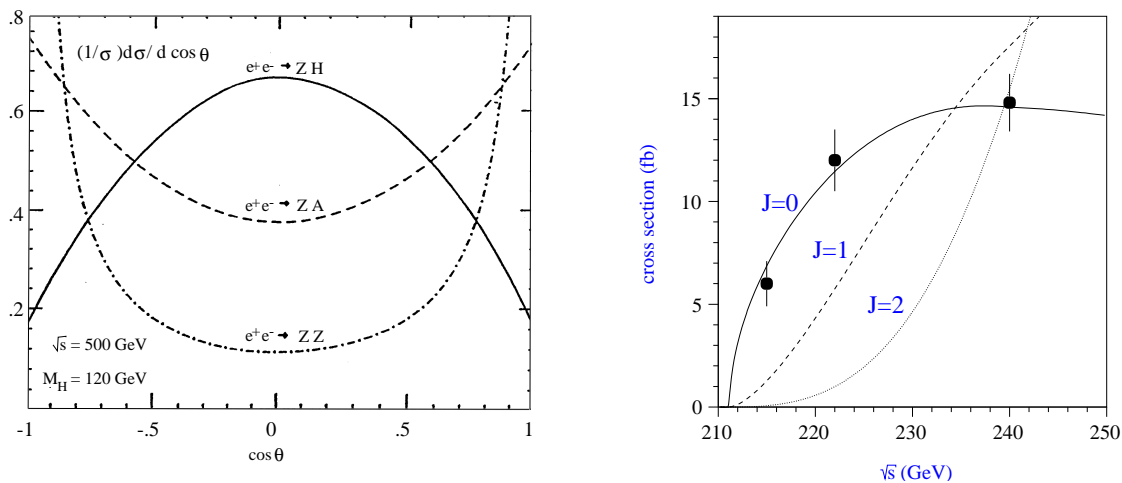


\vspace*{0.5cm}
\hspace*{0.0cm}
\epsfxsize=8cm \epsfbox{higgs_sm_spin_th.eps}
\vspace*{0.0cm}

\vspace*{-6.3cm}
\hspace*{9.0cm}
\epsfig{figure=spinexp.eps,bbllx=0,bblly=0,bburx=560,bbury=539,width=6.5cm,clip=}
\vspace*{0.0cm}

\caption[]{\it \label{fg:spinpar} Left: Angular distribution of $Z/H$ bosons in
Higgs-strahlung, compared with the production of pseudoscalar
particles and the $ZZ$ background final states; Ref.~\cite{41}.
Right: Threshold excitation of Higgs-strahlung which discriminates spin=0 
from other assignments, Ref. \cite{MMM,exp}.}
\end{figure}

\paragraph{(c) Spin-parity} ~\\[0.5cm]
The angular distribution of the $Z/H$ bosons
in the Higgs-strahlung process is 
sensitive to the spin and parity of the
Higgs particle \cite{41}. Since the production
amplitude is given by ${\cal A}(0^+) \sim \vec{\epsilon}_{Z^*} \cdot
\vec{\epsilon}_Z$, the $Z$ boson is produced in a state of
longitudinal polarization at high
energies -- in accordance  with the equivalence
theorem. As a result, the angular distribution
\begin{equation}
\frac{d\sigma}{d\cos\theta} \sim \sin^2 \theta + \frac{8M_Z^2}{\lambda s}
\end{equation}
approaches the spin-zero $\sin^2\theta$
law asymptotically. This may be contrasted
with the distribution $\sim 1 + \cos^2\theta$
for negative parity states, which follows
from the transverse polarization amplitude
${\cal A}(0^-) \sim \vec{\epsilon}_{Z^*} \times \vec{\epsilon}_Z \cdot
\vec{k}_Z$. It is also characteristically different
from the distribution of the background
process $e^+e^- \to ZZ$, which, as a result of $t/u$-channel $e$ exchange,
is strongly peaked in the forward/backward
direction, Fig.~\ref{fg:spinpar}a.\\

In a similar way, the zero-spin of the
Higgs particle can be determined from the
isotropic distribution of the decay
products. Moreover, the parity can be
measured by observing the spin correlations
of the decay products. According to the
equivalence theorem, the azimuthal angles
of the decay planes in $H\to ZZ\to (\mu^+\mu^-) (\mu^+\mu^-)$
are asymptotically uncorrelated, $d\Gamma^+/d\phi_* \to 0$,
for a $0^+$ particle; this is to be contrasted with 
$d\Gamma^-/d\phi_* \to 1-\frac{1}{4} \cos 2\phi_*$
for the distribution of the azimuthal angle
between the planes for the decay of a $0^-$
particle. The difference between the angular distributions
is a consequence of the different polarization
states of the vector bosons in the two
cases. While they approach states of 
longitudinal polarization for scalar Higgs
decays, they are transversely polarized
for pseudoscalar particle decays. \\

A different method to determine the spin of the Higgs boson is provided by 
scanning the onset of the excitation curve in Higgs-strahlung \cite{MMM} 
$e^+e^- \to ZH$. For Higgs spin $S_H = 0$ 
the excitation curve rises steeply at the 
threshold $\sim \sqrt{s - (M_H + M_Z)^2}$. This behavior is distinctly 
different from higher spin excitations which rise with a power $> 1$ of the 
threshold factor. An ambiguity for states with spin/parity 
$1^+$ and $2^+$ can be resolved by evaluating 
also the angular distribution of the Higgs and $Z$ boson in 
the Higgs-strahlung process. 
The experimental precision will be 
sufficient to discriminate the spin=0 assignment to the Higgs boson from 
other assignments as shown in Fig.~\ref{fg:spinpar}b.

\newpage
\paragraph{(d) Higgs couplings} ~\\[0.5cm]
Since fundamental particles acquire
mass through the interaction with the
Higgs field, the strength of the Higgs
couplings to fermions and gauge bosons
is set by the masses of the particles.
It will therefore be a crucial experimental  
task to measure these couplings, which
are uniquely predicted by the very
nature of the Higgs mechanism.\\

\begin{figure}[hbt]
\begin{center}
\epsfig{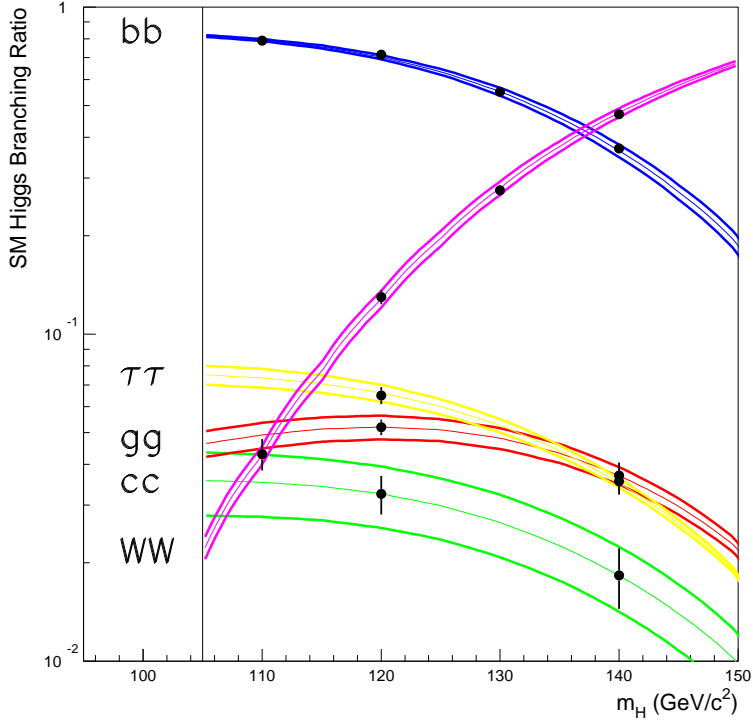}
\end{center}
\vspace*{-0.2cm}

\caption[]{\it \label{fg:brmeas} The predicted SM Higgs boson branching 
ratios. Points with error bars show the expected experimental accuracy, while 
the lines show the estimated uncertainties on the SM predictions. 
Ref.~\cite{13}.}
\end{figure}

The Higgs couplings to massive gauge
bosons can be determined from the
production cross sections in
Higgs-strahlung and $WW,ZZ$ fusion, with the
accuracy expected at the per cent level.
For heavy enough Higgs bosons the decay
width can be exploited to determine the
coupling to electroweak gauge bosons.
For Higgs couplings to fermions the
branching ratios $H\to b\bar b, c\bar c, \tau^+\tau^-$
can be used in the lower part of the
intermediate mass range, cf.~Fig.~\ref{fg:brmeas}; these observables
allow the direct measurement of the Higgs
Yukawa couplings.

A particularly interesting coupling
is the Higgs coupling to top quarks.
Since the top quark is by far the
heaviest fermion in the Standard Model,
irregularities in the standard picture
of electroweak symmetry breaking through
a fundamental Higgs field may become
apparent first in this coupling. Thus 
the $Ht t$ Yukawa coupling may eventually provide
essential clues to the nature of the
mechanism breaking the electroweak
symmetries.

\begin{figure}[hbt]
\vspace*{0.0cm}
\hspace*{2.0cm}
\epsfxsize=12cm \epsfbox{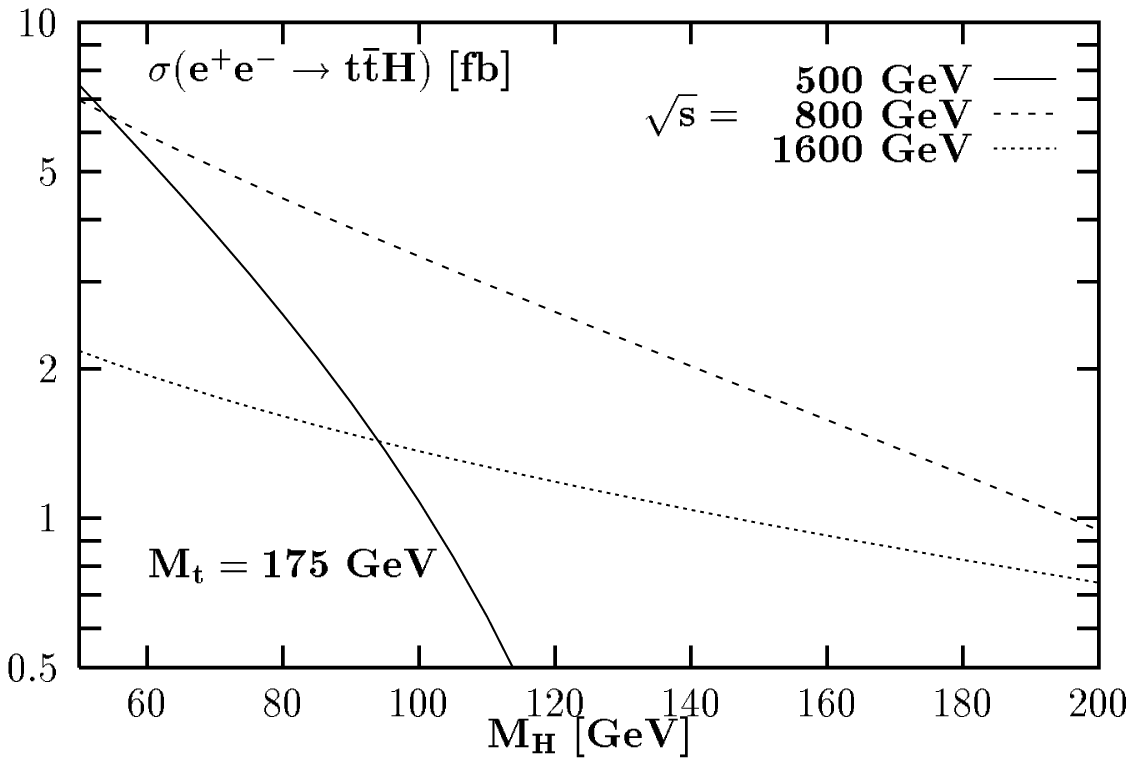}
\vspace*{-0.5cm}

\caption[]{\it \label{fg:eetth} The cross section for bremsstrahlung of SM
Higgs bosons off top quarks in the Yukawa
process $e^+e^-\to t\bar t H$.
[The amplitude for radiation off the
intermediate $Z$-boson line is small.] Ref. \cite{44}.}
\end{figure}

Top loops mediating the production
processes $gg\to H$ and $\gamma\gamma\to H$
(and the corresponding decay channels)
give rise to cross sections and partial
widths, which are proportional to the square of
the Higgs--top Yukawa coupling. This
Yukawa coupling can be measured directly,
for the lower part of the intermediate
mass range, in the bremsstrahlung
processes $pp\to t\bar t H$ and $e^+e^- \to t\bar t H$ \cite{44}.
The Higgs boson is radiated, in the first
process exclusively, in the second process
predominantly, from the heavy top quarks.
Even though these experiments are 
difficult because of  the small cross sections
[cf. Fig.~\ref{fg:eetth} for $e^+e^-$ collisions] 
and of the complex topology of
the $b\bar bb\bar bW^+W^-$ final state, this process 
is an important tool for exploring the
mechanism of electroweak symmetry breaking.
For large Higgs masses above the $t\bar t$
threshold, the decay channel $H\to t\bar t$
can be studied; in $e^+e^-$ collisions the cross section of
$e^+e^- \to t\bar t Z$ increases through the reaction
$e^+e^- \to ZH (\to t\bar t)$ \cite{45}. Higgs exchange between
$t\bar t$ quarks also affects the excitation curve
near the threshold at a level of a few per cent.

Only ratios of Higgs couplings to electroweak gauge bosons and fermions can be 
determined at LHC without auxiliary assumptions since only products of 
production cross sections and decay branching ratios can be measured. At 
linear colliders, however, the production cross sections can be measured 
independently of specific Higgs decay modes, {\it i.e.} inclusively 
in Higgs-strahlung, 
for instance. This can be exploited to measure the Higgs couplings to $Z$ or 
$W$ bosons relative to which all the other couplings are scaled; they 
are determined subsequently by the branching ratios. The expected 
accuracies for some of the couplings are collected in Table~\ref{tab:muehll}.

\begin{table}[h]
\begin{center}
{\small
\begin{tabular}{|lll|}
\hline
Coupling & $M_H=120$~GeV & 140~GeV \\
\hline
$g_{HWW}$ & $\pm 0.012$ & $\pm 0.020$ \\
$g_{HZZ}$ & $\pm 0.012$ & $\pm 0.013$ \\
\hline
$g_{Htt}$ & $\pm 0.030$ & $\pm 0.061$ \\
$g_{Hbb}$ & $\pm 0.022$ & $\pm 0.022$ \\
$g_{Hcc}$ & $\pm 0.037$ & $\pm 0.102$ \\
\hline
$g_{H\tau\tau}$ & $\pm 0.033$ & $\pm 0.048$ \\
\hline
\end{tabular}
}
\end{center}
\caption{Relative accuracy on the Higgs couplings assuming $\int\!{\cal L}=500$~fb$^{-1}$, $\sqrt{s}=500$~GeV ($\int\!{\cal L}=1$~ab$^{-1}$, $\sqrt{s}=800$~GeV for $g_{Htt}$).}
\label{tab:muehll}
\vspace*{-0.4cm}
\end{table}

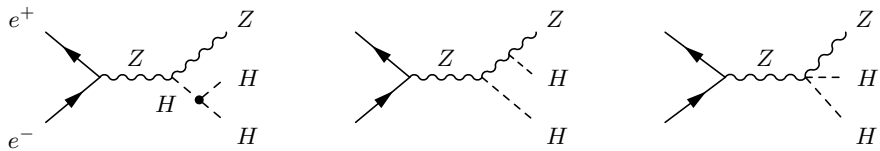
\begin{figure}[hbtp]
\begin{center}
\begin{fmffile}{fd}
{\footnotesize
\unitlength1mm
\begin{fmfshrink}{0.7}
\begin{fmfgraph*}(24,12)
  \fmfstraight
  \fmfleftn{i}{3} \fmfrightn{o}{3}
  \fmf{fermion}{i1,v1,i3}
  \fmflabel{$e^-$}{i1} \fmflabel{$e^+$}{i3}
  \fmf{boson,lab=$Z$,lab.s=left,tens=3/2}{v1,v2}
  \fmf{boson}{v2,o3} \fmflabel{$Z$}{o3}
  \fmf{phantom}{v2,o1}
  \fmffreeze
  \fmf{dashes,lab=$H$,lab.s=right}{v2,v3} \fmf{dashes}{v3,o1}
  \fmffreeze
  \fmf{dashes}{v3,o2}
  \fmflabel{$H$}{o2} \fmflabel{$H$}{o1}
  \fmfdot{v3}
\end{fmfgraph*}
\hspace{15mm}
\begin{fmfgraph*}(24,12)
  \fmfstraight
  \fmfleftn{i}{3} \fmfrightn{o}{3}
  \fmf{fermion}{i1,v1,i3}
  \fmf{boson,lab=$Z$,lab.s=left,tens=3/2}{v1,v2}
  \fmf{dashes}{v2,o1} \fmflabel{$H$}{o1}
  \fmf{phantom}{v2,o3}
  \fmffreeze
  \fmf{boson}{v2,v3,o3} \fmflabel{$Z$}{o3}
  \fmffreeze
  \fmf{dashes}{v3,o2}
  \fmflabel{$H$}{o2} \fmflabel{$H$}{o1}
\end{fmfgraph*}
\hspace{15mm}
\begin{fmfgraph*}(24,12)
  \fmfstraight
  \fmfleftn{i}{3} \fmfrightn{o}{3}
  \fmf{fermion}{i1,v1,i3}
  \fmf{boson,lab=$Z$,lab.s=left,tens=3/2}{v1,v2}
  \fmf{dashes}{v2,o1} \fmflabel{$H$}{o1}
  \fmf{dashes}{v2,o2} \fmflabel{$H$}{o2}
  \fmf{boson}{v2,o3} \fmflabel{$Z$}{o3}
\end{fmfgraph*}
\end{fmfshrink}
\\[0.5cm]
}
\end{fmffile}
\end{center}
\caption[]{\label{fg:wwtohh} \it Generic diagrams contributing to the 
double Higgs-strahlung process $e^+e^- \to ZHH$.}
\end{figure}

\newpage
\paragraph{(e) Higgs self-couplings} ~\\[0.5cm]
The Higgs mechanism, based on a non-zero
value of the Higgs field in the vacuum, must
finally be made manifest experimentally by
reconstructing the interaction potential
that  generates the non-zero field in
the vacuum. This program can be carried out
by measuring the strength of the  trilinear
and quartic self-couplings of the Higgs
particles:
\begin{eqnarray}
g_{H^3} & = & 3 \sqrt{\sqrt{2} G_F} M_H^2 \\ \non \\
g_{H^4} & = & 3 \sqrt{2} G_F M_H^2 ~.
\end{eqnarray}
This is a  difficult task since the
processes to be exploited are suppressed
by small couplings and phase space.
Nevertheless, the first step in this problem can be solved
at the LHC and in the high-energy phase
of the $e^+e^-$ linear colliders for sufficiently high
luminosities \cite{selfMMM}. The best-suited reaction
at $e^+e^-$ colliders for the measurement of the trilinear 
coupling for Higgs masses in the theoretically
preferred mass range of ${\cal O}(100~\mbox{GeV})$, is the
double Higgs-strahlung process
\begin{equation}
e^+e^- \to ZH^{\ast} \to ZHH
\end{equation}
in which, among other mechanisms, the two-Higgs
final state is generated by the 
exchange of a virtual Higgs particle so that this
process is sensitive to the trilinear $HHH$
coupling in the Higgs potential, Fig.~\ref{fg:wwtohh}. Since
the cross section is only a fraction of 1 fb, 
an integrated luminosity of
$\sim 1 ab^{-1}$ is needed to isolate the events at linear
colliders. Experimental accuracies close to 20\% can be expected in these 
measurements \cite{Rexp}. The quartic coupling $H^4$
seems to be accessible only through loop
effects in the foreseeable future.\\

{\it To sum up.} The essential elements of the
Higgs mechanism can be established experimentally
at the LHC and TeV $e^+e^-$ linear colliders.

\section{Higgs Bosons in Supersymmetric Theories}

\phantom{h}
Arguments  deeply rooted in the Higgs sector 
play an eminent role in introducing
supersymmetry as a fundamental symmetry
of nature \cite{14}. This is the only symmetry
that correlates bosonic with fermionic
degrees of freedom.
\paragraph{(a)}
The cancellation between bosonic and
fermionic contributions to the radiative
corrections of the light Higgs masses
in supersymmetric theories provides a
solution of the hierarchy problem in
the Standard Model. If the Standard Model
is embedded in a grand-unified theory,
the large gap between the high grand-unification
scale and the low scale of
electroweak symmetry breaking can be
stabilized in a natural way in boson--fermion symmetric
theories \cite{15,601}. Denoting
the bare Higgs mass by $M_{H,0}^2$,
the radiative corrections due to vector-boson
loops in the Standard Model by $\delta M_{H,V}^2$, 
and the contributions of supersymmetric
fermionic gaugino partners by $\delta M_{\tilde H,\tilde V}^2$, 
the physical Higgs mass is given by the sum
$M_H^2 = M_{H,0}^2 + \delta M_{H,V}^2 + \delta M_{\tilde H,\tilde V}^2$.
The vector-boson correction is quadratically
divergent, $\delta M_{H,V}^2 \sim \alpha [\Lambda^2 - M^2]$, 
so that for a cut-off scale $\Lambda \sim \Lambda_{GUT}$
extreme fine-tuning between  the
intrinsic bare mass and the radiative quantum fluctuations 
would be needed to
generate a Higgs mass of order $M_W$.
However, owing  to Pauli's principle, the
additional fermionic gaugino contributions
in supersymmetric theories are just
opposite in sign, $\delta M_{\tilde H,\tilde V}^2\sim -\alpha
[\Lambda^2-\tilde M^2]$,
so that the divergent terms cancel\footnote{Different statistics for bosons 
and fermions are sufficient for the cancellation of the divergencies; however, 
they are not necessary. Symmetry relations among couplings, as realized in 
Little Higgs Models, may also lead to cancellations individually between 
boson-boson or fermion-fermion amplitudes.}. Since
$\delta M_H^2\sim\alpha [\tilde M^2-M^2]$,
any fine-tuning is avoided for supersymmetric
particle masses $\tilde M \lessim {\cal O}(1$ TeV).
Thus, within this symmetry scheme the Higgs 
sector is stable in the low-energy range $M_H\sim M_W$
even in the context of high-energy GUT scales. This mechanism leads in a 
natural way to low-energy supersymmetry.

\paragraph{(b)}
The concept of supersymmetry is strongly
supported by the successful prediction
of the electroweak mixing angle in
the minimal version of this theory \cite{16}.
There, the extended particle spectrum 
 drives the evolution of the
electroweak mixing angle from the
GUT value 3/8 down to $\sin^2\theta_W = 0.2336 \pm 0.0017$,
the error including unknown threshold
contributions at the low and the
high supersymmetric mass scales.
The prediction coincides with the
experimentally measured value $\sin^2\theta_W^{exp} = 0.23120 \pm 0.00015$
within the theoretical uncertainty
of less than 2 per mille.

\paragraph{(c)}
Conceptually very interesting is 
 the interpretation
of the Higgs mechanism in supersymmetric
theories as a quantum effect \cite{50A}. The
breaking of the electroweak symmetry $SU(2)_L \times U(1)_Y$
can be induced radiatively while
leaving the electromagnetic gauge
symmetry $U(1)_{EM}$
and the color gauge symmetry $SU(3)_C$
unbroken for top-quark masses
between 150 and 200 GeV. Starting
with a set of universal scalar
masses at the high GUT scale, the
squared mass parameter of the Higgs
sector evolves to negative values
at the low electroweak scale, while
the squared squark and slepton
masses remain positive.\\

The Higgs sector of supersymmetric
theories differs in several aspects
from the Standard Model \cite{17}. To preserve
supersymmetry and gauge invariance,
at least two iso-doublet fields must
be introduced, leaving us with a
spectrum of five or more physical
Higgs particles. In the minimal
supersymmetric extension of the
Standard Model (MSSM) the Higgs
self-interactions are generated
by the scalar-gauge action, so that the
quartic couplings are related to
the gauge couplings in this scenario. This leads
to strong bounds \cite{19} of less than
about 140 GeV for the mass of 
the lightest Higgs boson [after including radiative corrections]. If the
system is assumed to remain
weakly interacting up to scales
of the order of the GUT or Planck
scale, the mass remains small, 
for reasons
quite analogous to those found in the Standard
Model,
even in more complex supersymmetric
theories involving additional Higgs fields and  
Yukawa interactions. The masses of the heavy 
Higgs bosons are expected to be
of the scale of electroweak symmetry
breaking up to order 1 TeV.

\subsection{The Higgs Sector of the MSSM}

\phantom{h}
The particle spectrum of the MSSM \cite{14} consists
of leptons, quarks and their scalar 
supersymmetric partners, and gauge
particles, Higgs particles and their
spin-1/2 partners. The matter and force fields are coupled
in supersymmetric and gauge-invariant
actions:
\begin{equation}
\begin{array}{lrcll}
S = S_V + S_\phi + S_W: \hspace*{1cm}
& S_V    & = & \frac{1}{4} \int d^6 z \hat W_\alpha \hat W_\alpha
\hspace*{1cm} & \mbox{gauge action} ~, \\ \\
& S_\phi & = & \int d^8 z \hat \phi^* e^{gV} \hat \phi
& \mbox{matter action} ~, \\ \\
& S_W    & = & \int d^6 z W[\hat \phi]
& \mbox{superpotential} ~.
\end{array}
\end{equation}
Decomposing the superfields into fermionic
and bosonic components, and carrying out
the integration over the Grassmann
variables in $z\to x$,
the following Lagrangians can be derived, which  
describe the interactions of the
gauge, matter and Higgs fields:
\begin{eqnarray*}
{\cal L}_V & = & -\frac{1}{4}F_{\mu\nu}F_{\mu\nu}+\ldots+\frac{1}{2}D^2 ~, \\ \\
{\cal L}_\phi & = & D_\mu \phi^* D_\mu \phi +\ldots+\frac{g}{2} D|\phi|^2  ~, \\ \\
{\cal L}_W & = & - \left| \frac{\partial W}{\partial \phi} \right|^2 ~. 
\end{eqnarray*}
The $D$ field is an auxiliary field that 
does not propagate in space-time and
 can be eliminated by applying  
the equations of motion: $D=-\frac{g}{2} |\phi|^2$.
Reinserted into the Lagrangian, the
quartic self-coupling of the scalar Higgs
fields is generated:
\begin{equation}
{\cal L} [\phi^4] = -\frac{g^2}{8} |\phi^2|^2 ~.
\end{equation}
Thus, the quartic coupling of the Higgs 
fields is given, in the minimal
supersymmetric theory, by the square
of the gauge coupling. Unlike the Standard
Model case, the quartic coupling is not a free parameter. Moreover,
this coupling is weak.\\

Two independent Higgs doublet fields $H_1$ and $H_2$
must be introduced into the superpotential: 
\begin{equation}
W = -\mu \epsilon_{ij} \hat H_1^i \hat H_2^j + \epsilon_{ij} [f_1 \hat H_1^i
\hat L^j \hat R + f_2 \hat H_1^i \hat Q^j \hat D +
f_2' \hat H_2^j \hat Q^i \hat U]
\end{equation}
to provide the down-type particles ($H_1$)
and the up-type particles ($H_2$) with mass.
Unlike the Standard Model, the second Higgs
field cannot be identified with the
charge conjugate of the first Higgs field
since $W$ must be analytic to preserve
supersymmetry. Moreover, the Higgsino
fields associated with a single Higgs
field would generate triangle anomalies;
they cancel if the two conjugate doublets
are added up, and the classical gauge
invariance of the interactions is not
destroyed at the quantum level. 
Integrating the superpotential over
the Grassmann coordinates generates
the supersymmetric Higgs self-energy
$V_0 = |\mu|^2 (|H_1|^2 + |H_2|^2)$.
The breaking of supersymmetry can be
incorporated in the Higgs sector by
introducing bilinear mass terms $\mu_{ij} H_i H_j$.
Added to the supersymmetric self-energy part $H^2$
and the quartic part $H^4$
generated by the gauge action, they
lead to the following Higgs potential
\begin{eqnarray}
V & = & m_1^2 H_1^{*i} H_1^i + m_2^2 H_2^{*i} H_2^i - m_{12}^2 (\epsilon_{ij}
H_1^i H_2^j + hc) \non \\ \non \\
& & + \frac{1}{8} (g^2 + g'^2) [H_1^{*i} H_1^i -
H_2^{*i} H_2^i]^2 + \frac{1}{2} |H_1^{*i} H_2^{*i}|^2 ~. 
\end{eqnarray}
The Higgs potential includes three
bilinear mass terms, while the strength
of the quartic couplings is set by the
$SU(2)_L$ and $U(1)_Y$
gauge couplings squared. The three mass
terms are free parameters.

The potential develops a stable minimum
for $H_1 \to [0,v_1]$ and $H_2\to [v_2,0]$,
if the following conditions are met:
\begin{equation}
m_1^2 +  m_2^2 >  2 | m^2_{12} |  \hspace*{0.5cm} \mbox{and}  \hspace*{0.5cm}
m_1^2    m_2^2  <  | m^2_{12} |^2 ~.
\end{equation}
Expanding the fields about the ground-state 
values $v_1$ and $v_2$,
\begin{equation}
\begin{array}{rclcl}
H_1^1 & = & & & H^+ \cos \beta + G^+ \sin \beta \\ \\
H_1^2 & = & v_1 & + & [H^0 \cos \alpha - h^0 \sin \alpha + i A^0 \sin \beta - i G^0
\cos \beta ]/\sqrt{2}
\end{array}
\end{equation}
and
\begin{equation}
\begin{array}{rclcl}
H_2^1 & = & v_2 & + & [H^0 \sin \alpha + h^0 \cos \alpha + i A^0 \cos \beta + i G^0
\sin \beta ]/\sqrt{2} \\ \\
H_2^2 & = & & & H^- \sin \beta - G^- \cos \beta ~, 
\end{array}
\end{equation}
the mass eigenstates are given by the
neutral states $h^0,H^0$ and $A^0$,
which are even and odd under ${\cal CP}$
transformations, and by the charged states $H^\pm$;
the $G$ states correspond to the Goldstone
modes, which are absorbed by the gauge
fields to build up the longitudinal
components. After introducing the three
parameters
\begin{eqnarray}
M_Z^2 & = & \frac{1}{2} (g^2 + g'^2) (v_1^2 + v_2^2) \non \\ \non \\
M_A^2 & = & m_{12}^2 \frac{v_1^2 + v_2^2}{v_1v_2} \non \\ \non \\
\tgb  & = & \frac{v_2}{v_1} ~, 
\end{eqnarray}
the mass matrix can be decomposed into
three $2\times 2$ blocks, which are easy to
diagonalize:
\begin{displaymath}
\begin{array}{ll}
\mbox{\bf pseudoscalar mass:} & M_A^2 \\ \\
\mbox{\bf charged mass:} & M_\pm^2 = M_A^2+M_W^2 \\ \\
\mbox{\bf scalar mass:} &
M_{h,H}^2 = \frac{1}{2} \left[ M_A^2 + M_Z^2 \mp \sqrt{(M_A^2+M_Z^2)^2
- 4M_A^2M_Z^2 \cos^2 2\beta} \right] \\ \\
& \displaystyle \tg 2\alpha = \tg 2\beta \frac{M_A^2 + M_Z^2}{M_A^2 - M_Z^2}
\hspace*{0.5cm} \mbox{with} \hspace*{0.5cm} -\frac{\pi}{2} < \alpha < 0
\nonumber
\end{array}
\end{displaymath}

From the mass formulae, two important
inequalities can readily be derived,
\begin{eqnarray}
M_h & \leq & M_Z, M_A \leq M_H \\ \nonumber \\
M_W & \leq & M_{H^\pm} ~, 
\end{eqnarray}
which, by construction, are valid in
the tree approximation. As a result,
the lightest of the scalar Higgs masses
is predicted to be bounded by the $Z$ mass,
{\it modulo} radiative corrections. These bounds
follow from the fact that the quartic
coupling of the Higgs fields is determined
in the MSSM by the size of the gauge
couplings squared. \\


\noindent
\underline{\it SUSY Radiative Corrections} \\[0.5cm]
The tree-level relations between the
Higgs masses are strongly modified
by radiative corrections that involve
the supersymmetric particle spectrum
of the top sector \cite{50B}; cf.~Ref.~\cite{66a,mhplot} for recent
summaries. These effects
are proportional to the fourth power
of the top mass and to the logarithm
of the stop mass. Their origin are
incomplete cancellations between virtual
top and stop loops, reflecting the
breaking of supersymmetry. Moreover,
the mass relations are affected by 
the potentially large mixing between
$\tilde t_L$ and $\tilde t_R$
due to the top Yukawa coupling.\\

To leading order in $M_t^4$
the radiative corrections can be
summarized in the parameter
\begin{equation}
\epsilon = \frac{3G_F}{\sqrt{2}\pi^2}\frac{M_t^4}{\sin^2\beta}\log
\frac{M_{\tilde t_1}M_{\tilde t_2}}{M_t^2} ~.
\end{equation}
In this approximation the light Higgs mass $M_h$ 
can be expressed by $M_A$ and $\tgb$
in the following compact form:
\begin{eqnarray*}
M^2_h & = & \frac{1}{2} \left[ M_A^2 + M_Z^2 + \epsilon \right.
\non \\
& & \left. - \sqrt{(M_A^2+M_Z^2+\epsilon)^2
-4 M_A^2M_Z^2 \cos^2 2\beta
-4\epsilon (M_A^2 \sin^2\beta + M_Z^2 \cos^2\beta)} \right]
\end{eqnarray*}
The heavy Higgs masses $M_H$ and $M_{H^\pm}$
follow from the sum rules
\begin{eqnarray*}
M_H^2 & = & M_A^2 + M_Z^2 - M_h^2 + \epsilon \non \\
M_{H^\pm}^2 & = & M_A^2 + M_W^2 ~.
\end{eqnarray*}
Finally, the mixing parameter $\alpha$,  
which diagonalizes the ${\cal CP}$-even mass
matrix, is given by the radiatively
improved relation:
\begin{equation}
\tg 2 \alpha = \tg 2\beta \frac{M_A^2 + M_Z^2}{M_A^2 - M_Z^2 +
\epsilon/\cos 2\beta} ~. 
\label{eq:mssmalpha}
\end{equation}

\begin{figure}[hbt]
\begin{center}
\hspace*{-0.3cm}
\epsfig{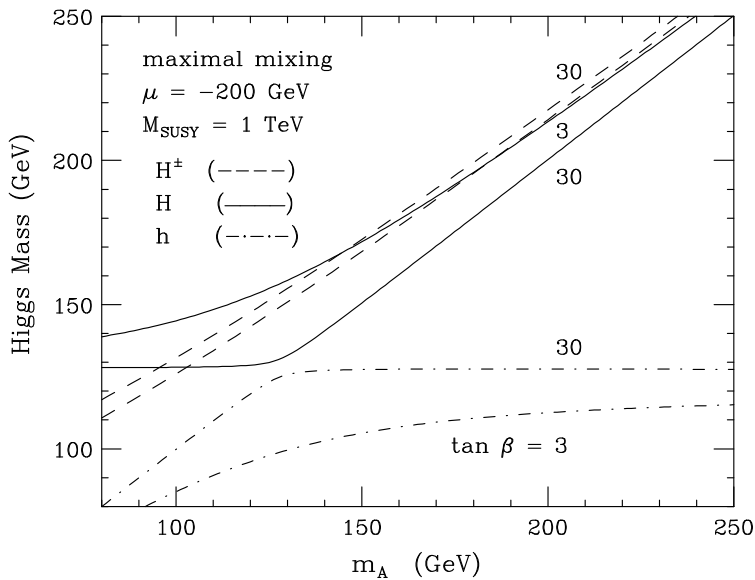}
\end{center}
\vspace*{-0.4cm}
\caption[]{\label{kdfig} \it The CP-even and charged MSSM Higgs boson 
masses as a function of $m_A$ for $\tan\beta=3$ and 30, including radiative
corrections. Ref.~\cite{66a}.}
\end{figure}

For large $A$ mass, the
masses of the heavy Higgs particles
coincide approximately, $M_A\simeq M_H \simeq M_{H^\pm}$,
while the light Higgs mass approaches
a small asymptotic value. The spectrum for large
values of $\tgb$ is quite  regular: for small $M_A$ one finds
$\{ M_h\simeq M_A; M_H  \simeq \mbox{const} \}$ \cite{intense}; 
for large $M_A$ the opposite relationship
$\{ M_h\simeq \mbox{const}, M_H \simeq M_{H^\pm}\simeq M_A \}$,
cf.~Fig.~\ref{kdfig} which includes radiative corrections.\\

While the non-leading effects of
mixing on the Higgs mass relations are
quite involved, the impact on the upper
bound of the light Higgs mass $M_h$ 
can be summarized in a simple way:
\begin{equation}
M_h^2 \leq M_Z^2 \cos^2 2\beta + \delta M_t^2 + \delta M_X^2 ~.
\end{equation}
The leading top contribution is related to
the parameter $\epsilon$,
\begin{equation}
\delta M_t^2 = \epsilon \sin^2\beta ~. 
\end{equation}
The second contribution
\begin{equation}
\delta M_X^2 = \frac{3G_F M_t^4}{2\sqrt{2}\pi^2} X_t^2 \left[ 2
h(M_{\tilde t_1}^2, M_{\tilde t_2}^2) + X_t^2~g(M_{\tilde t_1}^2,
M_{\tilde t_2}^2) \right]
\end{equation}
depends on the mixing parameter
\begin{equation}
M_t X_t = M_t \left[A_t - \mu~\ctgb \right] ~, 
\end{equation}
which couples left- and right-chirality
states in the stop mass matrix; $h,g$ are
functions of the stop masses:
\begin{equation}
h = \frac{1}{a-b} \log \frac{a}{b} \hspace*{0.5cm} \mbox{and} \hspace*{0.5cm}
g = \frac{1}{(a-b)^2} \left[ 2 - \frac{a+b}{a-b} \log \frac{a}{b} \right] ~.
\end{equation}
Subdominant contributions can essentially
be reduced to higher-order QCD effects.
They can effectively be incorporated by
interpreting the top mass parameter
$M_t \to M_t(\mu_t)$ as the $\overline{\rm MS}$
top mass evaluated at the geometric mean
between top and stop masses, $\mu_t^2 = M_t M_{\tilde t}$.\\

\begin{figure}[hbt]
\begin{center}
\hspace*{-0.3cm}
\epsfig{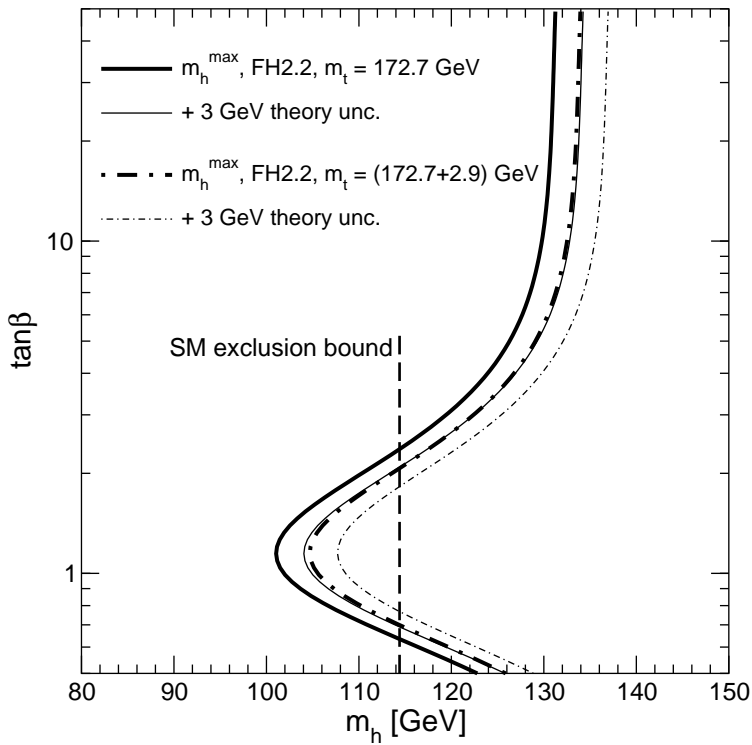}
\end{center}
\vspace*{-0.4cm}
\caption[]{\label{fg:mssmhiggs} \it Upper bounds on the light Higgs boson mass
as a function of $\tg\beta$ for various SUSY scenarios. Ref. \cite{mhplot}.}
\end{figure}

Upper bounds on the light Higgs mass are
shown in Fig.~\ref{fg:mssmhiggs} as a function of $\tg \beta$. The curves are 
the results of calculations with mixing effects. It turns out that the
general upper bound for maximal mixing is given by $M_h\lessim 140$ GeV, 
including large values of $\tgb$.  
The light Higgs sector could not entirely be
covered by the LEP2 experiments due to the increase of the mass limit with the 
top mass. 


\subsection{SUSY Higgs Couplings to SM Particles}

\phantom{h}
The size of MSSM Higgs couplings to quarks,
leptons and gauge bosons is similar to
the Standard Model, yet modified by the
mixing angles $\alpha$ and $\beta$.
Normalized to the SM values, they are
listed in Table \ref{tb:hcoup}. The pseudoscalar Higgs
boson $A$ does not couple to gauge bosons
at the tree level, but the coupling, 
compatible with  ${\cal CP}$ symmetry, can be
generated by higher-order loops.
The charged Higgs bosons couple to up   
and down fermions with the left- and
right-chiral amplitudes $g_\pm = -
\left[ g_t (1 \mp \gamma_5) + g_b (1 \pm \gamma_5) \right]/\sqrt{2}$, where
$g_{t,b} = (\sqrt{2} G_F)^{1/2} m_{t,b}$.
\begin{table}[hbt]
\renewcommand{\arraystretch}{1.5}
\begin{center}
\begin{tabular}{|lc||ccc|} \hline
\multicolumn{2}{|c||}{$\Phi$} & $g^\Phi_u$ & $g^\Phi_d$ &  $g^\Phi_V$ \\
\hline \hline
SM~ & $H$ & 1 & 1 & 1 \\ \hline
MSSM~ & $h$ & $\cos\alpha/\sin\beta$ & $-\sin\alpha/\cos\beta$ &
$\sin(\beta-\alpha)$ \\
& $H$ & $\sin\alpha/\sin\beta$ & $\cos\alpha/\cos\beta$ &
$\cos(\beta-\alpha)$ \\
& $A$ & $ 1/\tg\beta$ & $\tg\beta$ & 0 \\ \hline
\end{tabular}
\renewcommand{\arraystretch}{1.2}
\caption[]{\label{tb:hcoup}
\it Higgs couplings in the MSSM to fermions and gauge bosons [$V=W,Z$]
relative to SM couplings.}
\end{center}
\end{table}

The modified couplings incorporate the
renormalization due to SUSY radiative
corrections, to leading order in $M_t$, 
if the mixing angle $\alpha$ is related to
$\beta$ and $M_A$
through the corrected formula Eq.~(\ref{eq:mssmalpha}).
For large $M_A$, in practice $M_A\gsim 200$ GeV, 
the couplings of the light Higgs boson
$h$ to the fermions and gauge bosons
approach the SM values asymptotically.
This is the essence of the \underline{decoupling theorem} in the 
Higgs sector \cite{66AA}: 
Particles with large masses
must decouple from the light-particle
system as a consequence of the
quantum-mechanical uncertainty principle.

\subsection{Decays of Higgs Particles}

\phantom{h}
The lightest \underline{\it neutral Higgs boson} $h$ 
will decay mainly into fermion pairs
since the mass is smaller than $\sim 140$
GeV, Fig.~\ref{fg:mssmbr}a (cf. \cite{613A} for a
comprehensive summary). This is, in general,
also the dominant decay mode of the
pseudoscalar boson $A$. For values of $\tgb$
larger than unity and for masses less than
$\sim 140$ GeV, the main decay modes of the neutral
Higgs bosons are decays into $b\bar b$ and $\tau^+\tau^-$
pairs; the branching ratios are of order $\sim 90\%$ and $8\%$,
respectively. The decays into $c\bar c$
pairs and gluons are suppressed, especially
for large $\tgb$.  
For large masses, the top decay channels
$H,A \to t\bar t$ open up; yet for large $\tgb$
this mode remains suppressed and the
neutral Higgs bosons decay almost 
exclusively into $b\bar b$ and $\tau^+\tau^-$
pairs. If the mass is large enough, the
heavy ${\cal CP}$-even Higgs boson $H$
can in principle decay into weak gauge
bosons, $H\to WW,ZZ$.
Since the partial widths are proportional
to $\cos^2(\beta - \alpha)$,
they are strongly suppressed in general,
and the gold-plated $ZZ$ signal of the
heavy Higgs boson in the Standard Model
is lost in the supersymmetric extension.
As a result, the total widths of the Higgs
bosons are much smaller in supersymmetric
theories than in the Standard Model.
\begin{figure}[hbtp]

\vspace*{-2.5cm}
\hspace*{-4.5cm}
\begin{turn}{-90}%
\epsfxsize=16cm \epsfbox{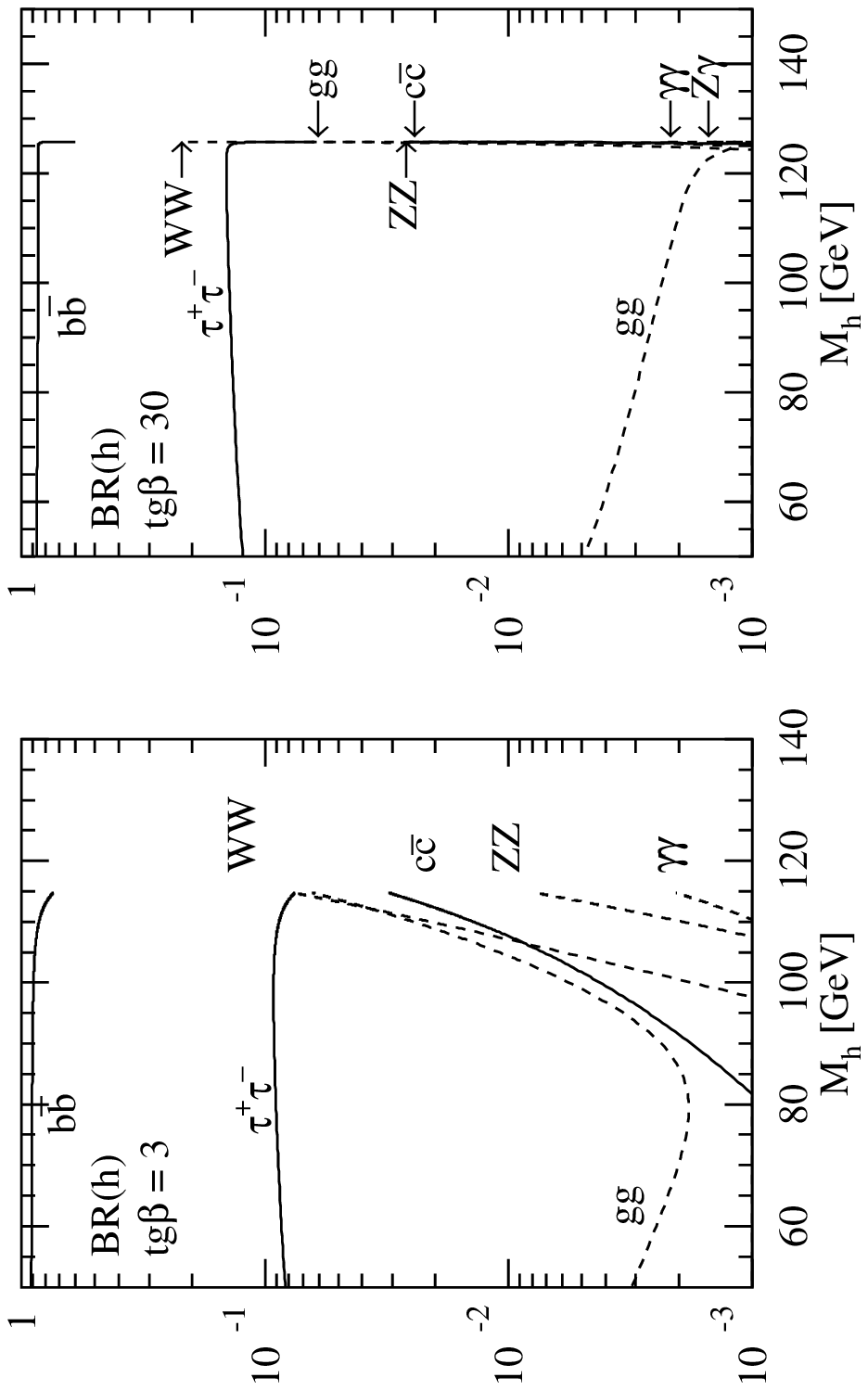}
\end{turn}
\vspace*{-4.2cm}

\centerline{\bf Fig.~\ref{fg:mssmbr}a}

\vspace*{-2.5cm}
\hspace*{-4.5cm}
\begin{turn}{-90}%
\epsfxsize=16cm \epsfbox{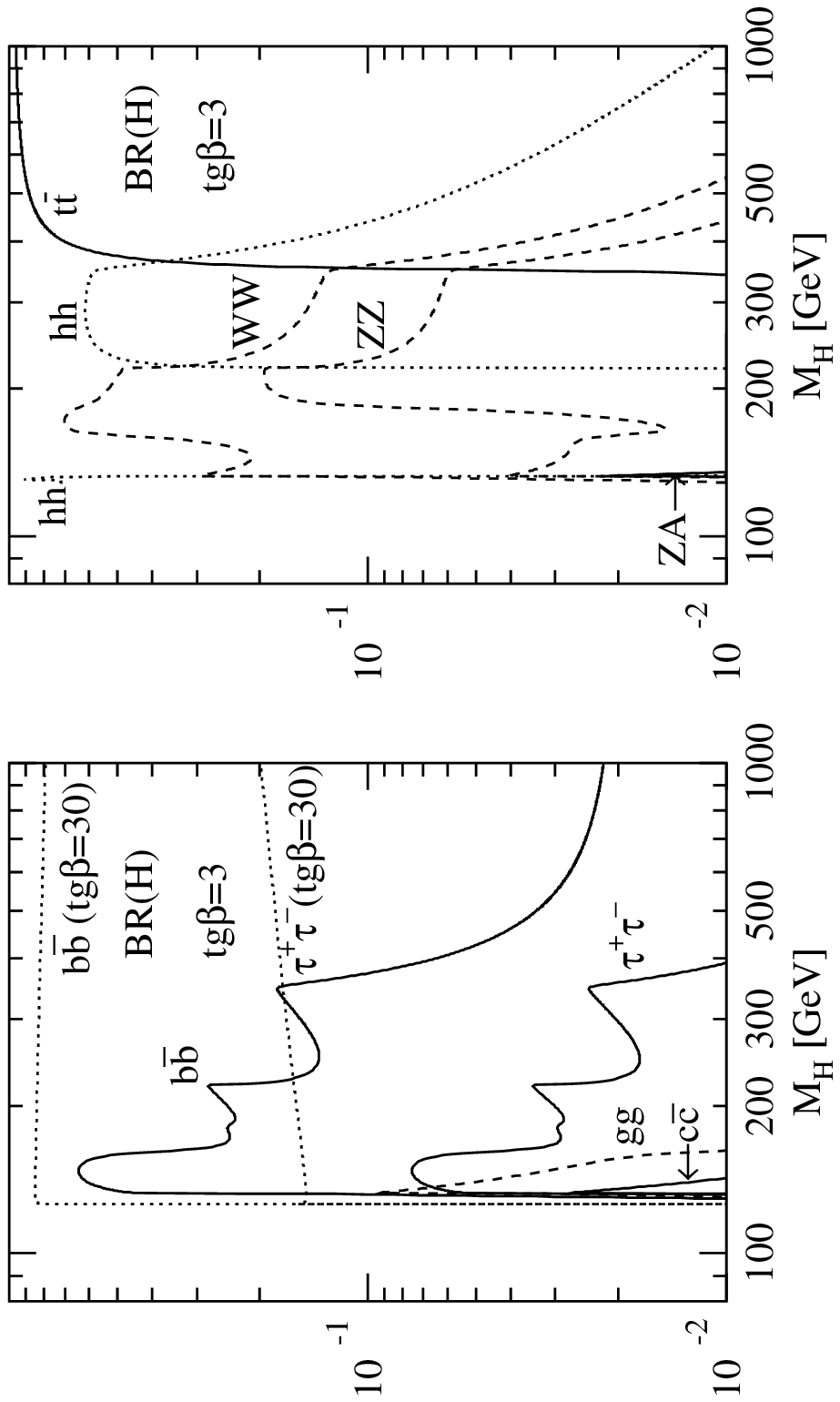}
\end{turn}
\vspace*{-4.2cm}

\centerline{\bf Fig.~\ref{fg:mssmbr}b}

\caption[]{\label{fg:mssmbr} \it Branching ratios of the MSSM Higgs bosons $h,
 H, A, H^\pm$ for non-SUSY decay modes as a function of the
masses for two values of $\tgb=3, 30$ and vanishing mixing. The common squark
mass has been chosen as $M_S=1$ TeV.}
\end{figure}
\addtocounter{figure}{-1}
\begin{figure}[hbtp]

\vspace*{-2.5cm}
\hspace*{-4.5cm}
\begin{turn}{-90}%
\epsfxsize=16cm \epsfbox{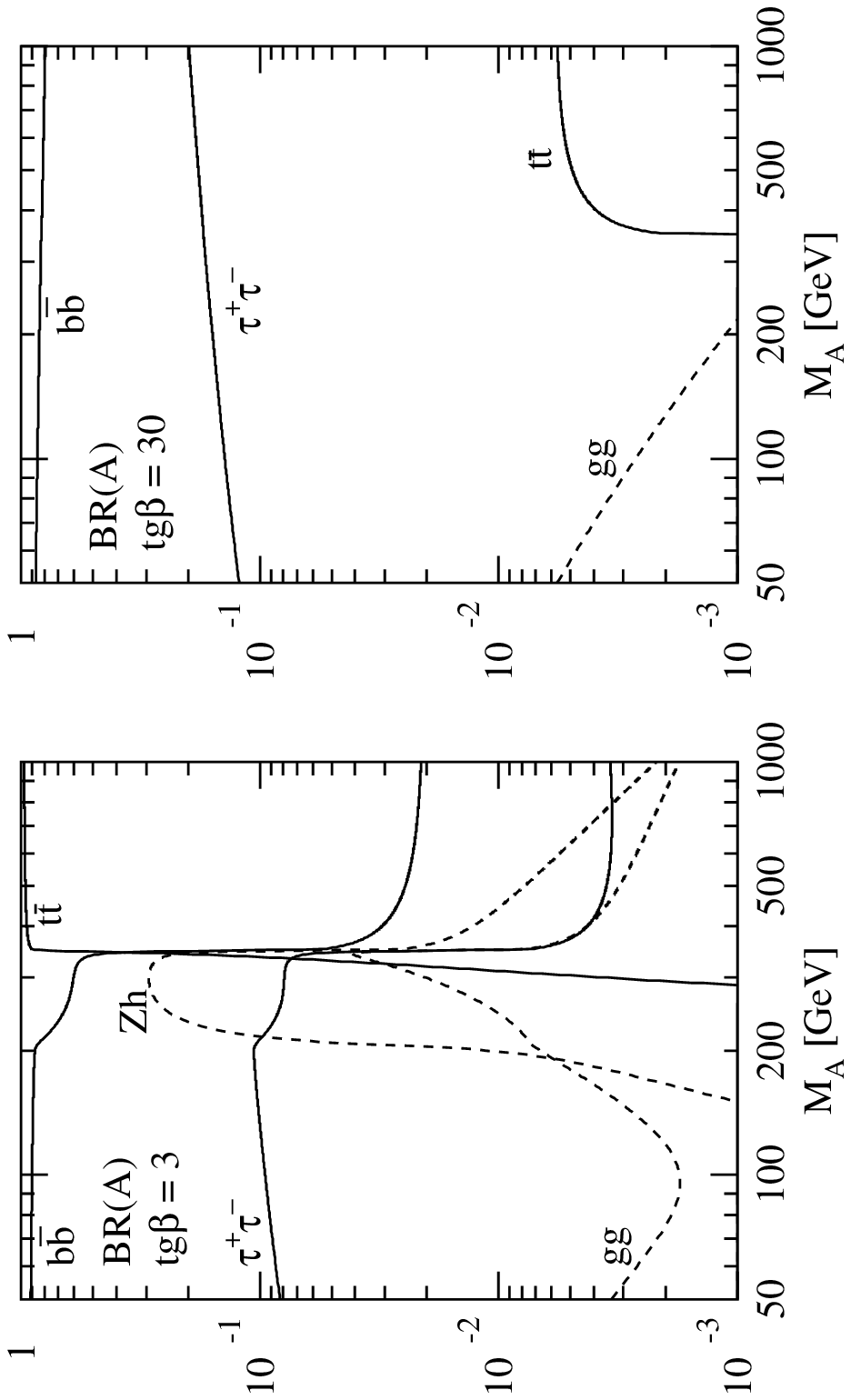}
\end{turn}
\vspace*{-4.2cm}

\centerline{\bf Fig.~\ref{fg:mssmbr}c}

\vspace*{-2.5cm}
\hspace*{-4.5cm}
\begin{turn}{-90}%
\epsfxsize=16cm \epsfbox{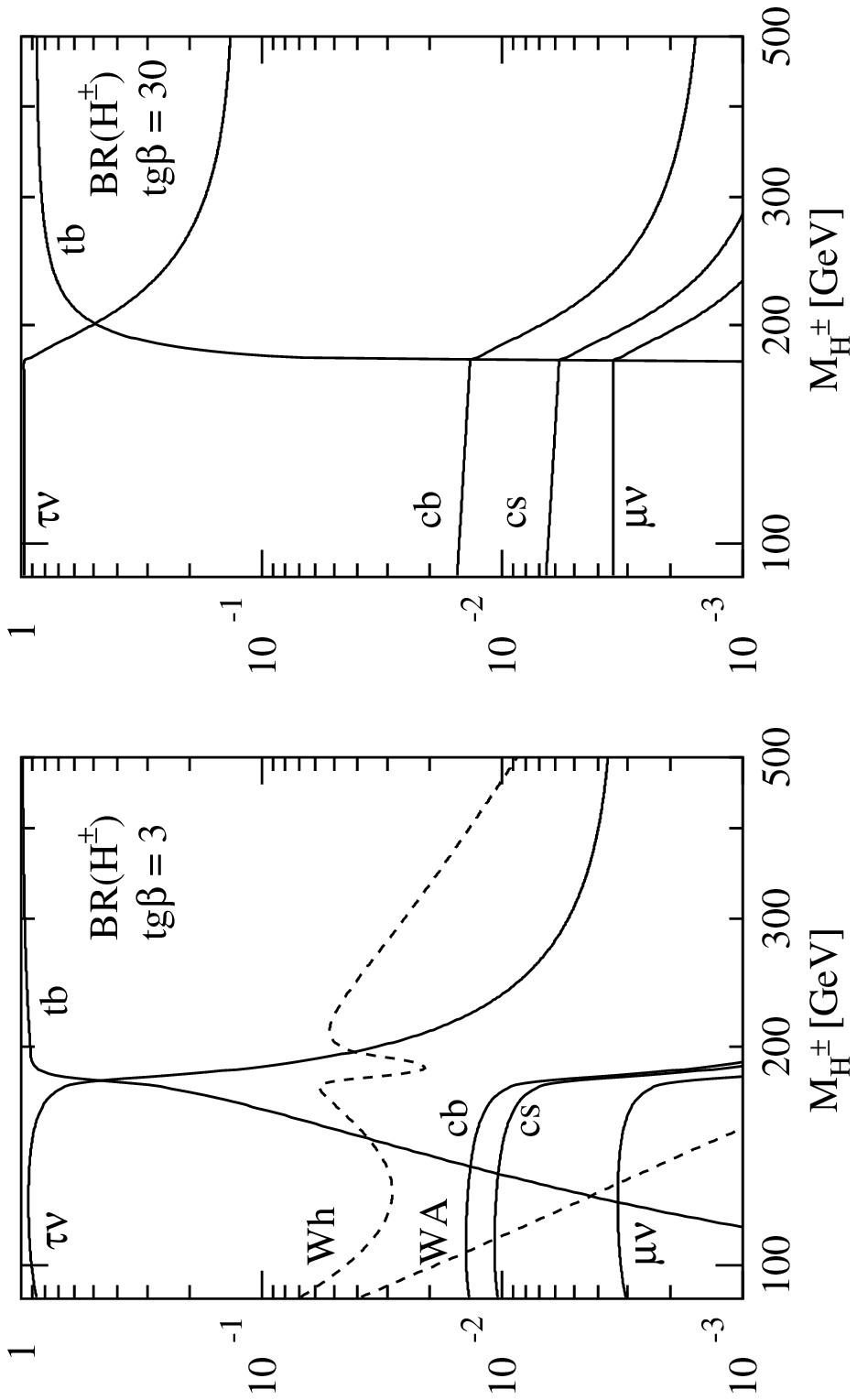}
\end{turn}
\vspace*{-4.2cm}

\centerline{\bf Fig.~\ref{fg:mssmbr}d}

\caption[]{\it Continued.}
\end{figure}

\begin{figure}[hbt]

\vspace*{-2.5cm}
\hspace*{-4.5cm}
\begin{turn}{-90}%
\epsfxsize=16cm \epsfbox{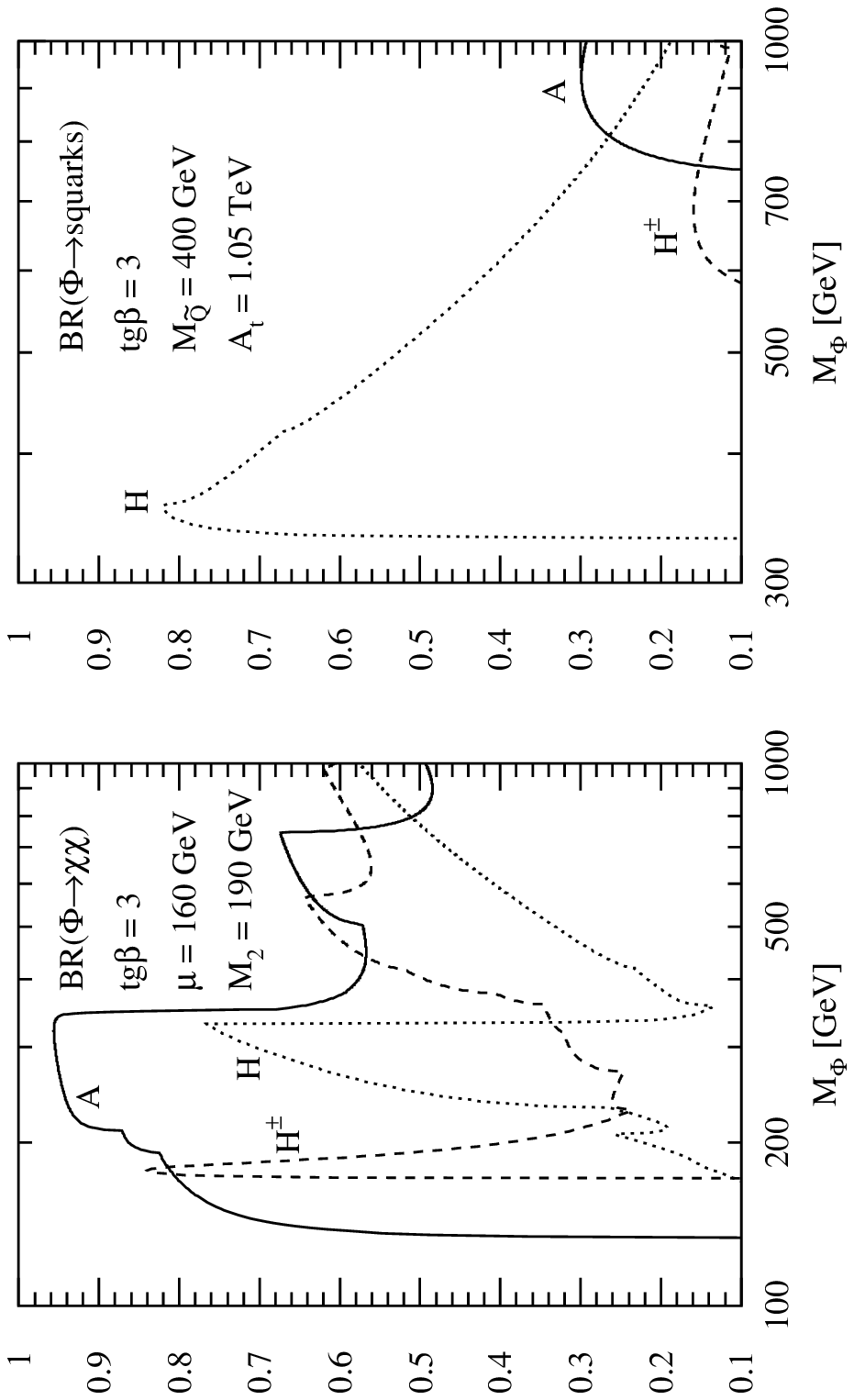}
\end{turn}
\vspace*{-4.2cm}

\caption[]{\label{fg:hcharneutsq} \it Branching ratios of the MSSM Higgs boson
$H,A,H^\pm$ decays into charginos/neutralinos and squarks as a function of their
masses for $\tgb=3$. The mixing parameters have been chosen as $\mu=160$ GeV,
$A_t=1.05$ TeV, $A_b=0$ and the squark masses of the first two generations as
$M_{\widetilde{Q}}=400$ GeV. The gaugino mass parameter has been set to
$M_2=190$ GeV.}
\end{figure}

The heavy neutral Higgs boson $H$
can also decay into two lighter Higgs bosons.
Other possible channels are Higgs cascade
decays and decays into supersymmetric
particles \citer{614,616}, Fig.~\ref{fg:hcharneutsq}. In addition to light
sfermions, Higgs boson decays into charginos
and neutralinos could eventually be important.
These new channels are kinematically accessible, 
at least for the heavy Higgs bosons $H,A$ and $H^\pm$;
in fact, the branching fractions can be very
large and they can even become dominant in some
regions of the MSSM parameter space. Decays of $h$
into the lightest neutralinos (LSP) are also
important, exceeding 50\% in some parts of
the parameter space. These decays
 strongly affect experimental search techniques.\\

The \underline{\it charged Higgs particles} decay into
fermions, but also, if allowed kinematically,
into the lightest neutral Higgs and a
$W$ boson. Below the $tb$ and $Wh$ thresholds,
the charged Higgs particles will 
decay mostly into $\tau \nu_\tau$ and $cs$
pairs, the former being dominant for $\tgb>1$.
For large $M_{H^\pm}$ values, the top--bottom decay mode
$H^+\to t\bar b$ becomes dominant. In some parts of
the SUSY parameter space, decays into
supersymmetric particles may exceed
50\%.\\

Adding up the various decay modes,
the width of all five Higgs bosons
remains very narrow, being of order
10 GeV even for large masses.

\subsection{The Production of SUSY Higgs Particles in $e^+e^-$ Collisions}

\phantom{h}
The search for the neutral SUSY Higgs bosons at \ee linear colliders will be
a straight\-forward ex\-tension of the search performed at LEP2, which
covered the mass range up to $\sim
100$~GeV for neutral Higgs bosons.  Higher
energies, $\sqrt{s}$ in excess of $250$~GeV, are required to sweep the
entire parameter space of the MSSM for moderate to large values
of $\tgb$.

\GS The main production mechanisms of \underline{\it neutral Higgs bosons} at
\ee colliders \cite{19, 615, 617} are the \Hs process and associated
pair production, as well as the fusion processes:
\begin{eqnarray}
(a) \ \ \mbox{Higgs--strahlung:} \hspace{1.4cm} \epem &
\stackrel{Z}{\longrightarrow} & Z+h/H \hspace{5cm}
\nonumber  \\
(b) \ \ {\rm Pair \ production:} \hspace{13.6mm} \epem &
\stackrel{Z}{\longrightarrow} & A+h/H 
\nonumber \\
(c) \ \ {\rm Fusion \ processes:} \hspace{10.7mm} \ \epem &
\stackrel{WW}{\longrightarrow} & \overline{\nu}_e \ \nu_e \ + h/H 
\hspace{3.3cm} \nonumber  \\
\epem & 
\stackrel{ZZ}{\longrightarrow} &  \epem + h/H  \nonumber
\end{eqnarray}
The ${\cal CP}$-odd Higgs boson $A$ cannot be produced in fusion
processes to leading order.  The cross sections for the four \Hs and
pair production processes can be expressed as
\begin{eqnarray}
\sigma(\epem \ra Z + h/H) & =& \sin^2/\cos^2(\beta-\alpha) \ \sigma_{SM}
\nonumber \\
\sigma(\epem \ra A + h/H) & =& \cos^2/\sin^2(\beta-\alpha) \
\bar{\lambda} \  \sigma_{SM} ~, 
\end{eqnarray}
where $\sigma_{SM}$ is the SM cross section for \Hs and the coefficient
$\bar{\lambda} \sim \lambda^{3/2}_{Aj} / \lambda^{\demi}_{Zj}$ accounts 
for the suppression of the $P$-wave 
$Ah/H$ cross sections near the threshold.

\STS The cross sections for  Higgs-strahlung and for  pair
production, much as those for the production of the
light and the heavy neutral Higgs bosons $h$ and $H$, are 
complementary, coming either with coefficients
$\sin^2(\beta-\alpha)$ or $\cos^2(\beta-\alpha)$.  As a result, since
$\sigma_{SM}$ is large, at least the lightest ${\cal CP}$-even Higgs
boson must be detected in $e^+e^-$ experiments.

\begin{figure}[hbtp]
\begin{center}
\vspace*{5mm}
\hspace*{5mm}
\epsfig{file=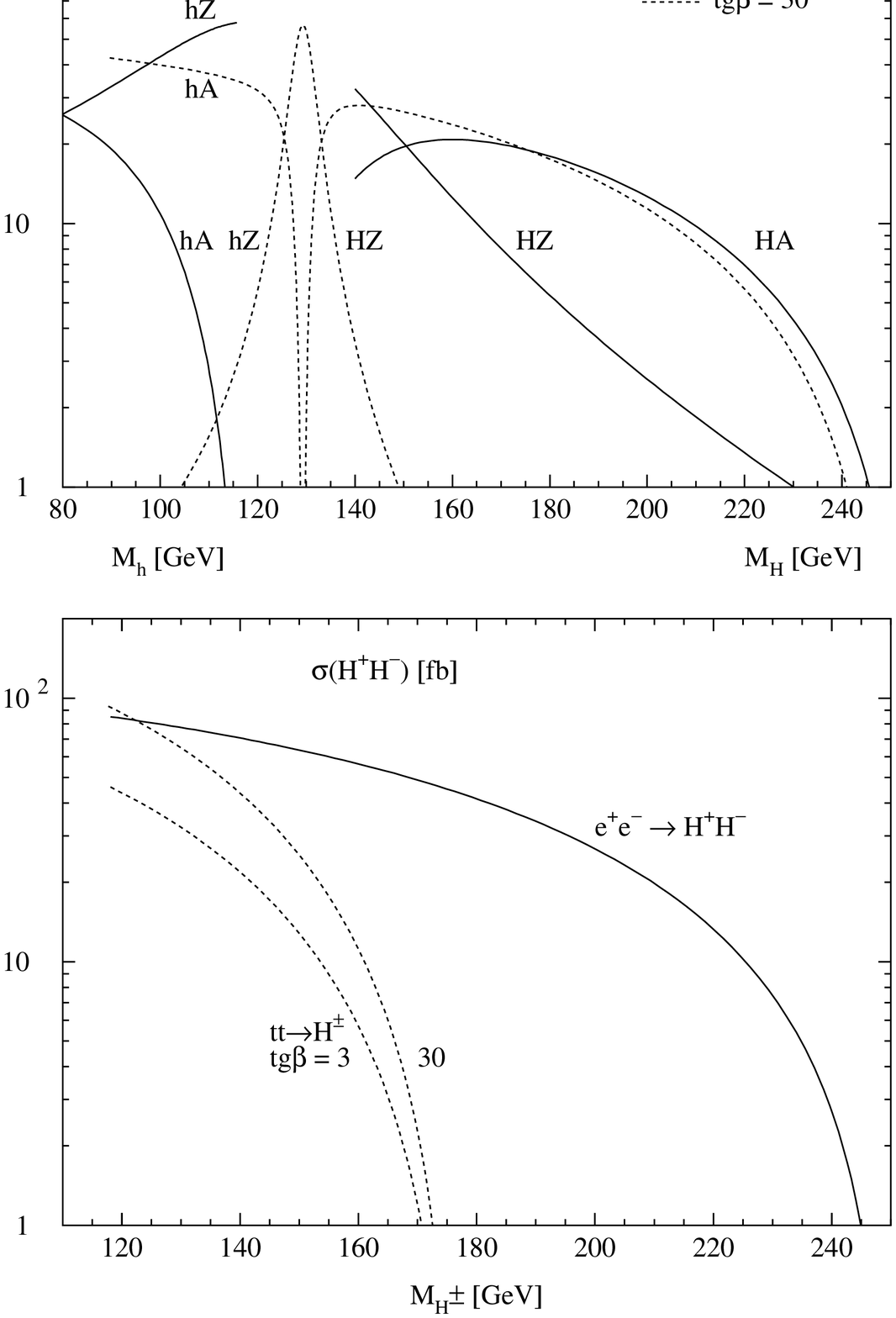,width=15cm}
\end{center}
\vspace{-2.5cm}
\caption[]{\it 
  Production cross sections of MSSM Higgs bosons at $\sqrt{s} =
  500$~GeV: Higgs-strahlung and pair production; upper part: neutral
  Higgs bosons, lower part: charged Higgs bosons.
  Ref. \protect\cite{613A}.  \protect\label{f604}\label{prodcs}}
\end{figure}
\STS Representative examples of cross sections for the production
mechanisms of the neutral Higgs bosons are exemplified in Fig.~\ref{f604}, 
as a function of the Higgs masses,  
 for $\tgb= 3$ and 30.  The cross
section for $hZ$ is large for $M_h$ near the maximum value allowed
for $\tgb$; it is of order 50~fb, corresponding to $\sim$ 2,500
events for an integrated luminosity of 50 fb$^{-1}$.  By contrast, the
cross section for $HZ$ is large if $M_h$ is sufficiently below the
maximum value  [implying small $M_H$].  For $h$
and for a low mass $H$, the signals consist of a $Z$ boson accompanied by
a $b\bar{b}$ or $\tau^+ \tau^-$ pair.  These signals are easy to separate
from the background,  which comes mainly from $ZZ$ production if the
Higgs mass is close to $M_Z$.  For the associated channels $\epem \to
Ah$ and $AH$, the situation is opposite to the previous case: the
cross section for $Ah$ is large for light $h$, whereas $AH$ pair
production is the dominant mechanism in the complementary region for
heavy $H$ and $A$ bosons.  The sum of the two cross sections
decreases from $\sim 50$ to 10~fb if $M_A$ increases from $\sim 50$ to
200~GeV at $\sqrt{s} = 500$~GeV.  In major parts of the parameter
space, the signals consist of four $b$ quarks in the final state,
requiring provisions for efficient $b$-quark tagging.  Mass
constraints will help to eliminate the backgrounds from QCD jets and
$ZZ$ final states.  For the $WW$ fusion mechanism, the cross sections
are larger than for Higgs-strahlung, if the Higgs mass is moderately small -- less
than 160~GeV at $\sqrt{s} = 500$ GeV.  However, since the final state
cannot be fully reconstructed, the signal is more difficult to
extract.  As in the case of the \Hs processes, the production of light
$h$ and heavy $H$ Higgs bosons complement each other in $WW$
fusion, too.

\GS The \underline{\it charged Higgs bosons}, if lighter than the top 
quark, can
be produced in top decays, $t \ra b + H^+$, with a branching ratio
varying between $2\%$ and $20\%$ in the kinematically allowed region.
Since the cross section for top-pair production is of order 0.5 pb at
$\sqrt{s} = 500$~GeV, this corresponds to 1,000 to 10,000 charged
Higgs bosons at a luminosity of 50~fb$^{-1}$.  Since, for $\tgb$ larger
than unity, the charged Higgs bosons will decay mainly into $\tau
\nu_\tau$, there is  a surplus of $\tau$ final states over $e,
\mu$ final states in $t$ decays, an apparent breaking of lepton
universality.  For large Higgs masses the dominant decay mode is the
top decay $H^+ \to t \overline{b}$.  In this case the charged Higgs
particles must be pair-produced in \ee colliders:
\[
              \epem \to H^+H^- ~.
\]
The cross section depends only on the charged Higgs mass.  It is of
order 100 fb for small Higgs masses at $\sqrt{s} = 500$~GeV, but it
drops very quickly due to the $P$-wave suppression $\sim \beta^3$
near the threshold.  For $M_{H^{\pm}} = 230$~GeV, the cross section
falls to a level of $\simeq 5\,$~fb. The \cs
is considerably larger for $\gamma \gamma$ collisions.

\GS \noindent
{\it Experimental Search Strategies} \\[0.5cm]
Search strategies have been described for neutral and charged
Higgs bosons in Ref. \cite{13}. The overall experimental situation
can be summarized as the following two points:

\STS
\noindent
{\bf (i)} The lightest ${\cal CP}$-even Higgs particle $h$ can be
detected in the entire range of the MSSM parameter space, either
via Higgs-strahlung  $\epem \to hZ$ or via pair
production $\epem \to hA$.  This conclusion holds true even at
a c.m. energy of 250 GeV, independently of the squark mass values; it
is also valid if decays to invisible neutralinos and other SUSY \ps
are realized in the Higgs sector.

\STS
\noindent
{\bf (ii)} The area in the parameter space where {\it all SUSY Higgs
bosons} can be discovered at \ee colliders is characterized by $M_H,
M_A \lessim \frac{1}{2} \sqrt{s}$, independently of $\tgb$.  The $h,
H$ Higgs bosons can be produced either via \Hs or in $Ah, AH$
associated production; charged Higgs bosons will be produced in
$H^+H^-$ pairs. \\

\begin{figure}[hbt]
\begin{center}
\hspace*{-0.3cm}
\epsfig{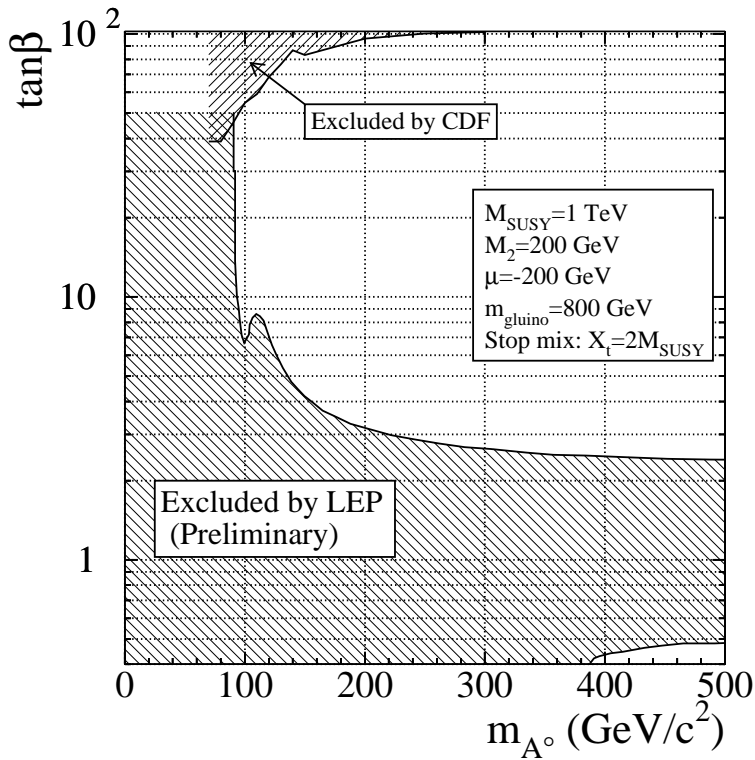}
\end{center}
\vspace*{-0.4cm}
\caption[]{\label{fig:igo} \it The 95\% CL bounds on $m_A$ and $\tan\beta$
for the $m_h-max$ benchmark scenario from LEP \cite{lep}. 
The exclusions at large 
$\tan\beta$ from CDF are also indicated. Ref. \cite{cdf}.}
\end{figure}

The search for the lightest neutral SUSY
Higgs boson $h$ had been one of the most important experimental
tasks at LEP2. Mass values of the
pseudoscalar boson $A$ of less about 90 GeV have 
been excluded, independently of $\tgb$, cf.~Fig.~\ref{fig:igo}. 
In MSSM scenarios
without mixing effects, the entire mass range
of the lightest Higgs particle $h$ has already
been covered for $\tgb$ less than about 1.6; however,
this conclusion does not hold true  for scenarios
with strong mixing effects \cite{mhplot}.

\subsection{The Production of SUSY Higgs Particles in Hadron Collisions}

\phantom{h}
The basic production processes of SUSY
Higgs particles at hadron colliders \cite{24A,32,620B}
are essentially the same as in the
Standard Model. Important differences
are  nevertheless generated by the
modified couplings, the extended particle
spectrum, and the negative parity of
the $A$ boson. For large $\tgb$ 
the coupling $hb\bar b$ is enhanced so that
the bottom-quark loop becomes competitive 
to the top-quark loop in the effective
$hgg$ coupling. Moreover squark loops
will contribute to this coupling \cite{sqloop}.\\

The partonic cross section $\sigma(gg\to \Phi)$
for the gluon fusion of Higgs particles
can be expressed by couplings $g$, in units
of the corresponding SM couplings, and
form factors $A$; to lowest order \cite{32,sqloopqcd}:
\begin{eqnarray}
\hat\sigma^\Phi_{LO} (gg\to \Phi) & = & \sigma^\Phi_0 M_\Phi^2 \times
BW(\hat{s}) \\
\sigma^{h/H}_0 & = & \frac{G_{F}\alpha_{s}^{2}(\mu)}{128 \sqrt{2}\pi} \
\left| \sum_{Q} g_Q^{h/H} A_Q^{h/H} (\tau_{Q})
+ \sum_{\widetilde{Q}} g_{\widetilde{Q}}^{h/H} A_{\widetilde{Q}}^{h/H}
(\tau_{\widetilde{Q}}) \right|^{2} \nonumber \\
\sigma^A_0 & = & \frac{G_{F}\alpha_{s}^{2}(\mu)}{128 \sqrt{2}\pi} \
\left| \sum_{Q} g_Q^A A_Q^A (\tau_{Q}) \right|^{2} \nonumber
\end{eqnarray}
While the quark couplings have been
defined in Table \ref{tb:hcoup}, the couplings of 
the Higgs particles to squarks are given by
\begin{eqnarray}
g_{\tilde Q_{L,R}}^{h} & = & \frac{M_Q^2}{M_{\tilde Q}^2} g_Q^{h}
\mp \frac{M_Z^2}{M_{\tilde Q}^2} (I_3^Q - e_Q \sin^2 \theta_W)
\sin(\alpha + \beta) \nonumber \\ \nonumber \\
g_{\tilde Q_{L,R}}^{H} & = & \frac{M_Q^2}{M_{\tilde Q}^2} g_Q^{H}
\pm \frac{M_Z^2}{M_{\tilde Q}^2} (I_3^Q - e_Q \sin^2 \theta_W)
\cos(\alpha + \beta)
\end{eqnarray}
Only ${\cal CP}$ non-invariance allows for non-zero
squark contributions to the pseudoscalar $A$
boson production. 
The form factors can be expressed
in terms of the scaling function $f(\tau_i=4M_i^2/M_\Phi^2)$,
cf. Eq. (\ref{eq:ftau}):
\begin{eqnarray}
A_Q^{h/H} (\tau) & = & \tau [1+(1-\tau) f(\tau)] \nonumber \\
A_Q^A (\tau) & = & \tau f(\tau) \nonumber \\
A_{\tilde Q}^{h/H} (\tau) & = & -\frac{1}{2}\tau [1-\tau f(\tau)] ~.
\end{eqnarray}
For small $\tgb$ the contribution of the top loop is
dominant, while for large $\tgb$
the bottom loop is strongly enhanced.
The squark loops can be significant
for squark masses below $\sim 400$ GeV \cite{sqloopqcd}.\\

Other production mechanisms for SUSY
Higgs bosons, vector boson fusion,
Higgs-strahlung off $W,Z$ bosons and
Higgs-bremsstrahlung off top and bottom
quarks, can be treated in analogy to
the corresponding SM processes.\\
\begin{figure}[hbtp]

\vspace*{0.3cm}
\hspace*{1.0cm}
\begin{turn}{-90}%
\epsfxsize=8.5cm \epsfbox{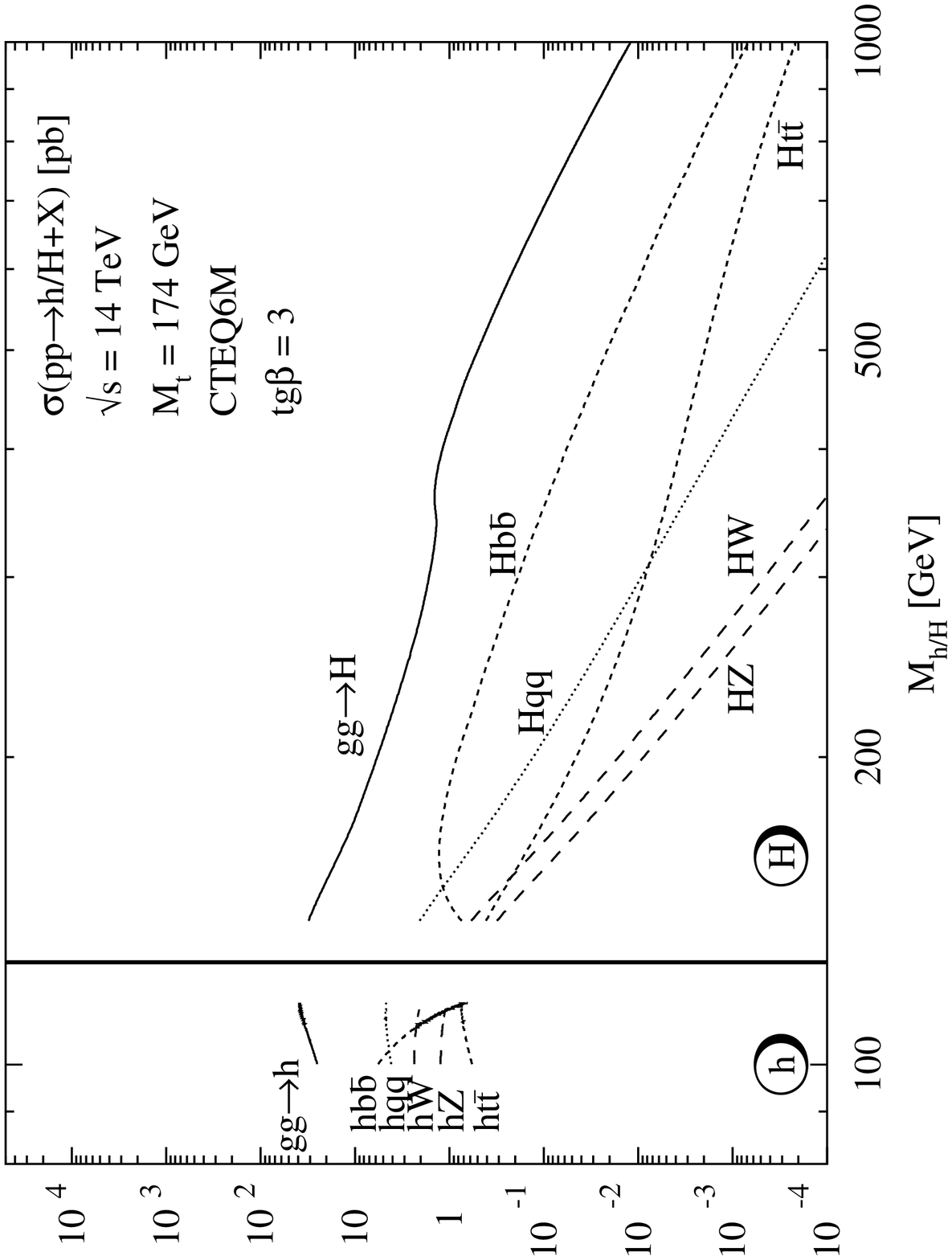}
\end{turn}
\vspace*{0.3cm}

\centerline{\bf Fig.~\ref{fg:mssmprohiggs}a}

\vspace*{0.2cm}
\hspace*{1.0cm}
\begin{turn}{-90}%
\epsfxsize=8.5cm \epsfbox{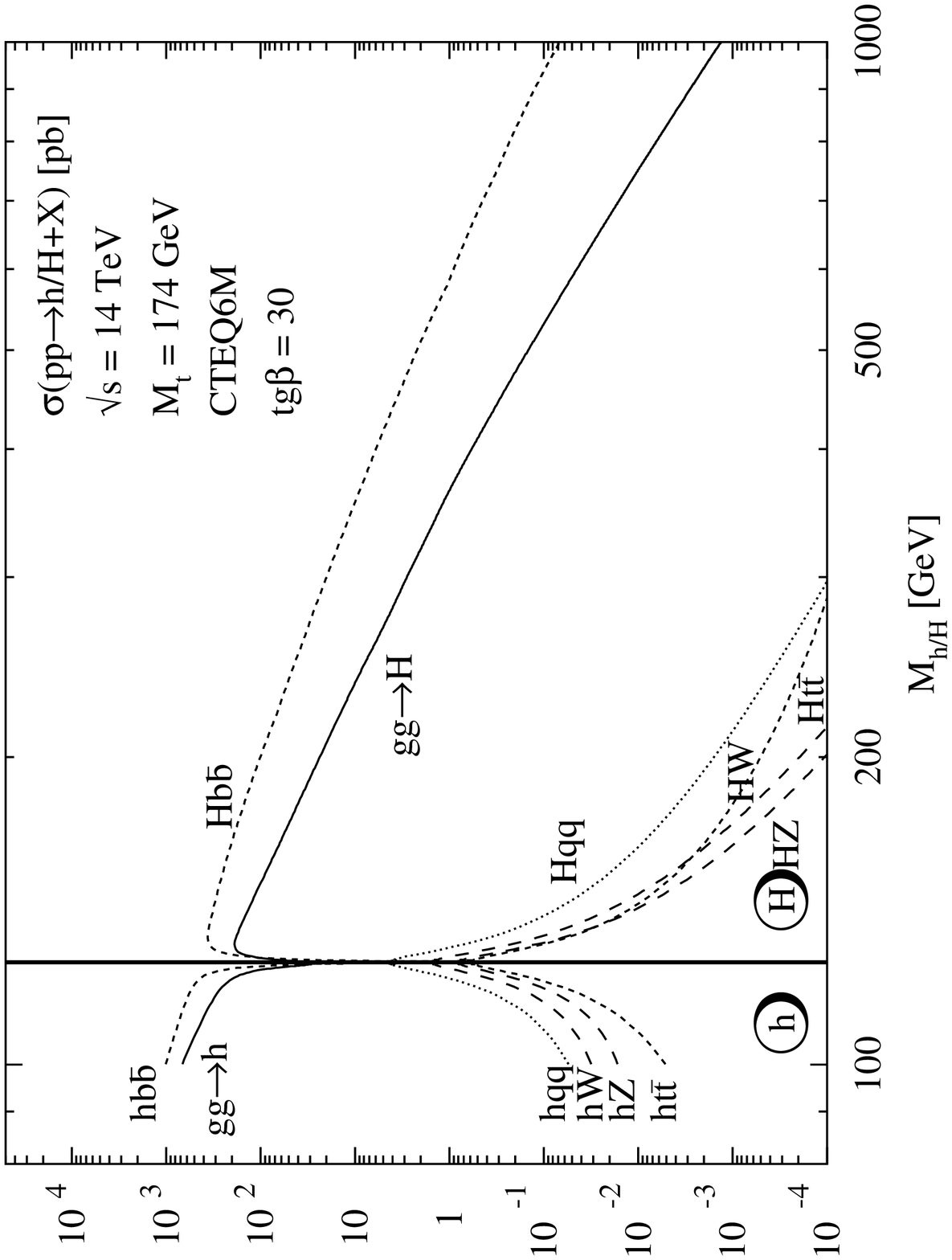}
\end{turn}
\vspace*{0.3cm}

\centerline{\bf Fig.~\ref{fg:mssmprohiggs}b}

\caption[]{\label{fg:mssmprohiggs} \it Neutral MSSM Higgs production cross
sections at the LHC  for gluon fusion $gg\to \Phi$,
vector-boson fusion $qq\to qqVV \to qqh/
qqH$, Higgs-strahlung $q\bar q\to V^* \to hV/HV$ and the associated
production $gg,q\bar q \to  b\bar b \Phi/ t\bar t \Phi$, including all known
QCD corrections. (a) $h,H$ production for $\tgb=3$, (b) $h,H$ production for
$\tgb=30$, (c) $A$ production for $\tgb=3$, (d) $A$ production for $\tgb=30$.}
\end{figure}
\addtocounter{figure}{-1}
\begin{figure}[hbtp]

\vspace*{0.3cm}
\hspace*{1.0cm}
\begin{turn}{-90}%
\epsfxsize=8.5cm \epsfbox{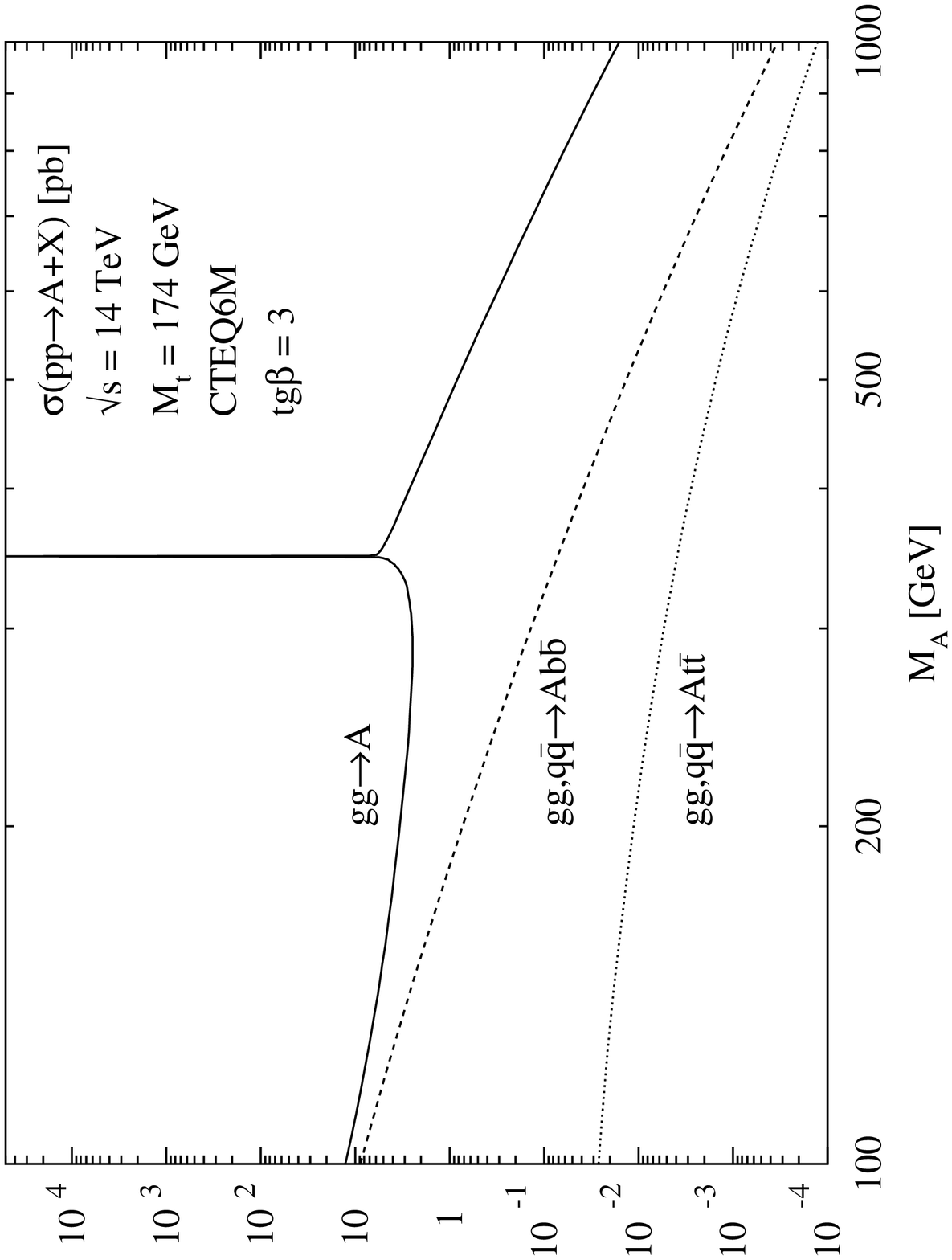}
\end{turn}
\vspace*{0.3cm}

\centerline{\bf Fig.~\ref{fg:mssmprohiggs}c}

\vspace*{0.2cm}
\hspace*{1.0cm}
\begin{turn}{-90}%
\epsfxsize=8.5cm \epsfbox{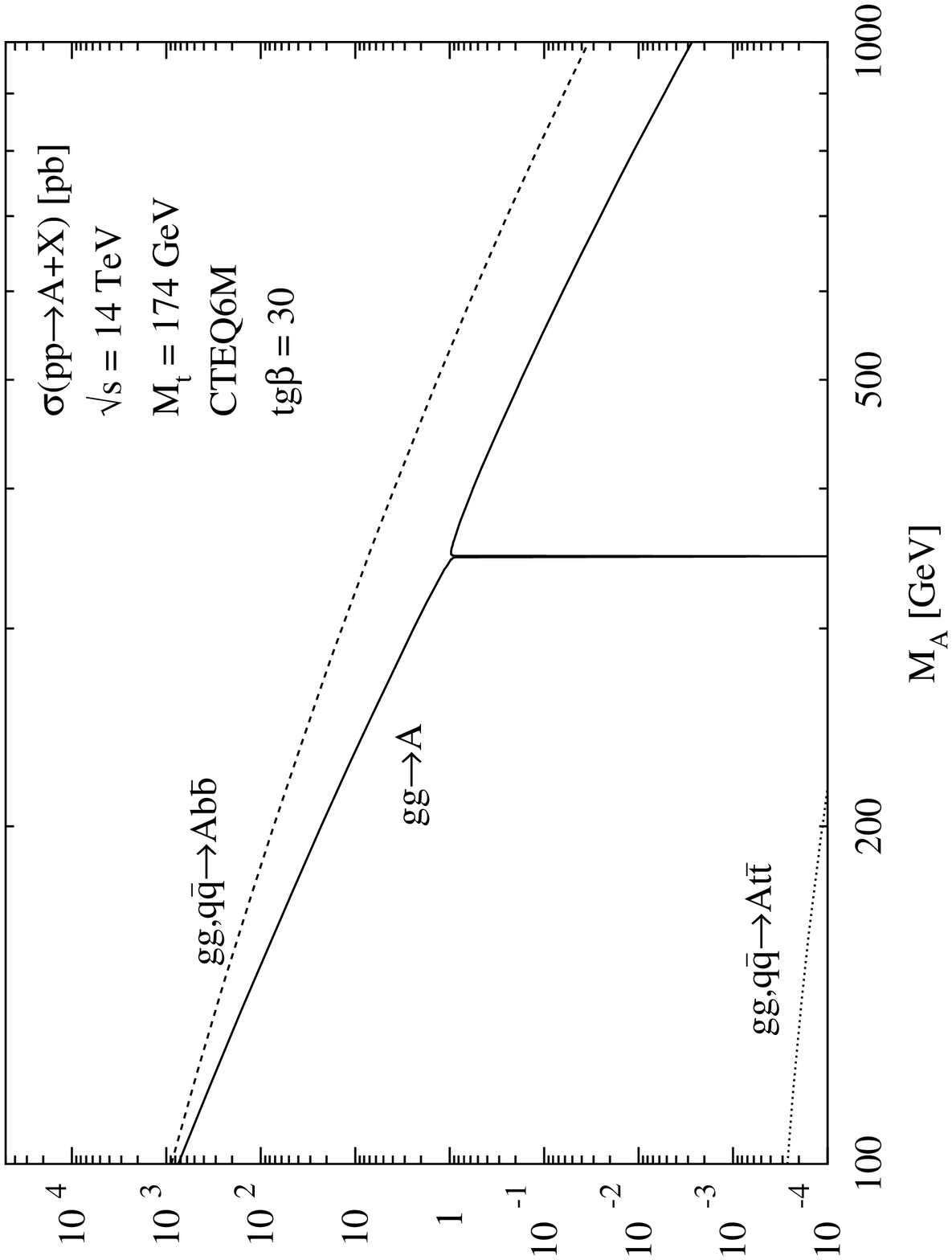}
\end{turn}
\vspace*{0.3cm}

\centerline{\bf Fig.~\ref{fg:mssmprohiggs}d}

\caption[]{\it Continued.}
\end{figure}

Data from the Tevatron in the channel $p \bar p \to b \bar b \tau^+ \tau^-$
 have been exploited \cite{63A} to exclude part of the supersymmetric Higgs
parameter space in the $[ M_A, \tgb]$ plane. In the
interesting range of $\tgb$ between 30 and 50, 
pseudoscalar masses $M_A$ of up to 150 to 190 GeV appear 
to be excluded.\\

The cross sections of the various MSSM Higgs production mechanisms at the LHC
are shown in Figs. \ref{fg:mssmprohiggs}a--d for two representative values of
$\tgb = 3$ and 30, as a function of the corresponding Higgs mass.
The CTEQ6M
parton densities have been adopted with $\alpha_s(M_Z)=0.118$; the top and
bottom masses have been set to $M_t=174$ GeV and $M_b=4.62$ GeV. For the
pseudoscalar Higgs bremsstrahlung off $t,b$ quarks, $pp \to Q\bar Q A +X$,  the
leading-order CTEQ6L1 parton densities have been used.
For small and moderate values of $\tgb\lessim 10$
the gluon-fusion cross section provides the dominant production cross section
for the entire Higgs mass region up to $M_\Phi\sim 1$ TeV. However, for large
$\tgb$, Higgs bremsstrahlung off bottom quarks, $pp\to b\bar b \Phi+X$, 
dominates
over the gluon-fusion mechanism since  the  bottom Yukawa
couplings are strongly enhanced in this case.\\

\begin{figure}[hbtp]
\begin{center}
\epsfig{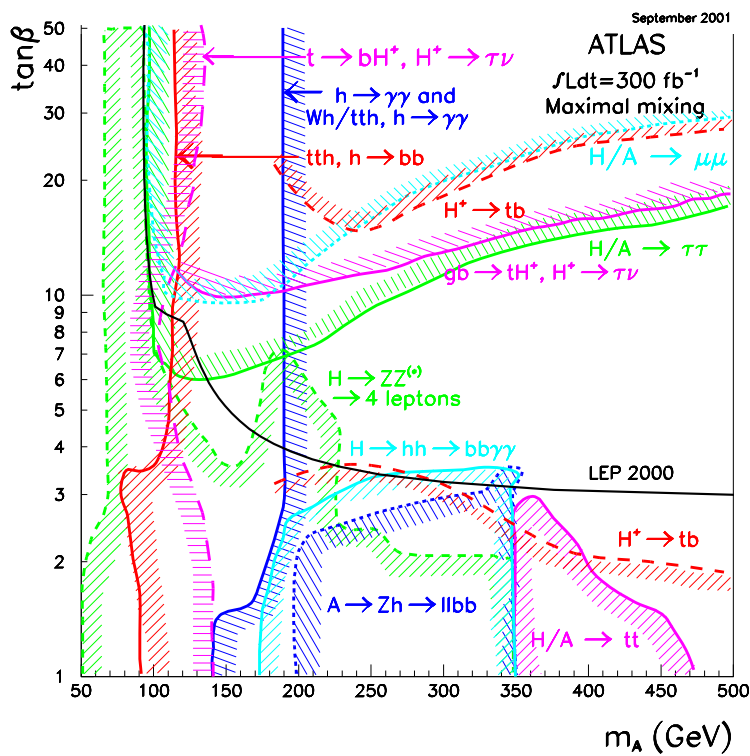}
\end{center}
\caption[]{\label{fg:atlascms} \it 
The ATLAS sensitivity for the discovery of the MSSM Higgs bosons in the 
case of maximal mixing. The 5$\sigma$ discovery curves are shown in the 
$(\tan\beta,m_A)$ plane for the individual channels and for an integrated 
luminosity of 300 fb$^{-1}$. The corresponding LEP limit is also shown. 
Ref.~\cite{richter}.}
\end{figure}

The MSSM Higgs search at the LHC will be more involved than the SM Higgs
search. 
%
The final summary is presented in Fig.~\ref{fg:atlascms}. It exhibits a 
difficult region for
the MSSM Higgs search at the
LHC. For $\tgb \sim 5$ and $M_A \sim 150$ GeV, the full
luminosity and the full data sample of both the ATLAS and CMS 
experiments at the
LHC are needed to cover the problematic parameter region \cite{richter}.
On the other hand, if no excess of Higgs events
above the SM background processes beyond 2 standard deviations will be found,
the MSSM Higgs bosons can be excluded at 95\% C.L. Even though the
entire supersymmetric Higgs parameter
space is expected to be finally covered by the
LHC experiments, the entire ensemble of individual Higgs
bosons is accessible only in part of
the parameter space. Moreover, the search
for heavy $H,A$ Higgs particles is very
difficult, because of the $t\bar t$
continuum background for masses  $\gsim 500$ GeV.\\

\subsection{Measuring the Parity of Higgs Bosons}

\phantom{h}
Once the  Higgs bosons  are
discovered, 
the properties of the particles must be established.
Besides the reconstruction of the supersymmetric Higgs potential \cite{66A},
which will be a very demanding effort, the external quantum
 numbers must be established, in particular the parity of the
heavy scalar and pseudoscalar Higgs particles $H$ and $A$ \cite{618}.\\[-0.1cm]

For large $H,A$ masses the decays $H,A\to t\bar t$ 
to top final states can be used to
discriminate between the different parity
assignments \cite{618}. For example, the $W^+$ and $W^-$
bosons in the $t$ and $\bar t$
decays tend to be emitted antiparallel
and parallel in the plane perpendicular
to the $t\bar t$ axis:
\begin{equation}
\frac{d\Gamma^\pm}{d\phi_*} \propto 1 \mp \left( \frac{\pi}{4} \right)^2
\cos \phi_*
\end{equation}
for $H$ and $A$ decays, respectively. \\[-0.1cm]

For light $H,A$ masses, $\gamma\gamma$
collisions appear to provide a viable
solution \cite{618}. The fusion of Higgs
particles in linearly polarized photon
beams depends on the angle between
the polarization vectors. For scalar $0^+$
particles the production amplitude
is non-zero for parallel polarization
vectors, while pseudoscalar $0^-$
particles require perpendicular
polarization vectors:
\begin{equation}
{\cal M}(H)^+  \sim  \vec{\epsilon}_1 \cdot \vec{\epsilon}_2  \hspace*{0.5cm}
\mbox{and} \hspace*{0.5cm}
{\cal M}(A)^-  \sim  \vec{\epsilon}_1 \times \vec{\epsilon}_2 ~.
\end{equation}
The experimental set-up for Compton
back-scattering of laser light can
be tuned in such a way that the
linear polarization of the hard-photon
beams approaches values close to 100\%.
 Depending on the $\pm$ parity 
of the resonance produced, the measured
asymmetry for photons of  parallel and perpendicular polarization, 
\begin{equation}
{\cal A} = \frac{\sigma_\parallel - \sigma_\perp}{\sigma_\parallel +
\sigma_\perp} ~, 
\end{equation}
is either positive or negative.

\subsection{Non-minimal Supersymmetric Extensions}

\phantom{h}
The minimal supersymmetric extension of the \SM may appear very
restrictive for supersymmetric theories in general, in particular in
the Higgs sector where the quartic couplings are identified with the
gauge couplings.  However, it turns out that the mass pattern of the
MSSM is quite typical if the theory is assumed to be valid up to the
GUT scale -- the motivation for supersymmetry {\it sui generis}.  This
general pattern has been studied thoroughly within the
next-to-minimal extension: the MSSM, incorporating two Higgs
isodoublets, is extended by introducing an additional isosinglet field $N$.
This extension leads to a model \citer{621,70A} that is generally
referred to as the NMSSM.

\STS The additional Higgs singlet can solve the so-called
$\mu$-problem [i.e. $\mu \sim$ order $M_W$] by
eliminating the $\mu$ higgsino parameter from the potential and by 
replacing this parameter  by the vacuum expectation value of the $N$ field,
which can  naturally be related to the usual vacuum expectation values
of the Higgs isodoublet fields.  In this scenario the superpotential
involves the two trilinear couplings $H_1 H_2 N$ and $N^3$.  The
consequences of this extended Higgs sector will be outlined  in
the context of (s)grand unification, including the universal soft breaking
terms of the supersymmetry \cite{622,70A}.\\

\begin{figure}[hbt]
\begin{center}
\hspace*{-0.3cm}
\epsfig{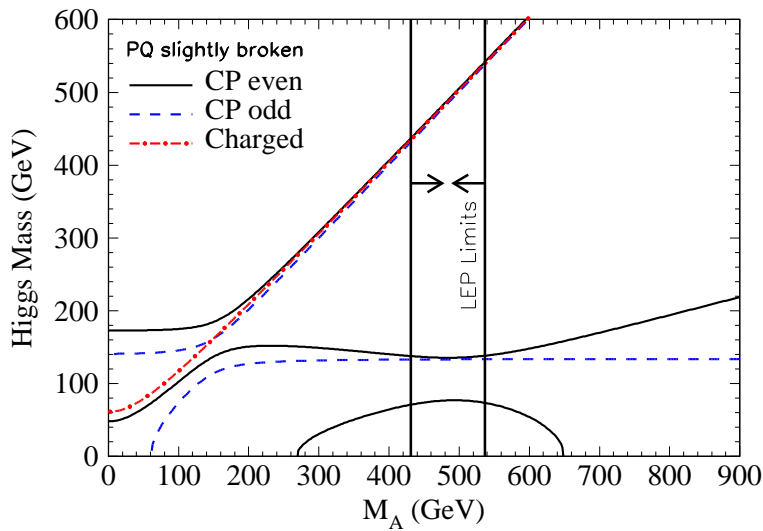}
\end{center}
\vspace*{-0.4cm}
\caption[]{\label{fig:26} \it The one-loop Higgs boson masses as a 
function of $M_A$ for $\lambda=0.3$, $\kappa=0.1$, $v_s=3v$, $\tan\beta=3$
and $A_\kappa=-100$~GeV. The arrows denote the region allowed by LEP searches
with 95\% confidence. Ref. \cite{70A}.}
\end{figure}

\GS The Higgs spectrum of the NMSSM includes, besides the minimal
set of Higgs particles, one additional scalar and pseudoscalar Higgs
particle.  The neutral Higgs \ps are in general mixtures of 
isodoublets, which couple to $W, Z$ bosons and fermions, and
the isosinglet, decoupled from the non-Higgs sector.
The trilinear self-interactions contribute to the masses of the Higgs
particles; for the lightest Higgs boson of each species: 
\begin{eqnarray}
M^2 (h_1) & \leq & M^2_Z \cos^2 2\beta + \lambda^2 v^2 \sin^2 2 \beta \\
M^2 (A_1) & \leq & M^2 (A)   \nonumber \\
M^2 (H^{\pm}) & \leq & M^2 (W) + M^2 (A) - \lambda^2 v^2 \nonumber
\end{eqnarray}
In contrast with the minimal model, the mass of the charged
Higgs \p could be smaller than the \W mass. An example of the mass spectrum
is shown in Fig.~\ref{fig:26}. Since the trilinear \cps
increase with energy, upper bounds on the mass of the lightest neutral
Higgs boson $h_1^0$ can be derived, in analogy to the Standard Model,
from the assumption that the theory be valid up to the GUT scale:
$m(h_1^0) \lessim 140 $~GeV.  Thus, despite the additional
interactions, the distinct pattern of the minimal extension remains
valid also in more complex \ssy \sces.  In fact, the mass
bound of 140~GeV for the lightest Higgs particle is realized in almost
all \ssy theories \cite{623}. If $h_1^0$ is (nearly) pure isosinglet, it
decouples from the gauge boson and fermion system and its role is
taken by the next Higgs \p with a large isodoublet component, implying
the validity of the mass bound again.\\

\STS If the Higgs \p $h_1^0$ is primarily isosinglet, the \cp
$ZZh_1^0$ is small and the \p cannot be produced by Higgs-strahlung.
However, in this case $h_2^0$ is generally light and couples with
sufficient strength to the $Z$ boson; if not, $h_3^0$ plays this role.\\

\STS {\it In summa}. Experiments at $e^+e^-$ colliders are in a `no-lose' 
situation 
\cite{L141A} for detecting the Higgs particles in general supersymmetric
theories, even for c.m. energies as low as $\sqrt{s} \sim 300$ GeV.


\section{Dynamical Symmetry Breaking}

\phantom{h}
\subsection{Little Higgs Models}

\phantom{h}
To interpret the Higgs boson as a (pseudo-)Goldstone boson has been a very 
attractive idea for a long time. The interest in this picture has been renewed 
within the Little Higgs scenarios \cite{2A} that have recently been developed 
to generate the electroweak symmetry breaking dynamically by new strong 
interactions.\\

Little Higgs models are based on a complex system of symmetries and symmetry 
breaking mechanisms; for a recent review see Ref.~\cite{littlest}. 
Three points are central in realizing the idea:
\begin{itemize}
\item[(i)]  The Higgs field is a Goldstone field associated with the breaking 
of a global symmetry $G$ at an energy scale of order $\Lambda_s \sim 4 \pi f 
\sim$ 10 to 30 TeV, with $f$ characterizing the scale of the symmetry 
breaking parameter;
\item[(ii)] In the same step, the gauge symmetry $G_0 \subset G$ is broken 
down to the gauge group $SU(2) \times U(1)$ of the Standard Model, generating 
masses for heavy vector bosons and fermions which cancel the standard 
quadratic divergencies in the radiative corrections to the light Higgs boson 
mass. Since the masses of these new particles are generated by the breaking of 
the gauge symmetry $G_0$ they are of the intermediate size $M \sim g f \sim 1$
 to 3 TeV;
\item[(iii)] The Higgs bosons acquires a mass finally through radiative 
corrections at the standard electroweak scale of order 
$v \sim g^2 f / 4 \pi \sim$ 100 to 300 GeV. 
\end{itemize}

Thus three characteristic scales are encountered in these models: the strong 
interaction scale $\Lambda_s$, the new mass scale $M$ and the electroweak 
breaking scale $v$, ordered in the hierarchical chain $\Lambda_s \gg M \gg v$. 
The light Higgs boson mass is protected at small value by requiring the 
collective breaking of two symmetries. In contrast to the boson-fermion 
symmetry that cancels quadratic divergencies in supersymmetry, 
the cancellation 
in Little Higgs models operates in the boson and fermion sectors individually, 
the cancellation ensured by the symmetries among the couplings of the SM 
fields and new fields to the Higgs field.\\

\noindent
{\it{Example: Littlest Higgs Model}}\\

\noindent
An interesting example in which these ideas are realized, is provided by the 
``Littlest Higgs Model'' \cite{91A,91B}. The model is formulated as a 
non-linear sigma model with a global $SU(5)$ symmetry group. This group is 
broken down to $SO(5)$ by the non-zero vacuum expectation value
\beq
\Sigma_0 = crossdiag\,[{\scriptstyle '}\!\mathbb{I},1,
{\scriptstyle '}\!\mathbb{I}]
\eeq
of the $\Sigma$ field. Assuming the subgroup $[SU(2) \times U(1)]^2$ to be 
gauged, the global symmetry breaking leads also to the breaking of this gauge 
group down to $[SU(2) \times U(1)]$. The global symmetry breaking generates 
$24 - 10 = 14$ Goldstone bosons, four of which are absorbed by the gauge 
bosons associated with the broken gauge group. The remaining 10 Goldstone 
bosons, incorporated in the $\Sigma$ field
\beq
\Sigma = exp[2i\Pi/f]: \quad \Pi = \left|\left| 
\begin{array}{ccc} 0 & h^\dagger/\sqrt{2} & 
\varphi^\dagger \\
h/\sqrt{2} & 0 & h^*/\sqrt{2} \\
\varphi & h^{\mathrm{T}}/\sqrt{2} & 0 
\end{array} \right|\right|
\eeq
are identified as an iso-doublet $h$ that will become the light Higgs field 
of the Standard Model, and a Higgs triplet $\varphi$ that will acquire a mass 
of order $M$.\\

The main construction principles of the model should be illustrated by 
analyzing the gauge and the Higgs sector qualitatively. The top sector, 
extended by a new heavy $[T_L, T_R]$ doublet, can be treated in a similar way 
after introducing the appropriate top-Higgs interactions.\\

\newpage
\noindent
{\it Vector Boson Sector}\\

\noindent
Inserting the $[SU(2) \times U(1)]^2$ gauge fields into the sigma Lagrangian, 
\beq
{\cal L} = \frac{1}{2} \frac{f^2}{4} \mathrm{Tr} | {\cal D}_\mu \Sigma |^2
\eeq 
with 
\beq
{\cal D}_\mu \Sigma = \partial_\mu \Sigma - i \sum_{j=1}^2 
[ g_j (W_j \Sigma + \Sigma W_j^\mathrm{T}) + \{U(1)\} ]
\eeq
the four vector bosons of the broken $[SU(2) \times U(1)]$ gauge symmetry 
acquire masses
\beq
M[W_H,Z_H,A_H] \sim g f
\eeq
where $W_H$ etc. denote the $W,Z$ and the heavy photon gauge fields.\\

Remarkably, the $W_H$ gauge bosons couple with the opposite sign to the square 
of the light Higgs boson compared with the standard $W$ bosons:
\beq
{\cal L} &=& + \frac{g^2}{4} W^2 \,\mathrm{Tr} h^\dagger h \nonumber \\
&& - \frac{g^2}{4} W'^2 \,\mathrm{Tr} h^\dagger h + ...
\eeq
The quadratic divergencies of the two closed $W$ and $W'$-loop diagrams 
attached to the 
light Higgs field, 
therefore cancel each other and, similarly to supersymmetric 
degrees of freedom, the new vector bosons should have masses not exceeding 1 
to 3 TeV to avoid excessive fine tuning.\\

The Standard Model gauge bosons remain still massless at this point; they 
acquire non-zero masses after the standard electroweak breaking mechanism is in 
operation.\\

\noindent
{\it Higgs Sector}\\

\noindent
Up to this level of the evolution of the theory, the global symmetries prevent 
a non-zero Higgs potential. Only if radiative corrections are switched on, the 
Coleman-Weinberg mechanism generates the Higgs potential that endows the Higgs 
bosons with masses and breaks the gauge symmetry of the Standard Model.\\

Casting the Higgs potential into the form
\beq
V = m_\varphi^2 \,\mathrm{Tr} \varphi^\dagger \varphi - \mu^2 h h^\dagger + 
\lambda_4 (h h^\dagger)^2
\eeq
the first term provides a non-zero mass to the $\varphi$ Higgs boson while the 
next two terms are responsible for the symmetry breaking in the gauge sector 
of the Standard Model. \\

\noindent
-- Cutting-off the quadratically divergent contributions to the 
Coleman-Weinberg potential at $\Lambda_s$, the masses squared of the [now] 
pseudo-Goldstone bosons $\varphi$ are of the order
\beq
m_\varphi^2 \sim g^2 (\Lambda_s / 4 \pi )^2 \sim g^2 f^2
\eeq
Thus the heavy Higgs bosons acquire masses of the size of the heavy vector 
bosons.\\

\noindent
-- The quartic coupling of the light Higgs boson is of order $g^2$. The 
coefficient $\mu^2$ however receives contributions only from one-loop 
logarithmically divergent and two-loop quadratically divergent parts in the 
Coleman-Weinberg potential:
\beq
\mu^2 = \mu_1^2 + \mu_2^2 : && \mu_1^2 \sim (\Lambda_s/4 \pi)^2 \log
\left(\Lambda_s^2/f^2\right)/16\pi^2 
\sim f^2 \log \left(\Lambda_s^2/f^2\right)/16\pi^2 \nonumber\\[0.3cm]
&& \mu_2^2 \sim \Lambda_s^2/(16\pi^2)^2 \sim f^2/16\pi^2 
\eeq
Both contributions are naturally of the order $f/4\pi$, i.e. they are an order 
of magnitude smaller than the intermediate scale $M$ of the heavy Higgs and 
vector masses.\\

{\it In summary.} A light Higgs boson with mass of order 100 GeV can be 
generated in 
Little Higgs models as a pseudo-Goldstone boson, 
and the light mass is protected against 
large radiative corrections individually in the boson and the fermion sectors. 
\\

\noindent
{\it{Phenomenology}}\\

\noindent
Such scenarios give rise to many predictions that can be checked 
experimentally.

Foremost, the spectrum of new heavy vector bosons and fermions should be 
observed with masses in the intermediate range of 1 to a few TeV at the LHC 
or TeV/multi-TeV $e^+e^-$ linear colliders.

\begin{figure}[hbt]
\begin{center}
\epsfig{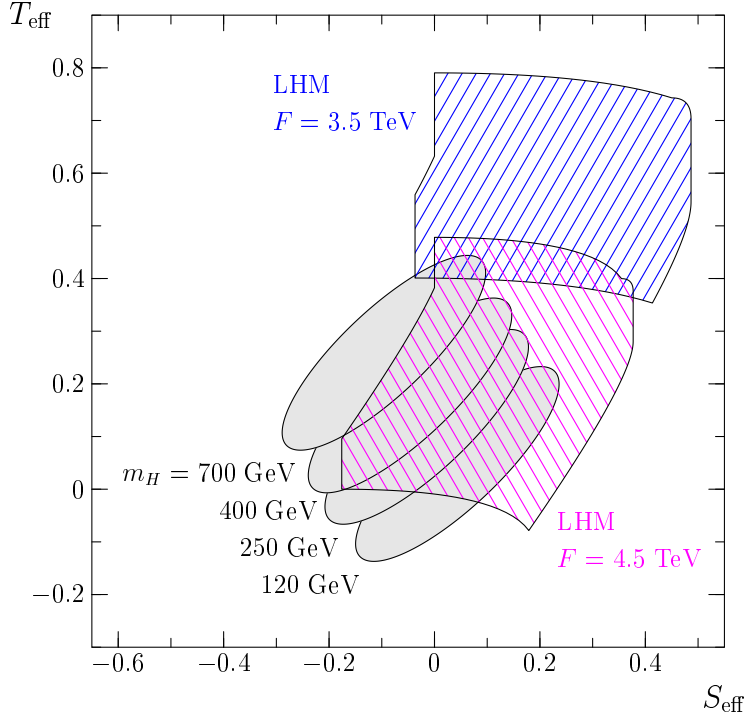}
\end{center}
\vspace*{-0.4cm}

\caption[]{\label{fig:kilian} \it Predictions of the $S,T$ precision parameters
for the Littlest Higgs 
model with standard $U(1)$ charge assigments. The shaded ellipses are the 
68 \% exclusion contours which follow from the electroweak precision data, 
assuming four different Higgs masses. The hatched areas are the allowed 
parameter ranges of the Littlest Higgs model for two different values of the 
scale $F$. The limits from contact interactions have been taken into account.
Ref. \cite{kilfigure}.}
\end{figure}

However, the model can already be checked by analyzing existing precision 
data from LEP and elsewhere. The impact of the new degrees of freedom on the 
Little Higgs models must be kept small enough not to spoil the success of the 
radiative corrections including just the light Higgs boson in the description 
of the data. This leads to a constraint of order 3 to 5 TeV on the parameter 
$f$, Fig.~\ref{fig:kilian}. Thus the theory is compatible with present 
precision data, but only marginally and the overlap appears already narrow.

\subsection{Strongly Interacting $W$ Bosons}

\phantom{h}
The Higgs mechanism is based on the
theoretical concept  of spontaneous symmetry
breaking \cite{1}. In the canonical 
formulation, adopted in the Standard
Model, a four-component {\it fundamental}
scalar field is introduced, which is
endowed with a self-interation such 
that the field acquires a non-zero
value in the ground state. The specific
direction in isospace, which is singled out by  the 
ground-state solution, breaks the
isospin invariance of the interaction
spontaneously\footnote{We retain the common language use also in the
context of gauge theories, although the gauge symmetry is not broken in
the strict sense.}. The interaction of the gauge fields with the 
scalar field in the ground state 
 generates the masses
of these fields. The longitudinal degrees
of freedom of the gauge fields are built
up by absorption of the Goldstone modes, 
which are associated with the spontaneous
breaking of the electroweak symmetries 
in the scalar field sector. Fermions
acquire masses through Yukawa interactions
with the ground-state field. While three
scalar components are absorbed by the 
gauge fields, one degree of freedom
manifests itself as a physical particle,
the Higgs boson. The exchange of this
particle in scattering amplitudes, including
longitudinal gauge fields and massive 
fermion fields, guarantees the unitarity
of the theory up to asymptotic energies.

In the alternative to this scenario 
based on a fundamental Higgs field, the
spontaneous symmetry breaking is generated
{\it dynamically} \cite{2}. A system of novel 
fermions is introduced, which interact
strongly at a scale of order 1 TeV. In
the ground state of such a system a scalar
condensate of fermion--antifermion pairs
may form. Such a process is  generally
expected to be realized in any non-Abelian gauge theory
of the novel strong interactions [and 
realized in QCD, for instance]. Since the
scalar condensate breaks the chiral
symmetry of the fermion system, Goldstone
fields will form, and these  can be absorbed
by the electroweak gauge fields to build
up the longitudinal components and the
masses of the gauge fields. Novel gauge
interactions must be introduced, which
couple the leptons and quarks of the
Standard Model to the new fermions in order
to generate lepton and quark masses
through interactions with the ground-state
fermion--antifermion condensate. In the
low-energy sector of the electroweak theory, 
the fundamental Higgs-field approach and
the dynamical alternative are equivalent.
However, the two theories are fundamentally
different at high energies. While the 
unitarity of the electroweak gauge theory
is guaranteed by the exchange of the scalar
Higgs particle in scattering processes, 
unitarity is restored in the dynamical
theory at high energies through the
non-perturbative strong interactions
between the particles. Since the longitudinal
gauge field components are equivalent to the
Goldstone fields associated with the microscopic
theory, their strong interactions at high
energies are transferred to the electroweak
gauge bosons. Since, by unitarity, the $S$-wave scattering
 amplitude of longitudinally polarized $W, Z$ bosons in the 
isoscalar channel $(2W^+W^- + ZZ) / \sqrt{3}$, 
$a^0_0 = \sqrt{2} G_F s/ 16 \pi$, is bounded by
1/2, the characteristic scale of the new strong interactions
must be close to 1.2 TeV. Thus near the critical energy of
1 TeV the $W, Z$ bosons interact strongly with each other.
Technicolor theories provide an elaborate form of such scenarios.

\subsubsection{Theoretical Basis\\ \\}
Physical scenarios of dynamical
symmetry breaking may be based on  new
strong interaction theories, which extend
the  spectrum of matter particles and of the interactions 
beyond the degrees of freedom realized in the
Standard Model. If the new strong interactions are 
invariant under transformations of a
chiral $SU(2) \times SU(2)$
group, the chiral invariance is generally  broken
spontaneously down to the diagonal custodial isospin group
$SU(2)$. This process is associated with the
formation of a chiral condensate in the
ground state and the existence of three
massless Goldstone bosons.

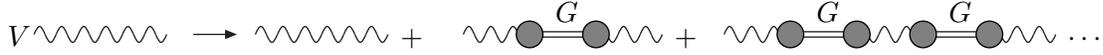
\begin{figure}[hbt]
\begin{center}
\begin{picture}(60,10)(90,30)
\Photon(0,25)(50,25){3}{6}
\LongArrow(60,25)(75,25)
\put(-10,21){$V$}
\end{picture}
\begin{picture}(60,10)(70,30)
\Photon(0,25)(50,25){3}{6}
\put(55,21){$+$}
\end{picture}
\begin{picture}(60,10)(55,30)
\Photon(0,25)(25,25){3}{3}
\Photon(50,25)(75,25){3}{3}
\Line(25,24)(50,24)
\Line(25,26)(50,26)
\GCirc(25,25){5}{0.5}
\GCirc(50,25){5}{0.5}
\put(80,21){$+$}
\put(35,30){$G$}
\end{picture}
\begin{picture}(60,10)(20,30)
\Photon(0,25)(25,25){3}{3}
\Photon(50,25)(75,25){3}{3}
\Photon(100,25)(125,25){3}{3}
\Line(25,24)(50,24)
\Line(25,26)(50,26)
\Line(75,24)(100,24)
\Line(75,26)(100,26)
\GCirc(25,25){5}{0.5}
\GCirc(50,25){5}{0.5}
\GCirc(75,25){5}{0.5}
\GCirc(100,25){5}{0.5}
\put(130,21){$\cdots$}
\put(35,30){$G$}
\put(85,30){$G$}
\end{picture}
\end{center}
\caption[]{\label{fg:gaugemass} \it Generating  gauge-boson masses (V)
 through the interaction with the Goldstone bosons (G).}
\end{figure}
The Goldstone bosons can be absorbed by
the gauge fields, generating longitudinal
states and non-zero masses of the gauge bosons, as shown in
Fig.~\ref{fg:gaugemass}. Summing up the geometric series
of vector-boson--Goldstone-boson transitions
in the propagator leads to in a shift of the  
mass pole:
\begin{eqnarray}
\frac{1}{q^2} & \to & \frac{1}{q^2} + \frac{1}{q^2} q_\mu \frac{g^2 F^2/2}{q^2}
q_\mu \frac{1}{q^2} + \frac{1}{q^2} \left[ \frac{g^2 F^2}{2} \frac{1}{q^2}
\right]^2 + \cdots \nonumber \\
& \to & \frac{1}{q^2-M^2}
\end{eqnarray}
The coupling between gauge fields and
Goldstone bosons has been defined as $ig F/\sqrt{2} q_\mu$.
The mass generated for the gauge field is related
to this coupling by
\begin{equation}
M^2 = \frac{1}{2} g^2 F^2 ~.
\end{equation}
The numerical value of the coupling $F$ must coincide
with $v=246$ GeV.\\

The remaining custodial $SU(2)$ symmetry
guarantees that the $\rho$
parameter, the relative strength between
$NC$ and $CC$ couplings, is one. Denoting the $W/B$ mass
matrix elements by
\begin{equation}
\begin{array}{rclcrcl}
\langle W^i | {\cal M}^2 | W^j \rangle & = & \displaystyle \frac{1}{2} g^2
F^2 \delta_{ij}
& \hspace*{1cm} & \langle W^3 | {\cal M}^2 | B \rangle & = & \langle B |
{\cal M}^2 | W^3 \rangle \\ \\
\langle B | {\cal M}^2 | B \rangle & = & \displaystyle \frac{1}{2} g'^2 F^2 &
& & = & \displaystyle \frac{1}{2} gg' F^2
\end{array}
\end{equation}
the universality of the coupling $F$ leads
to the ratio  $M_W^2/M_Z^2 = g^2/(g^2+g'^2) =
\cos^2\theta_W$ of the mass eigenvalues, equivalent to $\rho=1$.\\

Since the wave functions of longitudinally
polarized vector bosons grow with the
energy, the longitudinal field components are
the dominant degrees of freedom at high
energies. These states can, however, 
for asymptotic energies be identified with the
absorbed Goldstone bosons. This equivalence \cite{75} is apparent
in the 't Hooft--Feynman gauge where, for
asymptotic energies, 
\begin{equation}
\epsilon_\mu^L W_\mu \to k_\mu W_\mu \sim M^2 \Phi ~. 
\end{equation}
The dynamics of gauge bosons can therefore be
identified at high energies with the
dynamics of scalar Goldstone fields. An
elegant representation of the Goldstone
fields $\vec{G}$ in this context is provided by
the exponentiated form
\begin{equation}
U = \exp [-i \vec{G} \vec{\tau}/v ] ~, 
\end{equation}
which corresponds to an $SU(2)$ matrix field.\\

The Lagrangian of a system of strongly interacting
 bosons  consists in such a scenario 
of the Yang--Mills part ${\cal L}_{YM}$
and the interactions ${\cal L}_G$
of the Goldstone fields, 
\begin{equation}
{\cal L}={\cal L}_{YM}+{\cal L}_G ~. 
\end{equation}
The Yang--Mills part is written in the
usual form ${\cal L}_{YM} = -\frac{1}{4} {\rm Tr} [W_{\mu\nu} W_{\mu\nu} +
B_{\mu\nu} B_{\mu\nu} ]$.  
The interaction of the Goldstone fields can be systematically expanded
in chiral theories 
in the derivatives of the
fields, corresponding to expansions in
powers of the energy for scattering
amplitudes \cite{76}:
\begin{equation}
{\cal L}_G = {\cal L}_0 + \sum_{dim=4} {\cal L}_i + \cdots
\end{equation}
Denoting the SM covariant derivative of
the Goldstone fields by
\begin{equation}
D_\mu U = \partial_\mu U - i g W_\mu U + i g' B_\mu U ~, 
\end{equation}
the leading term ${\cal L}_0$, which is  
of dimension = 2, is given by
\begin{equation}
{\cal L}_0 = \frac{v^2}{4} {\rm Tr} [ D_\mu U^+ D_\mu U ] ~.
\end{equation}
This term generates the masses of the $W,Z$
gauge bosons: $M_W^2 = \frac{1}{4} g^2 v^2$ and
$M_Z^2 = \frac{1}{4} (g^2+g'^2) v^2$.
The only parameter in this part of the
interaction is $v$, which however is fixed
uniquely by the experimental value of the
$W$ mass; thus the amplitudes predicted by
the leading term in the chiral expansion
can effectively be considered as parameter-free.\\

The next-to-leading component in the expansion with
dimension = 4 consists of ten individual terms. If the
custodial $SU(2)$ symmetry is imposed, only two
terms are left, which do not affect propagators
and 3-boson vertices but only 4-boson vertices.
Introducing the vector field $V_\mu$ by 
\begin{equation}
V_\mu = U^+ D_\mu U
\end{equation}
these two terms are given by the interaction
densities
\begin{equation}
{\cal L}_4  =  \alpha_4 \left[Tr V_\mu V_\nu \right]^2 \hspace*{0.5cm}
\mbox{and} \hspace*{0.5cm}
{\cal L}_5  =  \alpha_5 \left[Tr V_\mu V_\mu \right]^2
\end{equation}

The two coefficients $\alpha_4,\alpha_5$
are free parameters that must be adjusted
experimentally from $WW$ scattering data.

Higher orders in the chiral expansion give
rise to an energy expansion of the scattering
amplitudes of the form ${\cal A} = \sum c_n (s/v^2)^n$.
This series  will diverge at energies for which
the resonances of the new strong interaction
theory can be formed in $WW$ collisions: $0^+$ `Higgs-like', 
$1^-$ `$\rho$-like' resonances, etc. The masses of these resonance
states are expected in the range $M_R \sim 4\pi v$ 
where chiral loop expansions diverge,
i.e. between about 1 and 3 TeV.

\subsubsection{An Example: Technicolor Theories\\ \\}
A simple example for such scenarios is provided by technicolor theories, 
see e.g. Ref.~\cite{94A}. They 
are built on a pattern similar to QCD but characterized by a scale 
$\Lambda_{TC}$ in the TeV range so that the interaction becomes strong
already at short distances of order $10^{-17}$~cm. \\

The basic degrees of freedom in the simplest version are a chiral set 
$[(U,D)_L;U_R,D_R]$ of massless fermions that interact with technicolor gauge 
fields. The chiral $SU(2)_L\times SU(2)_R$ symmetry of this theory is broken 
down to the diagonal $SU(2)_{L+R}$ vector symmetry by the formation of 
$\langle\bar{U} U\rangle =\langle\bar{D} D\rangle = {\cal O}(\Lambda^3_{TC})$ 
vacuum condensates. 
The breaking of the chiral symmetry generates three massless Goldstone bosons
$\sim \bar{Q} i \gamma_5 \stackrel{\to}{\tau} Q$, that can be 
absorbed by the gauge fields of the Standard Model to build the massive 
states with $M_W \sim 100$~GeV. From the chain
\beq
M_W = \frac{1}{2} g F \quad \mathrm{and} \quad F \sim \Lambda_{TC} / 4 \pi
\eeq 
the parameter $F$ is estimated to be 
of order 1 TeV while $\Lambda_{TC}$ should be 
in the 10 TeV range. \\

While the electroweak gauge sector can be formulated consistently in this 
picture, generating fermion masses leads to severe difficulties. Since gauge 
interactions couple only left-left and right-right field components, a 
helicity-flip left-right mass operator $\bar{f}_L f_R$ 
is not generated
for the fermions of 
the Standard Model. To solve this problem, new gauge 
interactions between the SM and TC fermions must be introduced 
[Extended Technicolor] so that the helicity can flip through the ETC 
condensate in the vacuum. The SM masses predicted this way are of order 
$m_f \sim g^2_E \Lambda^3_{ETC}/M_E^2$ with $g_E$ being the coupling in the 
extended technicolor gauge theory and $M_E$ the mass of the ETC gauge fields.
However, estimates of $M_E$ lead to a clash if one tries to reconcile the size
of the scale needed for generating the top mass, order TeV, with the 
suppression of flavor-changing  processes, like $K\bar{K}$ oscillations, 
which require a size of order PeV. \\

Thus, the simplest realization of the technicolor theories suffers from 
internal conflicts in the fermion sector. More involved theoretical models
are needed to reconcile these conflicting estimates \cite{94A}.
Nevertheless, the idea of generating 
electroweak symmetry breaking dynamically, is a theoretically attractive and 
interesting scenario in principle. 

\subsection{$WW$ Scattering at High-Energy Colliders}

\phantom{h}
Independently of specific realizations of dynamical symmetry breaking, 
theoretical tools have been developed which can serve to investigate these 
scenarios quite generally. The (quasi-) elastic 2--2 $WW$ scattering
amplitudes can be expressed at high
energies by a master amplitude
$A(s,t,u)$, which depends on the three
Mandelstam variables of the scattering processes:
\begin{eqnarray}
A(W^+ W^- \to ZZ) & = & A(s,t,u) \\
A(W^+ W^- \to W^+ W^-) & = & A(s,t,u) + A(t,s,u) \nonumber \\
A(ZZ \to ZZ) & = & A(s,t,u) + A(t,s,u) + A(u,s,t) \nonumber \\
A(W^- W^- \to W^- W^-) & = & A(t,s,u) + A(u,s,t) ~. \nonumber
\end{eqnarray} ~\\

To lowest order in the chiral expansion, ${\cal L} \to {\cal L}_{YM} +
{\cal L}_0$, the master amplitude is given, in a
parameter-free form, by the energy squared $s$:
\begin{equation}
A(s,t,u) \to \frac{s}{v^2} ~.
\end{equation}
This representation is valid for energies $s \gg M_W^2$
but below the new resonance region, i.e. in
practice at energies $\sqrt{s}={\cal O}(1~\mbox{TeV})$.
Denoting the scattering length for the
channel carrying isospin $I$ and angular
momentum $J$ by $a_{IJ}$,
the only non-zero scattering channels
predicted by the leading term of the
chiral expansion correspond to
\begin{eqnarray}
a_{00} & = & + \frac{s}{16\pi v^2} \\
a_{11}   & = & + \frac{s}{96\pi v^2} \nonumber \\
a_{20}   & = & - \frac{s}{32\pi v^2} ~.
\end{eqnarray}
While the exotic $I=2$ channel is repulsive,
the $I=J=0$ and $I=J=1$ channels are attractive,
indicating the formation of non-fundamental
Higgs-type and $\rho$-type resonances.\\

Taking into account the next-to-leading terms
in the chiral expansion, the master amplitude
turns out to be \cite{24}
\begin{equation}
A(s,t,u) = \frac{s}{v^2} + \alpha_4 \frac{4(t^2+u^2)}{v^4}
+ \alpha_5 \frac{8s^2}{v^4} + \cdots ~,
\end{equation}
including the two parameters $\alpha_4$ and $\alpha_5$.\\

Increasing the energy, the
amplitudes will approach the resonance area.
There, the chiral character of the
theory does not provide any more guiding principle
for constructing  the scattering
amplitudes. Instead, {\it ad-hoc} hypotheses must
be introduced to define the nature of the
resonances; see e.g. Ref. \cite{24a}. A typical example
is provided by the
\begin{eqnarray}
\hspace*{-0.5cm}\mbox{\bf{chirally coupled scalar resonance:}}\quad
A & = & \frac{s}{v^2} - \frac{g_s^2 s^2}{v^2} \frac{1}{s-M_S^2 - iM_S \Gamma_S}
\\
& & \mbox{with}~~~\Gamma_S = \frac{3g_s^2 M_S^3}{32 \pi v^2} \nonumber
\end{eqnarray}

For small energies, the scattering
amplitude is reduced to the leading chiral
form $s/v^2$.  In the resonance region it is
described by two parameters, the mass
and the width of the resonance. The
amplitudes interpolate between the
two regions in a smooth way.\\
 
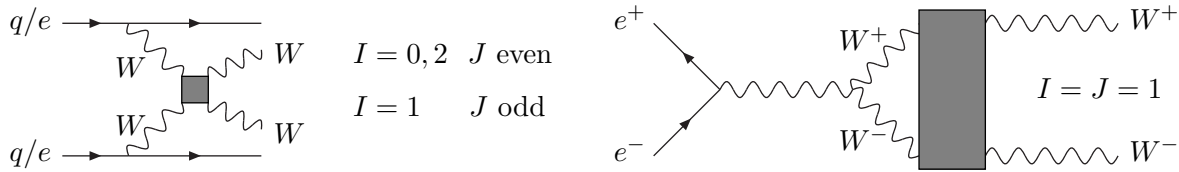
\begin{figure}[hbt]
\begin{center}
\begin{picture}(60,50)(140,0)
\ArrowLine(0,50)(25,50)
\ArrowLine(25,50)(75,50)
\ArrowLine(0,0)(25,0)
\ArrowLine(25,0)(75,0)
\Photon(25,50)(45,30){-3}{3}
\Photon(25,0)(45,20){3}{3}
\Photon(55,20)(75,10){-3}{3}
\Photon(55,30)(75,40){3}{3}
\GBox(45,20)(55,30){0.5}
\put(-20,48){$q/e$}
\put(-20,-2){$q/e$}
\put(20,8){$W$}
\put(20,30){$W$}
\put(80,35){$W$}
\put(80,5){$W$}
\put(110,35){$I=0,2~~J~\mbox{even}$}
\put(110,15){$I=1~~~~~J~\mbox{odd}$}
\end{picture}
\begin{picture}(60,50)(-20,0)
\ArrowLine(25,25)(0,50)
\ArrowLine(0,0)(25,25)
\Photon(25,25)(75,25){3}{5}
\Photon(75,25)(100,50){3}{4}
\Photon(75,25)(100,0){3}{4}
\Photon(125,0)(175,0){3}{5}
\Photon(125,50)(175,50){3}{5}
\GBox(100,-5)(125,55){0.5}
\put(-15,48){$e^+$}
\put(-15,-2){$e^-$}
\put(70,2){$W^-$}
\put(70,40){$W^+$}
\put(180,48){$W^+$}
\put(180,-2){$W^-$}
\put(145,22){$I=J=1$}
\end{picture}
\end{center}
\caption[]{\label{fg:qqtoqqww} \it $WW$ scattering and rescattering 
at high energies at the LHC
and TeV $e^+e^-$ linear colliders.}
\end{figure}

\begin{figure}[hbtp]
\begin{center}
\vspace*{-5.5cm}

\hspace*{-3.5cm}
\epsfig{file=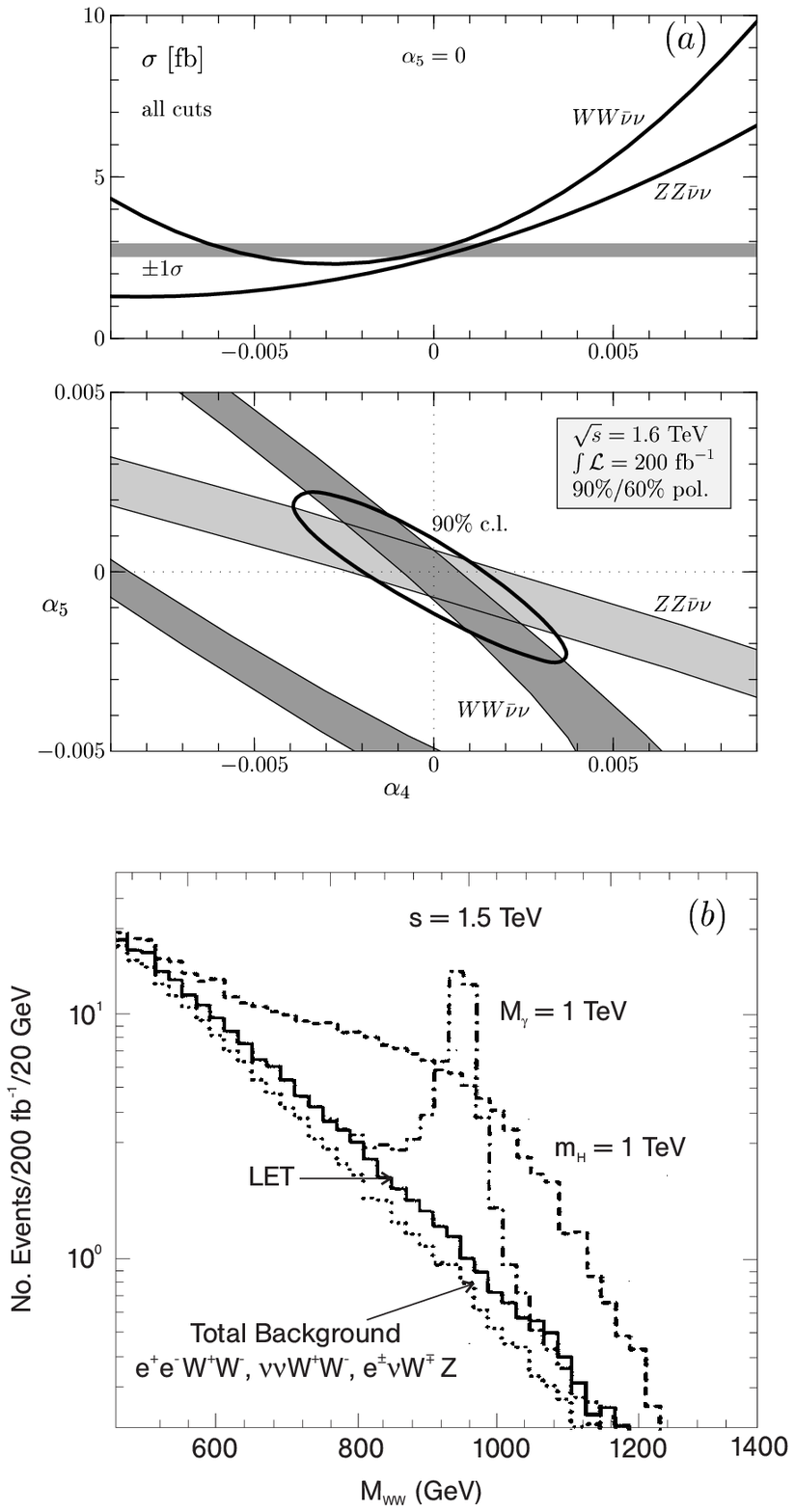,width=22.0cm,angle=0}
\vspace*{-8.5cm}

\end{center}
\caption[]{\it
  Upper part: Sensitivity to the expansion
  parameters in chiral electroweak models of $WW \to WW$ and $WW \to
  ZZ$ scattering at the strong-interaction threshold;
  Ref. \protect\cite{24}.
Lower part: The distribution of the $WW$ invariant energy in $e^+e^-
  \to \overline{\nu} \nu WW$ for scalar and vector resonance
  models [$M_H, M_V$ = 1 TeV];
  Ref. \protect\cite{24a}. 
\protect\label{17tt}\label{PKB}
}
\end{figure}
\begin{figure}[hbtp]

\vspace*{-2.0cm}
\hspace*{-2.0cm}
\epsfxsize=20cm \epsfbox{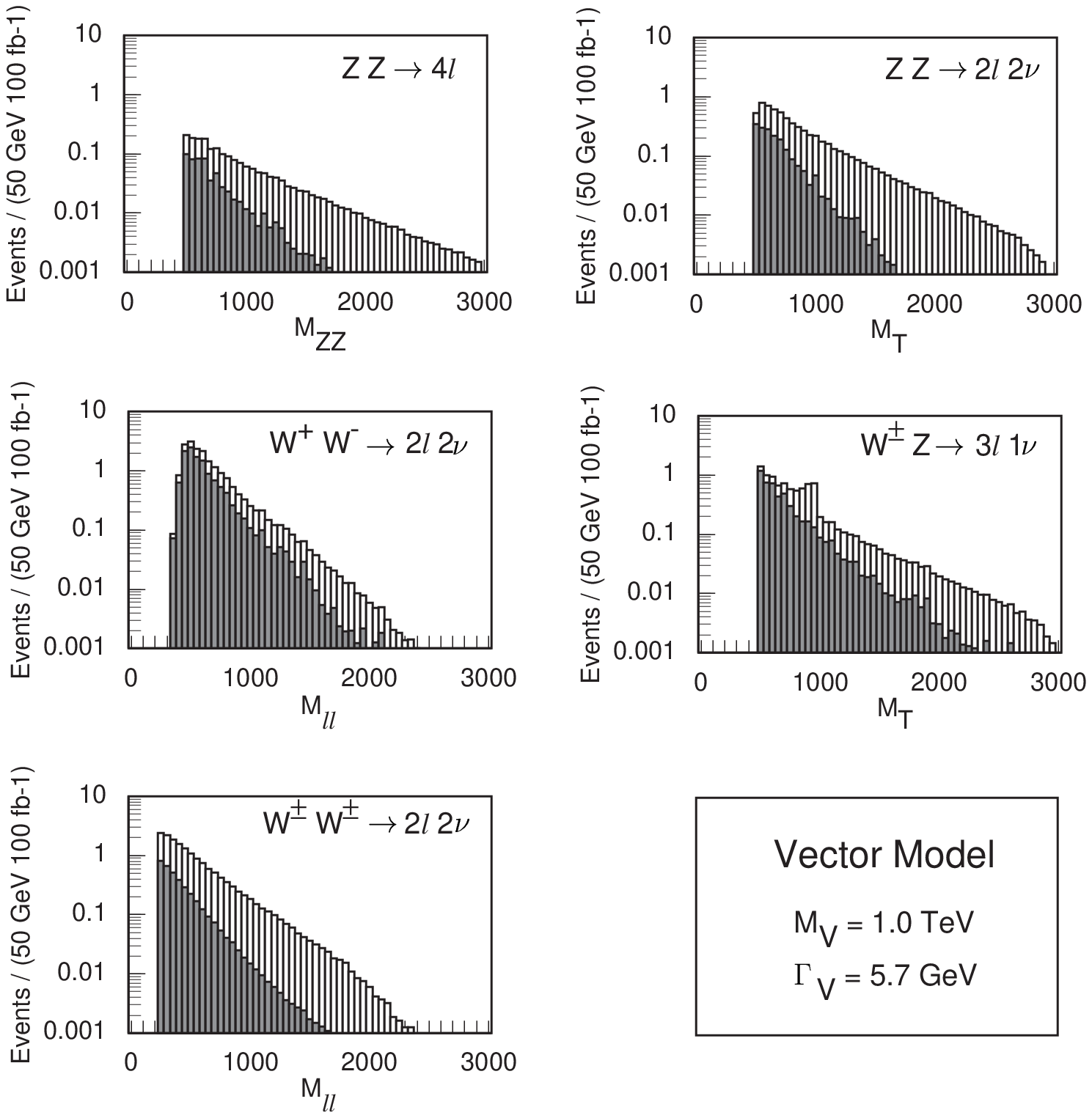}
\vspace*{-11cm}

\caption[]{\label{fg:vvto4l} \it Invariant mass distributions for the
gold-plated purely leptonic final states that arise from the processes
$pp\to ZZX \to 4\ell X, pp\to ZZX\to 2\ell 2\nu X, pp\to W^+W^-X, pp\to
W^\pm ZX$ and $pp\to W^\pm W^\pm X$, for the LHC (mass in  GeV).
The signal is plotted above the summed background. Distributions are shown
for a chirally coupled vector with $M_V=1$ TeV, $\Gamma_V=5.7$ GeV;
Ref. \protect\cite{23B}.}
\end{figure}
$WW$ scattering can be studied at the LHC
and at TeV $e^+e^-$ linear colliders. At high energies, 
equivalent $W$ beams accompany the quark
and electron/positron beams (Fig.~\ref{fg:qqtoqqww})
in the fragmentation processes $pp\to qq \to qqWW$ and
$ee\to \nu\nu WW$; the spectra of the longitudinally
polarized $W$ bosons have been given in Eq. (\ref{eq:xyz}). 
In the hadronic LHC  environment the final-state 
$W$ bosons can only be observed in
leptonic decays. Resonance reconstruction
is thus not possible for charged $W$ final
states. However, the clean environment of
$e^+e^-$ colliders will allow the reconstruction 
of resonances from $W$ decays to jet pairs.
The results of three experimental 
simulations are displayed in Fig.~\ref{PKB}.
In Fig.~\ref{PKB}a the sensitivity to the parameters
$\alpha_4,\alpha_5$ of the chiral expansion is shown for $WW$
scattering in $e^+e^-$ colliders \cite{24}. The results of this analysis
can be reinterpreted as sensitivity to the
parameter-free prediction of the chiral
expansion, corresponding to an error of
about 10\% in the first term of the master
amplitude $s/v^2$.  These experiments test the basic concept
of dynamical symmetry breaking through
spontaneous symmetry breaking. The production
of a vector-boson resonance of mass $M_V=1$ TeV
is exemplified in Fig.~\ref{PKB}b \cite{24a}. Expectations
for leptonic invariant energies of  $WW$ scattering final states
 at the LHC are compared in the vector model 
with the background in Fig.~\ref{fg:vvto4l} \cite{23B}.\\

A second powerful method measures the elastic 
$W^+W^- \to W^+W^-$ scattering in the $I=1, J=1$ channel. The
rescattering of $W^+W^-$ bosons produced in $e^+e^-$
annihilation, cf. Fig.~\ref{fg:qqtoqqww}, depends at high
energies on the $WW$ scattering phase $\delta_{11}$ 
\cite{78}. The production amplitude $F = F_{LO} \times R$
is the product of the lowest-order
perturbative diagram with the Mushkelishvili--Omn\`es rescattering amplitude
${\cal R}_{11}$,
\begin{equation}
{\cal R}_{11} = \exp \frac{s}{\pi} \int \frac{ds'}{s'}
\frac{\delta_{11}(s')}{s'-s-i\epsilon} ~,
\end{equation}
which is determined by the $I = J = 1$ $WW$ phase shift $\delta_{11}$.
The power of this method derives from the 
fact that 
the entire $e^+e^-$
collider energy is transferred to the $WW$ system
[while a major fraction of the energy
is lost in the fragmentation of $e \to \nu W$
if the $WW$ scattering is studied in the
process $ee\to \nu\nu WW$]. Detailed simulations \cite{78}
have shown that this process is sensitive
to vector-boson masses up to about $M_V \lessim 6$ TeV in technicolor-type 
theories. \\

The experimental analysis of the $\alpha$ parameters at the $e^+e^-$ linear 
collider in the first phase with energy up to $\sim 1$~TeV can be 
reinterpreted in the following way. Associating the parameters $\alpha$ 
with new 
strong interaction scales, $\Lambda_\star \sim M_W/\sqrt{\alpha}$, 
upper bounds on $\Lambda_\star$ 
of $\sim 3$~TeV can be probed
in $WW$ scattering. Thus this 
instrument allows to cover the entire threshold region $\lessim 4\pi v 
\sim 3$~TeV of the new strong interactions. In the $W^+W^-$ production channel 
of $e^+e^-$ collisions a range even up to order 10~TeV can be probed
indirectly. If a new scale $\Lambda_\star$ would be discovered below
$\sim 3$~TeV, novel $WW$ resonances could be searched for at the LHC while 
CLIC could investigate new resonance states potentially up to a mass close to 
5 TeV.

\newpage

\section{Summary}

\phantom{h}
The mechanism of electroweak symmetry
breaking can be established in the present
or the next generation of $p\bar p/pp$ and $e^+e^-$
colliders:
\begin{itemize}
\item[$\star$] Whether there exists a light fundamental Higgs boson;
\item[$\star$] The profile of the Higgs particle can be
  reconstructed, which reveals the physical
  nature of the underlying mechanism of
  electroweak symmetry breaking;
\item[$\star$] Analyses of strong WW scattering can be
  performed if the symmetry breaking is of
  dynamical nature and generated by novel
  strong interactions.
\end{itemize}
Moreover, depending on the experimental answer to these questions,
the electroweak sector will provide the
platform for extrapolations into physical
areas beyond the Standard Model:  either
to the low-energy  supersymmetry sector or, alternatively,
to a new strong interaction theory at a characteristic 
scale of order 1 TeV and beyond.
 
\section*{Acknowledgments}

  P.M.Zerwas is very thankful to the organisers, A.Bashir, J.Erler and
  M.Mondrag\'on, for the invitation to the XI Mexican School of Particles
  and Fields, Xalapa (Veracruz) 2004. He gratefully acknowledges the
  cooperation with his coauthors in writing this report, in particular
  with the Mexican students and Myriam Mondrag\'on who have prepared the
  Spanish version.\\\\
 Partially supported by the projects PAPIIT-IN116202 and Conacyt 42026-F.

\newpage

\appendix

\section{The O(3) $\sigma$ Model}

\phantom{h}
A transparent but, at the same time, sufficiently complex model to study all 
the aspects of electroweak symmetry breaking is the O(3) $\sigma$ model. By 
starting from the standard version, in a number of variants it allows to 
develop the idea of spontaneous symmetry breaking and the Goldstone theorem 
while gauging the theory leads to the Higgs phenomenon. This evolution will be
described step by step in the next three subsections. \\

The O(3) $\sigma$ model includes a triplet of field components:
\beq
\sigma = (\sigma_1,\sigma_2,\sigma_3)
\eeq
If the self-interaction potential of the field depends only on the overall 
field-strength, the theory, described by the Lagrangian
\beq
{\cal L} = \frac{1}{2} (\partial \sigma)^2 - V(\sigma^2)
\eeq
is O(3) rotationally invariant. These iso-rotations are generated by the 
transformation
\beq
\sigma \to e^{i\alpha t}\sigma \quad \mathrm{with} \quad
t_{ik}^j = -i \epsilon_{ijk}
\eeq
This transformation corresponds to a rotation about the axis $\alpha =
(\alpha_1,\alpha_2,\alpha_3)$. Choosing a quartic interaction for the 
potential, the theory is renormalizable and thus well-defined.

\subsection{``Normal'' Theory:}

\begin{figure}[hbt]
\begin{center}
\epsfig{figure=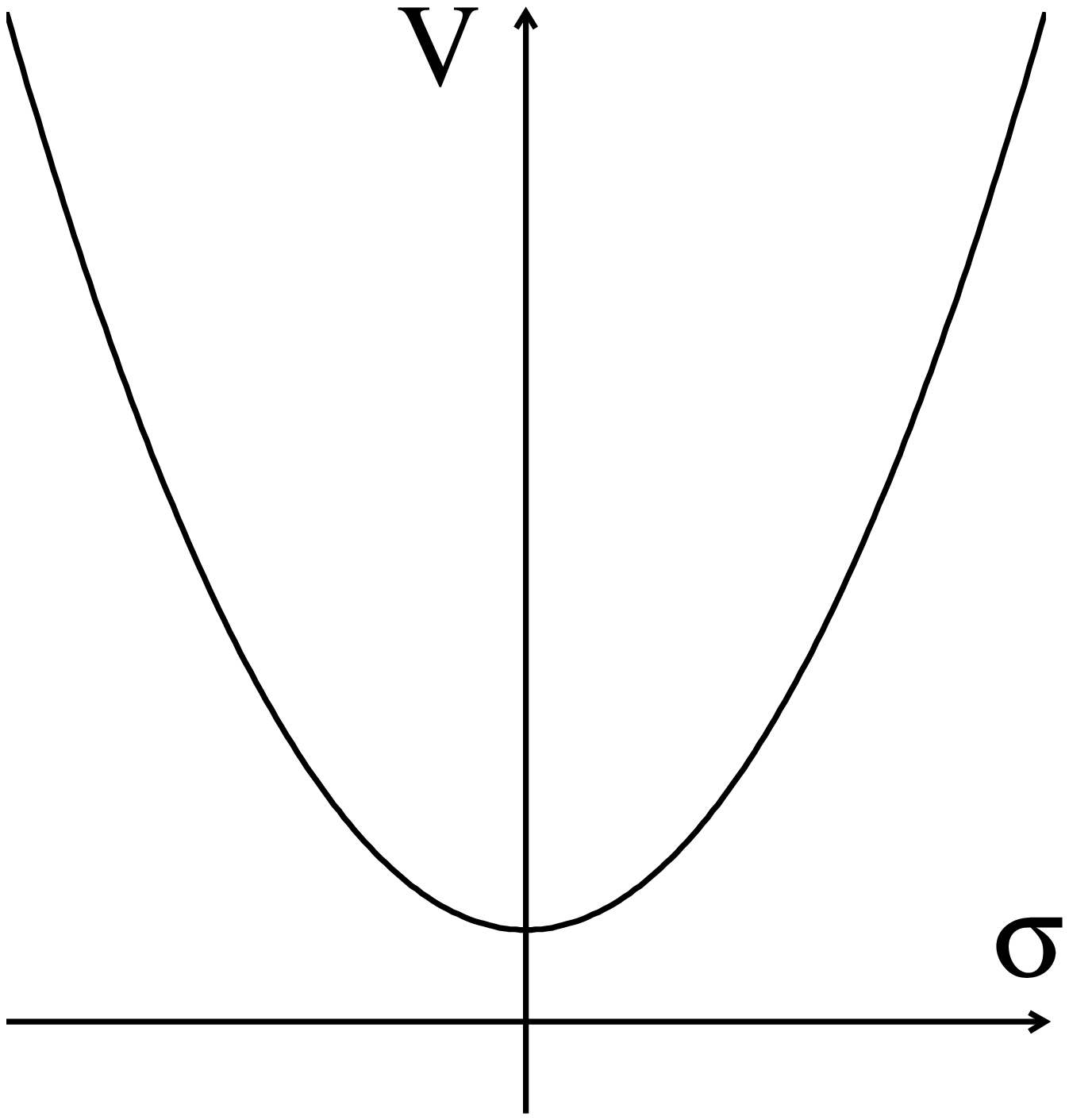,width=4cm}
\end{center}
\vspace*{-0.4cm}
\caption{\it \label{fig:pot} }
\end{figure}

\phantom{h}
If the quartic potential $V$ is chosen to be, cf.~Fig.~\ref{fig:pot},
\beq
V(\sigma^2) = \lambda^2 (\sigma^2 + \mu^2)^2
\eeq
the spectrum of particles and the interactions can easily be derived from the 
form
\beq
V(\sigma^2) = 2\lambda^2\mu^2\sigma^2 + \lambda^2 \sigma^4 + \mathrm{const.}
\eeq

\noindent
The bilinear field-term describes three degenerate masses
\beq
m(\sigma_1) = m(\sigma_2) = m(\sigma_3) = 2\lambda\mu
\eeq
corresponding to three physical particle degrees of freedom. The fields 
interact through the second quartic term. The ground state of the system is 
reached for zero field-strength:
\beq
\sigma^0 = (0,0,0)
\eeq
This theory describes a standard particle system in which the ground state 
preserves the rotational invariance of the Lagrangian. Thus the Lagrangian 
and the solution of the field equation obey the same degree of symmetry.

\subsection{Spontaneous Symmetry Breaking and Goldstone Theorem:}

\begin{figure}[hbt]
\begin{center}
\epsfig{figure=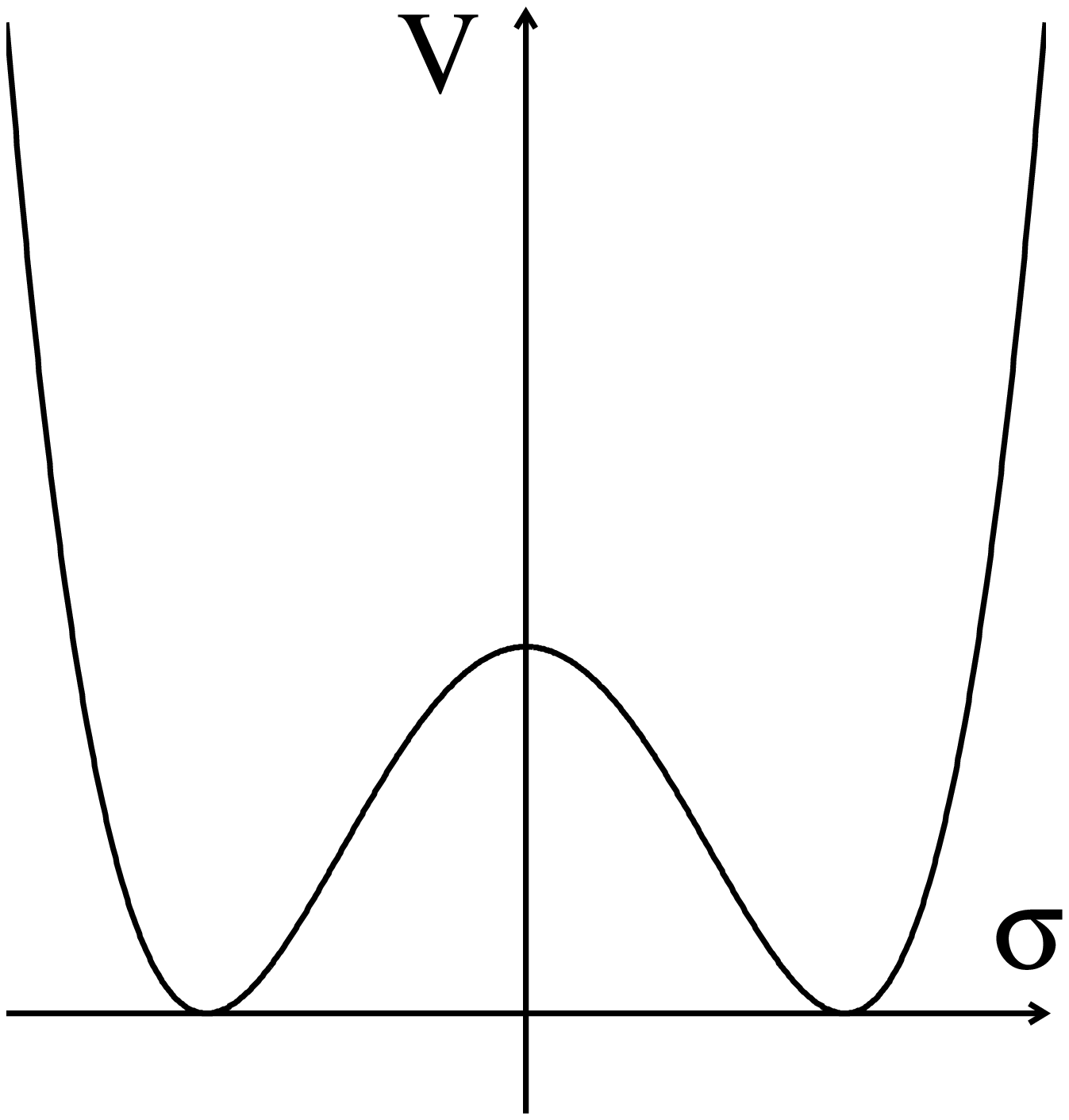,width=4cm}
\end{center}
\vspace*{-0.4cm}
\caption{\it \label{fig:pothiggs} }
\end{figure}

\phantom{h}
However, if the sign in the mass parameter in the potential flips to negative 
values,
\beq
V(\sigma^2) = \lambda^2(\sigma^2-\mu^2)^2
\eeq
the ground state is a state of non-zero field strength, 
cf.~Fig.~\ref{fig:pothiggs}. Fixing the axis of the ground state such that
\beq
\sigma^0 = (0,0,v) \quad \mathrm{with} \quad v=\mu
\eeq
the original O(3) rotational invariance of the Lagrangian is not obeyed any 
more by the ground-state solution which singles out a specific direction in 
iso-space. However, no principle determines the arbitrary direction of the 
ground state vector 
in iso-space. Such a phenomenon in which solutions of the field 
equations do not respect the symmetry of the Lagrangian, is generally termed 
``spontaneous symmetry breaking''.\\

Expanding the $\sigma$ field about the ground state,
\beq
\sigma = (\sigma_1', \sigma_2', v+\sigma_3')
\eeq
an effective theory emerges for the new dynamical degrees of freedom 
$\sigma_1', \sigma_2'$ and $\sigma_3'$. Evaluating the potential for the 
new fields,
\beq
V = 4v^2\lambda^2\sigma_3^{'2} + 4 v \lambda^2 \sigma_3' (\sigma_1^{'2} +
\sigma_2^{'2} + \sigma_3^{'2}) + \lambda^2 (\sigma_1^{'2} + \sigma_2^{'2} 
+ \sigma_3^{'2})^2 
\eeq
two massless particles plus one massive particle correspond to the bilinear 
field terms:
\beq
m(\sigma_1')=m(\sigma_2')=0 \nonumber\\
m(\sigma_3')=2\sqrt{2}\lambda v \neq 0
\eeq
The two massless particles are called Goldstone bosons, Ref.~\cite{appcite}.\\

The Goldstone bosons and the massive particle interact with each other through 
trilinear terms in the effective potential, in addition to the standard 
quartic terms. \\

The symmetry of the effective theory is reduced from the original O(3) 
rotational invariance to O(2) invariance restricted to rotations about the 
ground-state axis. \\

This $\sigma$ model is only a simple example of the general \\

{\it \noindent\underline{Goldstone theorem:}\\
If N is the dimension of the symmetry group of the basic Lagrangian, but the 
symmetry of the ground-state solution is reduced to M, then the theory 
includes (N-M) massless scalar Goldstone bosons.}\\

For each destroyed symmetry degree of freedom, a massless particle appears in 
the spectrum. A most famous example of this theorem are the three 
nearly massless pions which emerge from spontaneously broken chiral isospin 
symmetry in QCD.

\subsection{The Higgs mechanism}

\phantom{h}
The Higgs mechanism Ref.~\cite{1} 
provides the vector bosons in gauge theories with masses 
without destroying the renormalizability of the theory. Would masses be 
introduced by hand, the gauge invariance which ensures the renormalizability 
would be destroyed by the mass terms in the Lagrangian.\\

The global isospin symmetry of the O(3) $\sigma$ model can be extended to a 
local symmetry by introducing an iso-triplet $W$ of gauge fields coupled 
minimally to the $\sigma$ field. Introducing the covariant derivative
\beq
\partial_\mu \sigma \to \partial_\mu \sigma + igtW\sigma
\eeq
into the Lagrangian
\beq
{\cal L} = \frac{1}{2} [ (\partial + igtW)\sigma]^2 - V(\sigma^2) +
{\cal L}_{kin}(W)
\eeq
the theory is invariant under the local gauge transformation
\beq
\sigma \to e^{i\alpha t} \sigma \quad \mathrm{with} \quad
\alpha = \alpha(x)
\eeq
with the matter transformation complemented by the usual transformation of 
the non-abelian gauge field. The gauged Lagrangian includes the gauge kinetic 
part, the $\sigma$ kinetic part and the $\sigma$-gauge interaction, as well 
as the potential. 

\begin{itemize}
\item[--] If the $\sigma$ potential is just the standard potential, the 
theory is a 
non-abelian Yang-Mills gauge theory with a triplet of $\sigma$ particles 
interacting in the standard way with the $W$ gauge triplet fields.\\

\item[--] However, if the potential is chosen of the type 
$V=\lambda^2
(\sigma^2-\mu^2)^2$, which leads in the $\sigma$ model to spontaneous symmetry 
breaking, the physical field/particle content of the theory changes 
dramatically [a phenomenon similar to the non-gauged theory].
\end{itemize}

Parametrizing the $\sigma$ triplet-field through a rotation of the field about 
the ground-state axis,
\beq
\sigma = e^{i\Theta t/v} (\sigma^0 + \eta)
\eeq
with
\beq
\sigma^0 = (0,0,v)\,;\quad \eta = (0,0,\eta)\,; \quad
\Theta = (\Theta_1,\Theta_2,0)
\eeq
the $\Theta$ components of $\sigma$ perpendicular to the ground-state axis 
can be removed 
by the gauge transformation $\sigma \to exp[-i\Theta t/v] \sigma$ supplemented 
by the corresponding transformation of the gauge field. Keeping the original 
notation for the gauge-transformed fields, the new Lagrangian for the physical 
degrees of freedom is given by 
\beq
{\cal L} = \frac{1}{2} [(\partial + igWt)(\sigma^0 +\eta)]^2 - V([\sigma^0 +
\eta]^2) + {\cal L}_{kin}(W)
\eeq
After writing the resulting Lagrangian of the effective theory as
\beq
{\cal L} = {\cal L}_{kin}(W) + \frac{1}{4} g^2 v^2 ( W_1^2 + W_2^2) + 
\frac{1}{2} (\partial\eta)^2 - V + {\cal L}_{int} (\eta, W)
\eeq
the physical particle/field content becomes manifest:

\begin{itemize}
\item[--] a massless vector field $W_3$ corresponding to the residual
rotational invariance about the ground-state 3-axis;\\

\item[--] two massive $W$ fields $W_1$ and $W_2$ perpendicular to the 
ground-state 
axis with masses determined by the ground-state $\sigma$ field-strength $v$ 
and the gauge coupling $g$. These two massive fields correspond to the 
symmetry degrees of freedom that were broken spontaneously in the non-gauged 
$\sigma$ model;\\

\item[--] the Goldstone bosons have disappeared from the spectrum, absorbed to 
build up the longitudinal degrees of the massive gauge bosons;\\

\item[--] a real scalar Higgs boson $\eta$. 
\end{itemize}

This example can easily be extended, in parallel to the Goldstone theorem, to 
formulate the general\\

{\it \noindent \underline{Higgs mechanism}

\noindent
If N is the dimension of the symmetry group of the original Lagrangian, M the 
dimension of the symmetry group leaving invariant the ground state of the n 
scalar fields, then the physical theory consists of M massless vector fields, 
(N-M) massive vector fields, and n-(N-M) scalar Higgs fields.}

\newpage


\end{document}